\documentclass[useAMS,usenatbib,a4paper]{mn2e}
\usepackage{subfig,graphicx,amsmath, amsfonts, amssymb,footnote,color,dcolumn,rotating,blindtext,multicol,multirow,romannum,morefloats,float,lastpage,times}

\newcommand{\atlas}{ATLAS$^{\rm 3D}$}
\newcommand{\kms}{km s$^{-1}$}
\newcommand{\hi}{\hbox{H {\footnotesize I}}}
\newcolumntype{.}{D{.}{.}{-1}}
\newcolumntype{d}[1]{D{.}{.}{#1}}

\long\def\symbolfootnote[#1]#2{\begingroup%
\def\thefootnote{\fnsymbol{footnote}}\footnote[#1]{#2}\endgroup}

\title[The CARMA ATLAS$^{3D}$ CO survey of ETGs]{The \atlas\ project --  \Romannum{18}. CARMA CO imaging survey of early-type galaxies}
\author[K. Alatalo and the ATLAS$^{3D}$ team]
{Katherine Alatalo,$^{1,2}$\thanks{email: kalatalo@ipac.caltech.edu}
Timothy A. Davis,$^{3}$
Martin Bureau,$^{4}$
Lisa M. Young,$^5$
\newauthor Leo Blitz,$^{1}$
Alison F. Crocker,$^{6,7}$
Estelle Bayet,$^4$
Maxime Bois,$^8$
Fr\'ed\'eric Bournaud,$^9$
\newauthor Michele Cappellari,$^4$
Roger L. Davies,$^4$
P. T. de Zeeuw,$^{3,10}$
Pierre-Alain Duc,$^9$
\newauthor Eric Emsellem,$^{3,11}$
Sadegh Khochfar,$^{12}$
Davor Krajnovi\'c,$^3$
Harald Kuntschner,$^3$
\newauthor Pierre-Yves Lablanche,$^{3,11}$
Raffaella Morganti,$^{13,14}$
Richard M. McDermid,$^{15}$
\newauthor Thorsten Naab,$^{12}$
Tom Oosterloo,$^{13,14}$
Marc Sarzi,$^{16}$
Nicholas Scott,$^{17}$
\newauthor Paolo Serra,$^{13}$
and Anne-Marie Weijmans$^{18}$\thanks{Dunlap Fellow} \\
$^{1}$Department of Astronomy, Hearst Field Annex, University of California, Berkeley, CA 94720-3411, USA\\
$^{2}$NASA Herschel Science Center, California Institute of Technology, 770 S. Wilson Ave., Pasadena, CA 91125, USA\\
$^{3}$European Southern Observatory, Karl-Schwarzschild-Str. 2, 85748 Garching, Germany\\
$^4$Sub-department of Astrophysics, Department of Physics, University of Oxford, Denys Wilkinson Building, Keble Road, Oxford OX1 3RH, UK\\
$^5$Physics Department, New Mexico Institute of Mining and Technology, Socorro, NM 87801, USA\\
$^6$Department of Astronomy, Lederle Graduate Research Tower B 619E, University of Massachusetts, Amherst, MA 01003, USA\\
$^7$Physics Department, University of Toledo, 2801 W. Bancroft, Toledo, OH, USA\\
$^8$Observatoire de Paris, LERMA and CNRS, 61 Av. de l'Observatoire, F-75014 Paris, France\\
$^9$Laboratoire AIM Paris-Saclay, CEA/IRFU/SAp -- CNRS -- Universit\'e Paris Diderot, 91191 Gif-sur-Yvette Cedex, France\\
$^{10}$Sterrewacht Leiden, Leiden University, Postbus 9513, 2300 RA Leiden, The Netherlands\\
$^{11}$Universit\'e Lyon 1, Observatoire de Lyon, Centre de Recherche Astrophysique de Lyon and Ecole Normale Sup\'erieure de Lyon, \\
\hspace{1mm} 9 avenue Charles Andr\'e, F-69230 Saint-Genis Laval, France\\
$^{12}$Max Planck Institut f\"ur Extraterrestrische Physik, PO Box 1312, D-85478 Garching, Germany\\
$^{13}$Netherlands Institute for Radio Astronomy (ASTRON), Postbus 2, 7990 AA Dwingeloo, The Netherlands\\
$^{14}$Kapteyn Astronomical Institute, University of Groningen, Postbus 800, 9700 AV Groningen, The Netherlands\\
$^{15}$Gemini Observatory, Northern Operations Centre, 670 N. A`ohoku Place, Hilo, HI 96720, USA\\
$^{16}$Centre for Astrophysics Research, University of Hertfordshire, Hatfield, Herts AL1 9AB, UK\\ 
$^{17}$Centre for Astrophysics and Supercomputing, Swinburne University of Technology, Hawthorn, Victoria 3122, Australia\\ 
$^{18}$Dunlap Institute for Astronomy \& Astrophysics, University of Toronto, 50 St. George Street, Toronto, ON M5S 3H4, Canada}

\begin{document}

\pagerange{\pageref{firstpage}--\pageref{LastPage}} \pubyear{2012}
\pagenumbering{arabic}
\label{firstpage}

\maketitle
\clearpage 
\begin{abstract}
We present the Combined Array for Research in Millimeter Astronomy (CARMA) \atlas\ molecular gas imaging survey, a systematic study of the distribution and kinematics of molecular gas in CO-rich early-type galaxies. Our full sample of 40 galaxies (30 newly mapped and 10 taken from the literature) is complete to a $^{12}$CO(1--0) integrated flux of 18.5 Jy km s$^{-1}$\symbolfootnote[3]{}, and it represents the largest, best-studied sample of its type to date. A comparison of the CO distribution of each galaxy to the $g-r$ color image (representing dust) shows that the molecular gas and dust distributions are in good agreement and trace the same underlying interstellar medium. The galaxies exhibit a variety of CO morphologies, including discs (50\%), rings (15\%), bars+rings (10\%), spiral arms (5\%), and mildly (12.5\%) and strongly (7.5\%) disrupted morphologies. There appear to be weak trends between galaxy mass and CO morphology, whereby the most massive galaxies in the sample tend to have molecular gas in a disc morphology. We derive a lower limit to the total accreted molecular gas mass across the sample of $2.48\times10^{10}~M_\odot$, or approximately $8.3\times10^8~M_\odot$ per minor merger within the sample, consistent with minor merger stellar mass ratios.
\end{abstract}


\section{Introduction}
\label{intro}

\let\theOldFootnote\thefootnote\relax
\renewcommand{\thefootnote}{\fnsymbol{footnote}}\relax
\footnotetext[3]{18.5 Jy km s$^{-1}$ is drawn from the IRAM 30m survey, and as the interferometric data indicate that the 30m molecular gas masses are generally underestimates, the true limit is likely slightly higher.}
\let\thefootnote\theOldFootnote\relax

The bimodality in the morphological classification of galaxies has long been known \citep{hubble26}.  Late-type galaxies (LTGs) typically have exponential discs, spiral structure and blue colours.  Early-type galaxies (ETGs) are more ellipsoidal, with smoother isophotes and red colours.  One scenario for producing this galaxy bimodality is via mergers.  Simulations have shown that a merger between two LTGs often creates an ETG \citep{toomre72,springel+05}.  ETGs are also much more likely to be found on the ``red sequence'' portion of the color-magnitude diagram (CMD;  \citealt{baldry+04,faber+07}), presumably due to the loss of most of their star-forming material.

ETGs have been shown to be deficient in star formation relative to LTGs at a given mass \citep{visvan+76,bower+92}, and therefore should be molecular and atomic gas-poor \citep{lees+91}.  Recently, it has however become clear that they are not devoid of cold gas, containing reservoirs of dust (e.g. \citealt{hawarden+81,jura86,knapp+89}), neutral atomic gas (e.g. \citealt{knapp+85, sagewelch06, oosterloo+10}) and molecular gas (e.g. \citealt{sagewrobel89, welch+03, combes+07}).  

The \atlas\ sample is a complete, volume-limited sample of 260 local ETGs brighter than $M_K = -21.5$, covering distances to 42 Mpc with some restrictions on declination and Galactic latitude (\citealt{cappellari+11}, hereafter Paper \Romannum{1}).  The survey was designed to help us understand ETG formation and  evolution.  Optical integral-field spectroscopy with {\tt SAURON} on the William Herschel Telescope (WHT) has been obtained over the central $41''\times33''$ region of all \atlas\ galaxies, revealing their internal stellar kinematics, stellar population properties and ionized gas distributions and kinematics.  The \atlas\ sample has also been completely observed in CO J=1--0 and J=2--1 with the Institut de Radioastronomie Millimetrique (IRAM) 30m telescope (\citealt{young+11}, hereafter Paper \Romannum{4}), and 65\% of the sample has been observed in \hi\ with the Westerbork Synthesis Radio Telescope (WSRT; \citealt{morganti+06}; \citealt{oosterloo+10}; \citealt{serra+12}, hereafter Paper \Romannum{13}).  For studies of the cold and warm interstellar medium (ISM), the \atlas\ sample is thus one of the largest and best-observed samples of local ETGs available.  

Of the 260 ETGs in the \atlas\ sample, 56 were detected in CO (22\% detection rate; Paper \Romannum{4}).  However, despite emerging evidence that a non-negligible fraction of ETGs contain molecular gas, little is known about its origin and evolution, although various scenarios have been put forth (see e.g. \citealt{davis+11b}, hereafter Paper \Romannum{10}\footnote{The current paper presents in detail the observations used in Paper \Romannum{10} to discuss the origin of the CO.}).  The three most prominent scenarios are outlined here.  First, it is possible that the gas survived the galaxy's transformation into an ETG and was not completely consumed subsequently through secular evolution.  Second, the molecular gas may have accumulated internally through stellar mass loss \citep{fabergallagher76}.  In the \atlas\ sample, it is expected that each galaxy produces on average $0.1~M_\odot~{\rm yr}^{-1}$ of ISM through mass loss from the old stellar population alone \citep{ciotti+91}.  Third, it is possible that the molecular gas was accreted from an external source, through tidal stripping events, cold accretion and/or minor mergers with gas-rich companions.  The minor merger (1:4 -- 1:10 mass ratios) rate calculated by \citet{lotz+08,lotz+11} suggests that there are approximately 0.12 minor mergers per galaxy per Gyr, so the ETGs of the \atlas\ sample as a whole should have undergone a total of $\approx 30$ minor mergers in the last Gyr.  It is also possible that this molecular gas is leftover from major mergers that have taken place, but minor mergers dominate in number.  Determining the origin of the molecular gas therefore requires a detailed analysis of its distribution and kinematics (e.g. angular momentum), only available from interferometric maps, and a comparison to the stellar kinematics (as done in Paper \Romannum{10}).

How the properties of the molecular gas depend on the environment of the host galaxy is an open question.  For instance, galaxies in clusters may well be unable to acquire cold external gas, specifically atomic hydrogen (\hi), due to the presence of the hot intracluster medium (ICM) and associated ram pressure stripping  (\citealt{cayatte+90}; \citealt{bohringer+94}; Paper \Romannum{10}).  A first effort towards studying the molecular gas properties of ETGs with multiple molecular transitions (including $^{13}$CO, HCN and HCO$^+$ in addition to $^{12}$CO; \citealt{crocker+12}, hereafter Paper \Romannum{11}) shows that the molecular gas properties widely vary amoung ETGs, probably due to $^{12}$CO optical thickness variations linked to the dynamical state of the gas.

It is also not well determined whether the properties of individual galaxies are the determining factors in the behavior of the molecular gas, or whether the molecular gas in ETGs follows the same star-formation relations as in spirals (e.g. \citealt{ken98,bigiel+08,shapiro+10,wei+10,crocker+11}).  To completely understand star formation in ETGs, it is vital to study in a spatially-resolved way the molecular gas in an unbiased sample of molecular-gas rich ETGs.  Therefore, as part of the \atlas\ project, we performed the present imaging survey with the Combined Array for Research in Millimeter Astronomy (CARMA), the first of its kind capable of providing significant conclusions about the state of molecular gas in ordinary ETGs from interferometric imaging. 

In \S\ref{sample}, we describe the selection criteria used to define the CARMA sample.  In \S\ref{obs}, we describe the observational strategy, the data calibration and reduction, and the analysis techniques.  In \S\ref{disc}, we investigate the properties of the molecular gas in the sample, including the range of morphologies present and their relationships to other host galaxy properties.  We conclude briefly in \S\ref{conc}.  Appendix \ref{app:gals} describes the CO data of individual galaxies in the CARMA \atlas\ sample.  Appendix \ref{app:litgals} presents CO literature data of CO-imaged \atlas\ galaxies, that will be included in the official data release in a uniform manner.

\section{The CARMA sample}
\label{sample}
The sample of ETGs chosen for the CARMA survey was extracted from the \atlas\ survey.  All \atlas\ galaxies were observed with the IRAM 30m telescope in CO(1--0) and CO(2--1), mostly by \citet{combes+07} and Paper \Romannum{4}, though a few were observed by \citet{welch+03}.  The CARMA sample consists of 30 of the CO-detected \atlas\ galaxies (see Table~\ref{tab:sample}). NGC~2697 and NGC~4292, removed from the \atlas\ sample because of evidence of spiral structure in optical images and of a lack of {\tt SAURON} data, respectively, were also imaged with CARMA.  We present their maps in Appendix \ref{app:remgals}, but will not use them in any statistical work discussed in \S\ref{disc}.  We will refer to the 30 CO-rich \atlas\ galaxies with CARMA data as the CARMA \atlas\ sample.
 
Including 10 CO-detected sample galaxies that have already been interferometrically-mapped (see \S\ref{litdata}).  The total number of \atlas\ galaxies with interferometric observations discussed in this paper is 40.  This sample of CO-imaged \atlas\ galaxies is complete for the 33 brightest objects, down to an integrated CO(1--0) flux limit of 18.5 Jy \kms{} (corresponding to the flux of NGC~3182).  Two galaxies with fluxes below this threshold also have CARMA CO images, UGC~05408 and NGC~5173.  Five other CO-faint galaxies have CO maps in the literature (NGC~4550 by \citealt{crocker+09}; NGC~2768, NGC~4477 and NGC~3489 by \citealt{crocker+11}; NGC~2685 by \citealt{schinnerer+02}).  The faintest detections from Paper \Romannum{4} were not observed due to observing time restrictions.

Objects in the CARMA \atlas\ sample have total absolute $K$-band magnitudes ranging from $M_K = -21.57$ to $-25.09$, and distances from 13.4 Mpc to 45.8 Mpc\footnote{UGC~05408 was kept in the \atlas\ survey because its estimated distance is within 42 Mpc when errors in the distance estimate are taken into account (Paper \Romannum{1})}.  Six of the 30 galaxies belong to the Virgo cluster.  All galaxies but three (IC~719, NGC~1222 and NGC~7465) are regular rotators according to their stellar kinematics (\citealt{krajnovic+11}, hereafter Paper \Romannum{2}).  NGC~1222, classified as a non-regular rotator, is known to be undergoing a strong interaction with a neighbour.  NGC~7465 is undergoing an interaction with NGC~7464 and NGC~7463, and has a kinematically-decoupled core.  IC~719 does show ordered rotation in its stellar component, but is recognized as a $2\sigma$ (double-peaked) galaxy when the velocity dispersion map is considered, indicating two counter-rotating stellar discs.  Further properties of the sample galaxies are listed in Table \ref{tab:sample}.

\begin{table*}
 \centering
  \begin{minipage}{140mm}
  \caption{CARMA \atlas\ galaxy sample.}
  \label{tab:sample}
  \begin{tabular}{rccccccc}
  \hline \hline
  \multicolumn{1}{c}{\multirow{2}{*}{Name}} 
  & $\alpha$  & $\delta$ & $v_{\rm sys}$ & $i$   & $\phi_{\rm mol}$ &  $d$  & Virgo\\
  & (J2000) & (J2000)  & (\kms)        & (deg) &  (deg)           & (Mpc) & membership\\
  \hline
  IC 676    & 11~12~39.84 & $+$09~03~20.7 & 1429~ & 69 & \phantom{0}16.5~    & 24.6 & 0\\
  IC 719    & 11~40~18.52 & $+$09~00~35.6 & 1833~ & 74 & 229.0~              & 29.4 & 0\\
  IC 1024   & 14~31~27.07 & $+$03~00~30.0 & 1479~ & 72 & \phantom{0}24.5~    & 24.2 & 0\\
  NGC 1222   & 03~08~56.76 & $-$02~57~19.3 & 2422~ & 41 & \phantom{0}33.0$^c$ & 33.3 & 0\\
  NGC 1266   & 03~16~00.79 & $-$02~25~38.6 & 2170~ & 26 & 270\phantom{.0}~    & 29.9 & 0\\
  NGC 2764   & 09~08~17.44 & $+$21~26~35.8 & 2706~ & 65 & 202.5~              & 39.6 & 0\\
  NGC 2824   & 09~19~02.22 & $+$26~16~12.3 & 2758~ & 61 & 161.5~              & 40.7 & 0\\
  NGC 3182   & 10~19~33.02 & $+$58~12~21.0 & 2118~ & 35 & 331.5~              & 34.0 & 0\\
  NGC 3607   & 11~16~54.54 & $+$18~03~07.1 & \phantom{0}942~ & 34 & 302.5~              & 22.2 & 0\\
  NGC 3619   & 11~19~21.60 & $+$57~45~28.3 & 1560~ & 48 & \phantom{0}74.5~    & 26.8 & 0\\
  NGC 3626   & 11~20~03.78 & $+$18~21~25.6 & 1486~ & 67 & 169.5~              & 19.5 & 0\\
  NGC 3665   & 11~24~43.64 & $+$38~45~46.2 & 2069~ & 64 & 219.5~              & 33.1 & 0\\
  NGC 4119   & 12~08~09.60 & $+$10~22~44.7 & 1656~ & 69 & 296.0~              & 16.5 & 1\\
  NGC 4150   & 12~10~33.65 & $+$30~24~05.5 & \phantom{0}208~ & 54 & 146.0~              & 13.4 & 0\\
  NGC 4324   & 12~23~06.17 & $+$05~15~02.8 & 1665~ & 62 & 232.0~              & 16.5 & 1\\
  NGC 4429   & 12~27~26.56 & $+$11~06~27.3 & 1104~ & 68 & \phantom{0}82.0~    & 16.5 & 1\\
  NGC 4435   & 12~27~40.49 & $+$13~04~44.3 & \phantom{0}791~ & 52 & 201.0~              & 16.7 & 1\\
  NGC 4694   & 12~48~15.10 & $+$10~59~01.3 & 1171~ & 69 & 155.5~              & 16.5 & 1\\
  NGC 4710   & 12~49~39.36 & $+$15~10~11.7 & 1102~ & 86 & 207.0~              & 16.5 & 1\\
  NGC 4753   & 12~52~22.07 & $-$01~11~57.9 & 1163~ & 75 & \phantom{0}93.0~    & 22.9 & 0\\
  NGC 5173   & 13~28~25.29 & $+$46~35~29.9 & 2424~ & 24 & 100$^c$\phantom{.0} & 38.4 & 0\\
  NGC 5379   & 13~55~34.35 & $+$59~44~34.3 & 1774~ & 64 & \phantom{0}66\phantom{.0}~ & 30.0 & 0\\
  NGC 5866   & 15~06~29.60 & $+$55~45~48.0 & \phantom{0}755~ & 89 & 127\phantom{.0}~    & 14.9 & 0\\
  NGC 6014   & 15~55~57.39 & $+$05~55~54.7 & 2381~ & 22 & 139.5~              & 35.8 & 0\\
  NGC 7465   & 23~02~00.96 & $+$15~57~53.3 & 1960~ & 70 & 106.0~              & 29.3 & 0\\
  PGC 029321 & 10~05~51.18 & $+$12~57~40.7 & 2816~ & 38 & \phantom{0}76\phantom{.0}~ & 40.9 & 0\\ 
  PGC 058114 & 16~26~04.29 & $+$02~54~23.6 & 1507~ & 71 & \phantom{0}94.5~    & 23.8 & 0\\
  UGC 05408  & 10~03~51.86 & $+$59~26~10.2 & 2998~ & 31 & 315$^c$\phantom{.0} & 45.8 & 0\\
  UGC 06176  & 11~07~24.68 & $+$21~39~25.6 & 2677~ & 68 & 201.0~              & 40.1 & 0\\
  UGC 09519  & 14~46~21.12 & $+$34~22~14.2 & 1631~ & 41 & 177.5~              & 27.6 & 0\\
  \hline
  NGC 2697$^a$ & 08~54~59.40 & $-$02~59~15.2 & 1814$^b$ & 30 & 301.5~           & 22.0 & 0\\
  NGC 4292$^a$ & 12~21~16.49 & $+$04~35~44.3 & 2258$^b$ & 50 & 230\phantom{.0}~ & 29.8 & 0\\
  \hline \hline
  \end{tabular}
{\vspace{3pt} \\
\raggedright
$^a$ Removed from the \atlas\ sample, but the CO data are presented in Appendix \ref{app:remgals}.\\
$^b$ From the {\tt HYPERLEDA} catalog, due to the absence of {\tt SAURON} data.\\
$^c$ Not originally in Paper \Romannum{5}.\\
\textbf{Notes:}\\
\noindent Column (1): Principle designation from LEDA, used as the standard designation.\\
\noindent Column (2): Right ascension (J2000), from LEDA.\\
\noindent Column (3): Declination (J2000), from LEDA.\\
\noindent Column (4): Optical stellar heliocentric velocity, from Paper \Romannum{1}.\\
\noindent Column (5): Best-determined CO inclination angle, from Paper \Romannum{5}.\\
\noindent Column (6): Kinematic position angle of the molecular gas, from Paper \Romannum{10}.\\
\noindent Column (7): Average distance to each galaxy, from Paper \Romannum{1}, originally from \citet{tonry+01} and \citet{mei+07}.\\
\noindent Column (8): Virgo membership of each galaxy, from Paper \Romannum{1}.}
  \end{minipage}
\end{table*}

\section{Observations, Calibration and Data Reduction}
\label{obs}
\subsection{CARMA observations}
Observations for this survey were taken in the $^{12}$CO(1--0) line at CARMA over the course of five semesters, beginning in Autumn 2008 and finishing in November 2010.  Galaxies were always first observed with the CARMA D array, with 11 -- 150m baselines, corresponding to observable angular scales of 3.5 -- 48\arcsec\ at CO(1--0).  NGC~2697 and NGC~7465, that appeared to have significant flux resolved out, were followed up with the more compact E array (8 -- 66m baselines).  NGC~1266 was spatially unresolved in D array, and was followed up with the CARMA B and A arrays (a detailed discussion of NGC~1266 can be found in \citealt{alatalo+11}).  The CARMA \atlas\ survey had a surface brightness sensitivity (1$\sigma$ in 100 \kms\ linewidth) ranging from 12 to 369 $M_\odot~{\rm pc}^{-2}$, with median a sensitivity of 84 $M_\odot~{\rm pc}^{-2}$.
On average, $\approx$100 hours of observations were obtained each semester.  Observational parameters are listed in Table \ref{tab:obs}.

Upgrades to the CARMA correlator and receivers were taking place while the \atlas\ survey was ongoing, so data taken later in the programme have larger bandwidths and simultaneous observations of \hbox{$^{12}$CO(1--0)} and \hbox{$^{13}$CO(1--0)}.  A handful of \atlas\ galaxies thus also have $^{13}$CO maps that will be presented in an upcoming paper. The galaxies with large line widths \hbox{($\Delta v \gtrsim 420$ \kms)} were observed only after the CARMA correlator was upgraded, providing sufficient bandwidth and channel resolution to properly cover and sample the line.  We were able to reliably image 3mm continuum in three sources (NGC~1266, NGC~3665 and NGC~5866), and Table \ref{tab:data} lists those fluxes as well as upper limits for the other galaxies.

\begin{table*}
 \centering
  \begin{minipage}{150mm}
  \caption{Observational parameters of CARMA galaxies}
  \label{tab:obs}
  \begin{tabular}{rlrcccc}
\hline \hline
\multicolumn{1}{c}{\multirow{2}{*}{Name}} & \multicolumn{1}{c}{\multirow{2}{*}{Semester}} & \multicolumn{1}{c}{Gain} & Total & $\theta_{\rm maj}\times\theta_{\rm min}$ & $\Delta V$ & Bandwidth \\
&  & calibrator & \multicolumn{1}{c}{hours} & (arcsec) & (\kms) & (\kms) \\ 
\hline
 IC 676    & 2009B~ & 1058+015 & \phantom{0}3.75 & $3.8\times3.3$ & 10 & 410\\
 IC 719    & 2010A~ & 3C273 & 13.17 & $3.9\times2.8$ & 20 & 580\\
 IC 1024   & 2008B~ & 3C279 & \phantom{0}4.68 & $3.9\times3.0$ & 10 & 420\\
NGC 1222   & 2008B~ & 0339-017 & \phantom{0}5.20 & $3.6\times3.2$ & \phantom{0}5 & 420\\
NGC 1266   & 2008B~ & 0339-017 & \phantom{0}5.43 & $4.2\times3.3$ & 10 & 410\\
NGC 2764   & 2009B~ & 0854+201 & \phantom{0}4.43 & $3.5\times2.9$ & 10 & 410\\
NGC 2824   & 2008B~ & 0956+252 & \phantom{0}8.79 & $4.3\times4.0$ & 25 & 400\\
NGC 3182   & 2009A~ & 0927+390 & \phantom{0}7.42 & $5.2\times4.1$ & 30 & 390\\
NGC 3607   & 2010B~ & 1058+015 & \phantom{0}9.30 & $5.6\times5.0$ & 20 & 900\\
NGC 3619   & 2010B~ & 0958+655 & 24.44 & $4.4\times3.9$ & 40 & 880\\
NGC 3626   & 2009B~ & 1159+292 & \phantom{0}7.50 & $3.9\times3.7$ & 25 & 400\\
NGC 3665   & 2010B~ & 1159+292 & 16.59 & $4.3\times4.2$ & 10 & 910\\
NGC 4119   & 2008B~ & 3C273 & \phantom{0}3.67 & $5.0\times4.0$ & 10 & 410\\
NGC 4150   & 2010A~ & 1159+292 & \phantom{0}6.74 & $5.3\times4.1$ & 10 & 560\\
NGC 4324   & 2009A~ & 3C273 & 20.18 & $4.7\times3.9$ & 20 & 400\\
NGC 4429   & 2010B~ & 3C273 & \phantom{0}8.83 & $4.7\times3.7$ & 10 & 840\\
NGC 4435   & 2010B~ & 3C273 & \phantom{0}8.48 & $3.9\times3.4$ & 10 & 920\\
NGC 4694   & 2010A~ & 3C273 & \phantom{0}5.38 & $3.9\times3.1$ & 10 & 390\\
NGC 4710   & 2009A~ & 3C273 & \phantom{0}5.10 & $3.9\times3.2$ & 10 & 410\\
NGC 4753   & 2010B~ & 3C273 & 10.38 & $5.6\times4.1$ & 15 & 735\\
NGC 5173   & 2009B$^a$ & 1310+323 & \phantom{0}7.06 & $3.9\times3.5$ & 20 & 240\\
NGC 5379   & 2010A~ & 1642+689 & \phantom{0}5.33 & $5.2\times3.6$ & 10 & 410\\
NGC 5866   & 2010B~ & 1419+543 & \phantom{0}3.52 & $3.6\times3.1$ & 10 & 920\\
NGC 6014   & 2008B~ & 1751+096 & \phantom{0}3.85 & $4.2\times3.9$ & 10 & 350\\
\multirow{2}{*}{NGC 7465} & 2008B~ & \multirow{2}{*}{3C454.3} & \multirow{2}{*}{15.02} & \multirow{2}{*}{$6.6\times5.6$} & \multirow{2}{*}{10} & \multirow{2}{*}{410}\\
         & 2009A$^b$  &         &        &     &    &    \\
PGC 029321 & 2009B~ & 1058+015 & \phantom{0}4.52 & $3.8\times3.7$ & 10 & 410\\
PGC 058114 & 2009A~ & 1549+026 & \phantom{0}3.62 & $4.4\times3.8$ & 10 & 410\\
UGC 05408  & 2009A~ & 0927+390 & \phantom{0}2.00 & $4.6\times3.5$ & 25 & 196\\
UGC 06176  & 2009A~ & 1159+292 & \phantom{0}4.08 & $3.5\times2.8$ & 15 & 735\\
UGC 09519  & 2008B~ & 1310+323 & \phantom{0}3.75 & $3.4\times2.9$ & 10 & 410\\
\hline
\multirow{2}{*}{NGC 2697} & 2008B~ & \multirow{2}{*}{0927+390} & \multirow{2}{*}{14.33} & \multirow{2}{*}{$6.8\times5.1$} & \multirow{2}{*}{20} & \multirow{2}{*}{380}\\
        & 2009A$^b$ &       &       &                &    & \\
NGC 4292 & 2008B~     & 3C273 & 19.86 & $4.1\times3.3$ & 10 & 410\\
\hline \hline
\end{tabular}
{\vspace{3pt}\\
\raggedright
$^a$ Observed both in C- and D-arrays\\
$^b$ Observed in D-array in 2008B and E-array in 2009A.\\
\textbf{Notes:}\\
\noindent Column (1): Rrinciple designation from LEDA, used as the standard designation.\\
\noindent Column (2): Semester in which the observations were taken at CARMA.\\
\noindent Column (3): Gain calibrator used.\\
\noindent Column (4): Total on-source observing time in hours.\\
\noindent Column (5): FWHM of the major and minor axes of the synthesized beam.\\
\noindent Column (6): Final velocity channel width.\\
\noindent Column (7): Velocity bandwidth covering the line.}
  \end{minipage}
\end{table*}

\subsection{Calibration and imaging}
Raw CARMA visibility data were reduced in the usual way, using the Multichannel Image Reconstruction Image Analysis and Display ({\tt MIRIAD}) package \citep{sault+95}.  For each source and track, the raw data were first Hanning-smoothed in velocity.  Then the phase vs. time behaviour of the calibrator was checked to search for decorrelations over baselines, which were flagged out.  Next, the data were corrected for differences in the lengths of the fiber optic lines between the antennas and the correlator.  The bandpass was then determined using a high signal-to-noise ratio (S/N) observation of a bright calibrator, and used to correct for instrumental spectral fluctuations.  The atmospheric phase offsets present in the data were determined using a phase calibrator, usually a quasar within 20$^\circ$ of the source, observed at regular ($\approx15$ minutes) intervals.  Amplitude calibration was performed using the phase calibrator.  A catalog of the most up-to-date fluxes of each calibrator is maintained at CARMA, both through monitoring of calibrator fluxes during science tracks and through dedicated flux calibration tracks taken on a weekly basis, and is updated weekly using all tracks that include a primary calibrator (e.g. a planet), used to infer the secondary calibrator fluxes \citep{bauermeister+12}. Flux calibration uncertainties are assumed to be 20\%, which is standard for millimeter observations using planetary models (Petric et al., in prep).  After the data were satisfactorily processed, the gain solutions derived from the nearby calibrator were applied to the source.  The calibrated source data of all observations of each source were then combined into one visibility file for imaging.

We then used {\tt MIRIAD} to convert all visibility files to three-dimensional (3D) data cubes.  When possible, a zeroth order continuum fit to the $uv$-data was made using channels free of line emission, and was subtracted from all channels.  Only galaxies observed after the correlator was upgraded (bandwidth $>420$ \kms) had sufficient bandwidth for a reliable continuum fit.  Galaxies with narrower bandwidths were not continuum-subtracted (continuum emission generally adds $< 10$\% extra flux to each galaxy, much less than the total flux calibration uncertainties, and this is usually confined to the central beam in each galaxy).  The continuum-subtracted $uv$-datasets were then transformed into RA-Dec-velocity space (with velocities determined from the line being imaged, mainly $^{12}$CO but also occasionally $^{13}$CO).  Channel widths were chosen to achieve at least a 3$\sigma$ detection in each channel where flux was present, and were always larger (by at least a factor of 2) than the original spectral resolution of the $uv$-data.  Pixels of 1\arcsec$\times$1\arcsec\ were chosen as a compromise between spatial sampling and resolution, typically giving approximately 4 pixels across the beam major axis.  A constant pixel size was chosen to maintain uniform spatial sampling across the sample.  One arcsecond corresponds to a physical scale between 72 and 222 pc, depending on the distance of the source.  The areas imaged were generally chosen to be within the primary beam of the 10m antennas ($\approx 54''$), but in the cases of sources that extend beyond this (NGC~4324, NGC~4710 and NGC~5866), the area imaged was taken to be closer to the primary beam of the 6m antennas ($\approx 90''$).  In these cases, the {\tt MIRIAD} imaging task {\tt INVERT} was run with the mosaicking option, to properly scale the data and account for the different primary beam widths.

{\tt Robust=0} weighting was used by default, but was changed if the size and make-up of the source dictated it.  Natural weighting ({\tt robust = 1}) was used for sources where large scale features presented in the initial ({\tt robust = 0}) channel maps, and we suspected that some flux might be resolved out (NGC~3619, NGC~3626 and NGC~4324).  The dirty cubes were cleaned in regions of source emission to a threshold equal to the rms of the dirty channels.  The clean components were then added back and re-convolved using a Gaussian beam of full-width at half maximum (FWHM) equal to that of the dirty beam. This produced the final, reduced and fully calibrated data cube for each galaxy.

The 3D data cubes were then used to create moment maps of each galaxy: a zeroth moment (moment0) or integrated intensity map, and a first moment (moment1) or mean velocity map.  The data cubes were first Hanning-smoothed in velocity and Gaussian-smoothed spatially (with a FWHM equal to that of the beam), and masks were created by selecting all pixels above fixed flux thresholds, adjusted to recover as much flux as possible in the moment maps while minimising the noise (generally about 2--3 times the rms noise in the smoothed channels).  The moment maps were then created using the unsmoothed cubes within the masked regions only.

All data products were converted into Flexible Image Transport System (FITS) files using {\tt MIRIAD}.  Tables \ref{tab:obs} and \ref{tab:data} list some datacube properties for each of the \atlas\ CARMA galaxies.  Figure \ref{fig:r+co} shows the CO moment0 contours of each CARMA \atlas\ sample galaxy overlaid on an $r$-band image, either from the Sloan Digital Sky Survey (SDSS) or the Isaac Newton Telescope (INT) when SDSS imaging is unavailable (Scott et al. 2012, submitted, hereafter Paper \Romannum{21}).  Figure \ref{fig:moment1} displays the moment1 maps of the 30 sample galaxies.  Figure \ref{fig:dust} shows the CO moment0 contours overlaid on unsharp-masked optical images, to enhance dust features (see \S\ref{dustcomp}).  Figure \ref{fig:pvd} displays position-velocity diagrams (PVDs) of galaxies whose kinematics are not well represented with moment1 maps, due to multiple velocity components along the line-of-sight (NGC~2764, NGC~4710 and NGC~5866).  Appendix \ref{app:gals} contains multiple figures for each CARMA \atlas\ galaxy, including a $r$-band-moment0 overlay, moment0 map, moment1 map, a comparison of the CARMA and IRAM 30m spectra, a PVD along the kinematic position angle determined in Paper \Romannum{10}, and individual channel maps.  A more detailed analysis of the PVD, as well as a comparison to the stellar and ionized gas kinematics is available in Davis et al. (2012, hereafter Paper \Romannum{14}).

The continuum measurement or upper limit for each galaxy depended on the correlator configuration.  For narrow-band (bandwidth $\approx$ 420 \kms) galaxies that did not include $^{13}$CO in the lower sideband, the full lower sideband was used to calculate the continuum (bandwidth $\approx$ 186 MHz).  For narrow-band galaxies that included $^{13}$CO observations, the edges of the band in the lower sideband were used (bandwidth $\approx$ 124 MHz).   Galaxies that were observed with the upgraded correlator were treated individually, the continuum being modeled in line-free channels only, which varied from galaxy to galaxy.   The continuum upper limit was taken to be three times the rms noise of the dirty map produced by inverting the combined line-free channels.  Galaxies where a continuum source was detected (NGC~1266, NGC~3665 and NGC~5866), the flux was measured from the detected point source, and the rms noise was measured in the flux-free regions of the cleaned maps.

\begin{table*}
 \centering
  \begin{minipage}[b!]{7in}
  \caption{Data parameters of CARMA \atlas galaxies}
  \label{tab:data}
  \begin{tabular}{r.ld{1}@{ $\pm$ }lr@{ $\pm$ }lr@{ $\pm$ }lc}
 \hline \hline
  \multicolumn{1}{c}{\multirow{2}{*}{Name}} 
& \multicolumn{1}{c}{$\sigma_{\rm rms}$} 
& Continuum 
& \multicolumn{2}{c}{$F_{\rm CARMA}$} 
& \multicolumn{2}{c}{\multirow{2}{*}{$log\frac{M_{H_2}}{M_\odot}$}} 
& \multicolumn{2}{c}{$\underline{F_{\rm CARMA}}$} & Morph.\\
& \multicolumn{1}{c}{(mJy bm$^{-1}$)}   
& (mJy bm$^{-1}$)  
& \multicolumn{2}{c}{(Jy km s$^{-1}$)}       
& \multicolumn{2}{c}{}
& \multicolumn{2}{c}{$F_{\rm 30m}$}      
& class.\\
\hline
 IC 676& 30.6  & $<$ 3.09            &  66.3 & 2.5 & 8.71 & 0.04 & 1.2 & 0.05 & D\\
 IC 719&  6.63 & $<$ 0.351           &  20.4 & 1.0 & 8.34 & 0.05 & 1.2 & 0.12 & D\\
 IC 1024& 19.4  & $<$ 3.06            &  94.6 & 1.7 & 8.87 & 0.02 & 1.8 & 0.04 & D\\
NGC 1222& 23.1  & $<$ 1.97            & 147.6 & 1.6 & 9.33 & 0.01 & 1.8 & 0.04 & X\\
NGC 1266& 22.8  & \phantom{= } 7.49 $\pm$ 1.96 & 162.9 & 0.7 & 9.28 & 0.01 & 1.0 & 0.03 & D\\
NGC 2764& 12.9  & $<$ 1.85            &  85.4 & 1.5 & 9.23 & 0.02 & 1.1 & 0.04 & B+R\\
NGC 2824& 10.2  & $<$ 2.21            &  37.8 & 0.8 & 8.91 & 0.02 & 1.8 & 0.08 & D\\
NGC 3182&  8.16 & $<$ 1.68            &  16.4 & 0.7 & 8.41 & 0.04 & 1.2 & 0.12 & R\\
NGC 3607&  7.50 & $<$ 0.830           &  58.0 & 0.8 & 8.50 & 0.01 & 1.2 & 0.04 & D\\
NGC 3619&  3.23 & $<$ 0.327           &  15.2 & 0.4 & 8.28 & 0.03 & 0.7 & 0.12 & M\\
NGC 3626&  5.34 & $<$ 1.48            &  43.2 & 0.9 & 8.32 & 0.02 & 1.3 & 0.10 & D\\
NGC 3665& 10.2  & \phantom{= } 8.98 $\pm$ 0.42 &  94.3 & 1.2 & 9.11 & 0.01 & 1.6 & 0.06 & D\\
NGC 4119& 20.2  & $<$ 2.64            &  39.3 & 0.9 & 8.14 & 0.02 & 1.8 & 0.07 & R\\
NGC 4150&  9.41 & $<$ 1.87            &  20.5 & 0.4 & 7.82 & 0.02 & 0.7 & 0.08 & M\\
NGC 4324&  4.30 & $<$ 0.974           &  27.2 & 0.6 & 7.97 & 0.02 & 1.9 & 0.12 & R\\
NGC 4429&  9.53 & $<$ 0.375           &  70.8 & 0.9 & 8.39 & 0.01 & 2.2 & 0.08 & D\\
NGC 4435& 11.5  & $<$ 0.375           &  31.6 & 1.0 & 8.05 & 0.03 & 1.5 & 0.10 & R\\
NGC 4694&  9.30 & $<$ 0.512           &  21.7 & 0.6 & 8.01 & 0.03 & 0.7 & 0.07 & X\\
NGC 4710& 14.0  & $<$ 5.20            & 351.2 & 1.7 & 9.08 & 0.01 & 2.3 & 0.03 & B+R\\
NGC 4753& 11.1  & $<$ 0.521           &  74.7 & 1.1 & 8.70 & 0.01 & 1.4 & 0.06 & M\\
NGC 5173&  8.16 & $<$ 1.46            &  12.5 & 0.2 & 8.36 & 0.02 & 1.2 & 0.14 & X\\
NGC 5379& 10.8  & $<$ 0.450           &  18.5 & 0.8 & 8.53 & 0.04 & 1.6 & 0.15 & B+R\\
NGC 5866& 17.6  & \phantom{= } 4.02 $\pm$ 0.54 & 258.2 & 3.4 & 8.75 & 0.01 & 1.9 & 0.02 & B+R\\
NGC 6014& 22.4  & $<$ 3.61            &  34.5 & 1.0 & 8.77 & 0.03 & 1.0 & 0.07 & S\\
NGC 7465 &  9.71 & $<$ 1.39            &  93.6 & 0.4 & 9.02 & 0.01 & 1.7 & 0.04 & M\\
PGC 029321& 13.1  & $<$ 2.09            &  18.3 & 0.7 & 8.61 & 0.04 & 1.2 & 0.09 & D\\
PGC 058114& 15.7  & $<$ 2.21            &  73.7 & 0.9 & 8.75 & 0.01 & 1.4 & 0.04 & D\\
UGC 05408& 13.6  & $<$ 2.83            &   6.1 & 0.5 & 8.32 & 0.08 & 0.8 & 0.18 & D\\
UGC 06176& 10.8  & $<$ 0.524           &  28.5 & 0.8 & 8.76 & 0.03 & 1.5 & 0.10 & D\\
UGC 09519& 19.5  & $<$ 2.27            &  51.4 & 1.4 & 8.77 & 0.03 & 0.9 & 0.04 & M\\
\hline
NGC 2697   & 9.89  & $<$ 1.36            &  48.2 & 0.7 & 8.61 & 0.01 & 2.3 & 0.11 & R\\
NGC 4292   & 9.58  & $<$ 1.05            &  13.8 & 0.5 & 7.74 & 0.04 & 1.2 & 0.14 & R\\
\hline \hline
\end{tabular}
{\vspace{3pt}\\
\raggedright
\textbf{Notes:}\\
\noindent Column (1): Principle designation from LEDA, used as the standard designation.\\
\noindent Column (2): Root mean square noise level per channel, measured in regions devoid of line emission.\\
\noindent Column (3): Continuum emission or upper limit (see text).\\
\noindent Column (4): Total CO flux from CARMA.\\
\noindent Column (5): Total molecular gas mass from the CARMA observations using $X_{\rm CO} = 3\times10^{20}$ (K km s$^{-1})^{-1}$ cm$^{-2}$ (as in Paper \Romannum{4})\\
\noindent Column (6): Fraction of CO flux recovered by CARMA (with respect to the IRAM 30m telescope).\\
\noindent Column (7): CO morphological classification, based on the moment0 and moment1 maps: \\
\noindent D = disc, X = strongly disrupted, \hbox{R = ring}, M = mildly disrupted, B = bar/ring, S=spiral.}
  \end{minipage}
\end{table*}

\subsection{Measurements from the data}
\label{data_analysis}


To calculate the noise level, $\sigma_{\rm rms}$ of each data cube, we selected all the pixels within each cube outside the region known to contain flux, and calculated the standard deviation.  The total flux per channel was then calculated by using the region of the moment0 map that contained flux, then summing over all pixels in each channel located within that region.  The statistical noise per channel is calculated as $\sigma_{rms} * \sqrt{N_b}$ , where $N_b$ is the number of beam areas in the region with identified emission.  The CARMA integrated spectra presented in Appendix \ref{app:gals} have been constructed using this method.  The integrated fluxes in the channel maps were converted from the native Jansky beam$^{-1}$ units into Jansky by dividing out the total beam area in pixels, and are given in Table \ref{tab:data}.

PVDs were created by taking thin (5 pixel) slices through the data cubes, positioned to intersect the centre of the CO emission, and using the kinematic position angle determined in Paper \Romannum{10}.  Table \ref{tab:sample} also lists the kinematic position angles of galaxies not originally listed in Paper \Romannum{10}, but subsequently had kinematic position angles fitted using the same methods.


\subsection{Comparison to the 30m data}
The integrated spectra from CARMA and the IRAM 30m telescope are overlaid in Figs. \ref{fig:firstgal}--\ref{fig:lastgal}.  Millimeter measurements are assumed to have 20\% absolute flux uncertainties (errors reported on the measured parameters listed in Table \ref{tab:data} only include the random errors on the measurement itself, not the systematic flux calibration error), so flux ratios between 0.78 and 1.28 indicate fluxes in agreement within the uncertainties.  The spectra from the IRAM 30m agree with the CARMA ones for 15/30 galaxies.  In all cases where there is disagreement, CARMA actually recovers more flux than the 30m.  Most discrepant galaxies either extend beyond, or fill the majority of the single-dish beam, with two large ratios (NGC~4435 and UGC~06176) likely due to a slight off-center pointing of the single dish.  In fact, only 7 of the galaxies within the CARMA \atlas\ sample are well contained within the 21.6$''$ IRAM beam.

As discussed above, all galaxies detected with the IRAM 30m were also detected at CARMA.  However, the IRAM 30m selection for the current CARMA survey does imply that galaxies with molecular gas exclusively beyond the 30m beam (e.g. rings, arms and tidal features beyond a radius of $\approx 12''$) would have been missed.  It is impossible to estimate the incidence of these objects based on the CO data alone, but the Galaxy Evolution Explorer ({\em GALEX}) ultraviolet imaging survey of the {\tt SAURON} sample (tracing the associated star formation) suggests that they are very rare (of 34 ETGs, only one has such features, NGC~2974; \citealt{jeong+09}).  Lower density neutral hydrogen is however fairly common at large radii (Paper \Romannum{13}).

Because the molecular gas in many ($>75$\%) galaxies extends beyond the single-dish beam, it is likely that the CARMA flux more accurately reflects the total molecular gas mass in each galaxy, and therefore we adopt the CARMA derived molecular gas masses for the remainder of this paper.

\begin{figure*}
\raggedright
\includegraphics[width=7in]{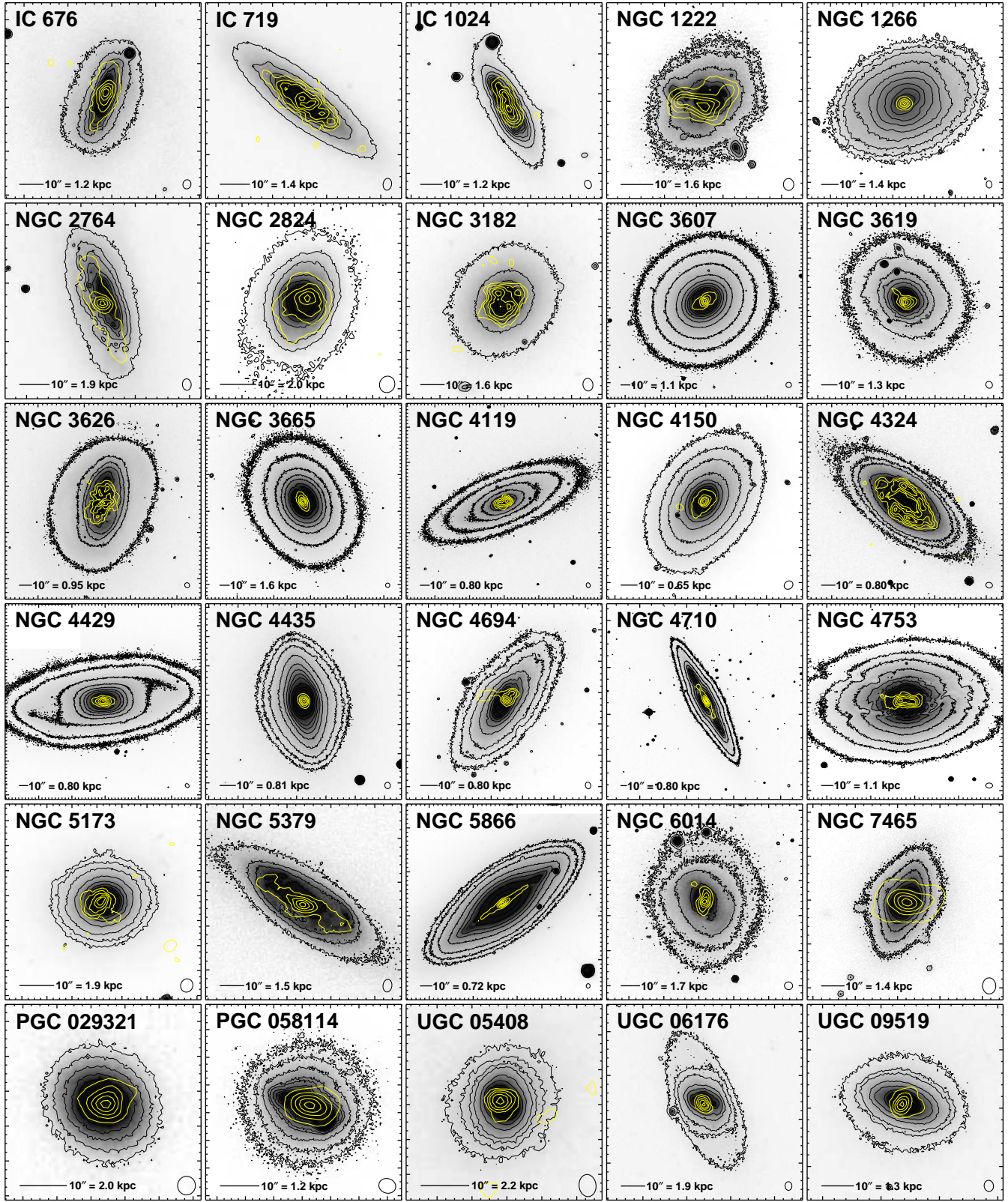}
\caption{$r$-band images (grayscale and black contours) overlaid with integrated CO(1--0) contours (yellow) at [10\%, 30\%, 50\%, 70\%, 90\%] of the peak intensity for the 30 galaxies in the \atlas\ CARMA sample.  The $r$-band images were either taken from the SDSS or  a dedicated program with the INT (Paper \Romannum{21})  The CO synthesized beam is shown in the bottom-right corner of each panel.  A bar in the bottom left corner of each panel indicates the scale of 10$''$, as well as the equivalent physical scale at the distance of the galaxy.}
\label{fig:r+co}
\end{figure*}

\begin{figure*}
\raggedright
\includegraphics[width=7in]{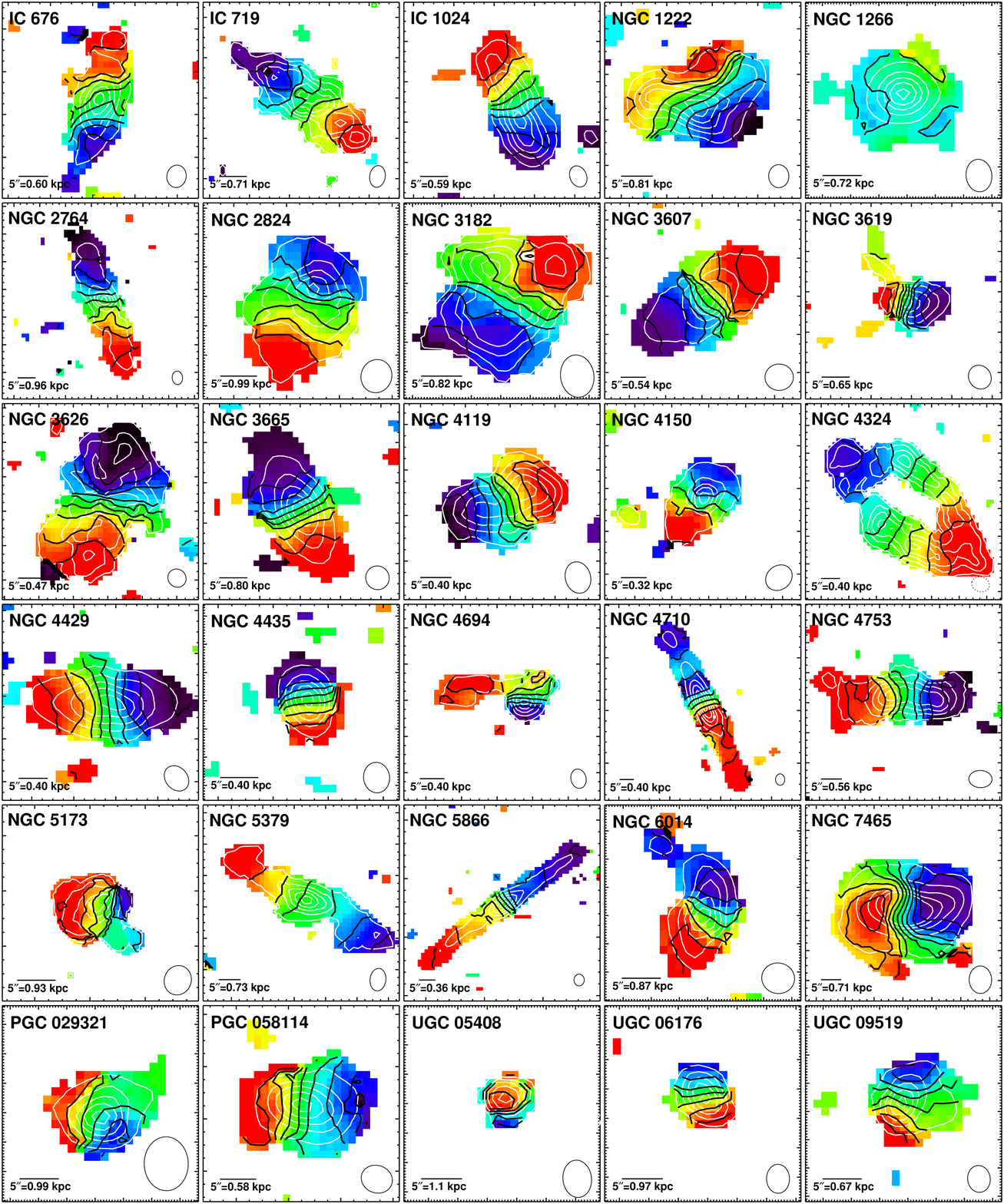}
\caption{Mean CO(1--0) velocity maps of the 30 galaxies of the \atlas\ CARMA sample.  Iso-velocity contours (black) are overlaid at 20 \kms\ intervals, as are contours from the moment0 map (white).  The CO synthesized beam is shown at the bottom-right corner of each panel.  A bar in the bottom left corner of each panel indicates the scale of 5$''$, as well as the equivalent physical scale at the distance of the galaxy.}
\label{fig:moment1}
\end{figure*}

\subsection{Literature data}
\label{litdata}
Twelve galaxies of the \atlas\ sample already have interferometric CO data available in the literature.  Those galaxies are: NGC~4710 \citep{wrobel+92}, NGC~2685 \citep{schinnerer+02}, and the ten galaxies that will be part of the \atlas\ data release: NGC~4476 \citep{young02}; NGC~3032, NGC~4150, NGC~4459 and NGC~4526 \citep{ybc08}; NGC~2768 \citep{crocker+08}; NGC~4550 \citep{crocker+09}; NGC~524, NGC~3489 and NGC~4477 \citep{crocker+11}.  Names, distances, Virgo membership, CO morphological classes and references of these galaxies are listed in Table \ref{tab:litgals}.  When required, data for each of the galaxies were provided by the original authors for use in this work.  Moment maps and dust comparisons appear in Figure \ref{fig:lit_gals} for the galaxies whose data will be included in the data release.  Both NGC~4710 and NGC~4150 were re-observed as part of the CARMA \atlas\ survey, and we only present these updated data in this paper.  

\subsection{Catalog of \atlas\ interferometric CO data}
\label{catalog}
The CARMA data archive\footnote{http://carma-server.ncsa.uiuc.edu:8181/} will include CARMA data for all 30 galaxies, as well as data for the 10 previously published galaxies and the two non-\atlas\ sources.  The data products provided will include data cubes and moment0 and moment1 maps in FITS format. The data will also be reachable from the \atlas project website\footnote{http://purl.org/atlas3d}.

\section{Discussion}
\label{disc}
Figures~\ref{fig:r+co} and \ref{fig:moment1} show the 30 ETGs imaged as part of the CARMA \atlas\ survey.  Before the survey was complete (which includes the 10 literature galaxies mentioned in \S\ref{litdata}), 16 ETGs had spatially-resolved CO maps \citep{wrobel+92,schinnerer+02,young02,young05,wei+10,ybc08,crocker+11}, most obtained as part of individual or small-sample studies.  The addition of the CO-imaged \atlas\ sources increases this amount to over 50, and provides a largely unbiased sample.  Combined with the maps from the {\tt SAURON} integral-field spectrograph, the CO-mapped \atlas\ galaxies thus provide the most robust picture of molecular gas-rich ETGs.

We now focus our analysis on the CO-rich \atlas\ ETGs only, adding the ten \atlas\ galaxies imaged in CO with the Berkeley-Illinois-Maryland Array (BIMA), the Owens Valley Radio Observatory (OVRO) and the Plateau de Bure Interferometer (PdBI) prior to the CARMA effort.  Between the 30 CARMA galaxies and 10 previously published galaxies (excluding two overlapping objects; see Table~\ref{tab:litgals}), we imaged 40 galaxies or 65\% of the CO detections in Paper \Romannum{4}, more than 90\% of the total molecular gas mass in those detections, and are complete down to a CO(1--0) integrated flux of 18.5 Jy km s$^{-1~}$.

\begin{table}
 \centering
 \begin{minipage}{80mm}
  \caption{Properties of the \atlas\ literature galaxies}
  \begin{tabular}{rcccc}
  \hline
\multicolumn{1}{c}{\multirow{2}{*}{Name}} & Distance & Morph. & Virgo & \multicolumn{1}{c}{\multirow{2}{*}{Reference}} \\
& (Mpc) & class.$^a$ & membership & \\
\hline
NGC 524& 23.3 & D & 0 & 6\\
NGC 2685& 16.7 & R & 0 & 1\\
NGC 2768& 21.8 & D & 0 & 4\\
NGC 3032& 21.4 & D & 0 & 3\\
NGC 3489& 11.7 & S & 0 & 6\\
NGC 4459& 16.1 & D & 1 & 3\\
NGC 4476& 17.6 & D & 1 & 2\\
NGC 4477& 16.5 & R & 1 & 6\\
NGC 4526& 16.4 & D & 1 & 3\\
NGC 4550& 15.5 & D & 1 & 5\\
\hline
\end{tabular}
\label{tab:litgals}

References: (1) \citet{schinnerer+02}; 
(2) \citet{young02};
(3) \citet{ybc08};
(4) \citet{crocker+08};
(5) \citet{crocker+09};
(6) \citet{crocker+11}\\
\textbf{Notes:}\\
$^a$CO morphological classification based on the moment0 and moment1 maps.  D = disc, X = strongly disrupted, R = ring,
M = mildly disrupted, B = bar/ring, S=spiral.
\end{minipage}
\end{table}

\begin{figure*}
\raggedright
\includegraphics[width=7in]{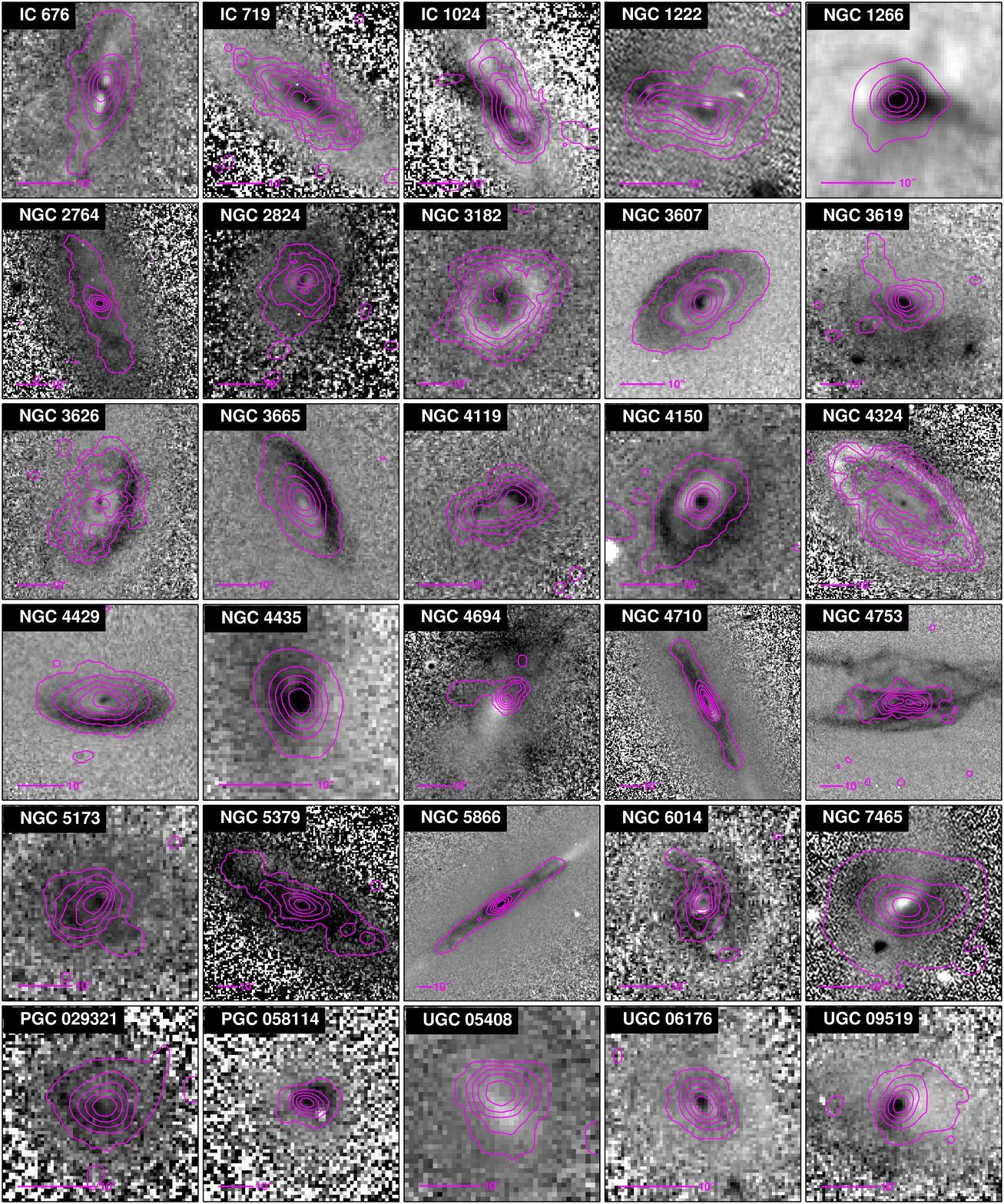}
\caption{Comparison of the integrated CO(1--0) distribution (magenta contours) and the dust distribution as represented by $g-r$ colour images (grayscale) for the 30 galaxies of the \atlas\ CARMA sample. It appears that galaxies with disrupted CO morphologies also exhibit disturbance in the dust distribution, and overall the molecular gas in the galaxies traces the filamentary dust.}
\label{fig:dust}
\end{figure*}

\subsection{Comparison to the dust}
\label{dustcomp}
Figure~\ref{fig:dust} shows the CO moment0 contours of each galaxy overlaid on a $g-r$ colour image.  The dust distributions are shown by creating $g-r$ colour SDSS images (and for NGC~1222 and NGC~7465, for which SDSS images are unavailable, on INT images, and for NGC~1266, on CTIO $b$ and $r$ band images from the SINGS survey; \citealt{kennicutt+03}).  In order to create the $g-r$ images, the Goddard IDL library task {\tt SKY} was used to determine the background sky level, then subtracted from the images.  The $g-r$ colour images were then created by dividing the sky-subtracted $g$-band image from the sky-subtracted $r$-band image.  The $g-r$ imaging enhances the visibility of dust features by directly tracing extinction and provides a qualitative view of the distribution of the filamentary dust in each galaxy of the \atlas\ CARMA sample.  This dust, for the most part, is a reasonable indicator of the distribution of the molecular gas.  

For simplicity we refer to the $g-r$ colour images as ``dust images'' for the remaining text, even if they are intrinsically only representations of stronger dust-obscured regions. Indeed, the CO is nearly always (28/29 instances) associated with evidence of dust obscuration.  It is evident that the converse is not true, as a Hubble Space Telescope (HST) survey of nearby ellipticals revealed that they contain dust in nearly half the cases \citep{tran+01,lauer+05}, much more than the CO detection rate of \atlas.
Additional dust is seen beyond the limits of the detected CO emission in most galaxies, but notably in NGC~1266, NGC~3619, NGC~4150, NGC~4694 and NGC~4753.  However, there is no noticeable dust beyond the detected extent of the CO in others.  It thus appears that in most cases where the dust and CO are co-spatial, dust provides a good, high spatial resolution rendering of the molecular gas structure.  Where dust is more extended than the CO, it could simply be that the outer CO is below our detection threshold, or that the ISM is not molecular at all (for instance, it could be atomic or warm).  With this caveat in mind, we use the dust images as a useful guide to the morphological structures in which the cold ISM resides.  We exploit this fact below when the CO morphological class is difficult to determine (e.g. IC~676 and NGC~4435).

\subsection{Morphologies of the molecular gas}
The CO-imaged \atlas\ survey illustrates that the molecular gas in ETGs comes in a large variety of morphologies.  We have identified six different morphologies, including smooth discs (D), spirals (S), rings (R), bars+rings (B+R), as bars are always found with rings, never alone, mildly disrupted objects (M), and strongly disrupted objects (X).  These classifications are strictly for the molecular gas, as all galaxies are optically-classified as ETGs.  The CO morphological classification of each galaxy was based on an analysis of its moment0 and dust maps (Fig.~\ref{fig:dust}), moment1 map (Fig.~\ref{fig:moment1}) and major-axis PVD (Figs.~\ref{app:gals}1-30).  In many cases the moment0 and moment1 map allowed us to unambiguously classify a galaxy.  For ambiguous cases, the PVD was used to search for distinct kinematic signatures. Finally, the dust map was used as a higher spatial resolution proxy for the molecular gas.

The criteria for each CO morphological class are as follows.  Galaxies classified as discs (D) show regular rotation and a smooth elliptical shape.  The PVD of D-class objects also show smooth velocity variations with no discontinuity.  Galaxies classified as spirals (S) show regular rotation, but the moment0 and/or dust maps exhibit discrete spiral arms.  Galaxies classified as rings (R), like discs, exhibit regular rotation, but unlike discs, the PVD of R-class objects include a solid-body component and/or a discontinuity in velocity. Well-resolved rings show a central hole in the moment0 map.  In the most ambiguous cases, the dust map was consulted to search for a dust ring (e.g. NGC~4119 and NGC~4435).  Bar+Ring systems (B+R) were identified either directly from the moment1 map, when the bar and ring are both distinctly visible, or using the PVD and searching for an X-shaped pattern (in the case of edge-on systems; see \S\ref{disc:bars}).  Galaxies classified as mildly disrupted (M) have mainly regular disc morphologies, but include small deviations in their moment0 and moment1 maps.  Finally, a galaxy is classified as strongly disrupted (X) for showing strong irregularities in the CO kinematics.  .

\begin{table}
 \centering
 \begin{minipage}{70mm}
  \caption{CO morphologies of the \atlas\ interferometric sample.}
  \begin{tabular}{cc..}
  \hline
Morphological & \multicolumn{1}{c}{\multirow{2}{*}{Symbol}} & \multicolumn{1}{c}{Number} & \multicolumn{1}{c}{Fraction} \\
configuration & & \multicolumn{1}{c}{(out of 40)} & \multicolumn{1}{c}{(\%)} \\
\hline
Disc & D & 20 & 50\\
Spiral & S & 2 & 5\\
Ring & R & 6 & 15\\
Bar+Ring & B+R & 4 & 10\\
Mild disruption & M & 5 & 12.5\\
Strong disruption & X & 3 & 7.5\\
\hline
\end{tabular}
\label{tab:morph}
\end{minipage}
\end{table}

\begin{figure*}
\raggedright
\subfloat{\includegraphics[height=1.7in,clip,trim=0.2cm 1.4cm 0.4cm 1cm]{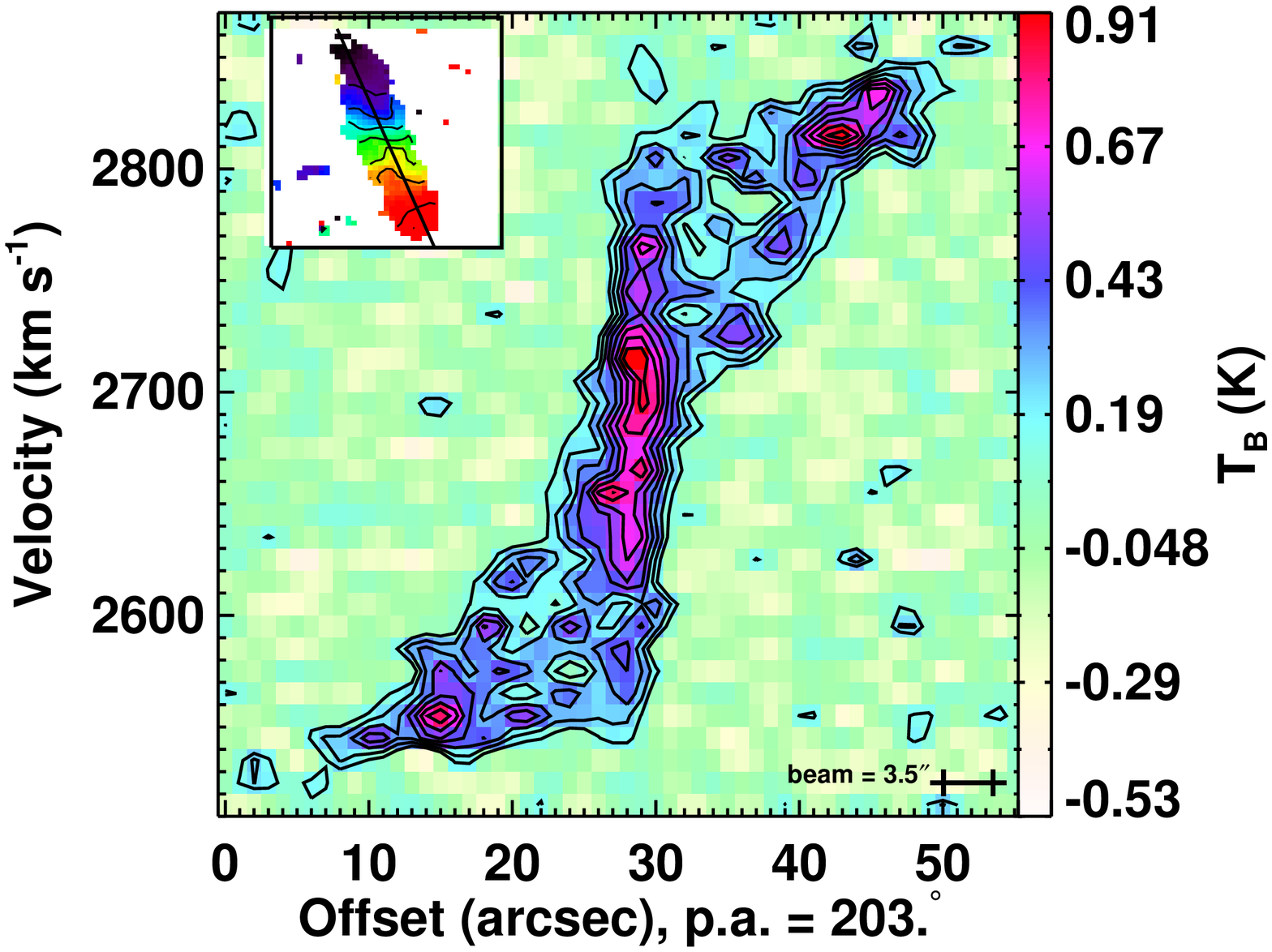}}
\subfloat{\includegraphics[height=1.7in,clip,trim=0.2cm 1.4cm 0.4cm 1cm]{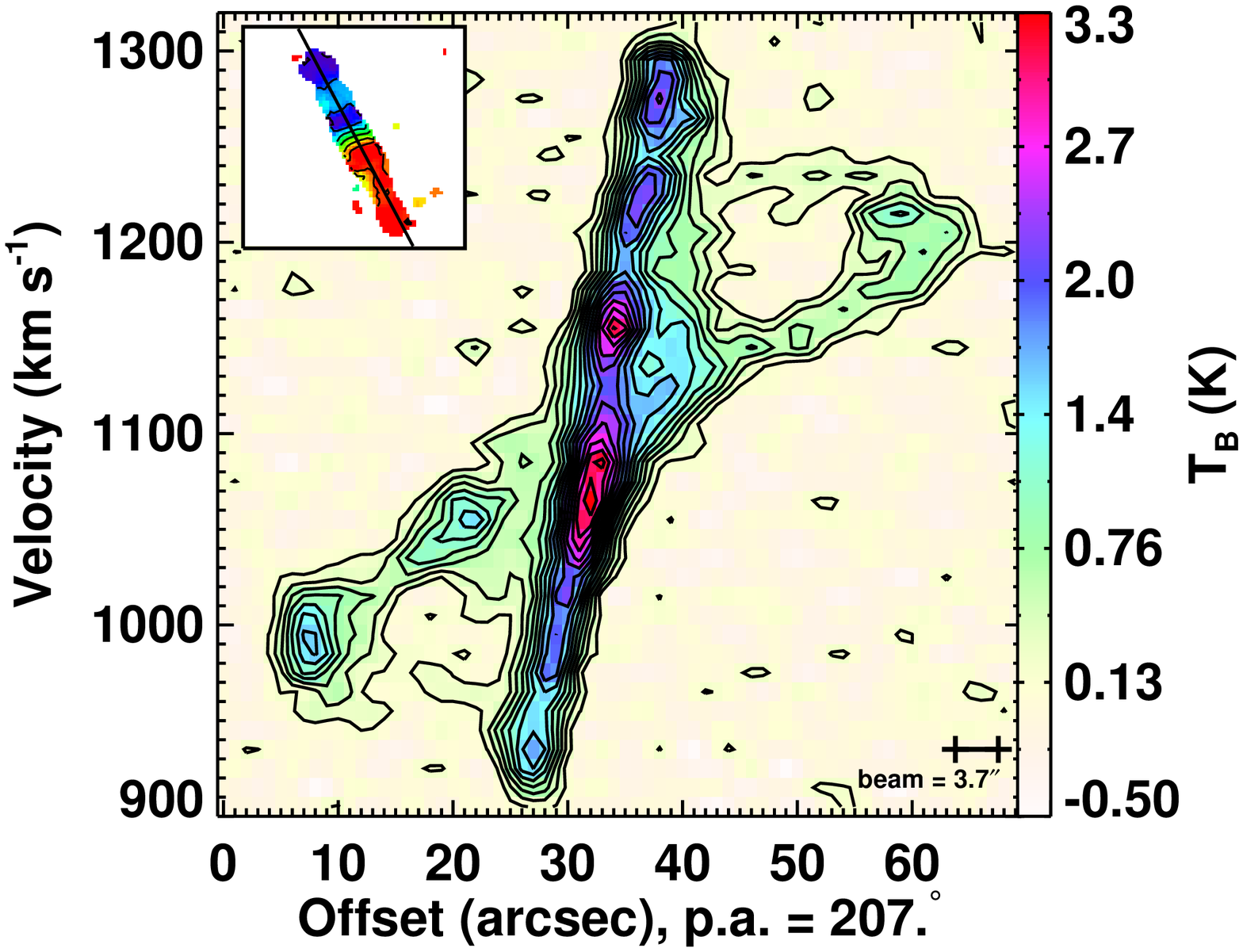}}
\subfloat{\includegraphics[height=1.7in,clip,trim=0.2cm 1.39cm 0.2cm 1cm]{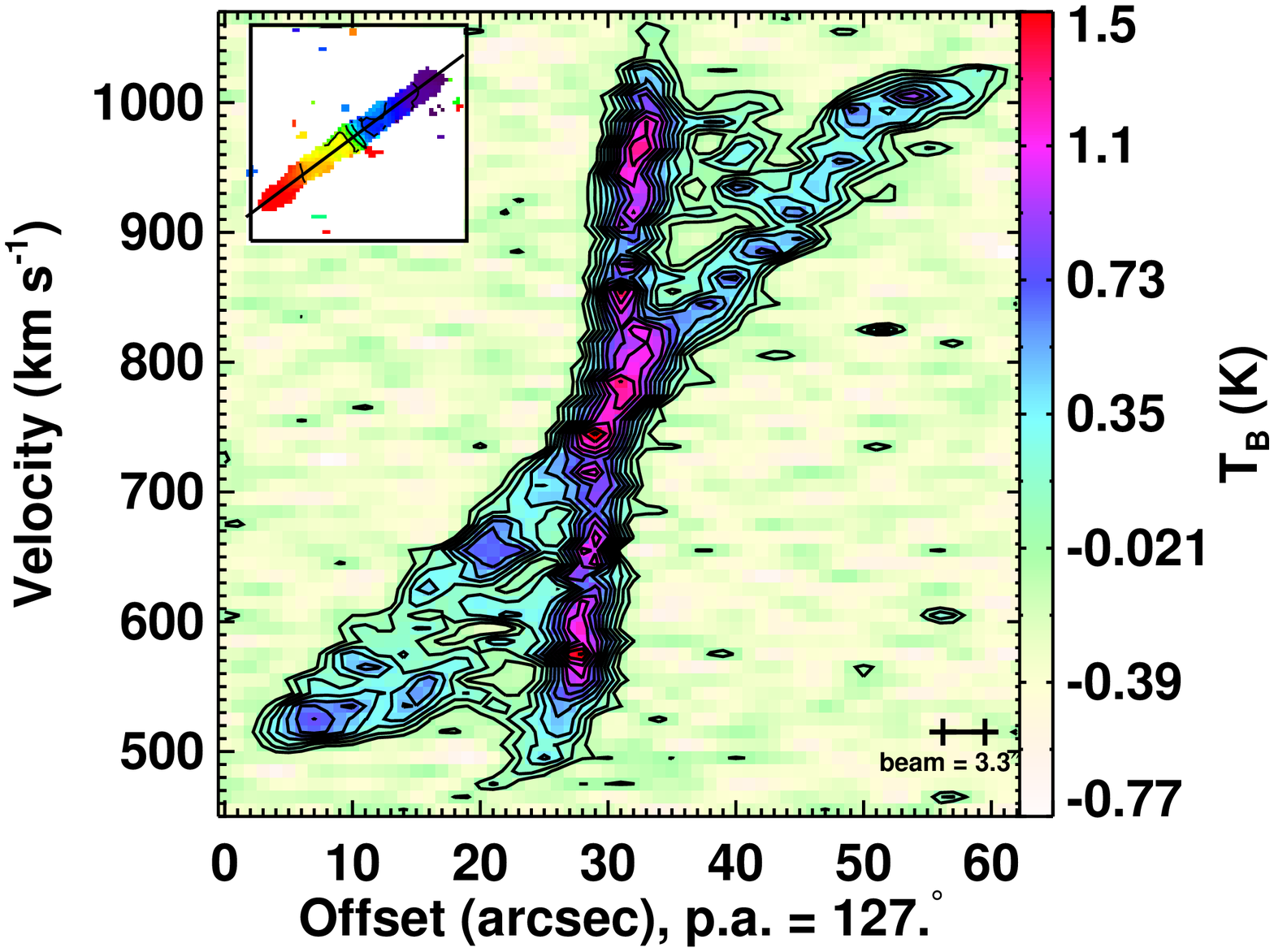}}
\vspace{-1mm}
\caption{Position-velocity diagrams of the three CARMA \atlas\ ETGs with clear two-component velocity profiles, that are not properly represented by the moment1 maps.  The galaxies are NGC~2764 (left), NGC~4710 (middle) and NGC~5866 (right).  All 3 galaxies are edge-on or almost edge-on and show the characteristic X-shaped PVD signature of an edge-on bar (see \S\ref{disc:bars})}
\label{fig:pvd}
\end{figure*}

\subsubsection{Strongly and mildly disrupted objects}
\label{disc:m+x}
The origin of the strongly disrupted (X) objects is the most easily explained.  These galaxies show highly irregular CO distributions and kinematics, that are normally characteristic of interacting systems, a fact often confirmed by observations at other wavelengths.  In the case of NGC~1222 and NGC~4694, it is clear that they acquired their gas via external accretion (see \citealt{beck+07} and \citealt{duc+07} for details on each galaxy).  NGC~5173 is known to host a large misaligned \hi\ disk (Paper \Romannum{13}).  

Mildly disrupted CO distributions (M) also appear to be in less relaxed systems than the rest.  In some cases, there is strong evidence that the gas is being acquired externally, such as in NGC~7465 \citep{liseaquist94}, which is accreting from an external \hi\ reservoir via a close interaction with other galaxies.  NGC~3619 and UGC~09519 are known to host large misaligned \hi\ discs (Paper \Romannum{13}), providing a likely source for the molecular gas.  The irregular dust filaments in NGC~4753 are the main reason that the CO is classified as an M morphology.  The NGC~4753 dust structure is well explained by a precessing twisted disc after the accretion of a gas-rich dwarf companion \citep{steiman+92}.  The original BIMA observations of NGC~4150 \citep{ybc08} show a pronounced low-surface density elongated tail that was not recovered by CARMA, both due to a smaller field of view and higher spatial resolution.  A study of the gas-phase metallicity of NGC~4150 also show it to be sub-Solar (Z $\approx 0.3-0.5$ Z$_\odot$), therefore the molecular gas is likely to originate from an external accretion event \citep{crockett+11}.

It appears that field galaxies are far more likely to exhibit disrupted morphologies than Virgo cluster galaxies, with $24\pm8$\% (7/29) of field galaxies classified as either M or X as opposed to $9\pm9$\% (1/11) for Virgo galaxies. This is not surprising given the results of Paper \Romannum{10}, which suggested that galaxies virialised within the cluster are unable to replenish their stores of molecular gas via external accretion, and thus only longer-lived phenomena survive, such as bars and rings.  Overall, the M- and X-class objects seem to host gas that has an external origin, and they are probably dynamically younger than the other objects, explaining the dearth of such objects within Virgo.

\subsubsection{Molecular rings and bars}
\label{disc:bars}
Rings (R) and bars+rings (B+R) also make up a non-negligible fraction of the \atlas\ molecular gas morphologies, representing about a fifth (8/40) of the CO sample.  It is of note that cold gas in barred galaxies accumulates in rings associated with the major resonances (e.g. \citealt{butacombes96}), so it is possible that bar+ring systems are simply in an early stage of an evolutionary sequence, whereby the bar funnels gas into the centre (e.g. \citealt{bournaud+02}), evacuating the central (but not nuclear) regions of molecular gas, and later leaving the ring structure intact, without a molecular gas-rich bar.  This scenario would explain why bars are only seen with rings, but not the converse.

Figure~\ref{fig:pvd} shows the PVDs of three galaxies (NGC~2764, NGC~4710 and NGC~5866) that possess two velocity components along the line-of-sight. All three galaxies are edge-on, and the characteristic X-shaped PVDs observed are normally associated with the kinematic signature of a bar in an edge-on disc. The material in the inner rapidly-rising component of each PVD is associated with gas on $x_2$ orbits (elongated perpendicular to the bar and located within the inner Lindblad resonance, ILR), while the material in the outer slowly-rising component is associated with gas on nearly circular orbits beyond corotation. The intermediate region has been largely swept free of gas by the bar, due to inflows toward the ILR (see \citealt{sellwood+93} for a review of bar dynamics). The signature is analogous to that of the Milky Way bar in a Galactic longitude-velocity diagram. It was first proposed in external galaxies by \citet{kuijken+95}, later refined by \citet{bureau+99} and \citet{athana+99}, who both identified multiple examples observationally \citep{merrifield+99,bureaufreeman99}.  The bar signatures in NGC~2764 and NGC~5866 are observed here for the first time, but that in NGC~4710 was previously observed in CO by \citet{wrobel+92}. The link between bars and boxy-shaped bulges (such as that in NGC~4710) has long been established, but the presence of a bar in NGC~5866 is rather unexpected given its large classical bulge (see \citealt{athana05} for the different bulge types and the bar-boxy bulge relationship).

The fraction of molecular gas-rich ETGs (6/40) that contain a ring structure is also a factor of $\approx 2$ larger than that of LTGs, based on the BIMA Survey of Nearby Galaxies (SONG) sample (3/39; \citealt{sheth+02,helfer+03}) modified to match the average noise and redshift of the \atlas\ sample (as in Paper \Romannum{14} and \S\ref{sec:d+s}).  Also, unlike for LTGs, the molecular gas morphology does not necessarily follow the stellar morphology, most notably stellar bars.  Galaxies with stellar bars identified in Paper \Romannum{2} are as likely to have a disc-like (D) CO morphology as R and B+R morphologies.

It also appears that Virgo cluster galaxies are slightly more likely to have R and B+R structures than field galaxies, with $45\pm15$\% (5/11) of Virgo galaxies classified as either R or B+R, as opposed to $17\pm7$\% (5/29) for field galaxies.  Paper \Romannum{10} has shown that the molecular gas in ETGs is preferentially kinematically-aligned with the stars in Virgo galaxies, so it is perhaps not surprising to find relatively more bars and rings there, where the gas has been under the influence of secular processes without external gas interference for some time.  This may also point to these morphologies being long-lived, as opposed to the M and X morphologies discussed above. It is also quite possible that the slightly higher detection rate of R and B+R structures in Virgo with respect to the field galaxies is due to a spatial resolution effect, as Virgo galaxies are closer on average ($D \approx 16.5$ Mpc; \citealt{mei+07}) than the field galaxies in the sample ($D \approx 28$ Mpc). If we redshift the Virgo galaxies to the average distance of field galaxies, the detection rate of R and B+R structures drops from 45 to 18\% and is then statistically consistent with the field rate (NGC~4119, NGC~4435 and NGC~4477 become unresolved).  Higher resolution and deeper imaging is required to determine whether the higher rate in Virgo is truly a resolution effect or an intrinsic property.

\subsubsection{Molecular discs and spirals}
\label{sec:d+s}
Smooth molecular discs make up the largest fraction of the CO morphologies that are seen in the CO-imaged \atlas\ galaxies, while spirals make up the smallest fraction.  This relative dearth of discrete spirals as compared to smooth discs seems to indicate that there is something fundamentally different about the way in which molecular gas behaves within an ETG as opposed to a LTG.  A caveat to this scenario is that galaxies with dominant molecular gas spirals (and associated star formation) would likely be classified as LTGs, and thus be excluded from the sample.

It is possible that molecular gas in ETGs remains in a smooth distribution and lacks distinct spiral arms for dynamical reasons.  For example, the steep potential wells in the central parts of ETGs (where the molecular gas resides) yield high epicyclic frequencies, and may thus prevent gravitational instabilities linked to spiral arms from developing (Toomre $Q$ parameter; e.g. \citealt{toomre81}).

It is also possible that some of the D morphologies would have been classified otherwise with observations of better sensitivity and/or spatial resolution.  For instance, much higher spatial resolution CO observations of NGC~1266 presented in \citet{alatalo+11} reveal a disrupted nuclear disc and a molecular outflow, both of which are unresolved here.  In addition, if only the peak of a spiral structure were detected, and convolved with a large beam, it could well be classified as a disc.  In fact, as much as half of BIMA SONG spirals \citep{helfer+03} would be classified as D if they were located at the average distance of the \atlas\ sample, with similar noise properties and beam sizes (see Paper \Romannum{14} for a full discussion of properly redshifting the BIMA SONG data to match \atlas).  However, even assuming that half of the current CO-rich ETGs classified as D actually harbour spatially-unresolved spirals, the spiral fraction of \atlas\ galaxies would be $30\pm7$\%, still much lower than the spiral fraction of BIMA SONG galaxies ($62\pm10$\%).

While obtaining higher resolution molecular gas maps is the only rigorous way to directly break the S/D degeneracy, we use here the unsharp-masked dust images as proxies.  As mentioned in \S\ref{dustcomp}, the molecular gas morphology follows the dust morphology faithfully in the majority of cases.  The unsharp-masked images therefore may be utilized to provide the higher spatial resolution and sensitivity required to separate the S and D classifications.  The dust comparison indicates that in the majority of cases (with the possible exceptions of NGC~3619, NGC~3626 and NGC~3665), and contrary to LTGs, a disc morphology without distinct spiral arms is indeed the most common morphology for the molecular gas in ETGs.

\subsubsection{Summary of the gas morphologies}
Table \ref{tab:morph} lists the total number of \atlas\ ETGs within each CO morphological class.  Overall, it thus appears that the CO in ETGs is most likely to be in a settled morphology, either in regular discs (comprising 50\% of the galaxies) or ring, bar+ring and spiral morphologies (comprising another 30\% of the sample).  This means that the majority (80\%) of ETGs have a dynamically settled molecular gas morphology.  The dynamically unsettled (M and X) systems comprise the remaining 20\%, and just over half of those are only mildly disrupted.  Of course, the CO structures observed here are rather compact spatially, with an average extent of only 1 kpc (Paper \Romannum{14}), and have correspondingly small dynamical timescales ($10^7-10^8$ yrs).  The CO distributions are thus expected to relax and reach equilibrium rapidly.

\begin{figure}
\centering
\includegraphics[width=3.3in,clip,trim=1.2cm 0.5cm 0.8cm 0.8cm]{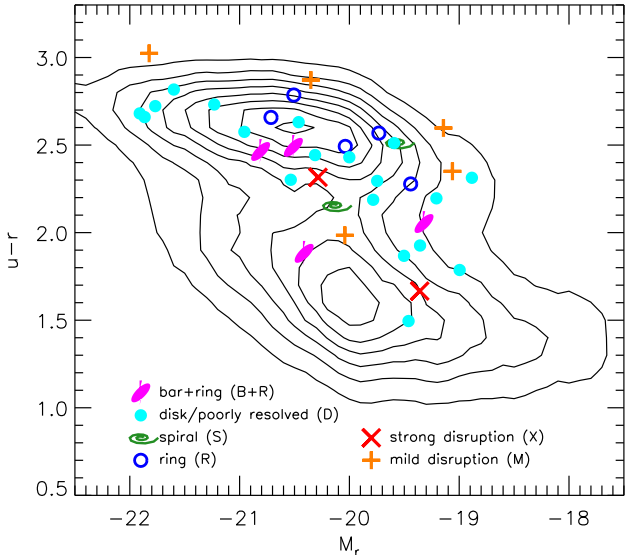}
\caption{Color-magnitude diagram of the CO-imaged \atlas\ galaxies, sorted based on the CO morphological classifications.  The contours (black) are from the SDSS Data Release 8 (DR8; \citealt{sdss8}), and represent all galaxies with redshift $z > 0.08$.  Data for the \atlas\ ETGs are either from the SDSS DR8 $u$, $r$ and $M_r$ catalog products or the INT (NGC~524, NGC~3489 and NGC~7465; Paper \Romannum{21}).  Colour information is unavailable for NGC~1222 and NGC~1266 due to poor $u$-band photometry, for PGC~058114 because of the presence of a bright star less than a degree from the galaxy, and for NGC~4710 due to erroneous SDSS photometry.}
\label{fig:cmd}
\end{figure}

Figure~\ref{fig:cmd} shows the $u-r$ vs. $M_r$ color-magnitude diagram (CMD) of the CO-imaged \atlas\ sample, from SDSS and INT data, overlaid upon the SDSS Data Release 8 distribution of all galaxies with redshift $z > 0.08$ \citep{sdss8}.  There is no clear separation of the morphological groups, although some trends do appear.  The brightest objects ($M_r < -21$) tend to be classified as D galaxies, with the exception of NGC~4753, which was discussed in \S\ref{disc:m+x}.  The CO discs in these massive galaxies are also generally well-resolved, limiting the chance that these objects are poorly resolved rings or spirals.  It also appears that although B+R galaxies can be found throughout the CMD, galaxies classified as R only seem to cluster on the red sequence, perhaps indicating a later evolutionary stage.

\subsection{CO spectral asymmetries}
It is worth mentioning that a fraction of the CO spectra exhibit asymmetries, and have representatives in many CO morphological classes (as opposed to solely M and X-class objects). Although asymmetries detected from a single-dish spectrum could be due to a slightly offset pointing, interferometric identifications of spectral asymmetries are more robust. Asymmetric spectra have long been known within \hi\ discs \citep{baldwin+80}. Within the Westerbork \hi\ sample of Spiral and Irregular Galaxies (WHISP) at least 80\% of all galaxies contain some amount of asymmetry \citep{vaneymeren+11}. In fact, asymmetries are found at small radii in at least 60\% of \hi\ discs from the WHISP survey.  Despite the different dynamical timescales at the radii where the CO has been found, finding a few asymmetric objects within the CARMA \atlas\ sample is unsurprising.

Asymmetric \hi\ spectra generally trace lopsided spatial distributions, and lopsided molecular gas discs are likely a result of the molecular gas responding to a change in the gravitational potential. In particular, long-lived lopsided \hi\ discs are well reproduced in models where the gas is off-centre with respect to the galactic halo (e.g. \citealt{levine+98}). Other possible causes for lopsidedness include responding to an interaction (e.g. \citealt{jog+combes09}), or possibly an internal instability that perturbs both the $m = 1$ (lopsided) and $m = 2$ (barred) modes \citep{masset+97,bournaud+05,jog+combes09}. In the strongest case of asymmetry (IC~1024), it is likely that the molecular gas disc is responding to a tidal interaction, which is also seen in the optical data (Paper \Romannum{2}). IC~1024, despite being lopsided, is dynamically relaxed (Paper \Romannum{14}) and shows regular kinematics and a CO disc morphology in the moment0 map. This may suggest that the tidal interaction happened sufficiently long ago for any detectable morphological and kinematic disturbance of the molecular gas to subside, while preserving the spectral asymmetry.

In the context of \hi, the structures are well resolved, extended, and clearly ``lopsided'', and the \hi\ cloud size is small compared to the global structure, and thus inferences can be made in most cases that the driver of the asymmetries are global, on a long timescale.  Structures traced by the CO distributions are much less extended compared to the rest of the galaxy (Paper \Romannum{14}), and thus the interpretation of CO asymmetries is less straightforward.  A more comprehensive study of asymmetries and lopsidedness in CO distributions within galaxies would likely address these issues, as well as uncover many other yet unstudied properties of molecular gas distributions in galaxies, but is beyond the scope of this paper.

\subsection{Mass of externally-acquired molecular gas in ETGs}
\label{extgas}

Paper \Romannum{10} discusses the origins of the molecular gas in ETGs, and identifies galaxies that likely have externally-acquired gas.  Using the criteria set forth in Paper \Romannum{10}, that the kinematic major axis of the ionized gas must be misaligned with the kinematic major axis of the stars by at least 30$^\circ$, at least thirteen CO-imaged \atlas\ galaxies are identified as having externally-acquired gas: IC~719, NGC~1266, NGC~2685, NGC~2768, NGC~3032, NGC~3626, NGC~4694, NGC~5173, NGC~7465, PGC~029321, PGC~058114 and UGC~09519.  NGC~3619 is classified as M in this work, exhibiting an unsettled CO morphology, but the origin of the CO is difficult to determine, with a kinematic misalignment angle of only $24\pm3^\circ$, right at the boundary of being classified as external (Paper \Romannum{10}), and the galaxy contains a kinematically-misaligned \hi\ disc (Paper \Romannum{13}).  Although it is possible that the CO indeed has an external origin, there is no conclusive measurement to classify it as such, and NGC~3619 will not be considered as having external CO in the following discussion.  The CO of NGC~4150 is included here as being external in origin, as studies show that its gas-phase metallicity is sub-Solar, and thus is very likely from the accretion of a gas-rich companion in a minor merger \citep{crockett+11}.  NGC~4753 is classified as M in this work, and \citet{steiman+92} make a compelling argument that the molecular disc is unsettled, and must come from an external source.  
For this reason, we include here NGC~4753 in the category of gas acquired externally.  The disrupted stellar and gas kinematics of NGC~1222 (CO class X), which is currently undergoing a strong three-way interaction \citep{beck+07}, also suggest that the molecular gas is likely to be of external origin.
Including these three additions, 15 (38\%) of the 40 gas-rich CARMA ETGs contain external gas, in good agreement with the fraction of externally-acquired gas estimated in Paper \Romannum{10} (36\%).

The fact that the majority of galaxies with unsettled gas morphologies are identified as having an external gas origin is unsurprising, as a parcel of gas falling into a galaxy takes time to virialise, which is not the case for internal gas (stellar mass loss).  The rest of the galaxies with external gas have disc morphologies (IC~719, NGC~3032, NGC~3626, PGC~029321 and  PGC~058114), represented at about the same rate as in the sample as a whole.  NGC~2685 and NGC~2768 are both known polar rings/discs, and are discussed in detail in \citet{schinnerer+02} and \citet{crocker+08}, respectively.

We use the H$_2$ masses from CARMA, where available, and find that the sum of the molecular gas masses in the 15 sources with external CO is $8.26\times10^9~M_\odot$, assuming $X_{\rm CO}$ = $3\times10^{20}$ (K km s$^{-1})^{-1}$ cm$^{-2}$ \citep{dickman+86}, and accounts for 44\% of the total molecular gas mass in our sample of 40 CO-rich galaxies with interferometric maps.  There is reasonable agreement between the external number (38\%) and mass (44\%) fractions.  We note that these fractions are lower limits, due to the fact that we can only unambiguously assign galaxies with misaligned stellar and gas kinematics as external accretors, while some fraction of aligned galaxies are also likely to be so.

Galaxies that were originally detected in molecular gas (Paper \Romannum{4}) but were not mapped interferometrically have a higher rate of ionised gas kinematically-aligned with the stars, compared to those actually mapped here (i.e. brighter CO; see Paper \Romannum{10}).  This likely means that the un-imaged CO-detected ETGs contribute negligibly ($<6$\%) to the total external molecular gas mass.  It is also likely that undetected galaxies do not contribute significantly  to the total gas mass, with an upper limit to the contribution of $\approx3\times10^9~M\odot$, assuming that 42.5\% upper limits in undetected field galaxies are due to an external phenomenon (Paper \Romannum{10}).

Assuming the molecular gas is depleted through star formation and reasonable gas depletion timescales ($\approx 1$ Gyr; \citealt{saintonge+11}) and star formation histories (allowing us to adopt the mean current-to-initial molecular gas mass correction factor calculated by \citealt{kaviraj+11} for ETGs with prominent dust lanes), we can infer the amount of molecular gas each galaxy acquired before star formation activity consumed some of the molecular gas.  The total amount of molecular gas originally acquired externally within the \atlas\ sample is then approximately $2.48\times10^{10}~M_\odot$.  At the rate of 30 minor mergers in the sample detailed in \S\ref{intro} and from the rate of \citet{lotz+08}, this would mean approximately $8.3\times10^8~M_\odot$ of H$_2$ per minor merger per molecular gas depletion time (\hbox{1 Gyr}; \citealt{saintonge+11}).  We derive an average stellar mass ratio for these minor mergers by first converting the molecular gas mass to an atomic mass using the 1:3 H$_2$-to-\hi\ ratio seen in the Milky Way \citep{wolfire+03}.  We then use the average \hi{}-to-stellar mass fraction of minor merger companions from \citet{kannappan04} of 20\% to calculate the average stellar mass of the accreted companions, $1.25\times10^{10}~M_\odot$.  For the mass of the accreting galaxy, we take the average stellar mass of the non-Virgo ETGs in the CO-imaged \atlas\ sample, $\approx 6.5\times10^{10}~M_\odot$.  This results in a merger stellar mass ratio of 1:5.2, consistent with the minor merger stellar mass ratios we expect.  We therefore conclude that our results are in good agreement with the origin of the external gas being due to minor mergers.

\section{Conclusions}
\label{conc}
We have presented millimeter-wave data products as part of the CARMA \atlas\ survey of galaxies, an imaging survey of the CO J = 1--0 emission of 30 nearby CO-rich ETGs, part of the \atlas\ survey.  The main conclusions drawn from these data are:\\

\indent 1. The CO-imaged \atlas\ survey is the largest CO imaging survey of ETGs to date, and provides the most detailed view of the nature of molecular gas in ETGs.\\
\indent 2. The molecular gas found in ETGs appears to be co-spatial with dust features (as traced by $g-r$ colour images), though the dust features often appear to be more extended, likely because we are unable to trace small molecular gas surface densities.\\
\indent 3. The molecular gas present in ETGs comes in a variety of morphologies: 50\% of the objects have their CO in inclined gas discs or is poorly resolved structures (D), 5\% show spiral arms or spiral structure (S), 15\% have resolved rings (R), 10\% have bar+ring systems (B+R), 12.5\% have mildly disrupted distributions (M) and 7.5\% have strongly disrupted distributions (X).\\
\indent 4. The CO morphology does not show a strong correlation with the $u-r$ color or total magnitude of the galaxy, although there appear to be weak trends.  For example, the most massive galaxies in this sample tend to have D CO morphologies, and R morphologies tend to lie on the red sequence whereas B+R morphologies tend to also be present in the blue cloud.\\
\indent 5. We currently observe a total of $8.26\times10^{9}~M_\odot$ of molecular gas that is likely of external origin in 15/40 of our galaxies.  Using the correction factor derived in \citet{kaviraj+11}, we estimate that the total externally acquired molecular gas prior to the consumption of some of it by star formation was  $\approx2.48\times10^{10}~M_\odot$. In conjunction with the minor merger rate of \citet{lotz+08} and assumptions about gas mass fractions, this amount of externally accreted molecular gas is consistent with a minor merger origin.\\
\indent 6. The CARMA \atlas\ (as well as 9 additional literature galaxy) data products including data cubes and moment0 and moment1 maps in FITS format, will be publicly released and hosted through the CARMA data archive\footnote{http://carma-server.ncsa.uiuc.edu:8181/} and will be reachable from our project website\footnote{http://purl.org/atlas3d}.

\section*{Acknowledgments}

We thank the anonymous referee for insightful comments and suggestions.  K.A. thanks C. Heiles, K. Nyland, G. Graves, R. Plambeck, E. Rosolowsky and J. Carpenter for useful conversations.  K.A. also wants to thank Lisa Wei in particular for collaboration on NGC~5173.  The research of K. Alatalo is supported by the NSF grants AST-0838258 and AST-0908572.  The research of K. Alatalo is also supported by NASA grant HST-GO-12526.  MC acknowledges support from a STFC Advanced Fellowship (PP/D005574/1) and a Royal Society University Research Fellowship.  PS is a NWO/Veni fellow.  RMcD is supported by the Gemini Observatory, which is operated by the Association of Universities for Research in Astronomy, Inc., on behalf of the international Gemini partnership of Argentina, Australia, Brazil, Canada, Chile, the United Kingdom, and the United States of America.  RLD and MB are supported by the rolling grants `Astrophysics at Oxfordâ' PP/E001114/1 and ST/H002456/1 from the UK Research Councils. RLD acknowledges travel and computer grants from Christ Church, Oxford and support from the Royal Society in the form of a Wolfson Merit Award 502011.K502/jd.  TN acknowledges support from the DFG Cluster of Excellence: "Origin and Structure of the Universe."  Support for CARMA construction was derived from the states of California, Illinois, and Maryland, the James S. McDonnell Foundation, the Gordon and Betty Moore Foundation, the Kenneth T. and Eileen L. Norris Foundation, the University of Chicago, the Associates of the California Institute of Technology, and the National Science Foundation. Ongoing CARMA development and operations are supported by the National Science Foundation under a cooperative agreement, and by the CARMA partner universities.  This paper is partly based on observations carried out with the IRAM 30m telescope. IRAM is supported by INSU/CNRS (France), MPG (Germany) and ING (Spain).  We acknowledge use of the HYPERLEDA database (http://leda.univ-lyon1.fr) and the NASA/IPAC Extragalactic Database (NED) which is operated by the Jet Propulsion Laboratory, California Institute of Technology, under contract with the National Aeronautics and Space Administration.

\appendix
\onecolumn

\section{CO CARMA data of \atlas\ galaxies}
\label{app:gals}
For each galaxy of the CARMA \atlas\ sample, we provide six figures:\\

\noindent {\bf Top-left:} $r$-band image (grayscale) either from SDSS or from the INT, overlaid with black contours.  The CO integrated intensity (moment0) map is also overlaid (cyan contours corresponding to 20, 40, 60, 80 and 100\% of the peak).  The cyan box corresponds to the size of the individual moment maps in the middle-left and middle-centre panels.  \\

\noindent{\bf Top-right:} Comparison of the single-dish IRAM 30m CO(1--0) integrated flux (black) and the CARMA integrated flux (red).  The total integrated CARMA flux is obtained by summing all flux above a given threshold (see \S\ref{data_analysis}).  The grey dashed vertical lines indicate the total spectral range of CARMA.  Both heliocentric velocities and velocities with respect to systemic (taken from Paper \Romannum{1}) are indicated.  The rms noise is shown as a single error bar in black for the IRAM 30m and red for CARMA.\\

\noindent{\bf Middle-left:}  CO(1--0) integrated intensity (moment0) map (colorscale), overlaid with black and grey contours (at equal increments of 20\% of the peak).  The colour table on the right provides the integrated flux scale.  The dot-dashed gray circle corresponds to the 21\farcs6 primary beam of the IRAM 30m telescope.  The CARMA synthesized beam is shown in the bottom-right corner (black ellipse).\\

\noindent{\bf Middle-center:}  CO(1--0) mean velocity (moment1) map (colorscale and black contours at 20 \kms\ intervals unless otherwise stated).  The colour table on the right provides the velocity scale with respect to the systemic velocity (taken from Paper \Romannum{1} and indicated in the top-left corner of the panel).  The dot-dashed gray circle corresponds to the 21\farcs6 primary beam of the IRAM 30m telescope. The CARMA synthesized beam is shown in the bottom-right corner (black ellipse).\\

\noindent{\bf Middle-right:} CO(1--0) position-velocity diagram (PVD) taken over a one synthesized beam width (5 pixels) slice at the position angle indicated in the top-right corner inset and at the bottom of the panel (see \S\ref{data_analysis}).  Velocities are heliocentric.  The synthesized beam along the PV slice direction is indicated in the bottom-right corner.  Contours are spaced by $1\sigma$ starting at $3\sigma$ (unless otherwise stated), where $\sigma$ is the rms noise in an individual channel multiplied by $\sqrt{5}$, to account for the summation over the synthesized beam width.\\

\noindent{\bf Bottom:}  CO(1--0) channel maps.  The panels for velocity channels with flux are color-coded based on their velocity with respect to systemic (taken from Paper \Romannum{1}).  The mean velocity offset is listed in the top-left corner of each panel.  The cross indicates a fixed reference position in all panels.  Contours are spaced by 1$\sigma$ beginning at $\pm2\sigma$ (unless otherwise stated), where $1\sigma$ is the rms noise in an individual channel.  Grey contours are negative.  The CARMA synthesized beam is shown in the lower-right corner of the first panel (black ellipse).

\clearpage
\begin{figure*}
\centering
\subfloat{\includegraphics[height=2in,clip,trim=2.2cm 5.2cm 0cm 4.8cm]{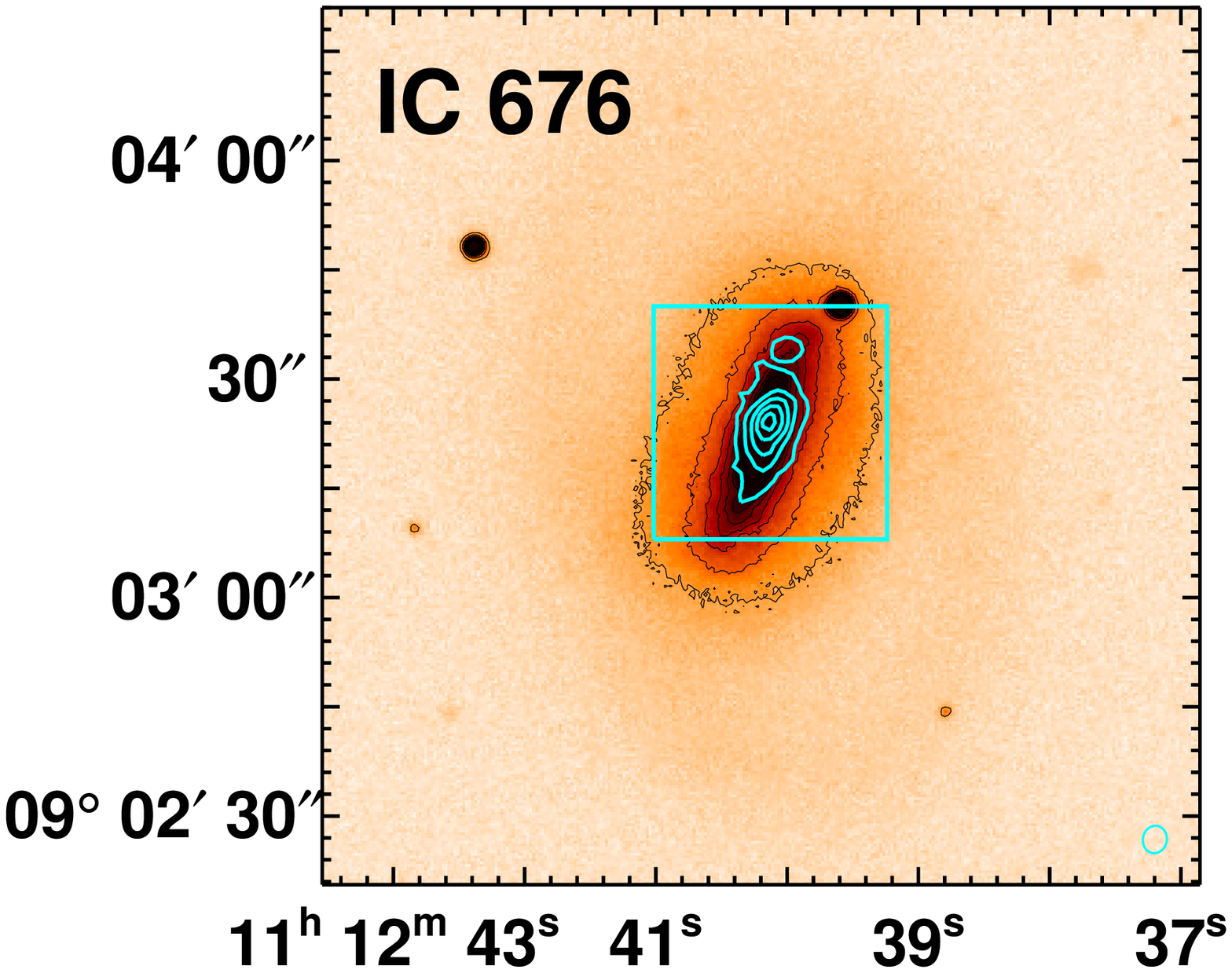}}
\subfloat{\includegraphics[height=2in,clip,trim=0.9cm 0.5cm 0cm 0.2cm]{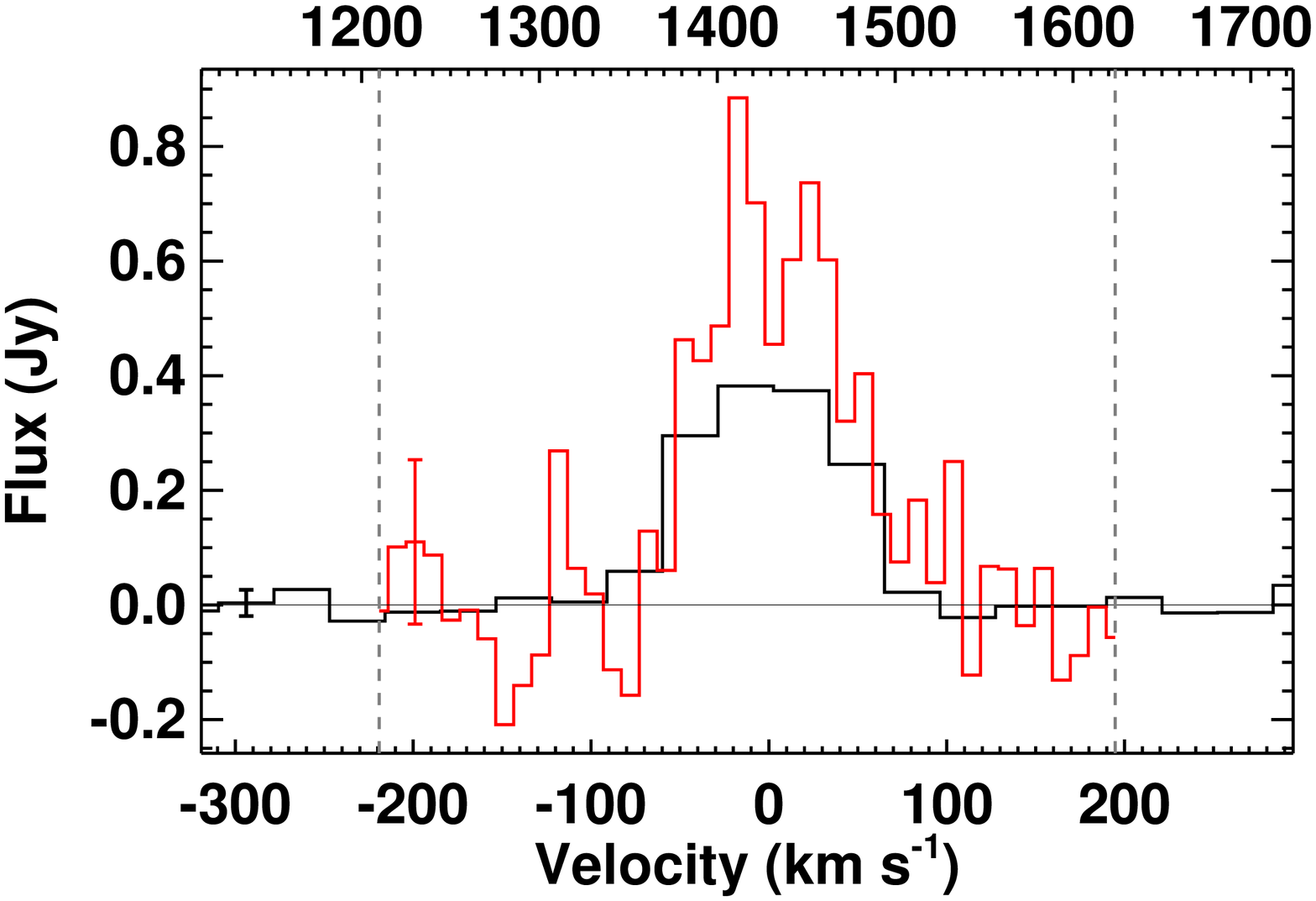}}
\end{figure*}
\begin{figure*}
\subfloat{\includegraphics[height=1.6in,clip,trim=0cm 1.5cm 0cm 2.5cm]{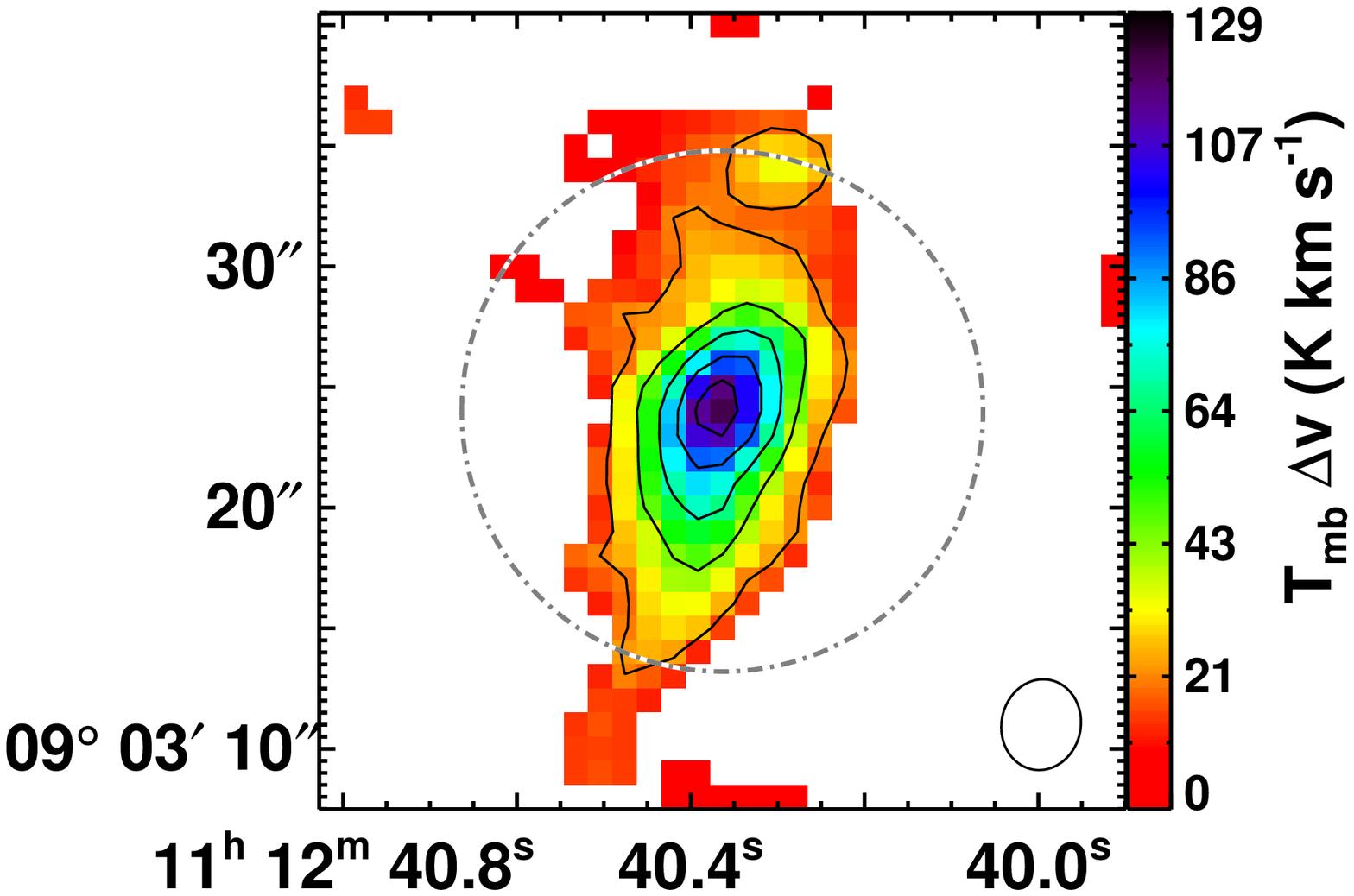}}
\subfloat{\includegraphics[height=1.6in,clip,trim=0.2cm 1.5cm 0cm 2.5cm]{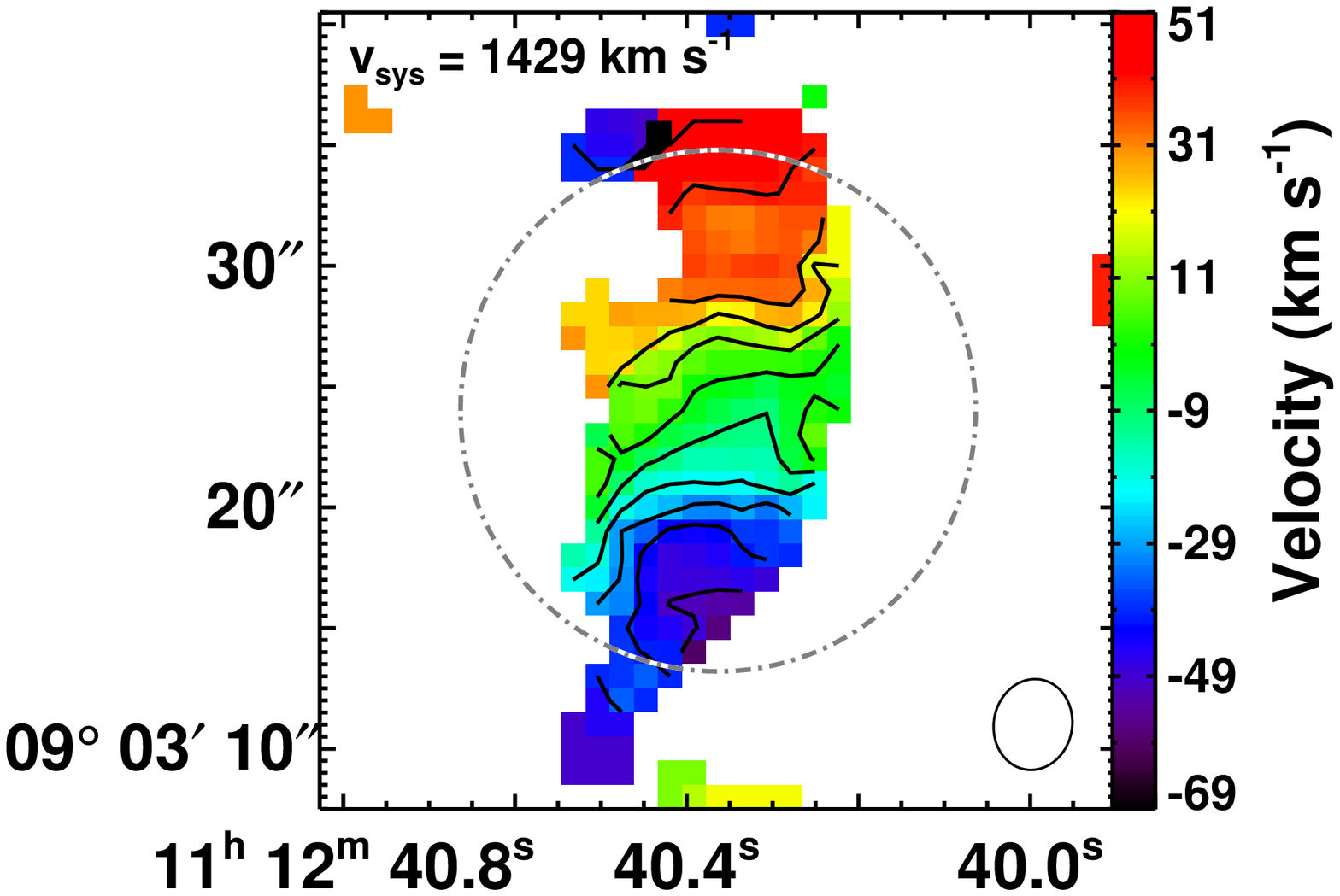}}
\subfloat{\includegraphics[height=1.6in,clip,trim=0cm 1.5cm 0cm 0.9cm]{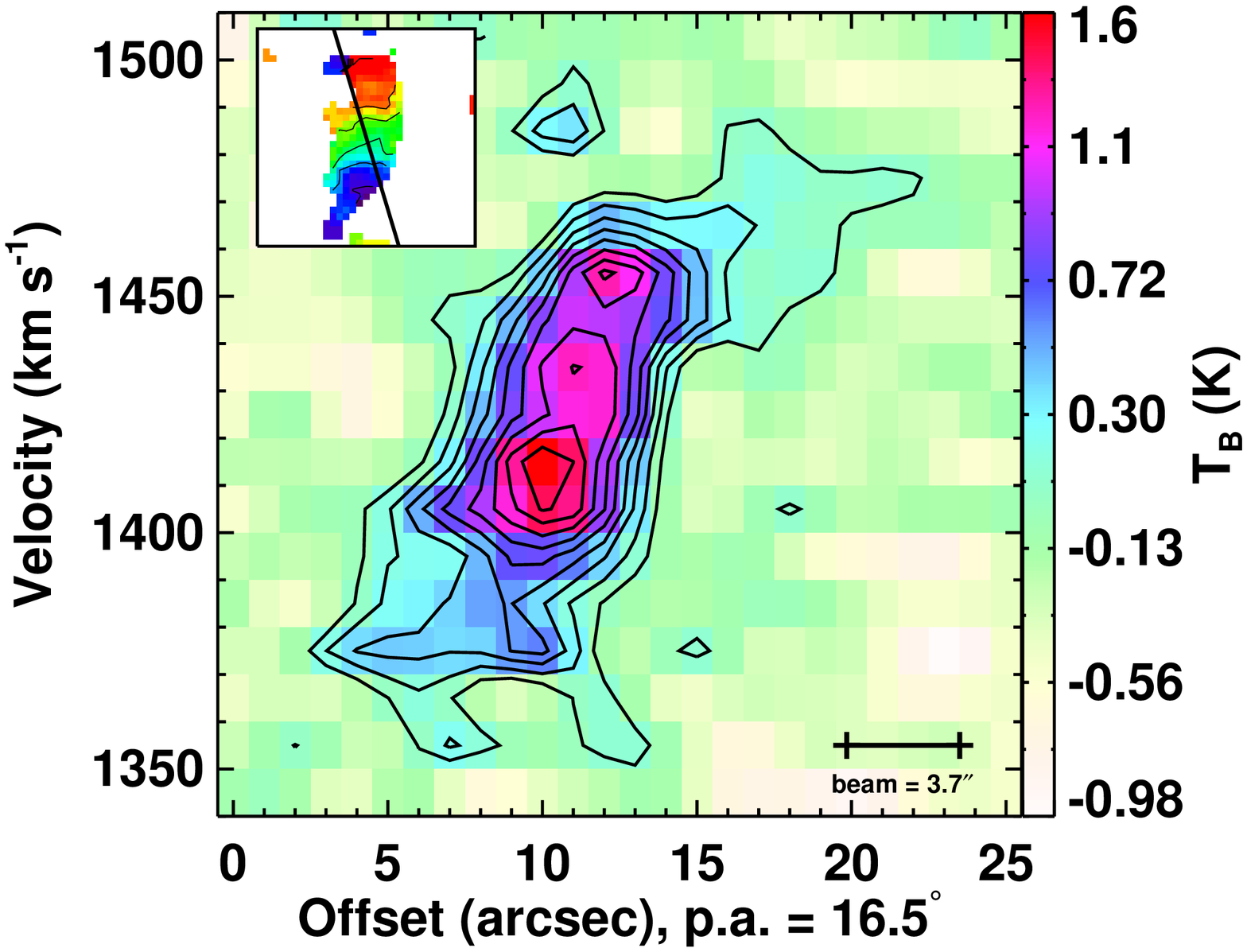}}
\end{figure*}
\begin{figure*}
\subfloat{\includegraphics[width=7in,clip,trim=6cm 4cm 8.7cm 5.3cm]{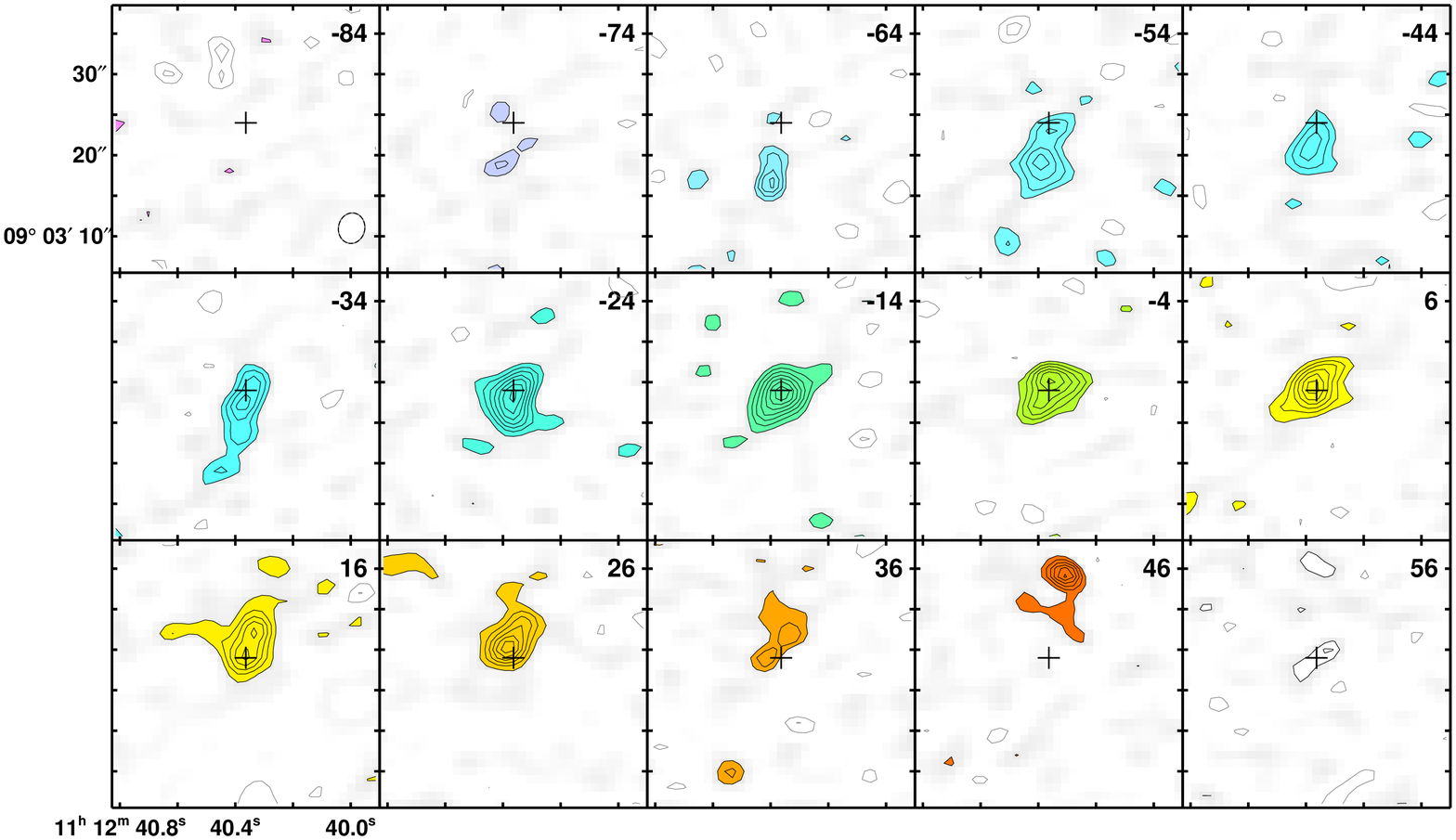}}
\caption{{\bf IC~676} is a field regular rotator ($M_K$ = -22.27) with a bar and ring stellar morphology.  It has a dust filament.  The moment0 peak is 17 Jy beam$^{-1}$ \kms.  The moment1 contours are placed at 10\kms\ intervals and the PVD contours are placed at $1.5\sigma$ intervals.}  
\label{fig:firstgal}
\end{figure*}

\clearpage
\begin{figure*}
\centering
\subfloat{\includegraphics[height=2in,clip,trim=2.2cm 5.2cm 0cm 4.8cm]{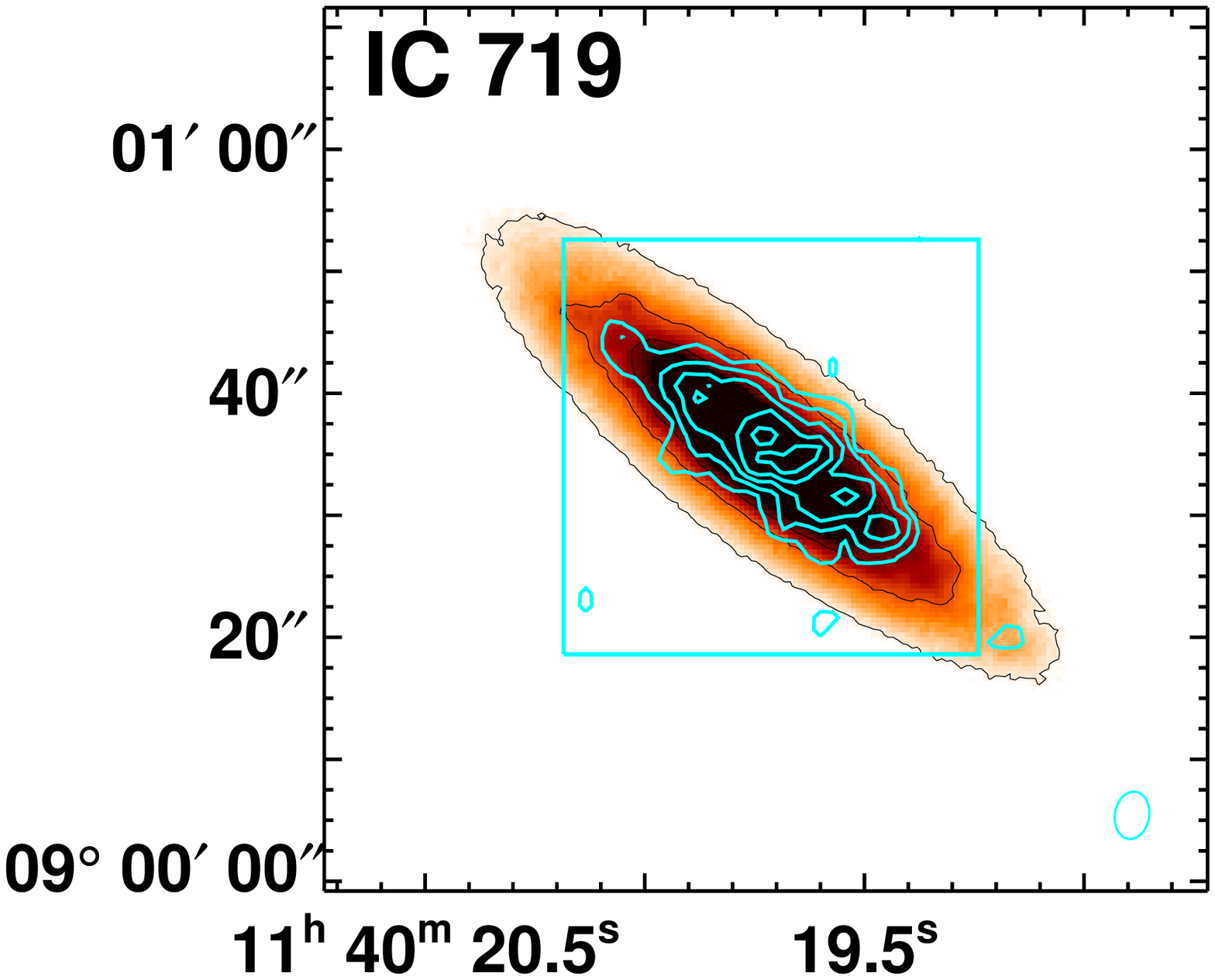}}
\subfloat{\includegraphics[height=2in,clip,trim=0.5cm 0.5cm 0cm 0.2cm]{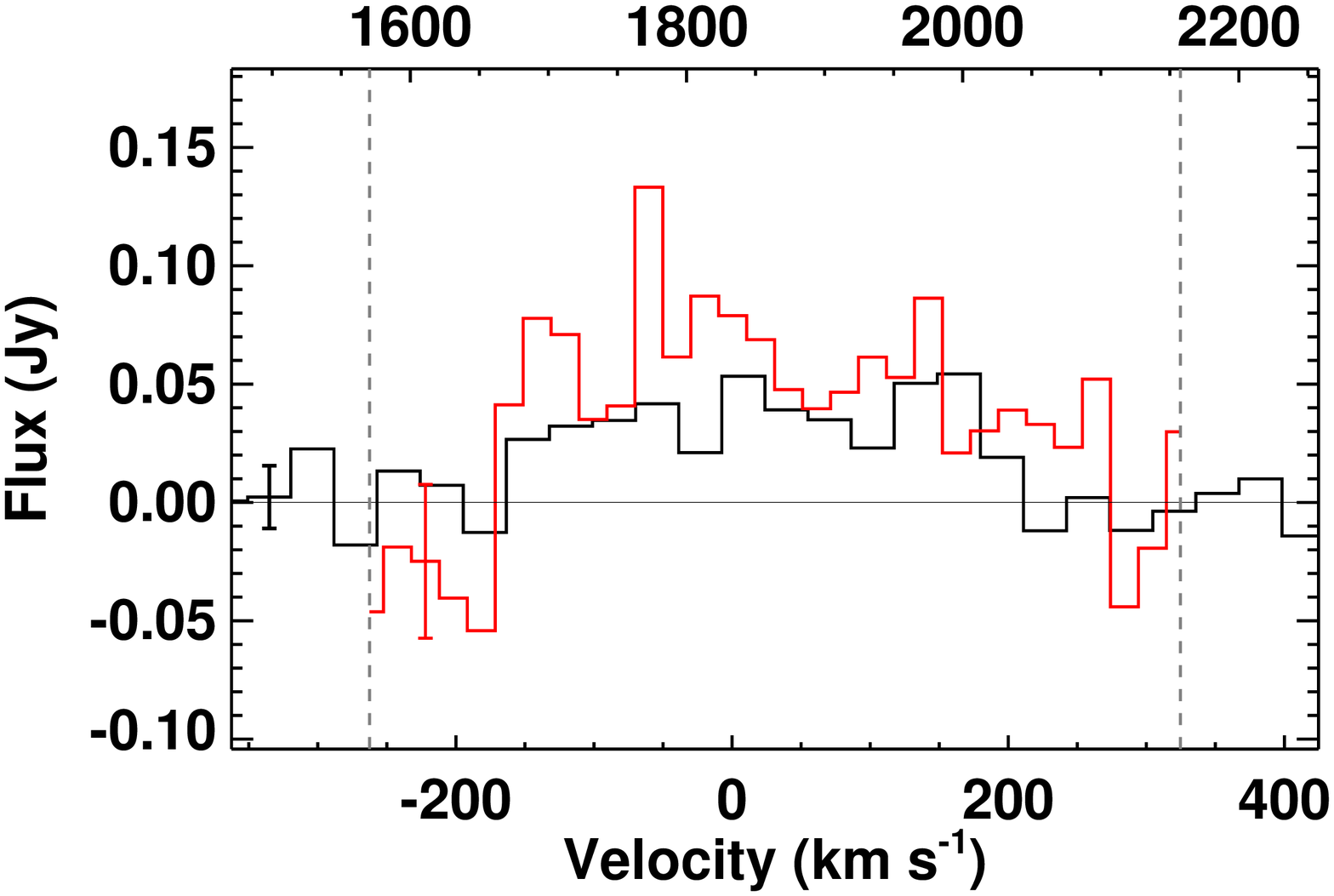}}
\end{figure*}
\begin{figure*}
\subfloat{\includegraphics[height=1.6in,clip,trim=0cm 1.5cm 0cm 2.5cm]{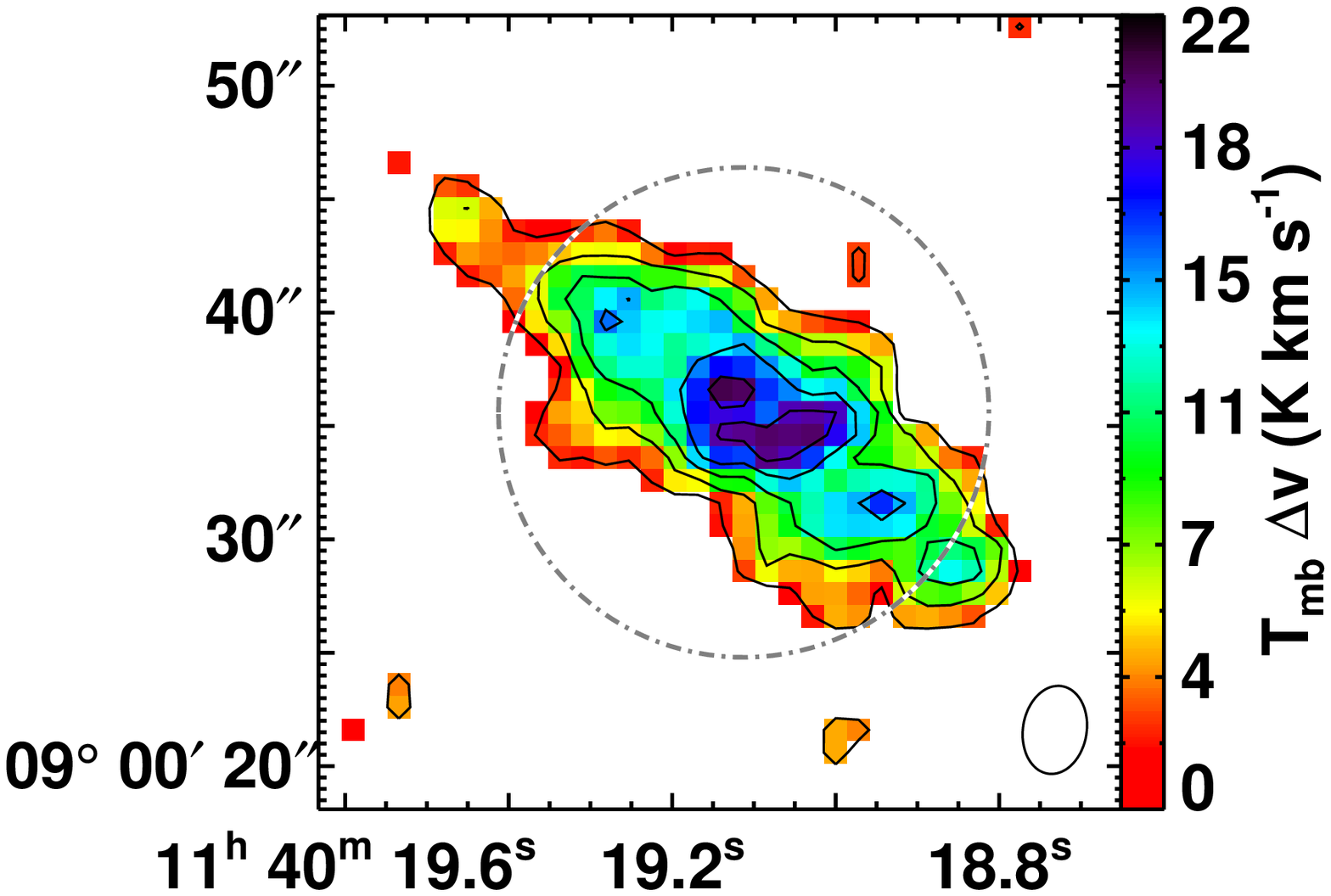}}
\subfloat{\includegraphics[height=1.6in,clip,trim=0.2cm 1.5cm 0cm 2.5cm]{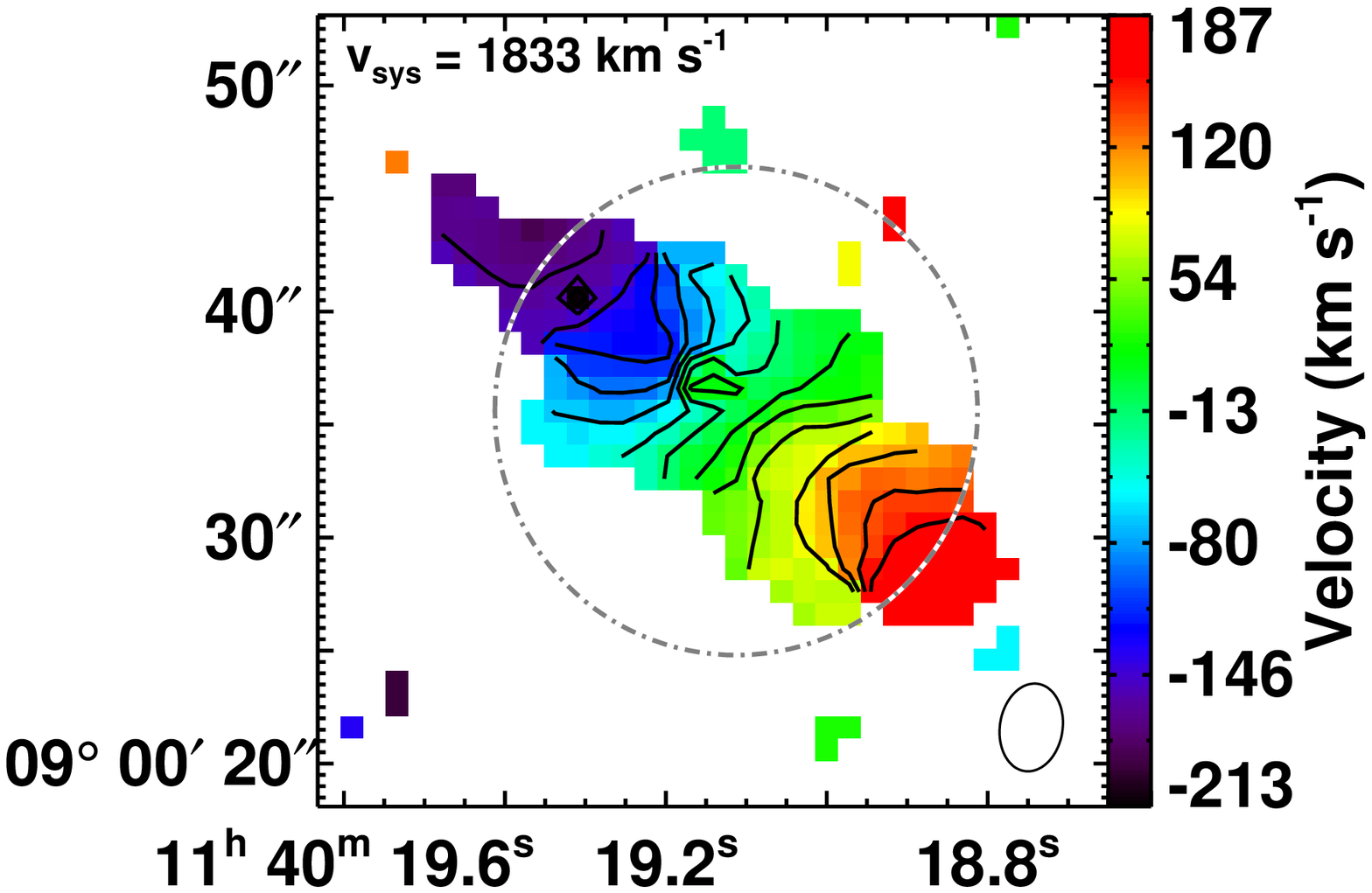}}
\subfloat{\includegraphics[height=1.6in,clip,trim=0cm 1.4cm 0cm 0.9cm]{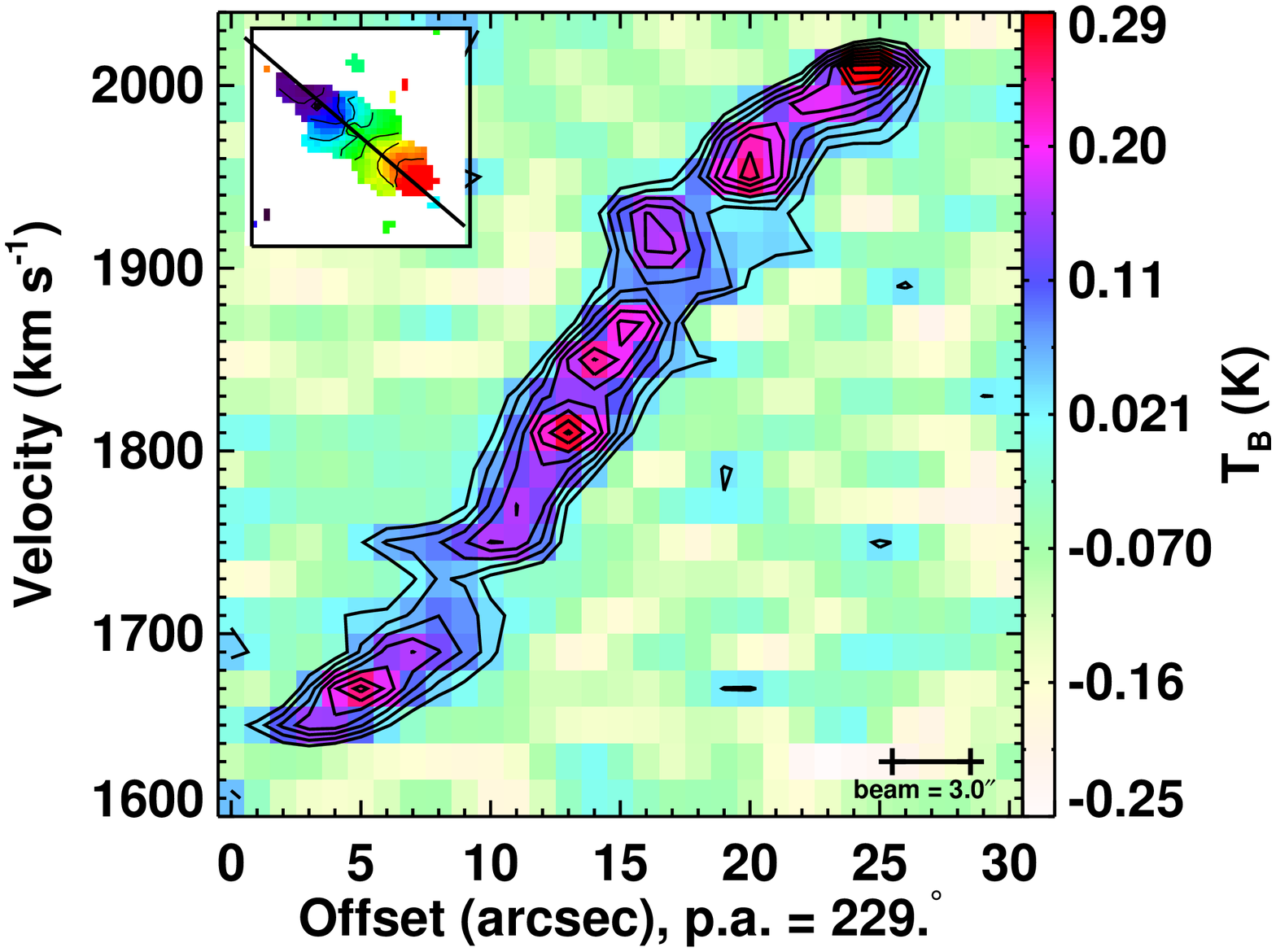}}
\end{figure*}
\begin{figure*}
\subfloat{\includegraphics[width=7in,clip,trim=1.5cm 1.6cm 2.5cm 2.5cm]{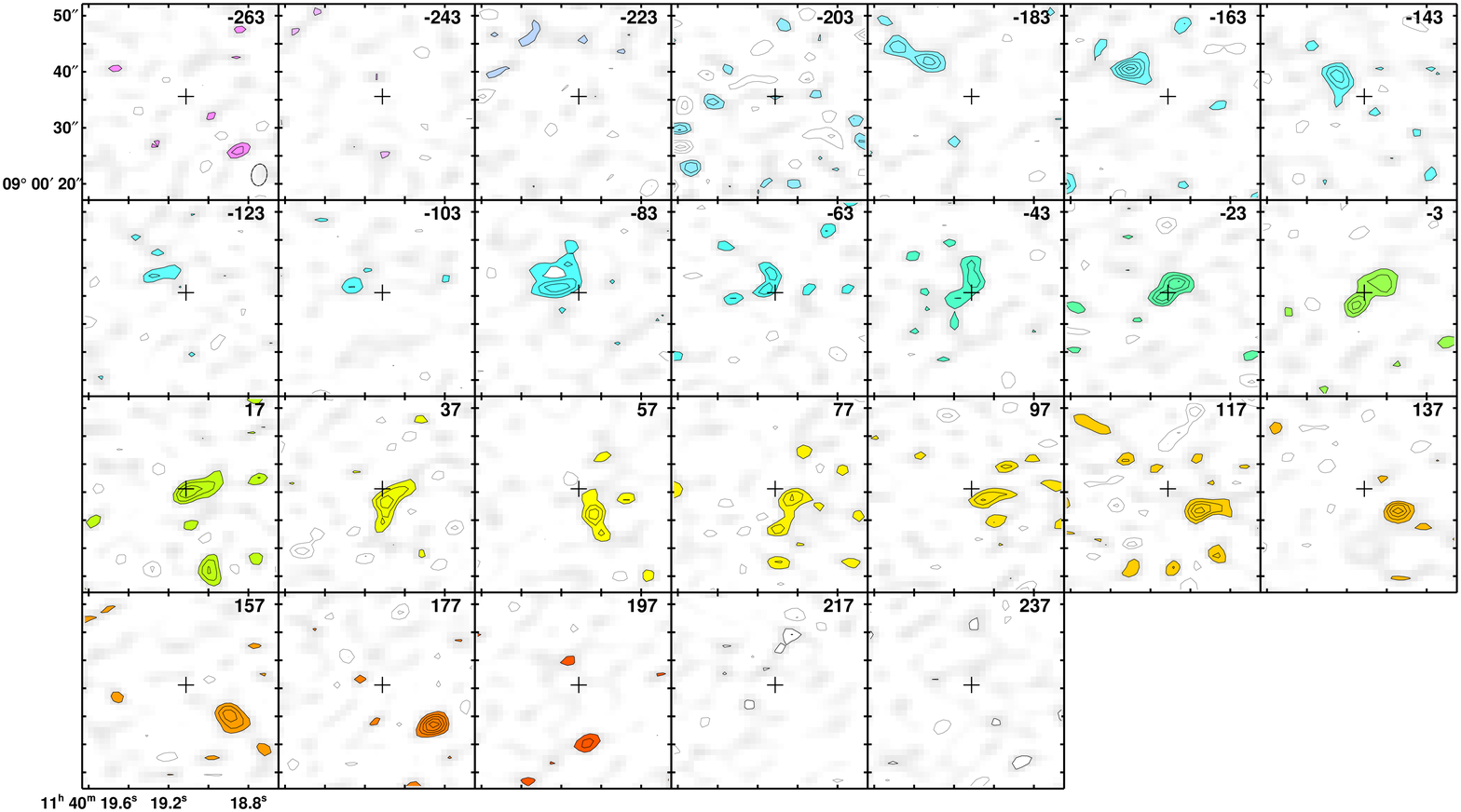}}
\caption{{\bf IC~719} is a field 2$\sigma$ non-regular rotator ($M_K$ = -22.70) with normal stellar morphology.  It contains a dusty disc.  The 2$\sigma$ peak is the signature of two counter-rotating stellar discs with opposite angular momenta.  The molecular gas is co-rotating with a kinematically decoupled core, and counter-rotating with respect to the dominant stellar component. The moment0 peak is 2.6 Jy beam$^{-1}$ \kms.  The moment1 contours are placed at 25\kms\ intervals.}
\end{figure*}

\clearpage
\begin{figure*}
\centering
\subfloat{\includegraphics[height=2.2in,clip,trim=0.8cm 2.7cm 0cm 2.3cm]{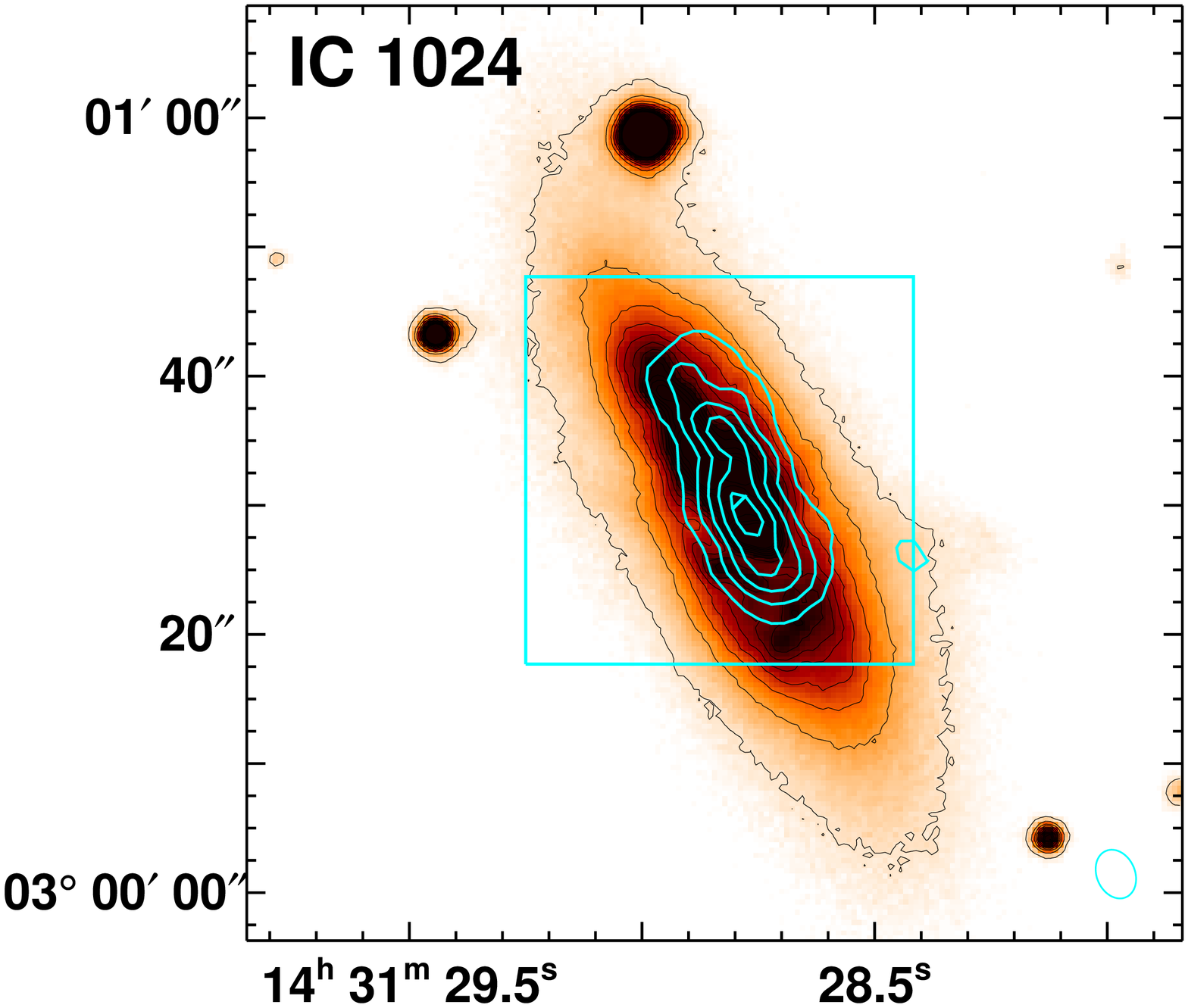}}
\subfloat{\includegraphics[height=2.2in,clip,trim=0.5cm 0.5cm 0cm 0.2cm]{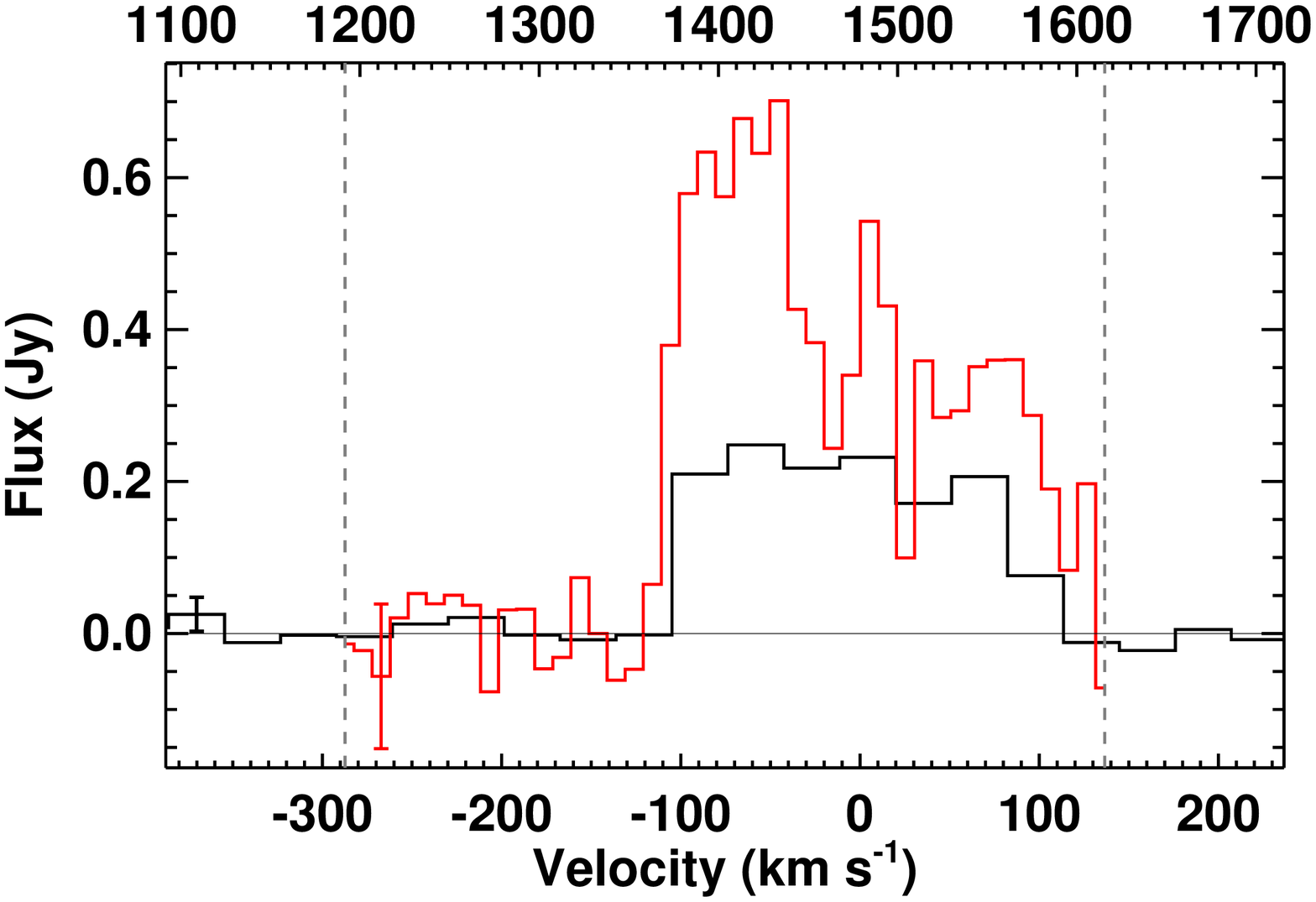}}
\end{figure*}
\begin{figure*}
\subfloat{\includegraphics[height=1.6in,clip,trim=0cm 1.5cm 0cm 1.5cm]{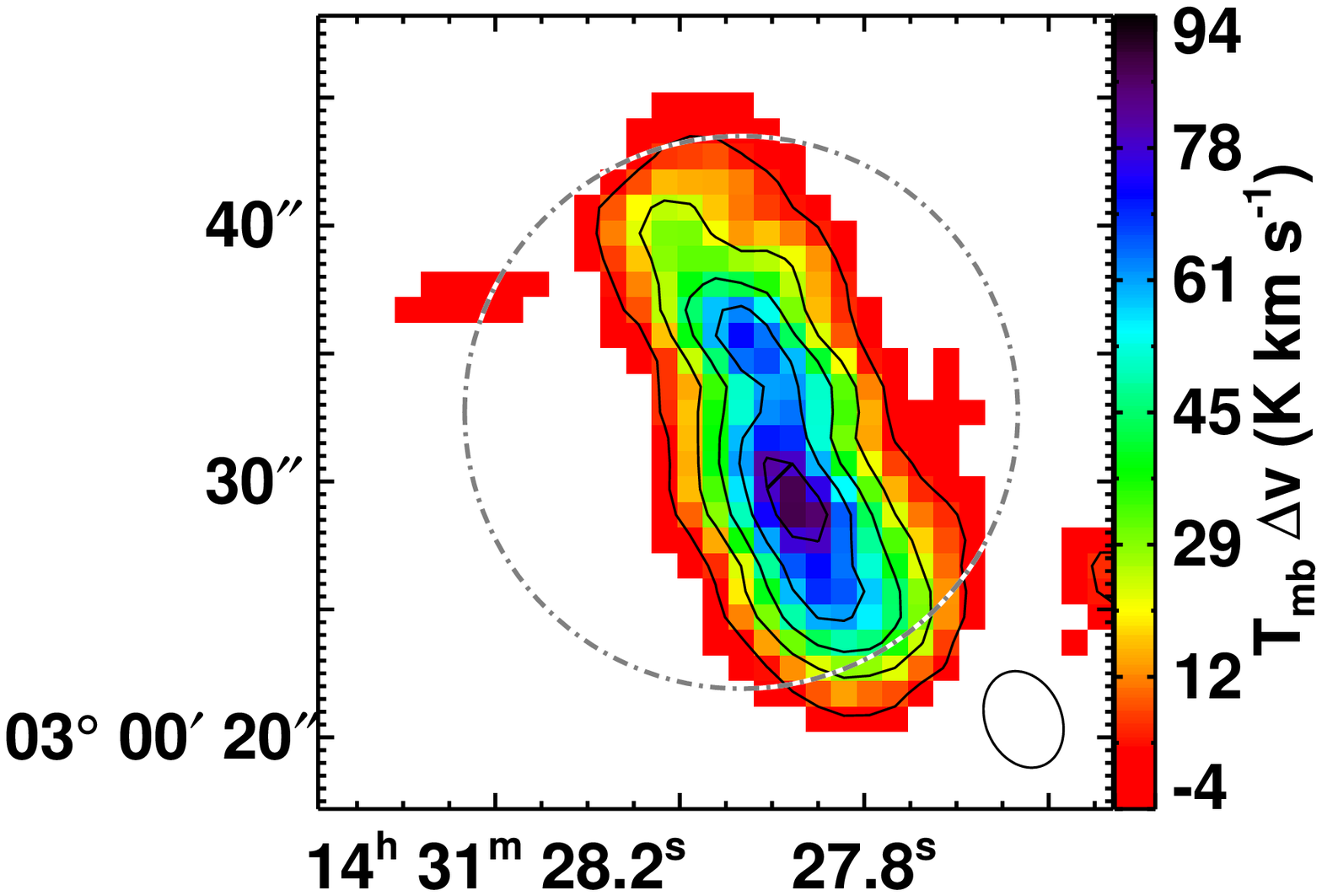}}
\subfloat{\includegraphics[height=1.6in,clip,trim=0cm 1.5cm 0cm 1.5cm]{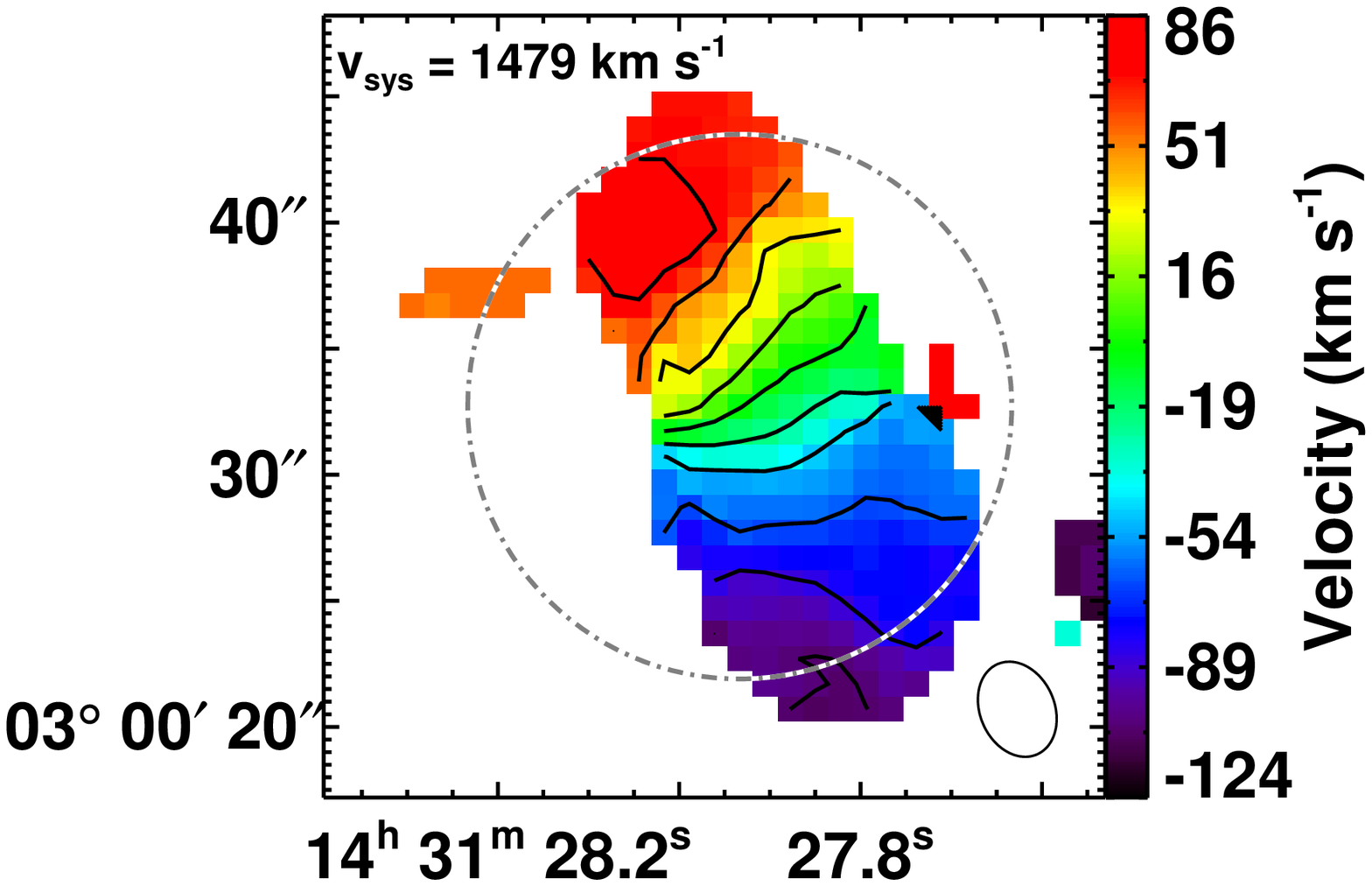}}
\subfloat{\includegraphics[height=1.6in,clip,trim=0cm 1.4cm 0cm 0.9cm]{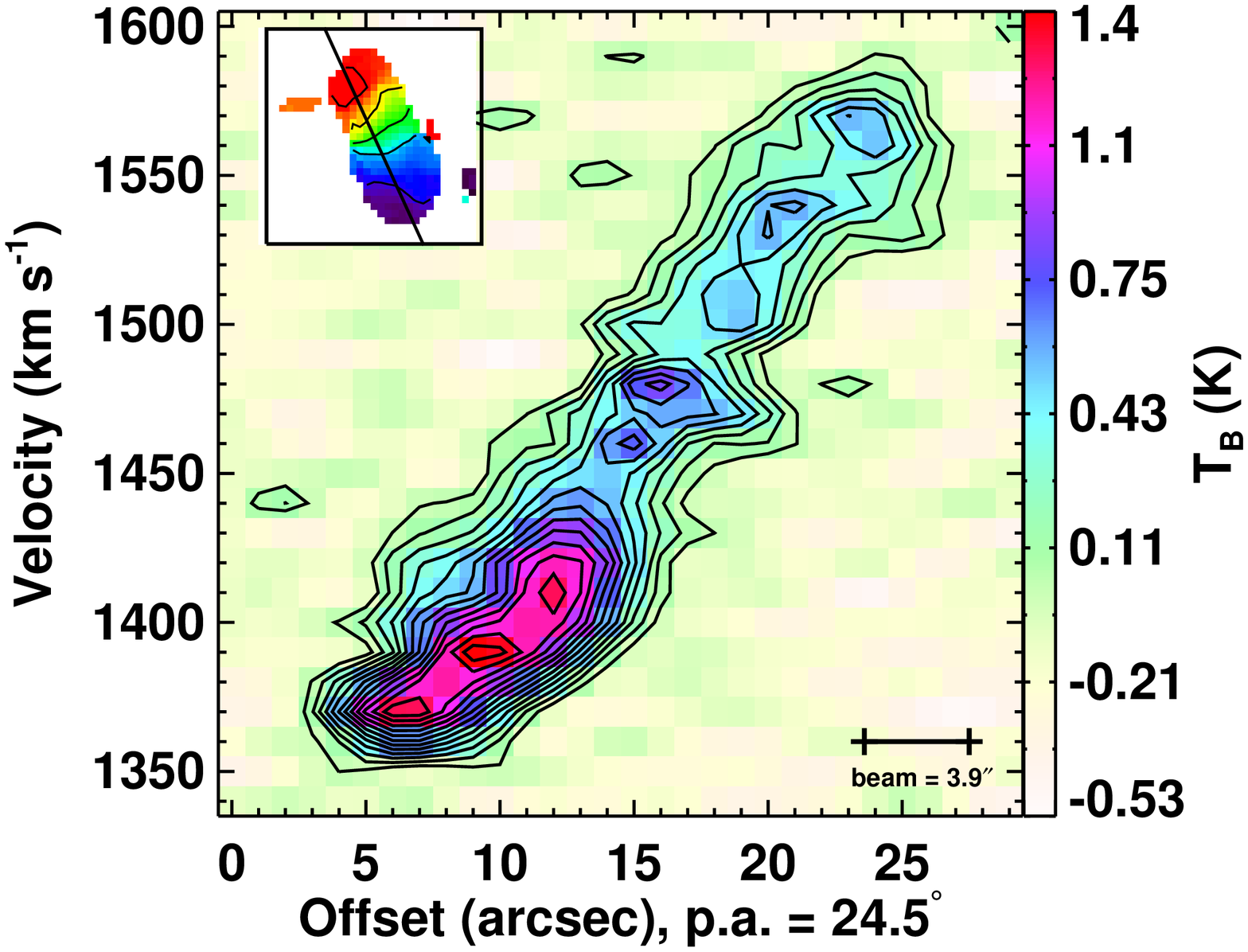}}
\end{figure*}
\begin{figure*}
\subfloat{\includegraphics[width=7in,clip,trim=1.5cm 1.6cm 2.5cm 2.5cm]{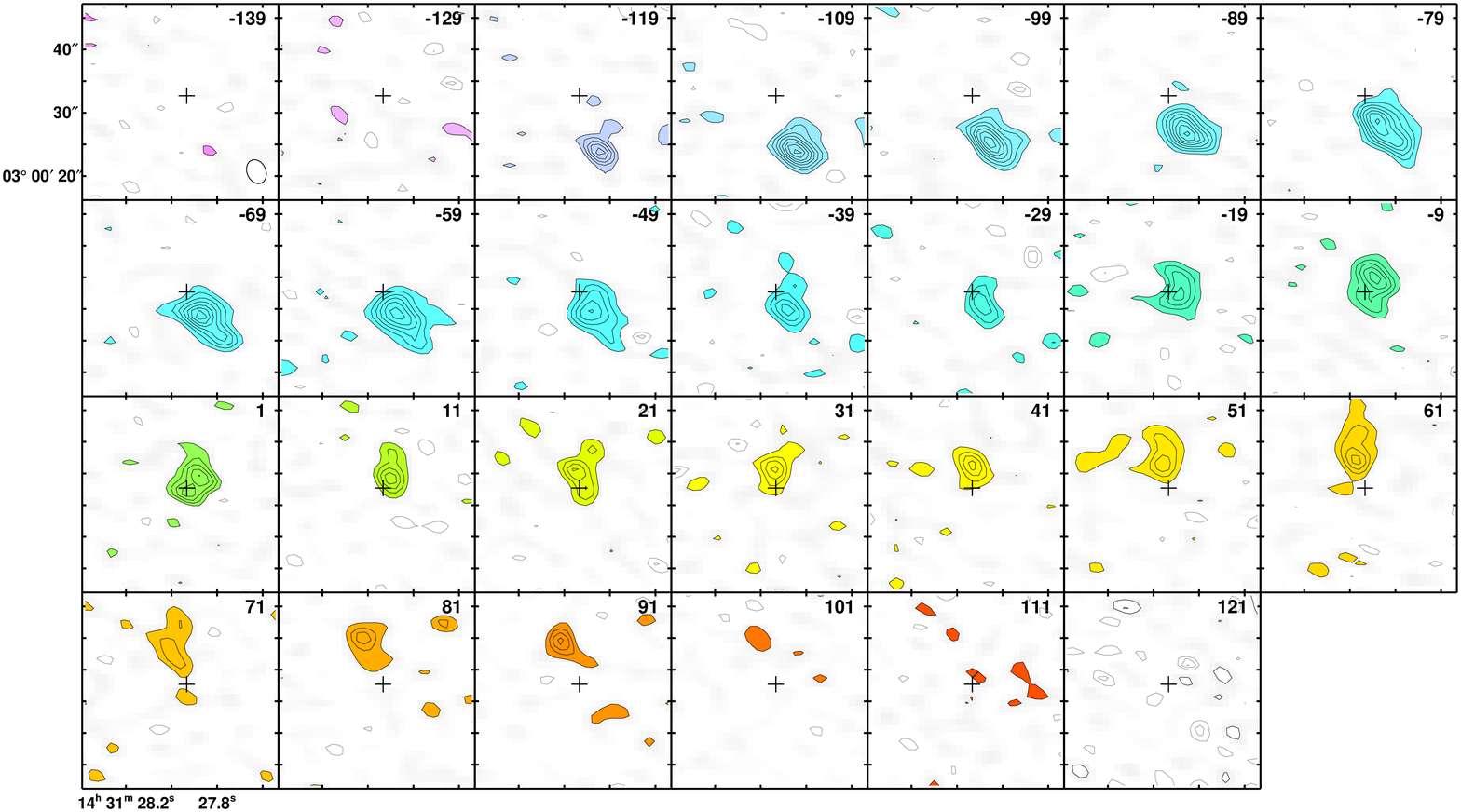}}
\caption{{\bf IC~1024} is a field regular rotator ($M_K$ = -21.85) with stellar morphology that shows interaction features as well as a dust filament.  The moment0 peak is 14 Jy beam$^{-1}$ \kms.  The channel map and PVD contours are placed at $1.5\sigma$ intervals.}
\end{figure*}

\clearpage
\begin{figure*}
\centering
\subfloat{\includegraphics[height=2.5in,clip,trim=2cm 3cm 0cm 2.8cm]{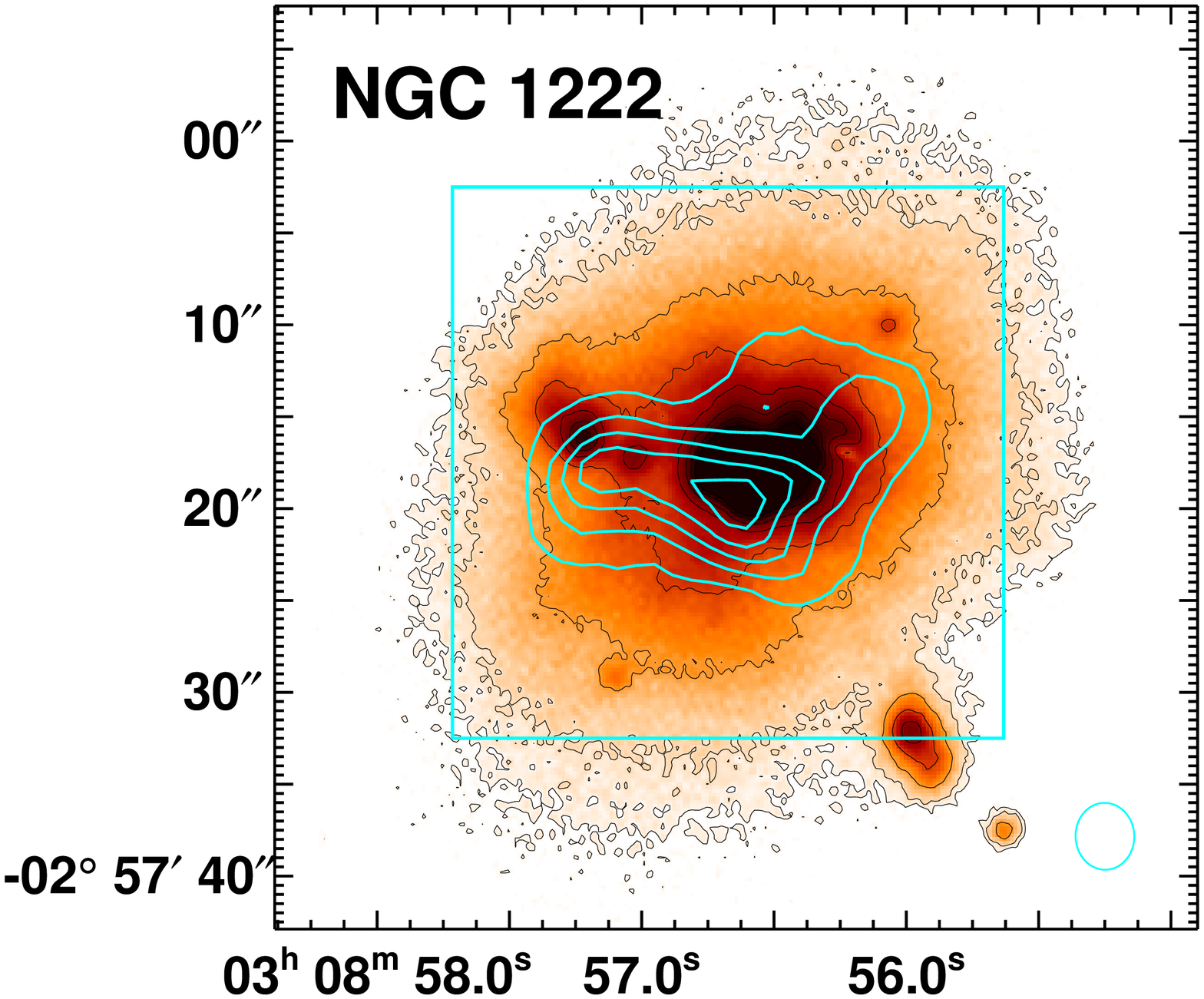}}
\subfloat{\includegraphics[height=2.5in,clip,trim=0cm 0.5cm 0.8cm 0.4cm]{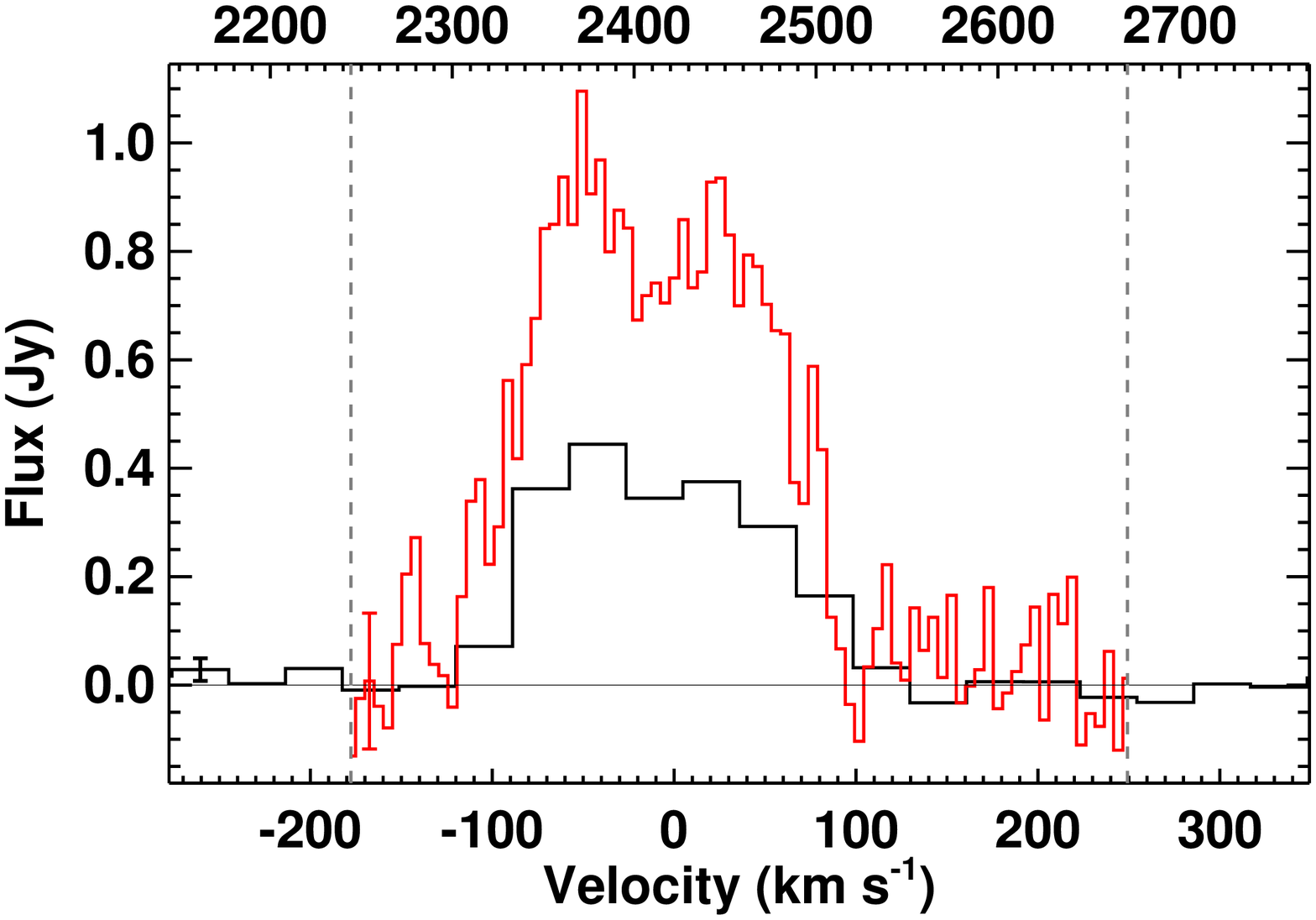}}
\end{figure*}
\begin{figure*}
\subfloat{\includegraphics[height=1.6in,clip,trim=0cm 1.4cm 0.4cm 2.4cm]{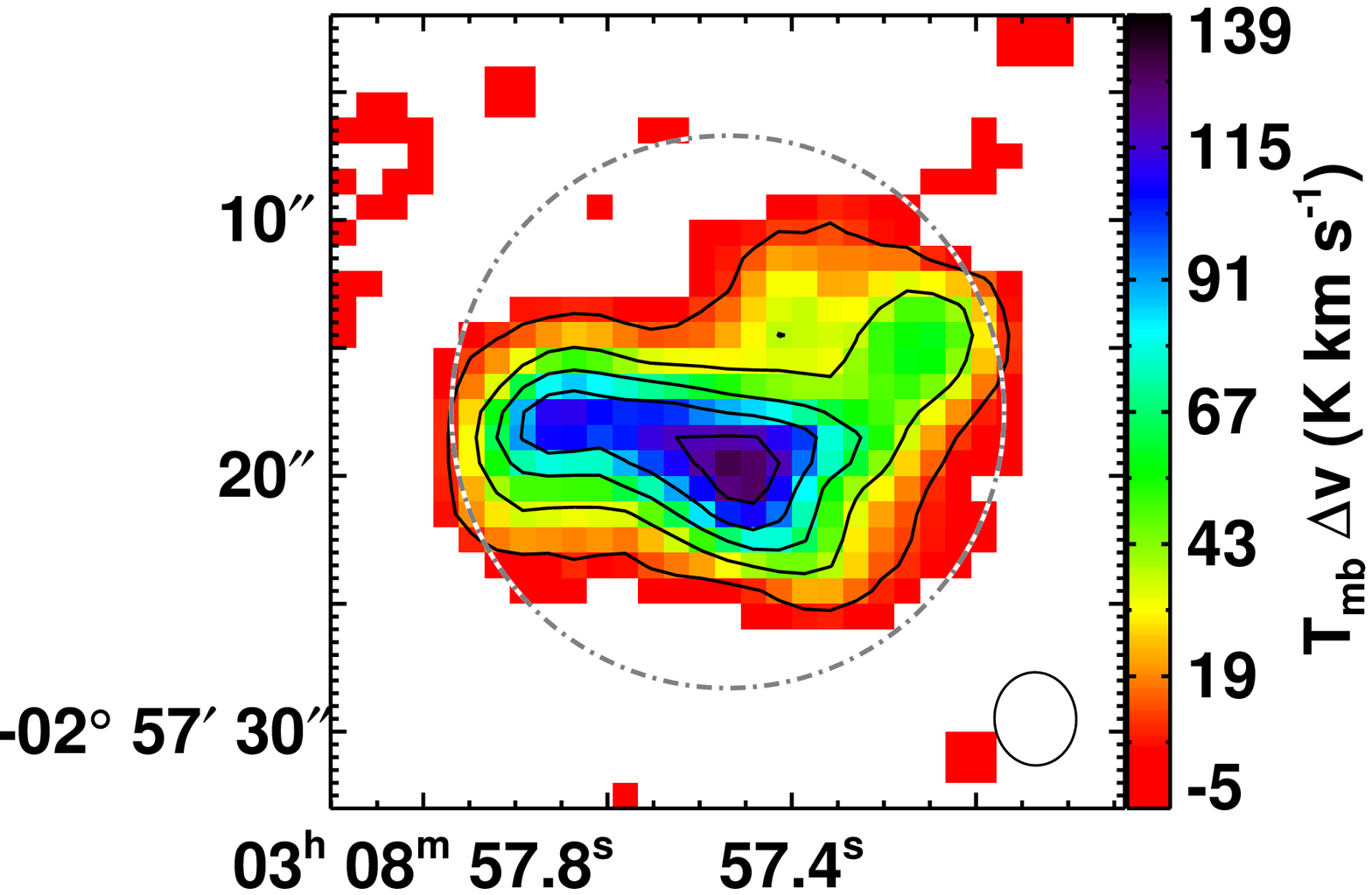}}
\subfloat{\includegraphics[height=1.6in,clip,trim=0cm 1.4cm 0.4cm 2.2cm]{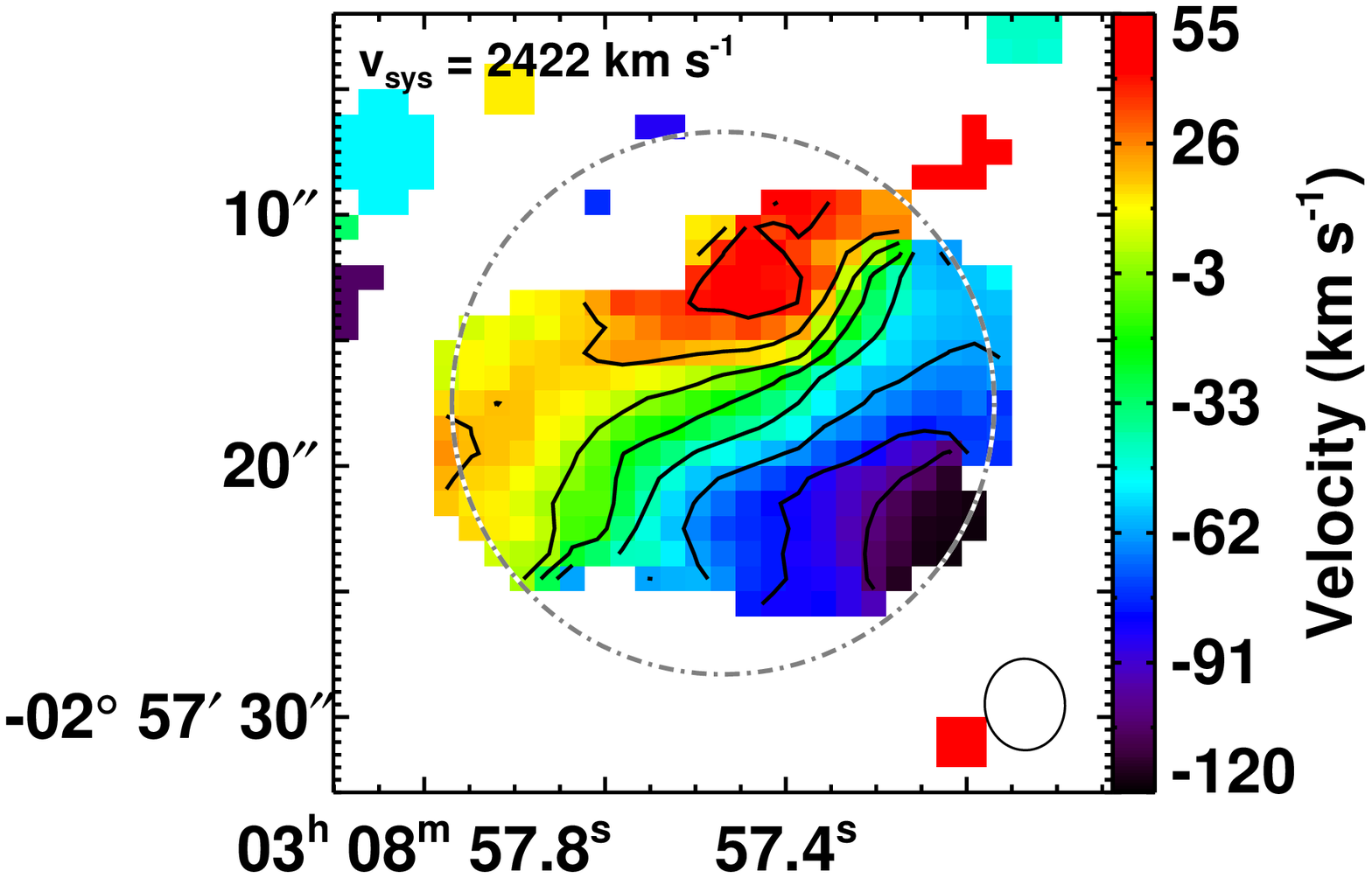}}
\subfloat{\includegraphics[height=1.6in,clip,trim=0cm 1.4cm 0cm 0.9cm]{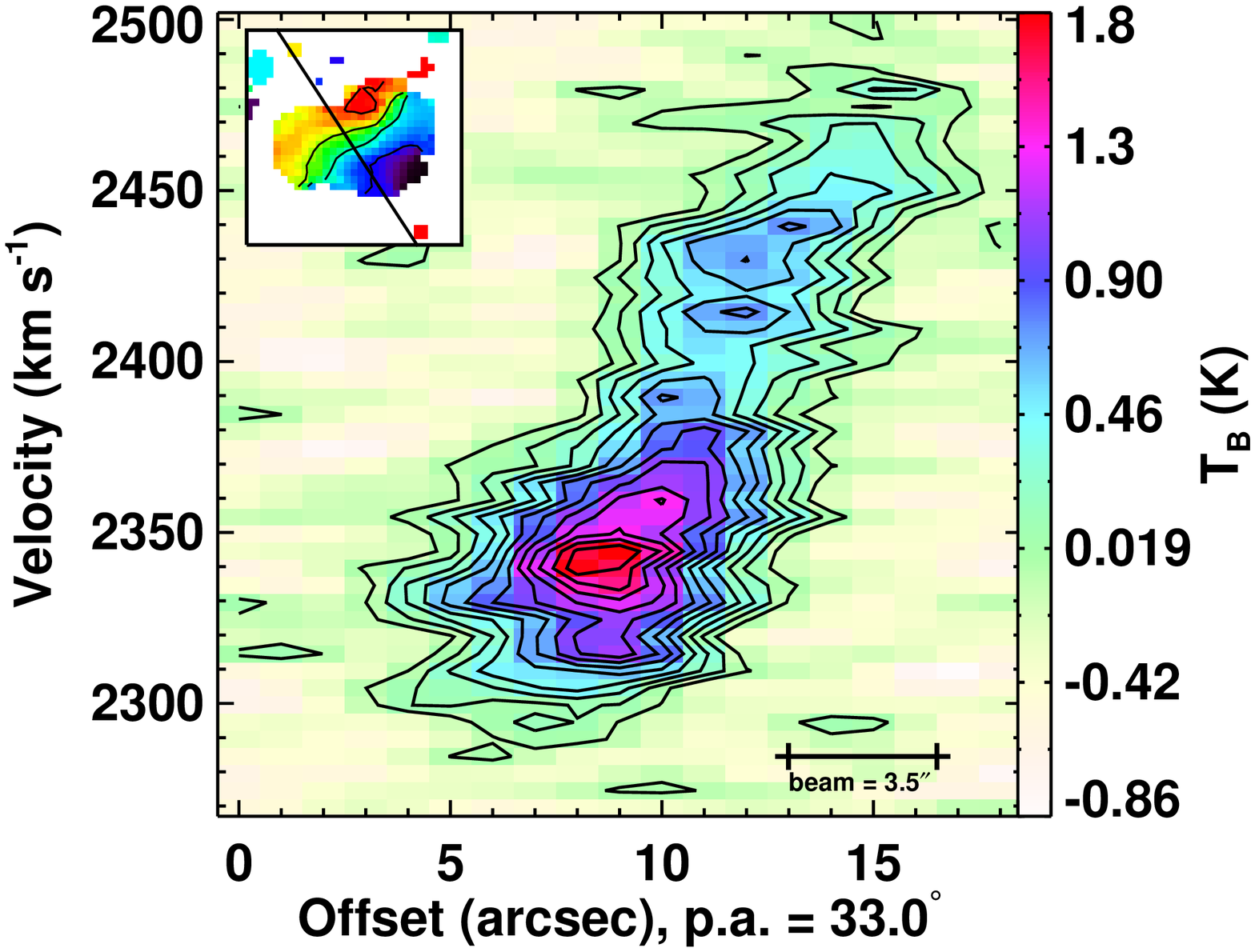}}
\end{figure*}
\begin{figure*}
\subfloat{\includegraphics[width=7in,clip,trim=1.4cm 5.3cm 6cm 1cm]{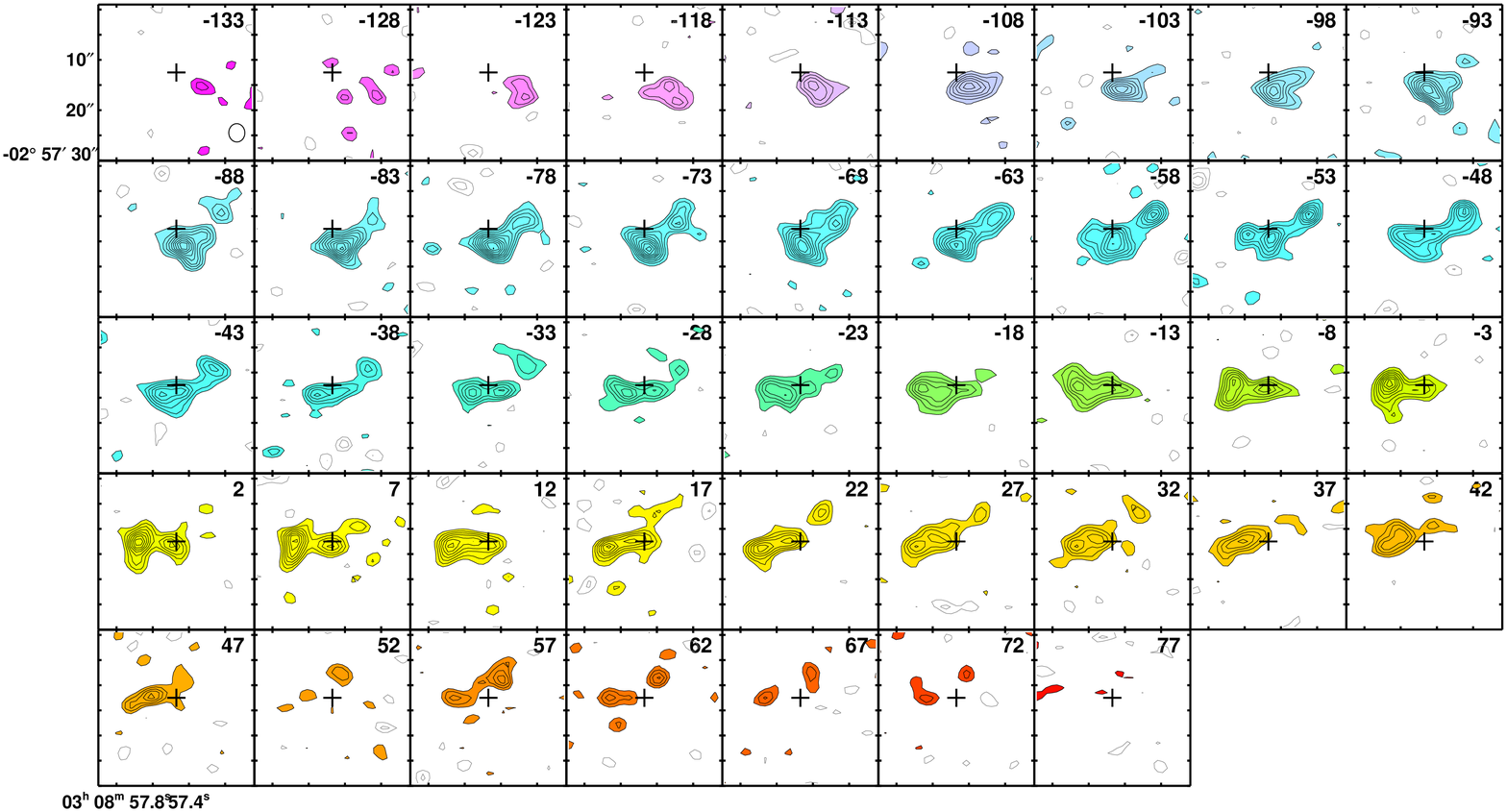}}
\caption{{\bf NGC~1222} is a field non-regular rotator ($M_K$ = -22.71) with stellar morphology that shows interaction features as well as dust filaments.  It is a well-known interaction, and cross-identified as Markarian 603 \citep{beck+07}.  The moment0 peak is 18 Jy beam$^{-1}$ \kms.  The PVD contours are placed at $1.5\sigma$ intervals.}
\end{figure*}

\clearpage
\begin{figure*}
\centering
\subfloat{\includegraphics[height=2.2in,clip,trim=0.4cm 3.1cm 0.2cm 2.8cm]{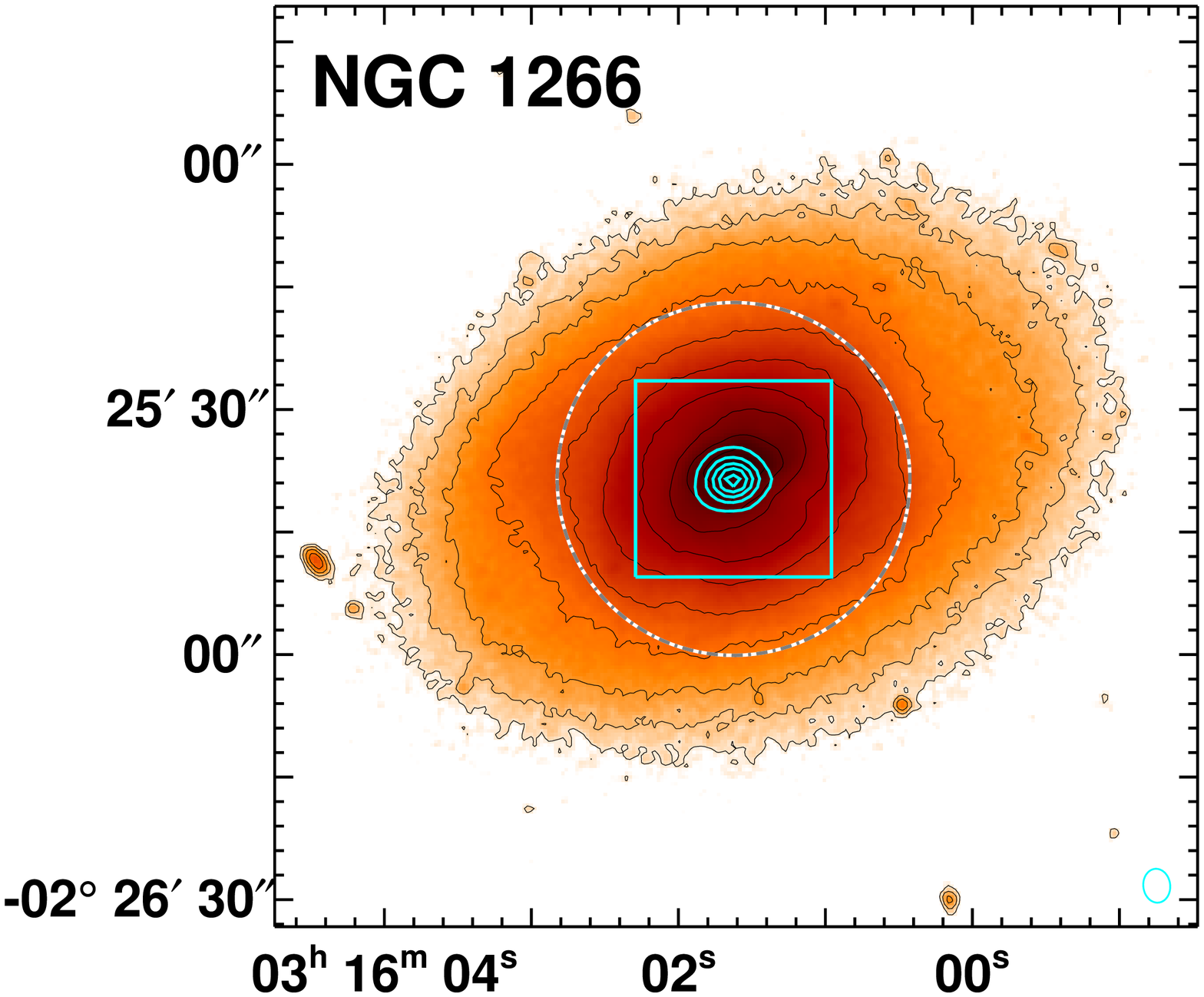}}
\subfloat{\includegraphics[height=2.2in,clip,trim=0cm 0.5cm 0.4cm 0.2cm]{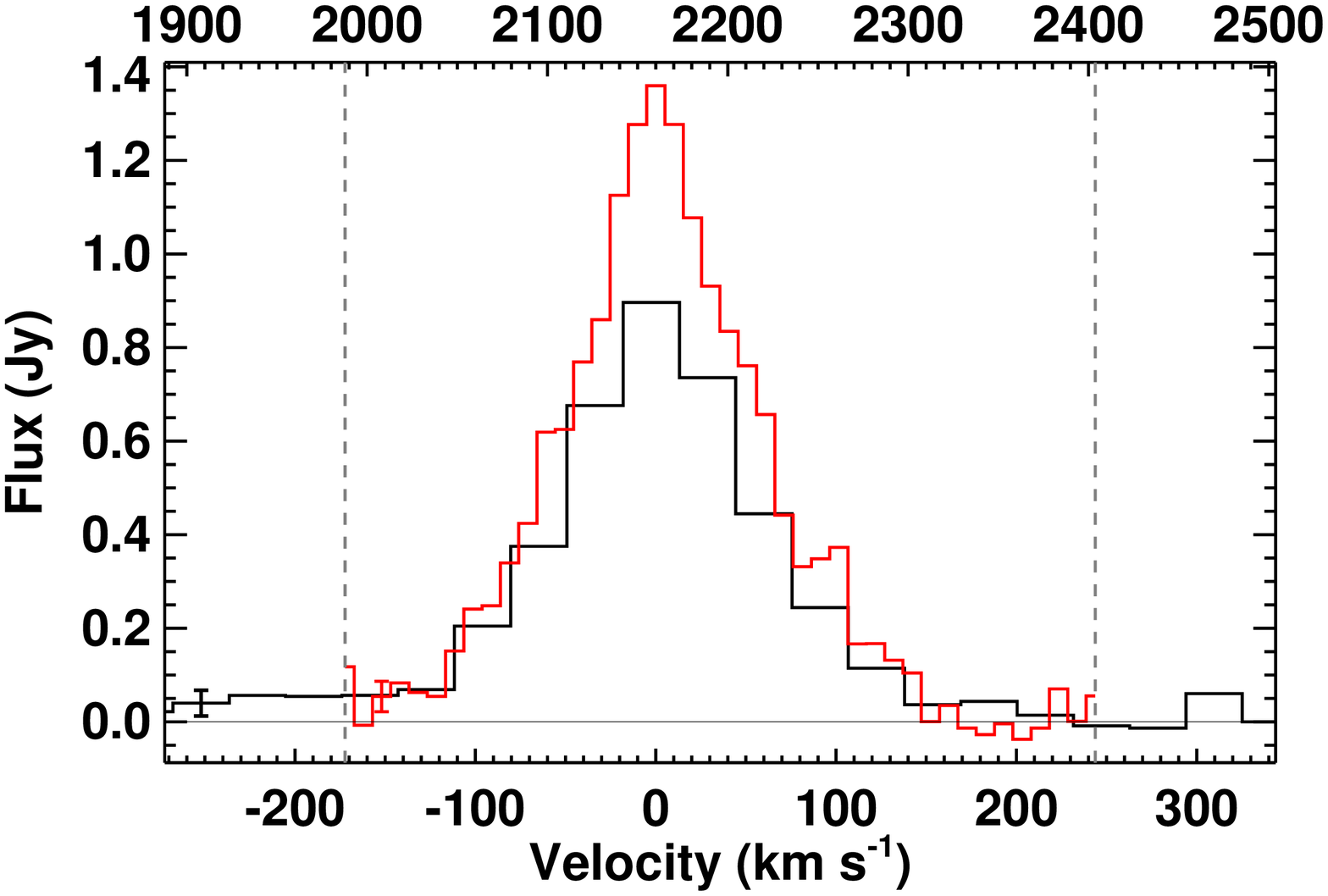}}
\end{figure*}
\begin{figure*}
\subfloat{\includegraphics[height=1.6in,clip,trim=0cm 1.4cm 0.4cm 2.4cm]{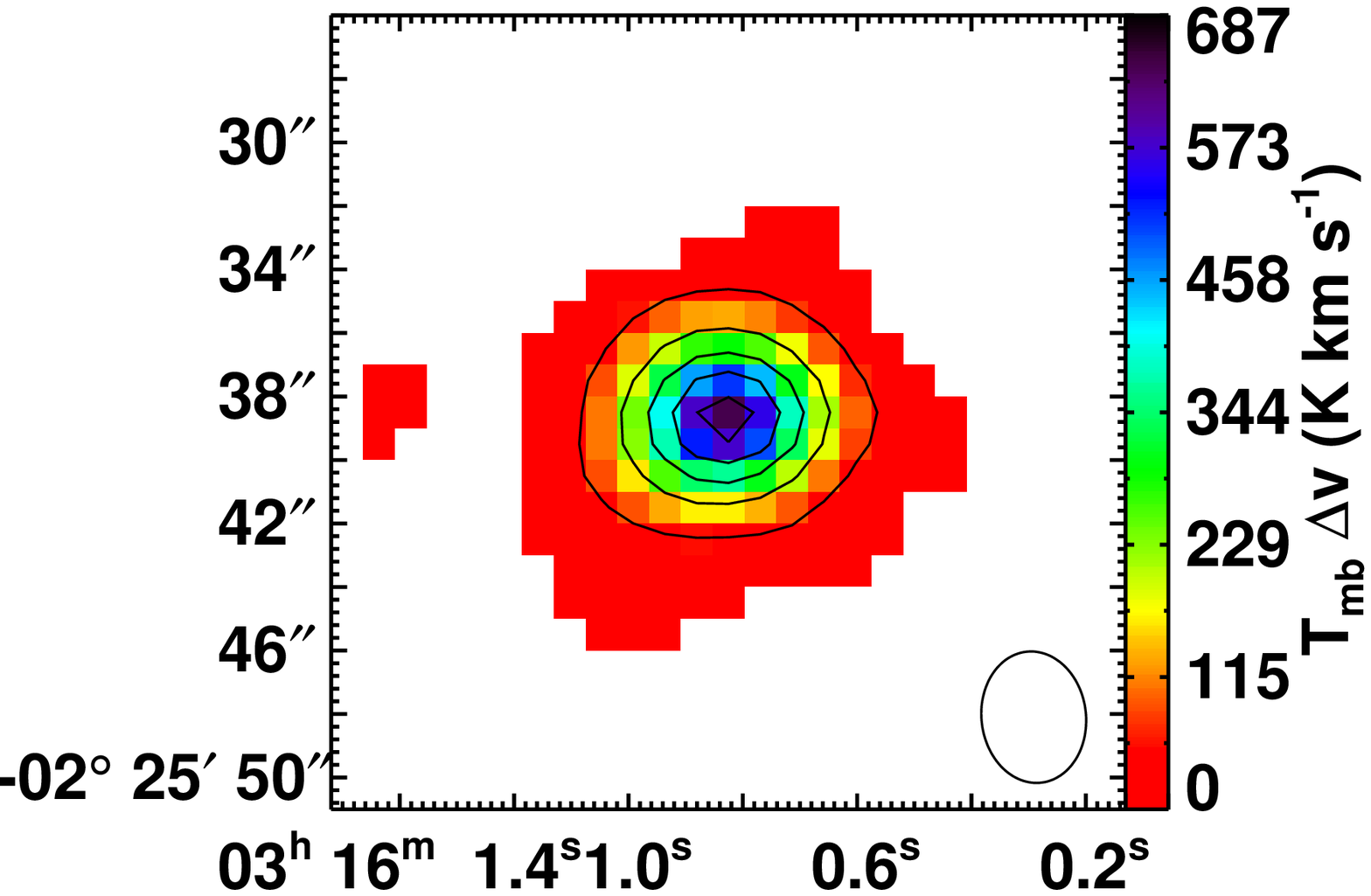}}
\subfloat{\includegraphics[height=1.6in,clip,trim=0cm 1.4cm 0.4cm 2.4cm]{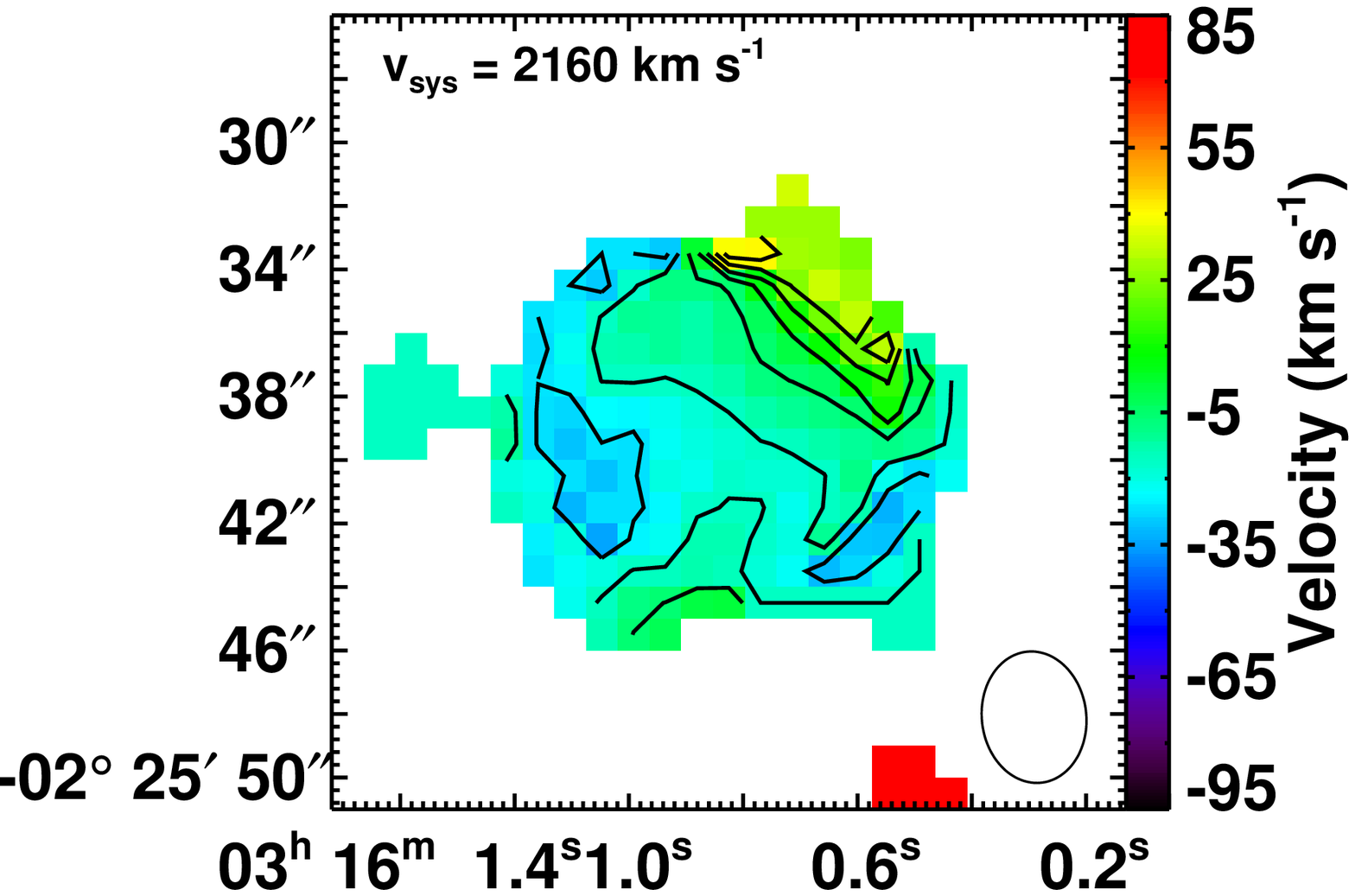}}
\subfloat{\includegraphics[height=1.6in,clip,trim=0cm 1.4cm 0cm 1cm]{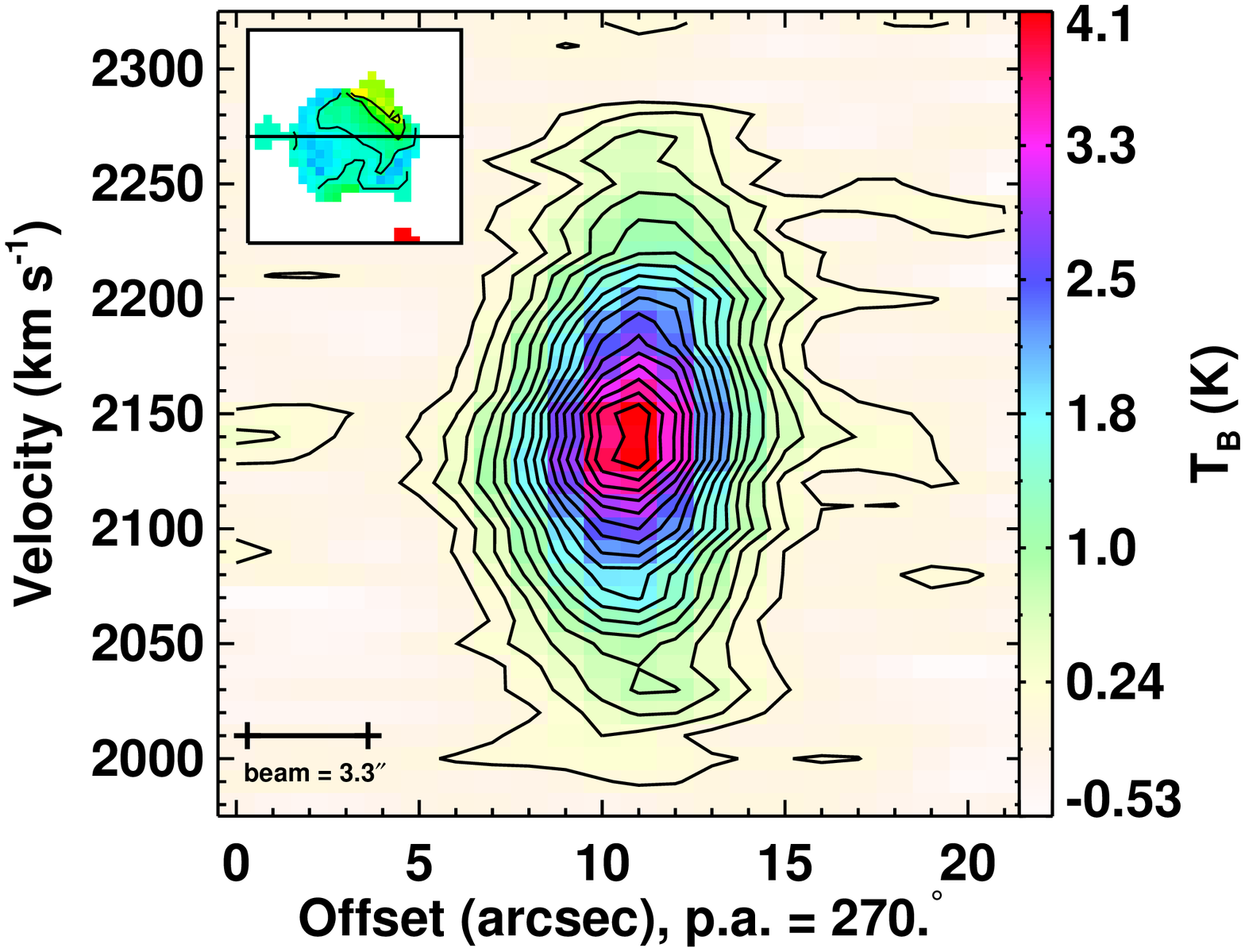}}
\end{figure*}
\begin{figure*}
\subfloat{\includegraphics[width=7in,clip,trim=1.4cm 5.3cm 6cm 1cm]{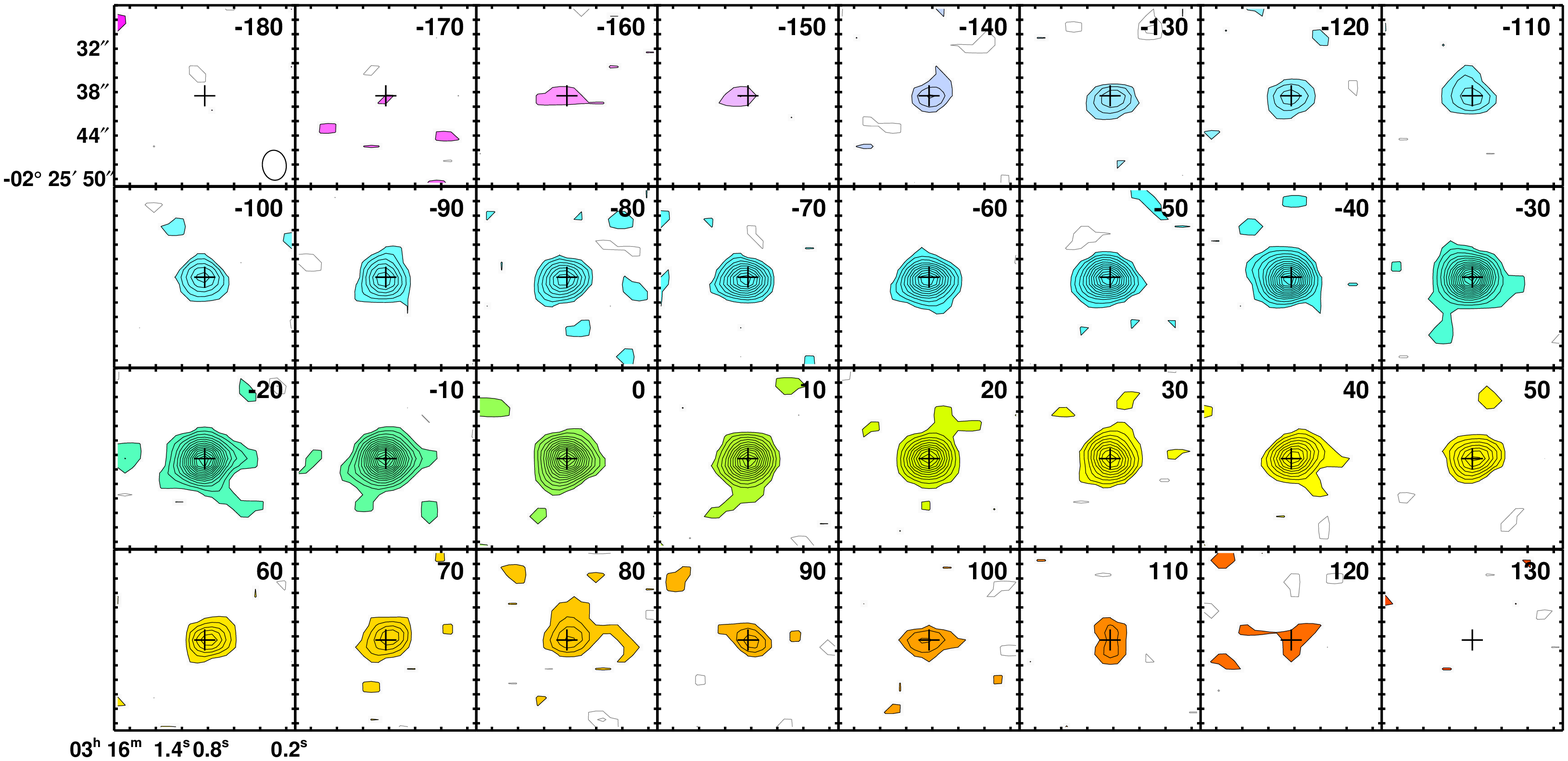}}
\vspace{-2cm}
\caption{{\bf NGC~1266} is a field regular rotator ($M_K$ = -22.93) with normal stellar morphology.  It is also observed to have a dust filament.  Not only was it the brightest detection in the sample, but it had the largest linewidths observed.  It was initially unresolved in CARMA D-array, thus was observed in the much higher resolution arrays.  NGC~1266 hosts a massive molecular outflow that appears to be driven by an AGN.  Further details can be found in \citealt{alatalo+11}.  The moment0 peak is 100 Jy beam$^{-1}$.  The moment1 contours are placed at 10\kms\ intervals.  The channel map contours are placed at $2\sigma$ intervals and the PVD contours are placed at $3\sigma$ intervals.}
\end{figure*}

\clearpage
\begin{figure*}
\centering
\subfloat{\includegraphics[height=2.2in,clip,trim=0.4cm 3.8cm 0cm 3.3cm]{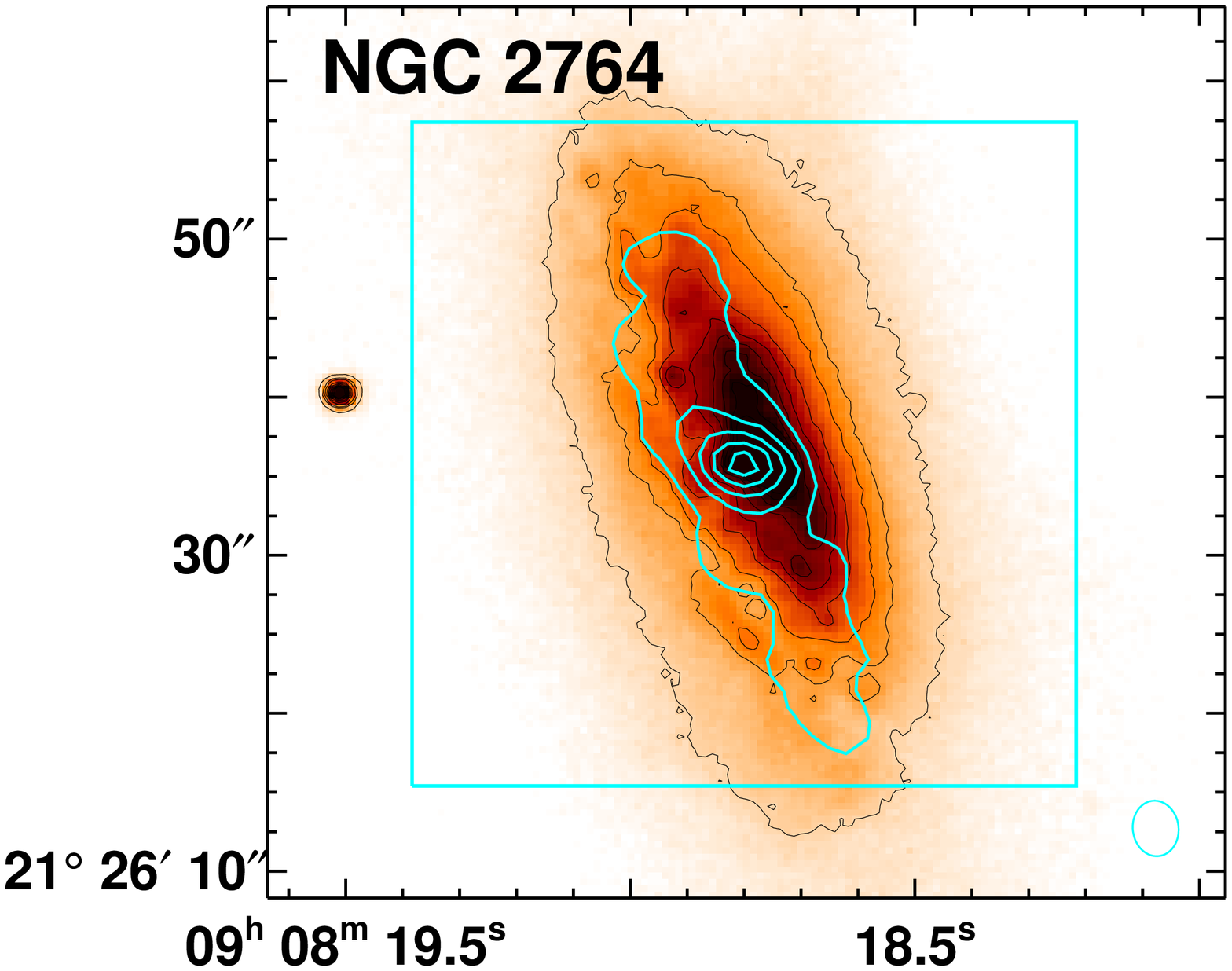}}
\subfloat{\includegraphics[height=2.2in,clip,trim=0.5cm 0.5cm 0.4cm 0.2cm]{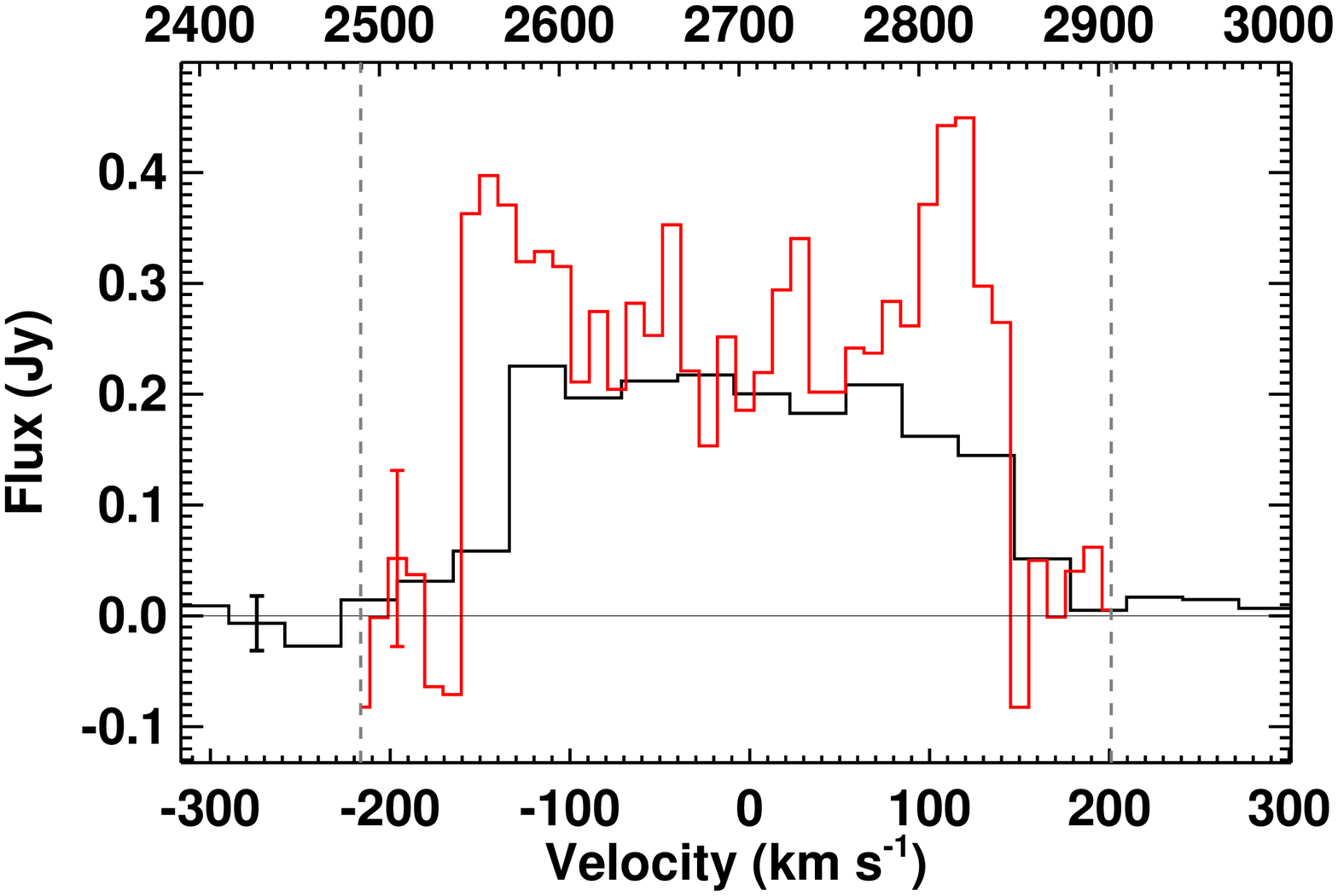}}
\end{figure*}
\begin{figure*}
\subfloat{\includegraphics[height=1.6in,clip,trim=0cm 1.4cm 0cm 2.3cm]{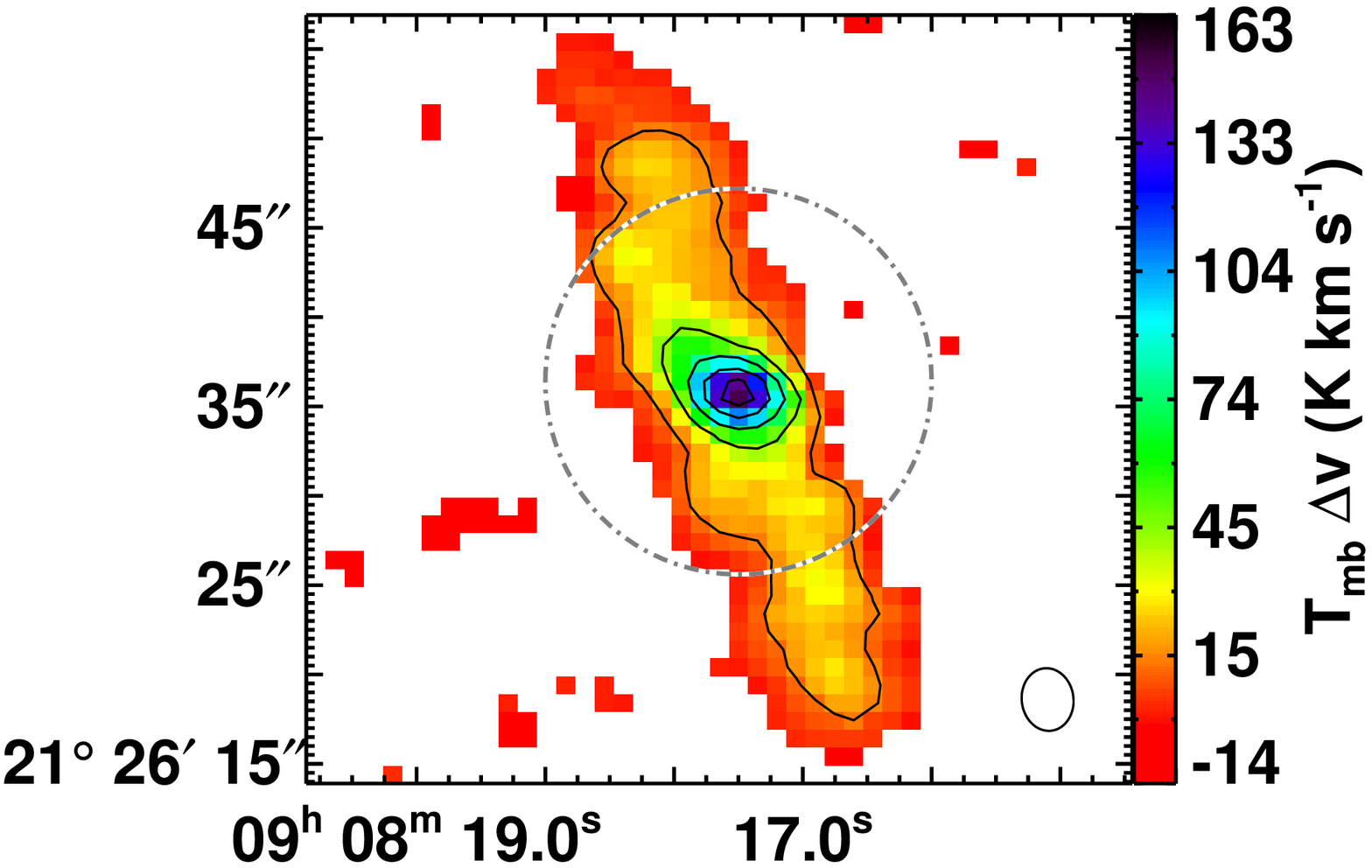}}
\subfloat{\includegraphics[height=1.6in,clip,trim=0cm 1.4cm 0cm 2.4cm]{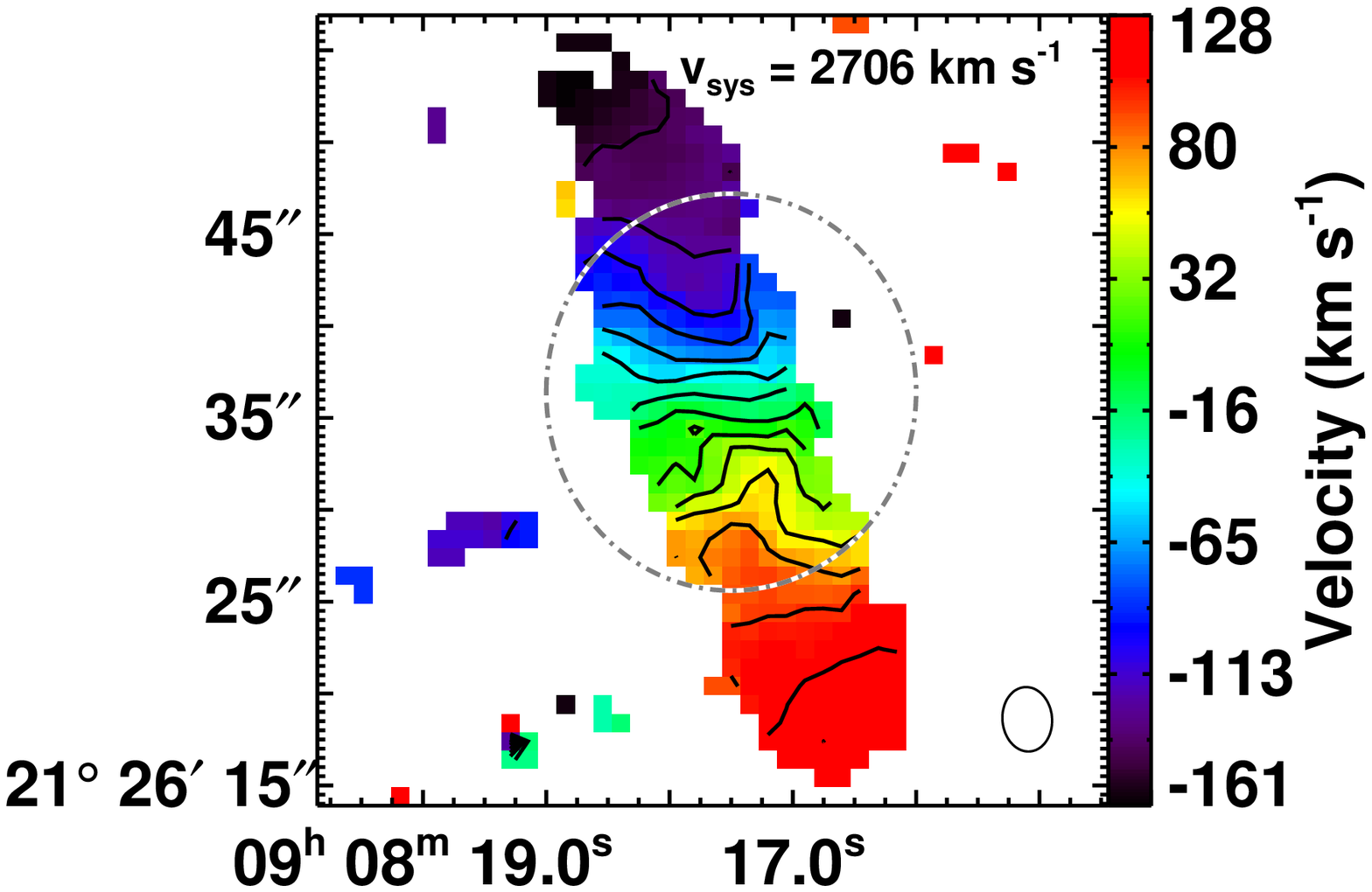}}
\subfloat{\includegraphics[height=1.6in,clip,trim=0cm 1.4cm 0cm 0.9cm]{figures/ngc2764_pv.eps}}
\end{figure*}
\begin{figure*}
\subfloat{\includegraphics[width=7in,clip,trim=1.5cm 6.2cm 2.5cm 0.6cm]{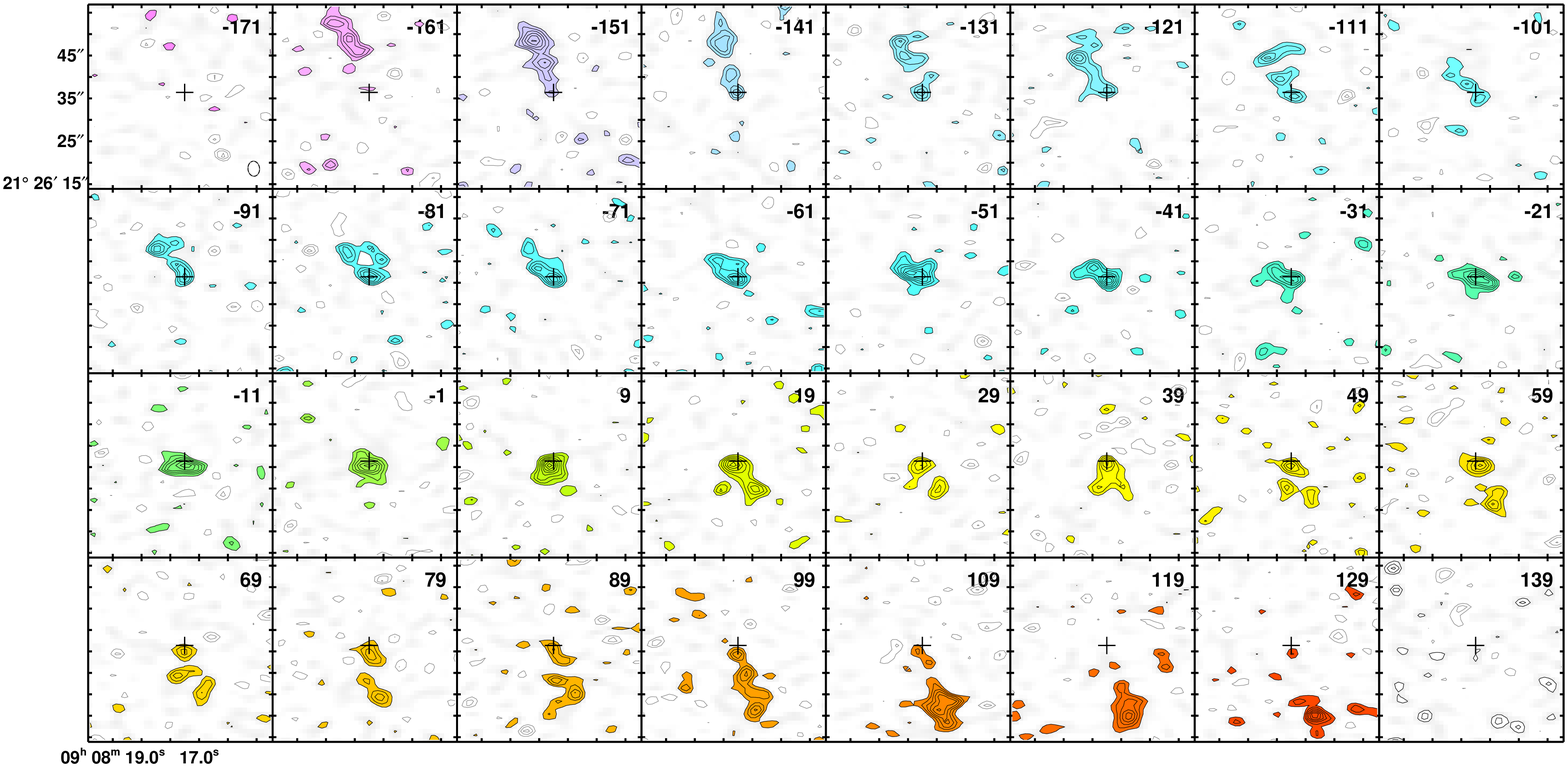}}
\caption{{\bf NGC~2764} is a field regular rotator ($M_K$ = -23.19) with stellar morphology that indicates interaction.  It is also observed to contain a blue nucleus and dust filaments.  The moment0 peak is 18 Jy beam$^{-1}$ \kms.  The PVD contours are placed at $1.5\sigma$ intervals.}
\end{figure*}

\clearpage
\begin{figure*}
\centering
\subfloat{\includegraphics[height=2.2in,clip,trim=0.4cm 4.1cm 0cm 3.7cm]{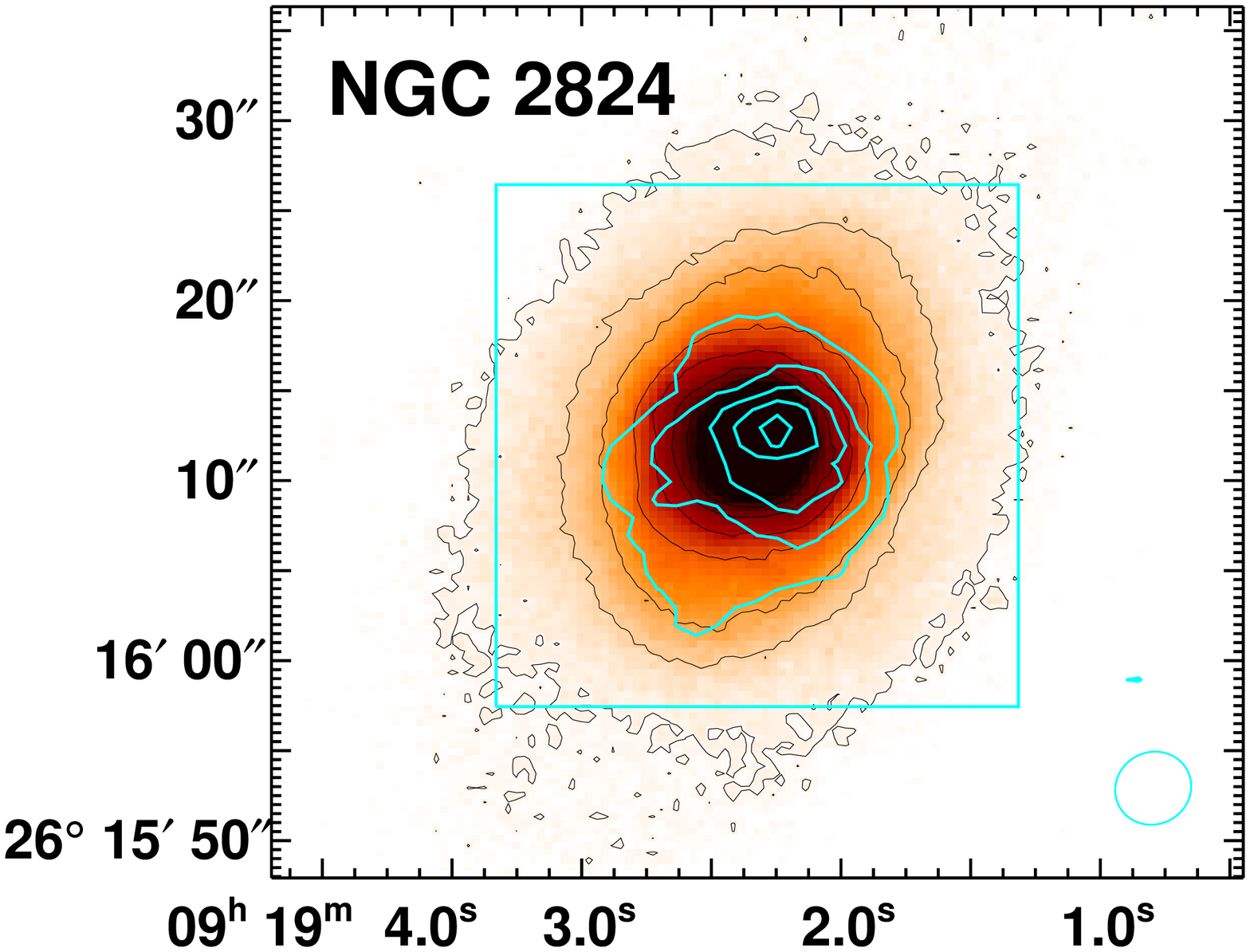}}
\subfloat{\includegraphics[height=2.2in,clip,trim=0cm 0.5cm 0.4cm 0.2cm]{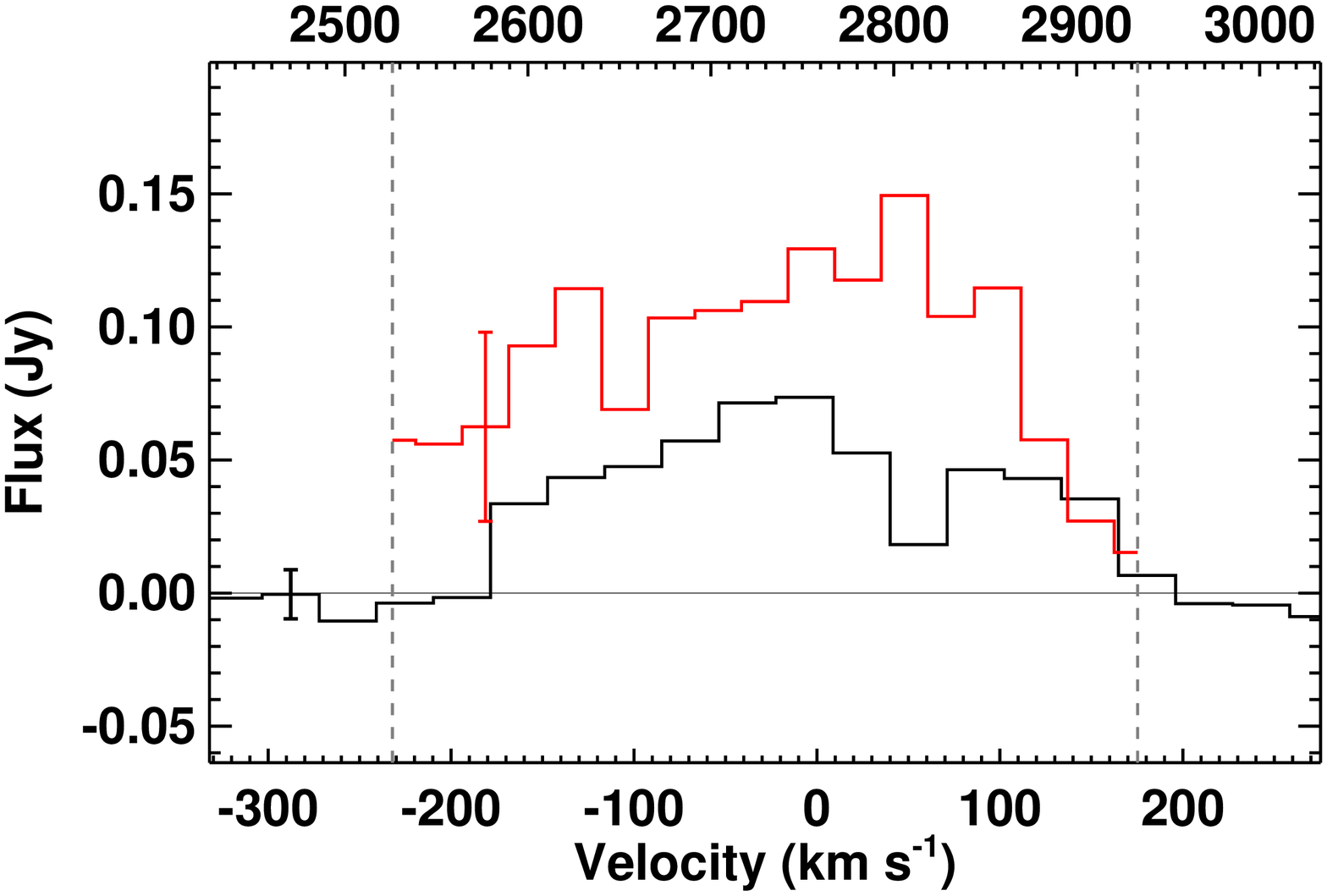}}
\end{figure*}
\begin{figure*}
\subfloat{\includegraphics[height=1.6in,clip,trim=0cm 1.4cm 0cm 2.2cm]{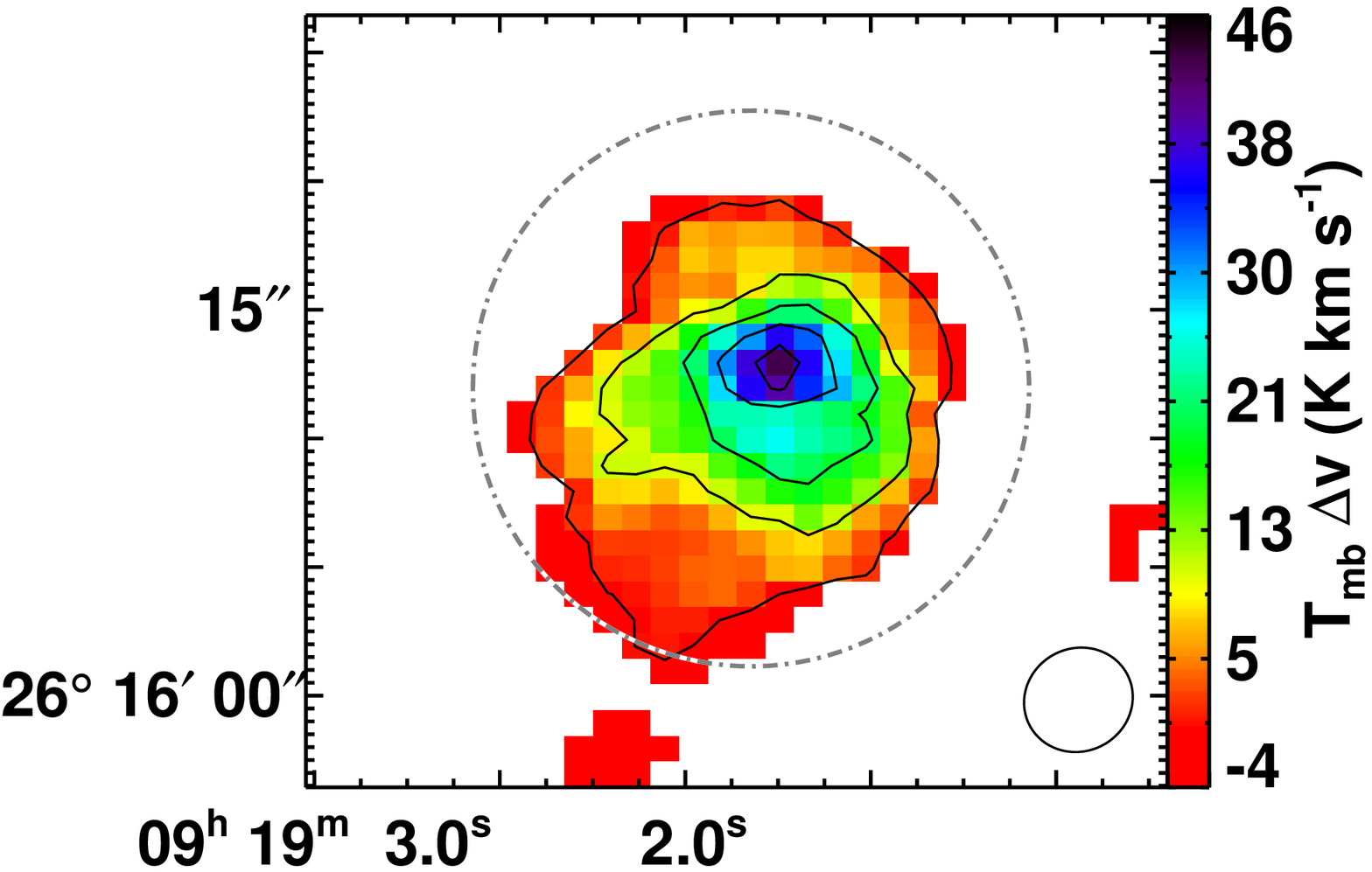}}
\subfloat{\includegraphics[height=1.6in,clip,trim=0cm 1.4cm 0cm 2.5cm]{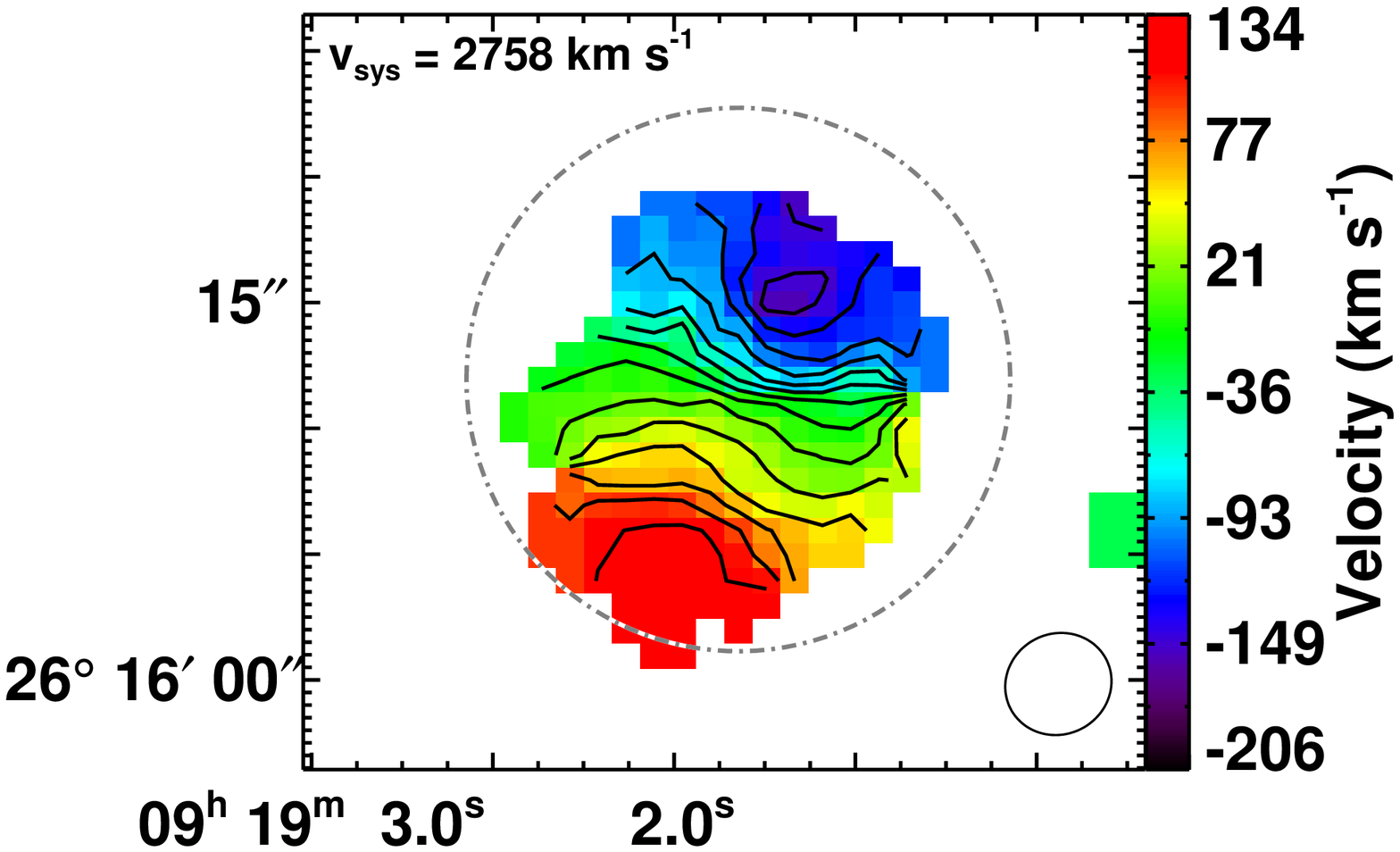}}
\subfloat{\includegraphics[height=1.6in,clip,trim=0cm 2.2cm 0cm 0.9cm]{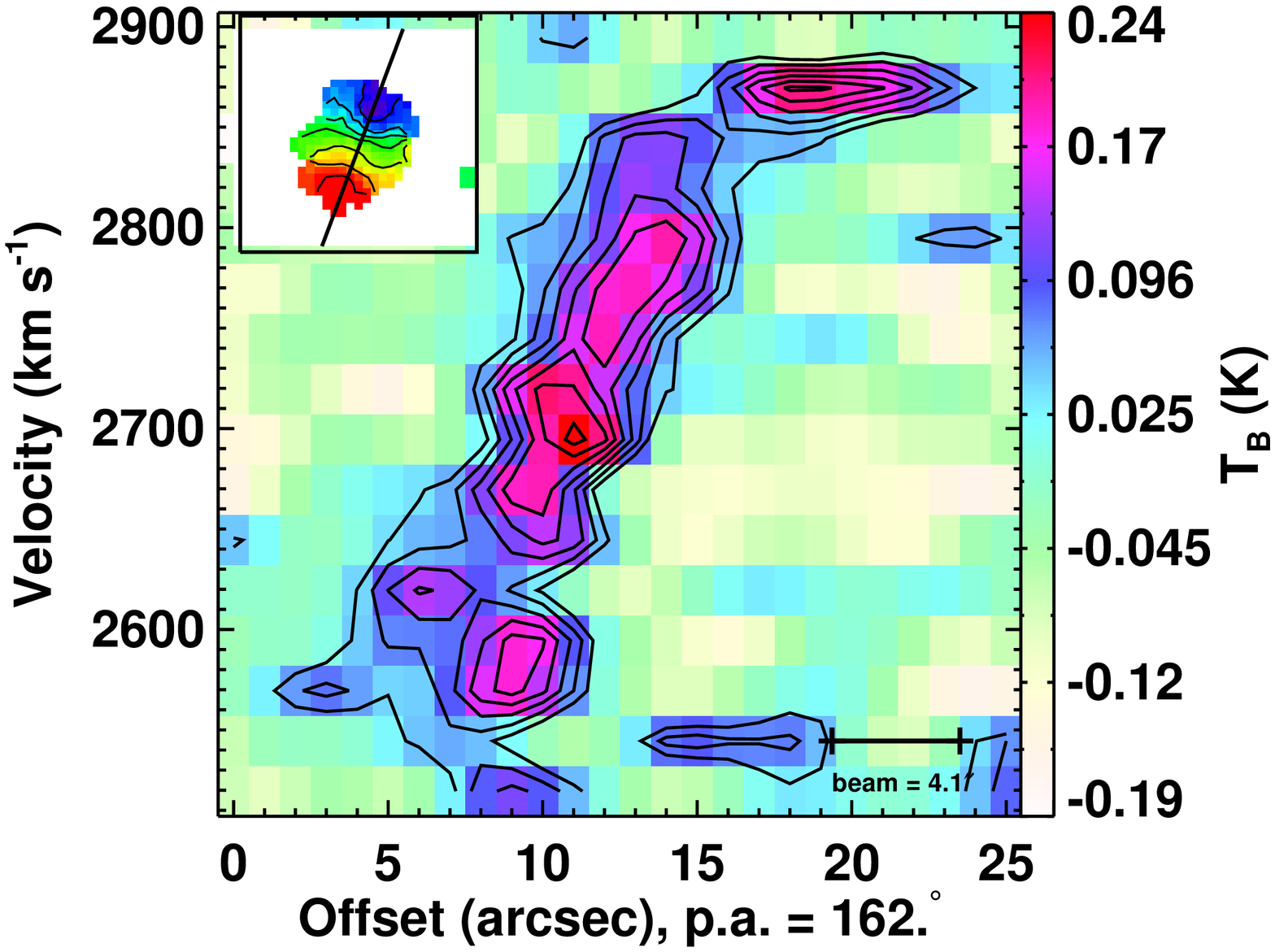}}
\end{figure*}
\begin{figure*}
\subfloat{\includegraphics[width=7in,clip,trim=1.5cm 1.5cm 1.2cm 4.7cm]{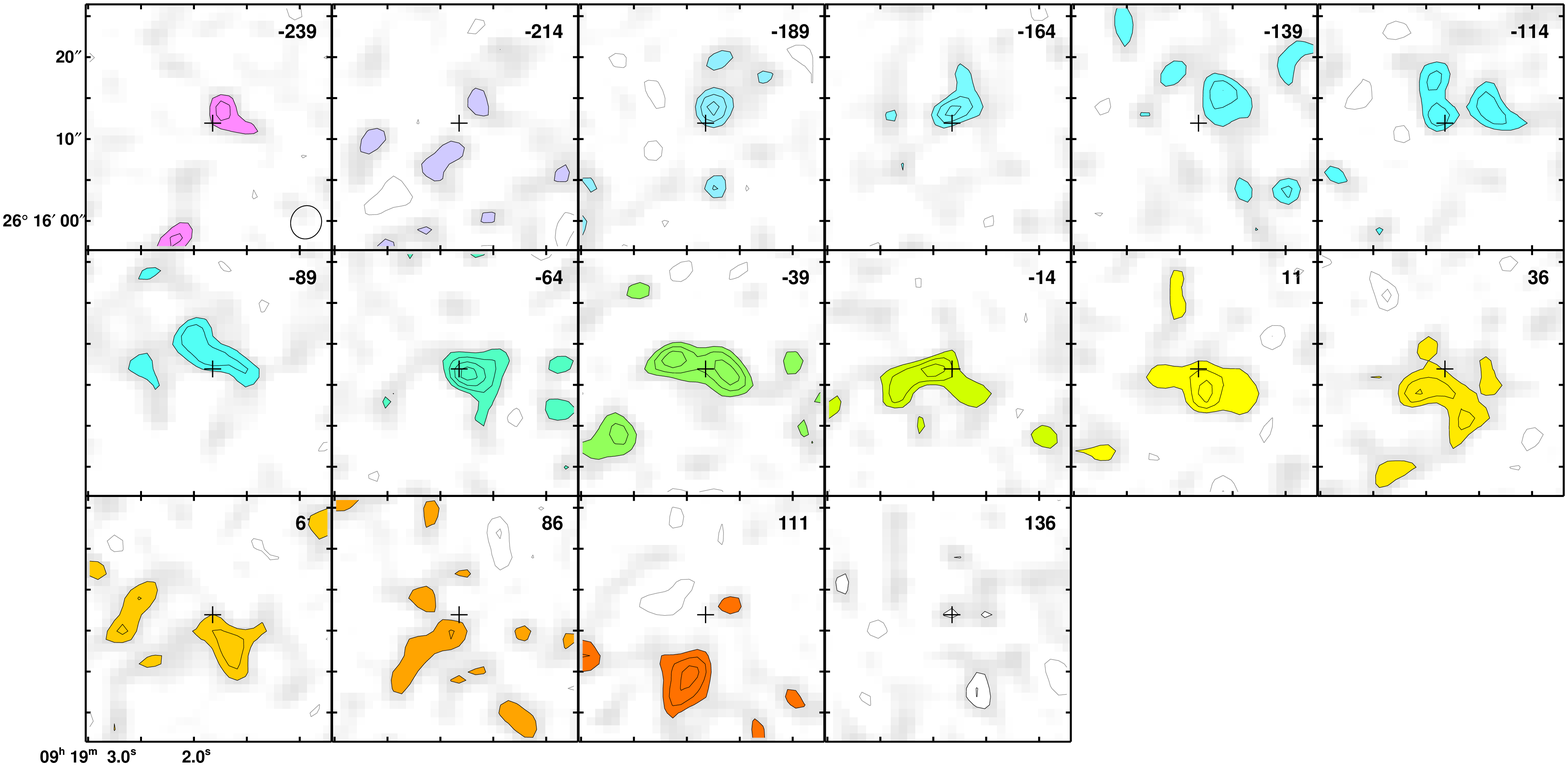}}
\caption{{\bf NGC~2824} is a field regular rotator ($M_K$ = -22.93) with a ring stellar morphology. It contains a dust disc.  The moment0 peak is 8.6 Jy beam$^{-1}$ \kms.}
\end{figure*}

\clearpage
\begin{figure*}
\centering
\subfloat{\includegraphics[height=2.2in,clip,trim=0.4cm 3.2cm 0cm 2.8cm]{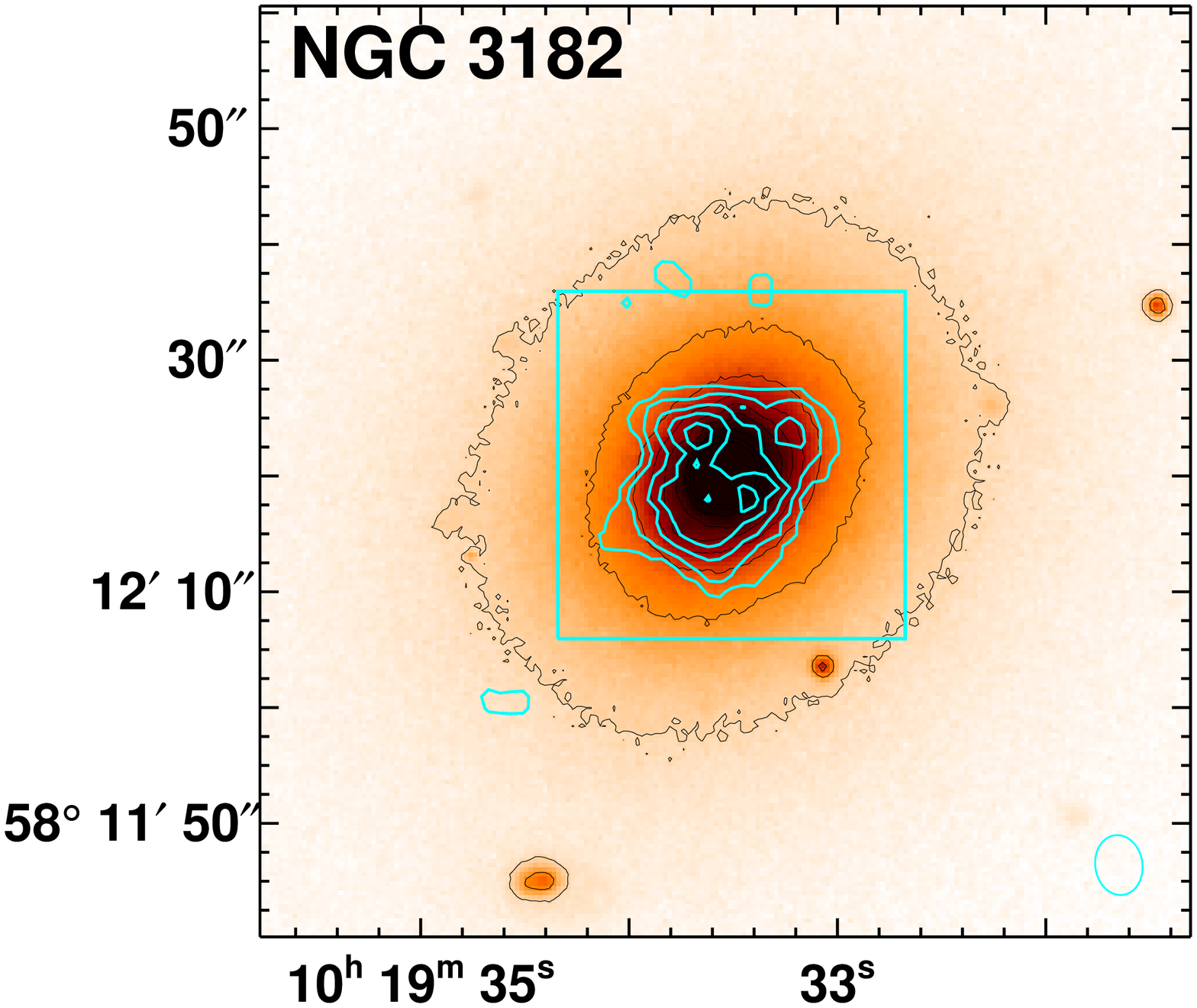}}
\subfloat{\includegraphics[height=2.2in,clip,trim=0cm 0.5cm 0.4cm 0.2cm]{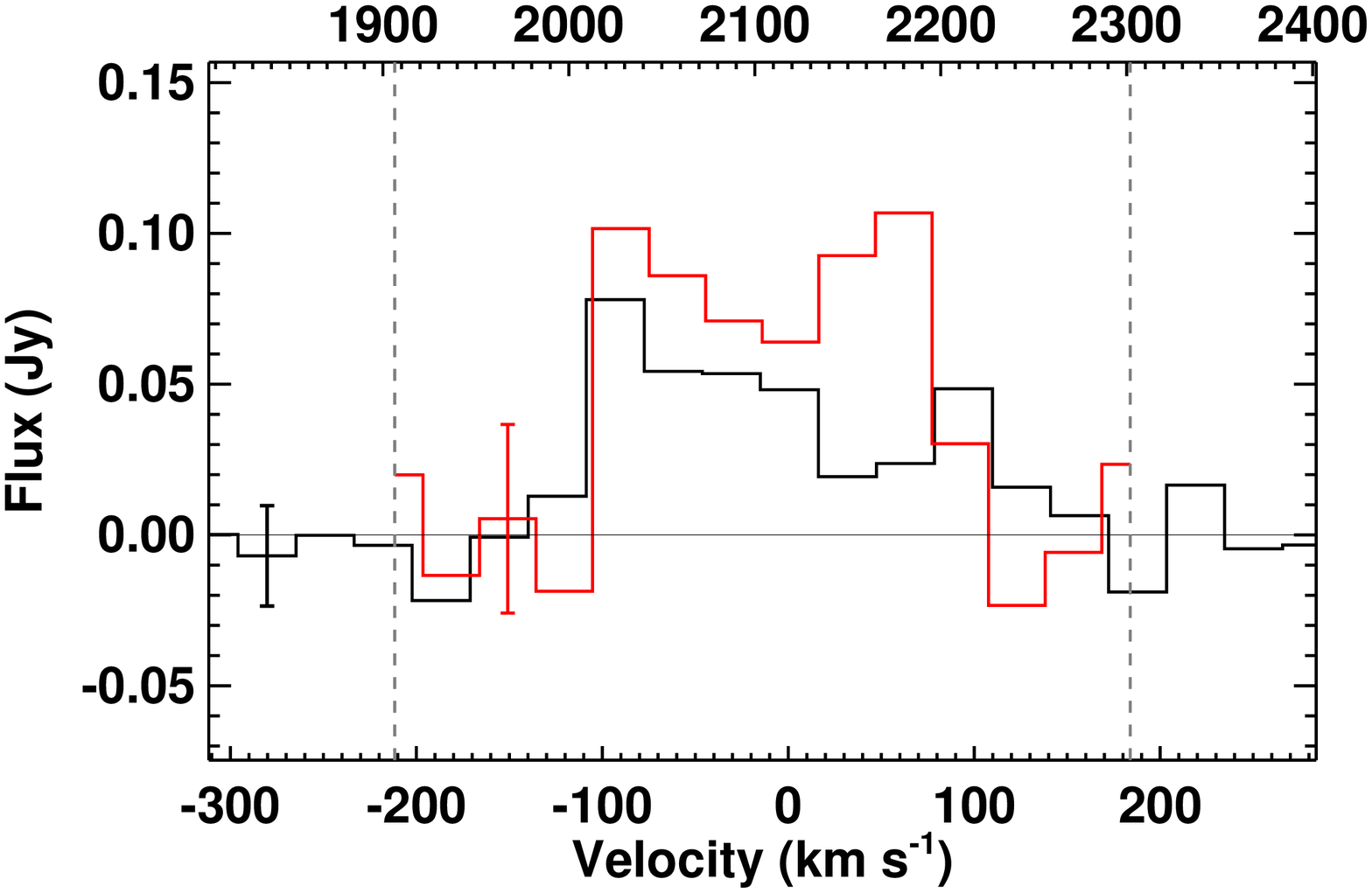}}
\end{figure*}
\begin{figure*}
\subfloat{\includegraphics[height=1.6in,clip,trim=0cm 1.4cm 0cm 2.5cm]{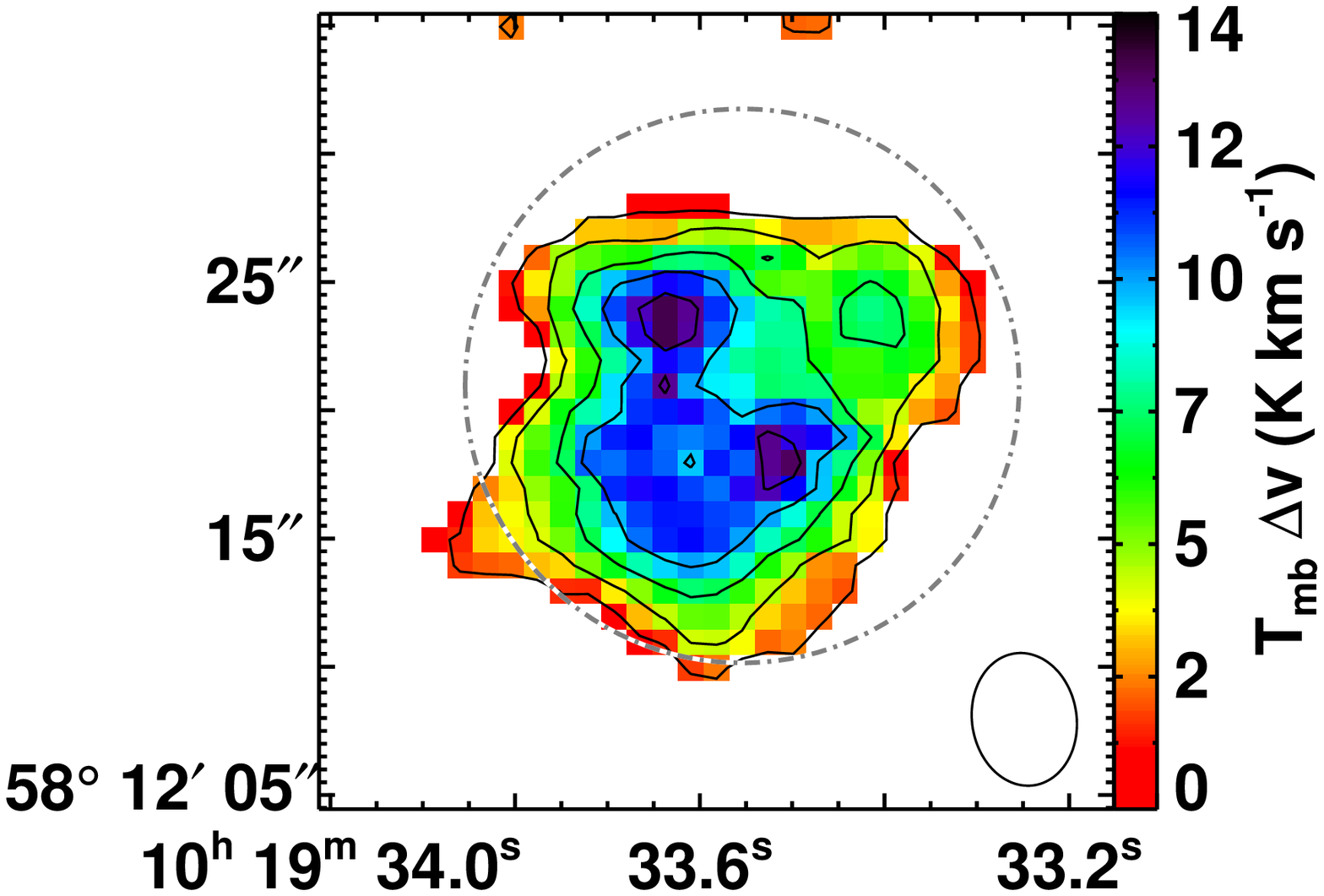}}
\subfloat{\includegraphics[height=1.6in,clip,trim=0cm 1.4cm 0cm 2.5cm]{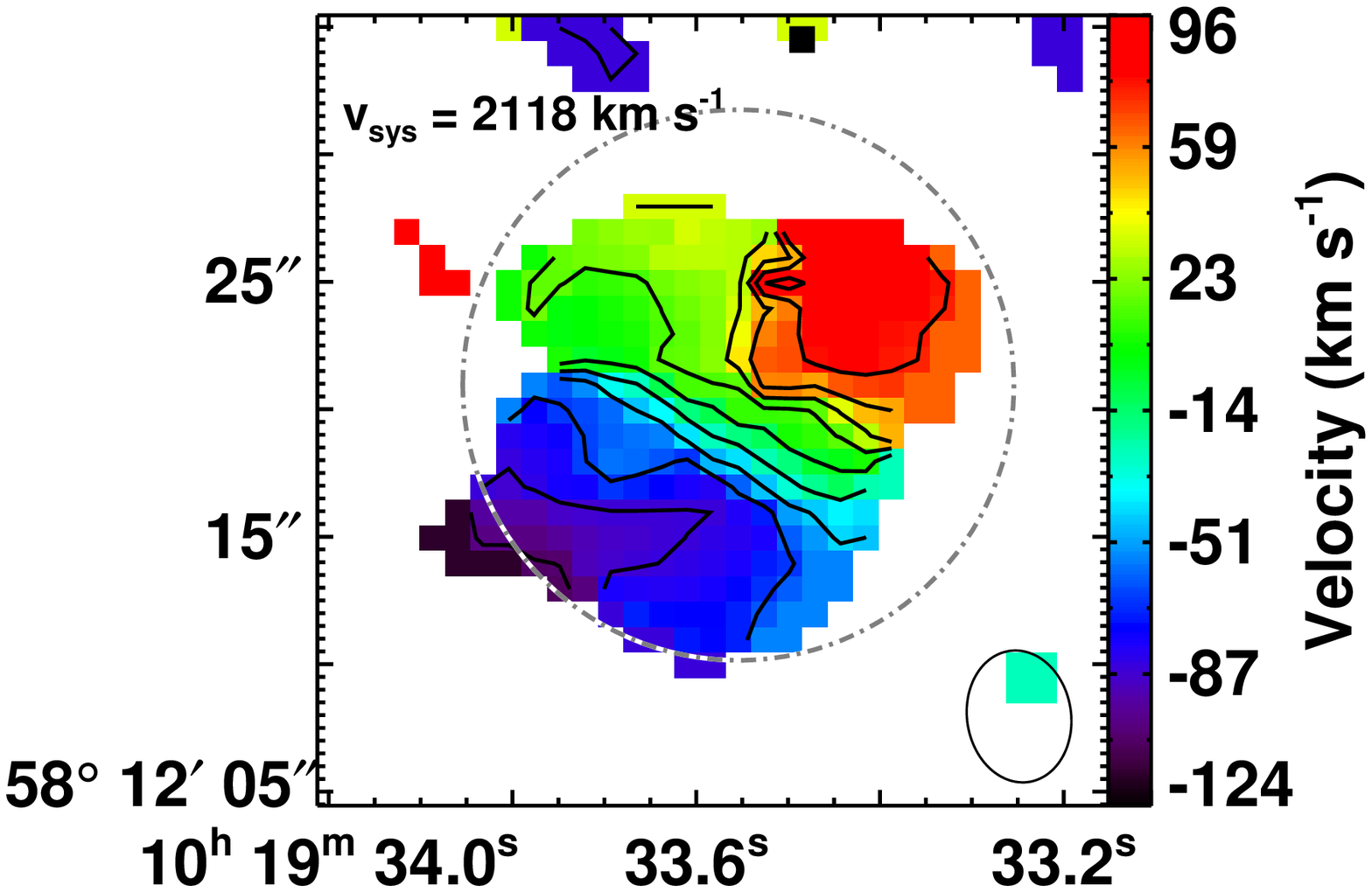}}
\subfloat{\includegraphics[height=1.6in,clip,trim=0cm 1.4cm 0cm 0.9cm]{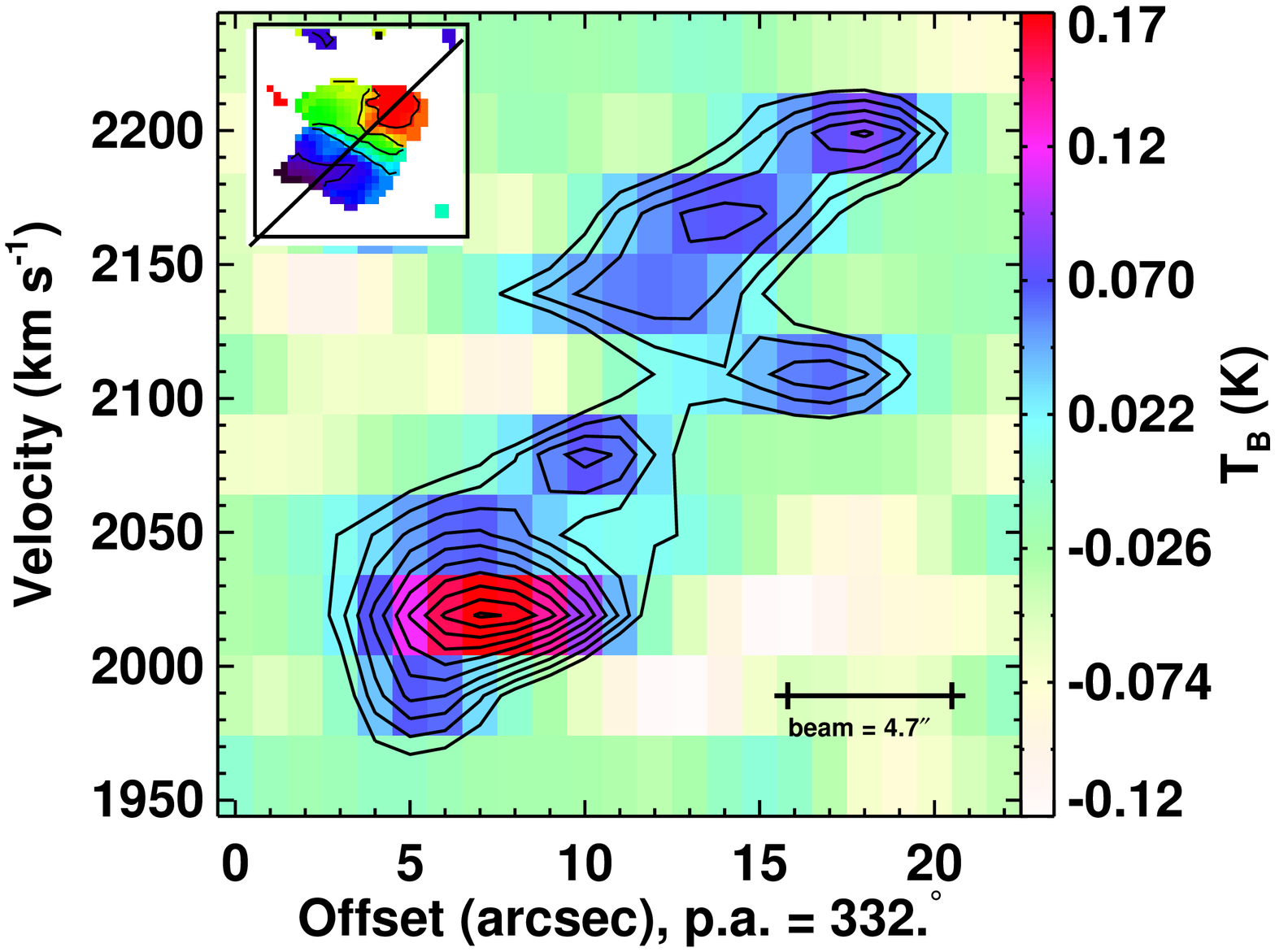}}
\end{figure*}
\begin{figure*}
\subfloat{\includegraphics[width=7in,clip,trim=1.6cm 8.8cm 4cm 2.3cm]{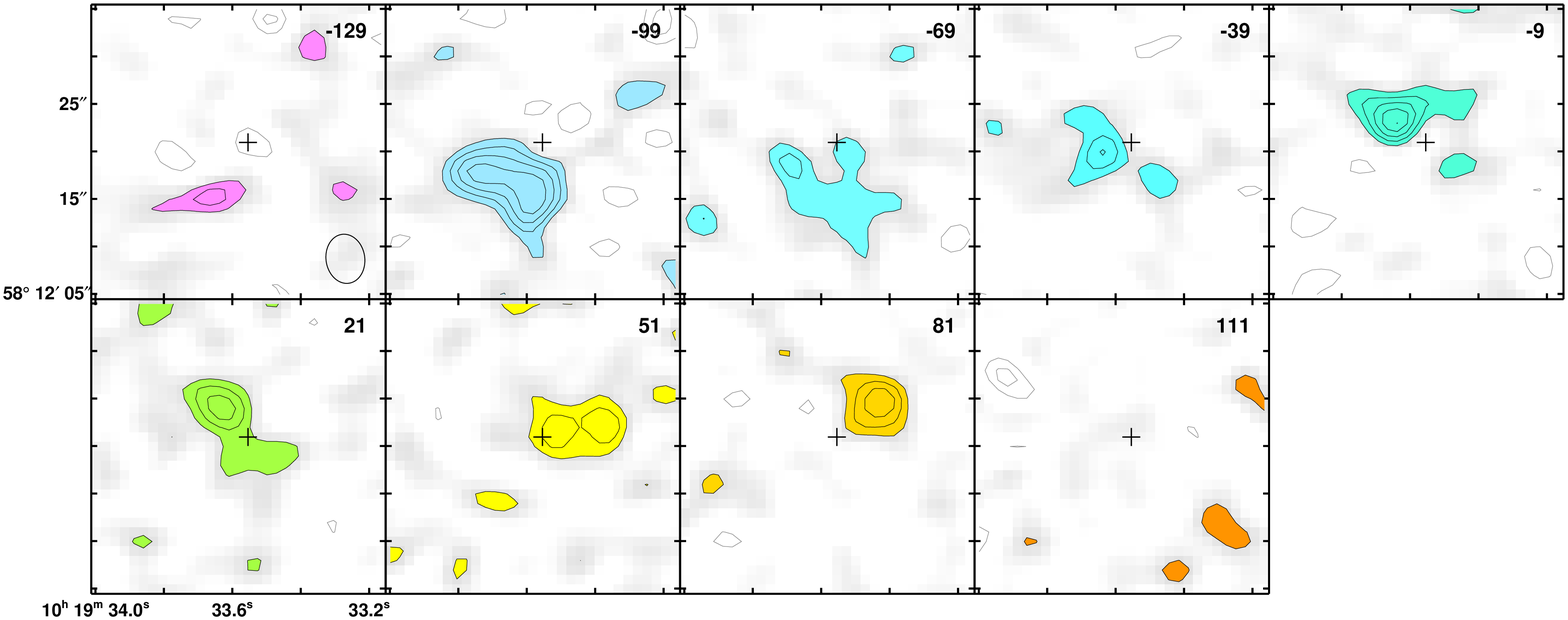}}
\caption{{\bf NGC~3182} is a field regular rotator ($M_K$ = -23.19) with normal stellar morphology and contains a dust bar and ring.  It is one of the faintest detections within the sample, thus it is likely much of the CO emission in this system is below the noise in the channel maps. The moment0 peak is 3.3 Jy beam$^{-1}$ \kms.}
\end{figure*}

\clearpage
\begin{figure*}
\centering
\subfloat{\includegraphics[height=2.2in,clip,trim=3.8cm 3.6cm 0cm 3.2cm]{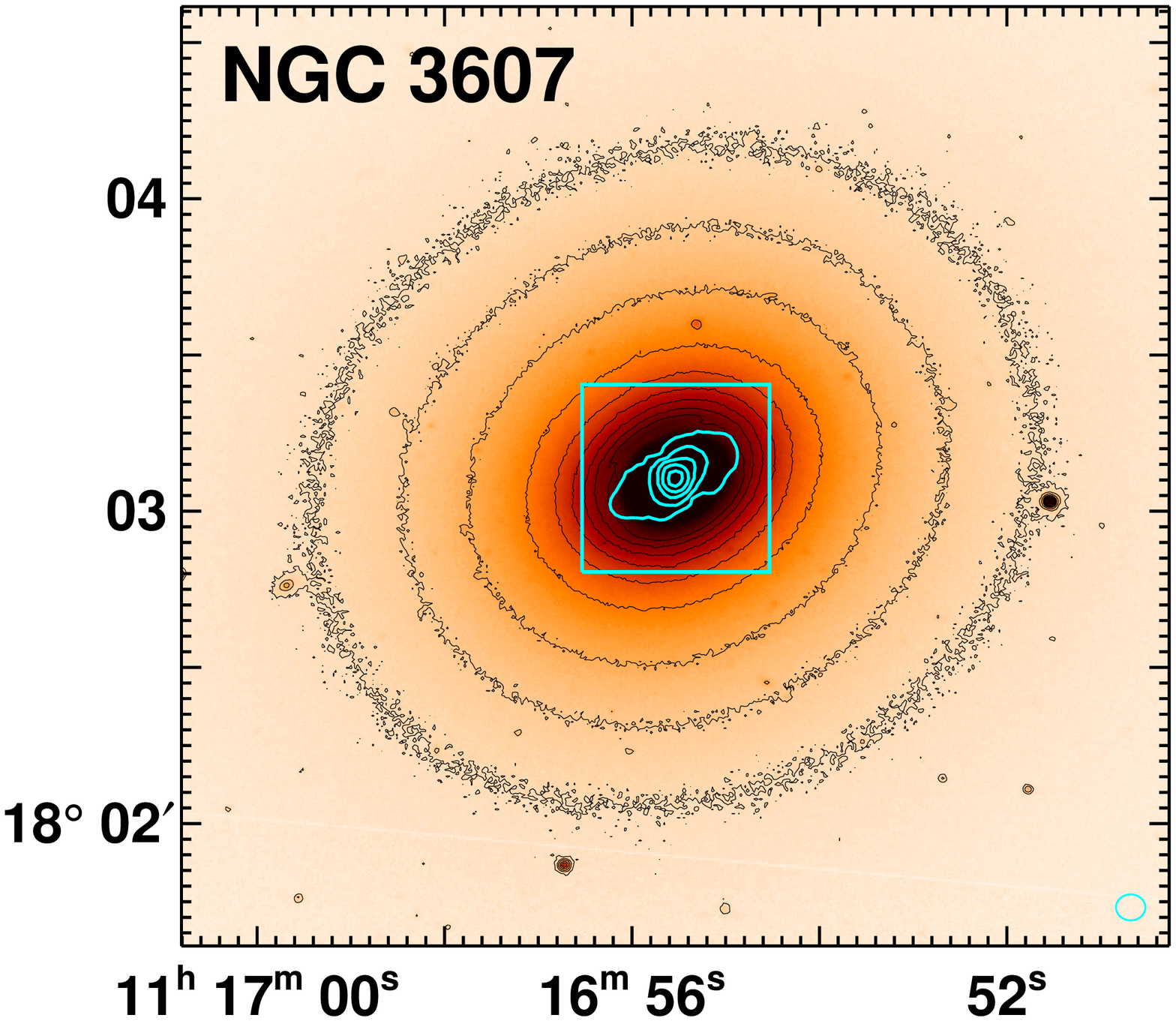}}
\subfloat{\includegraphics[height=2.2in,clip,trim=0cm 0.6cm 0.8cm 0.4cm]{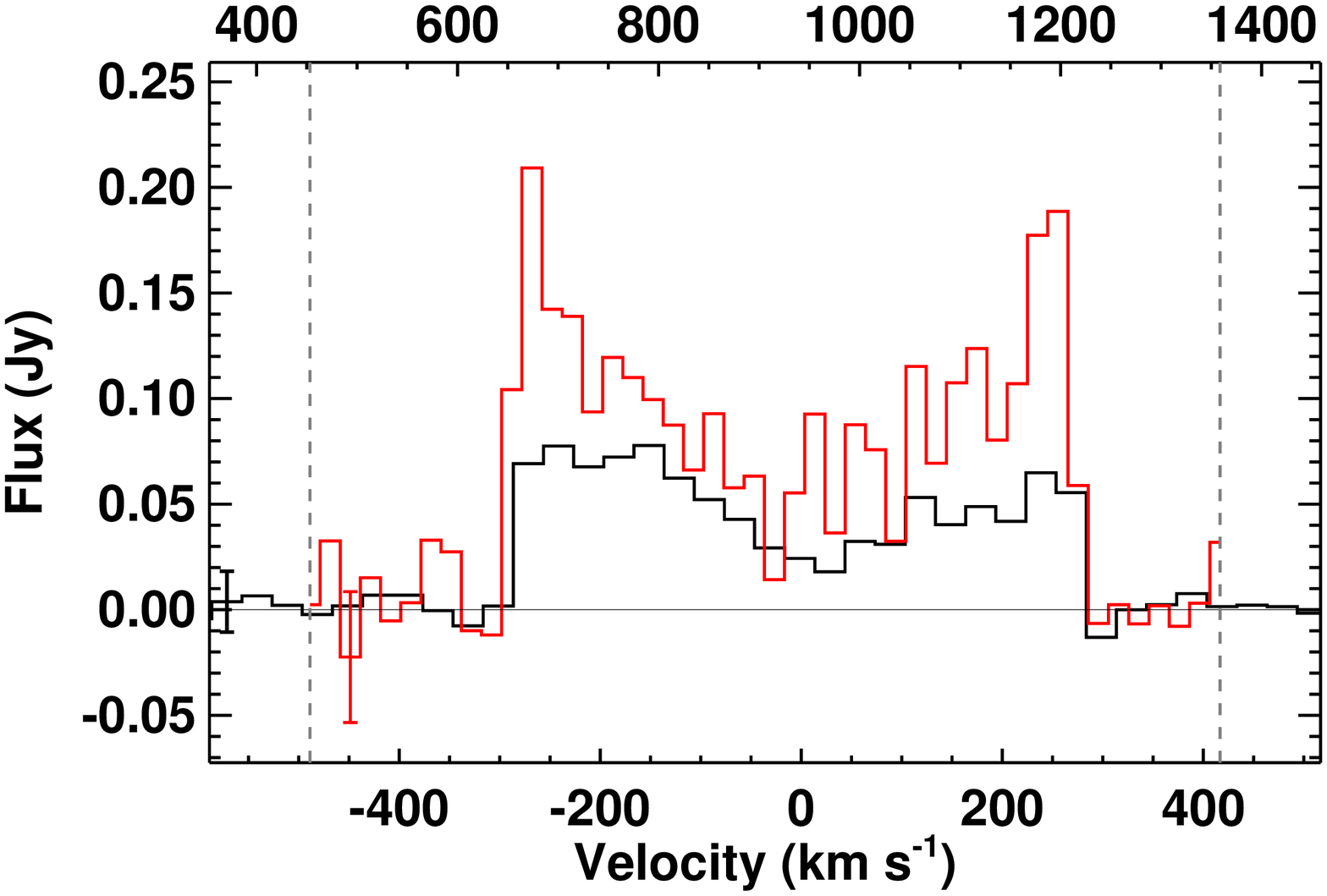}}
\end{figure*}
\begin{figure*}
\subfloat{\includegraphics[height=1.6in,clip,trim=0.1cm 1.4cm 0cm 2.5cm]{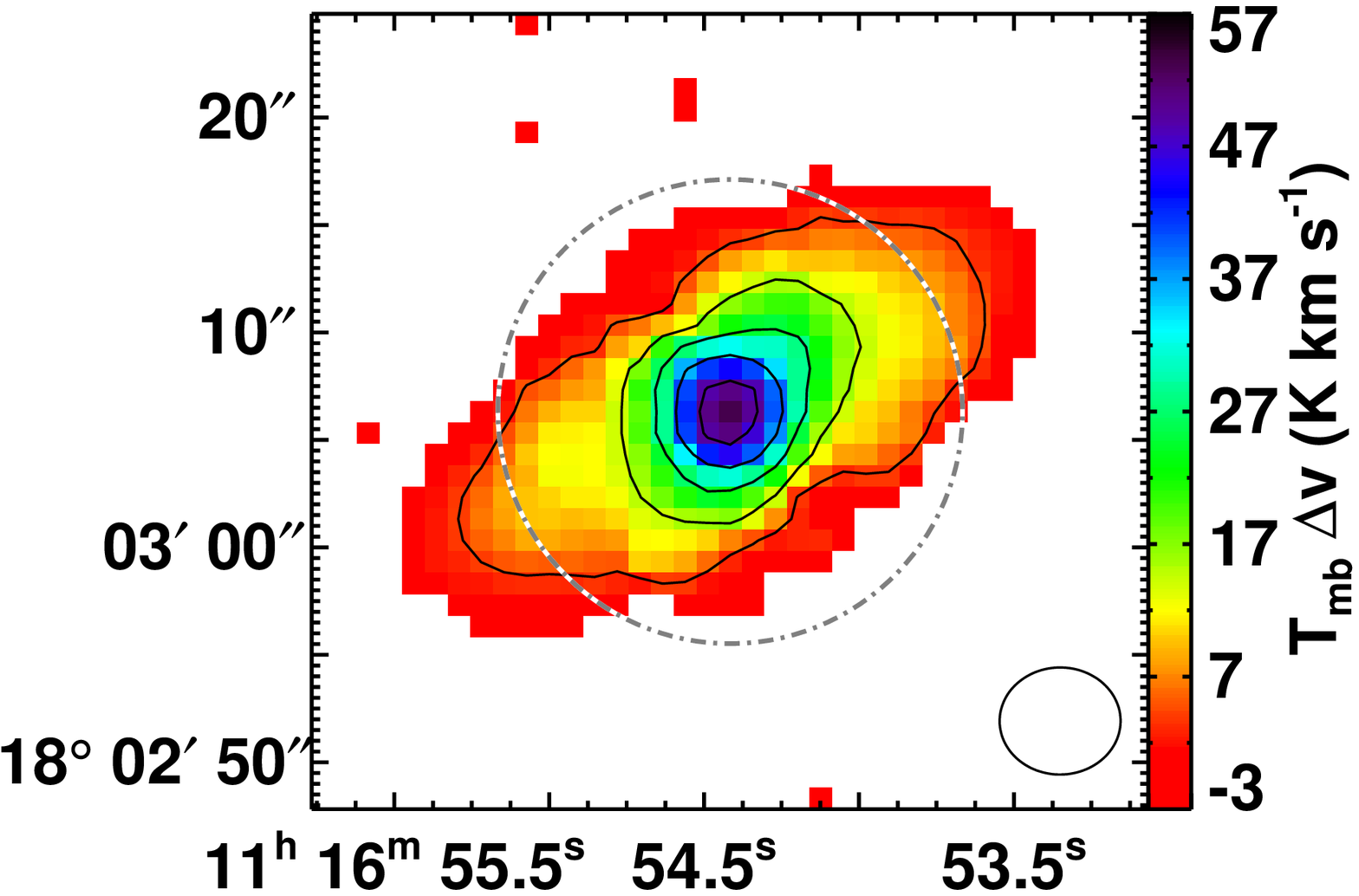}}
\subfloat{\includegraphics[height=1.6in,clip,trim=0cm 1.4cm 0cm 2.5cm]{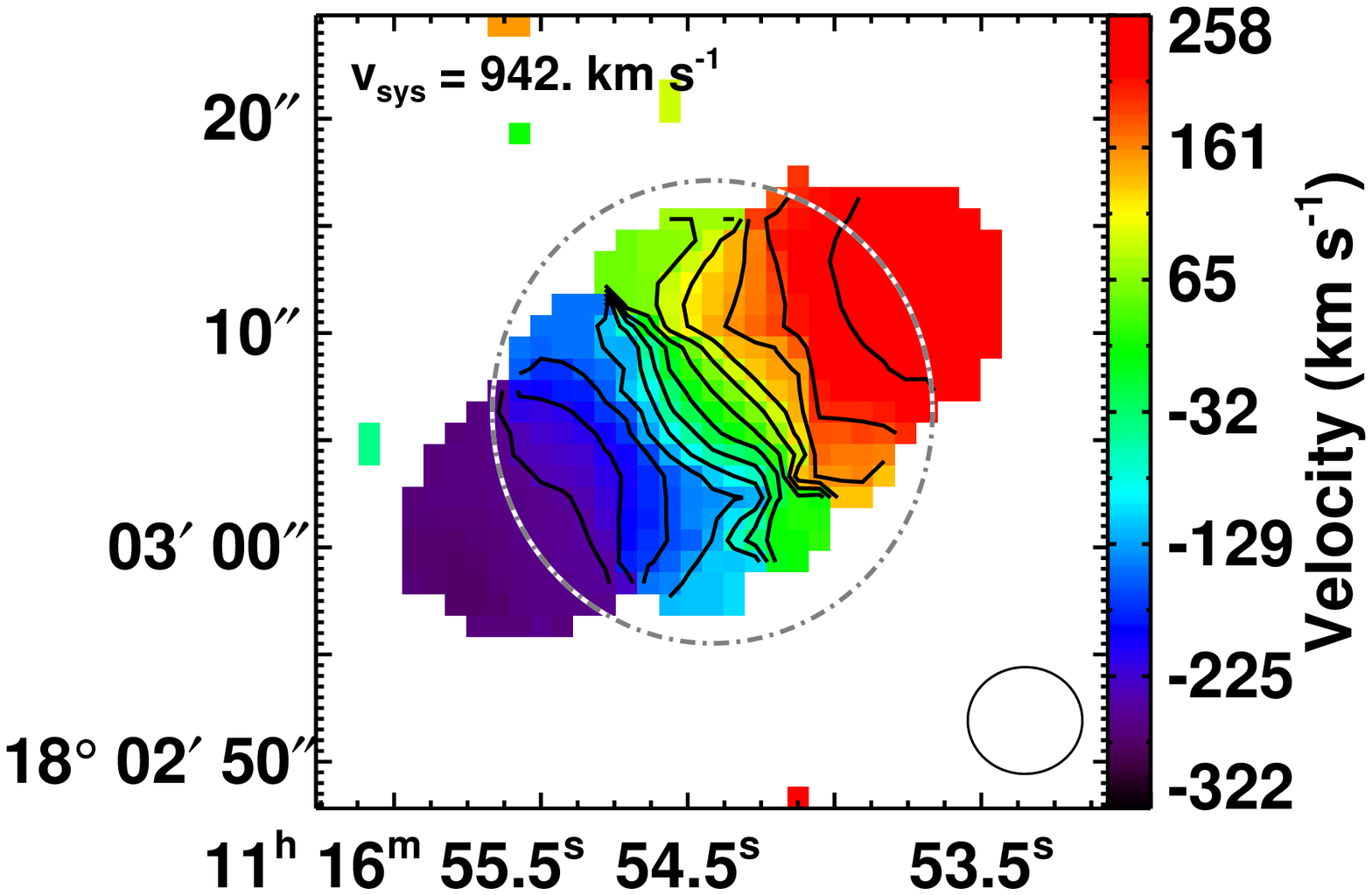}}
\subfloat{\includegraphics[height=1.6in,clip,trim=0cm 1.4cm 0cm 0.9cm]{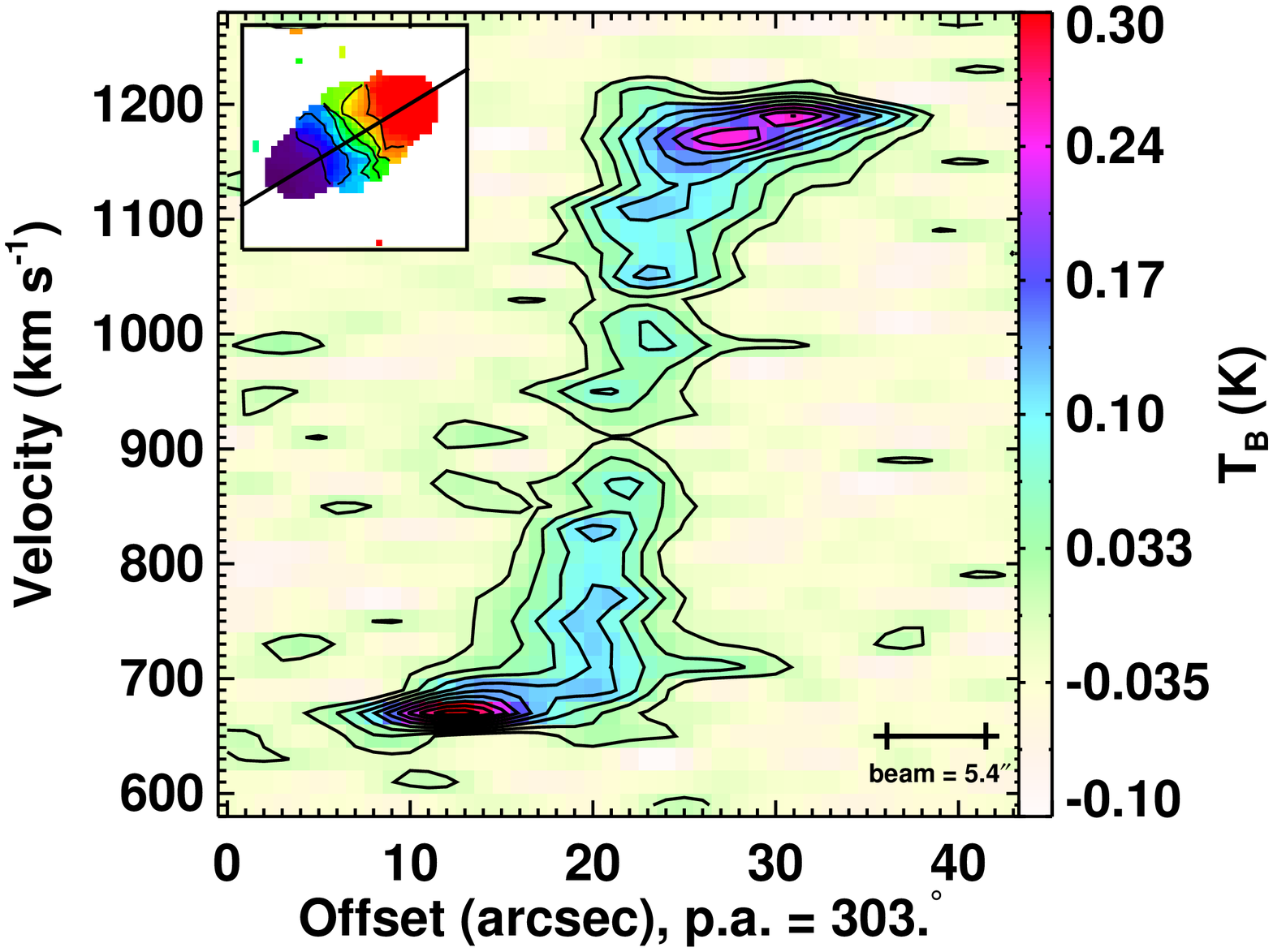}}
\end{figure*}
\begin{figure*}
\subfloat{\includegraphics[width=7in,clip,trim=1.6cm 9.2cm 9.5cm 1cm]{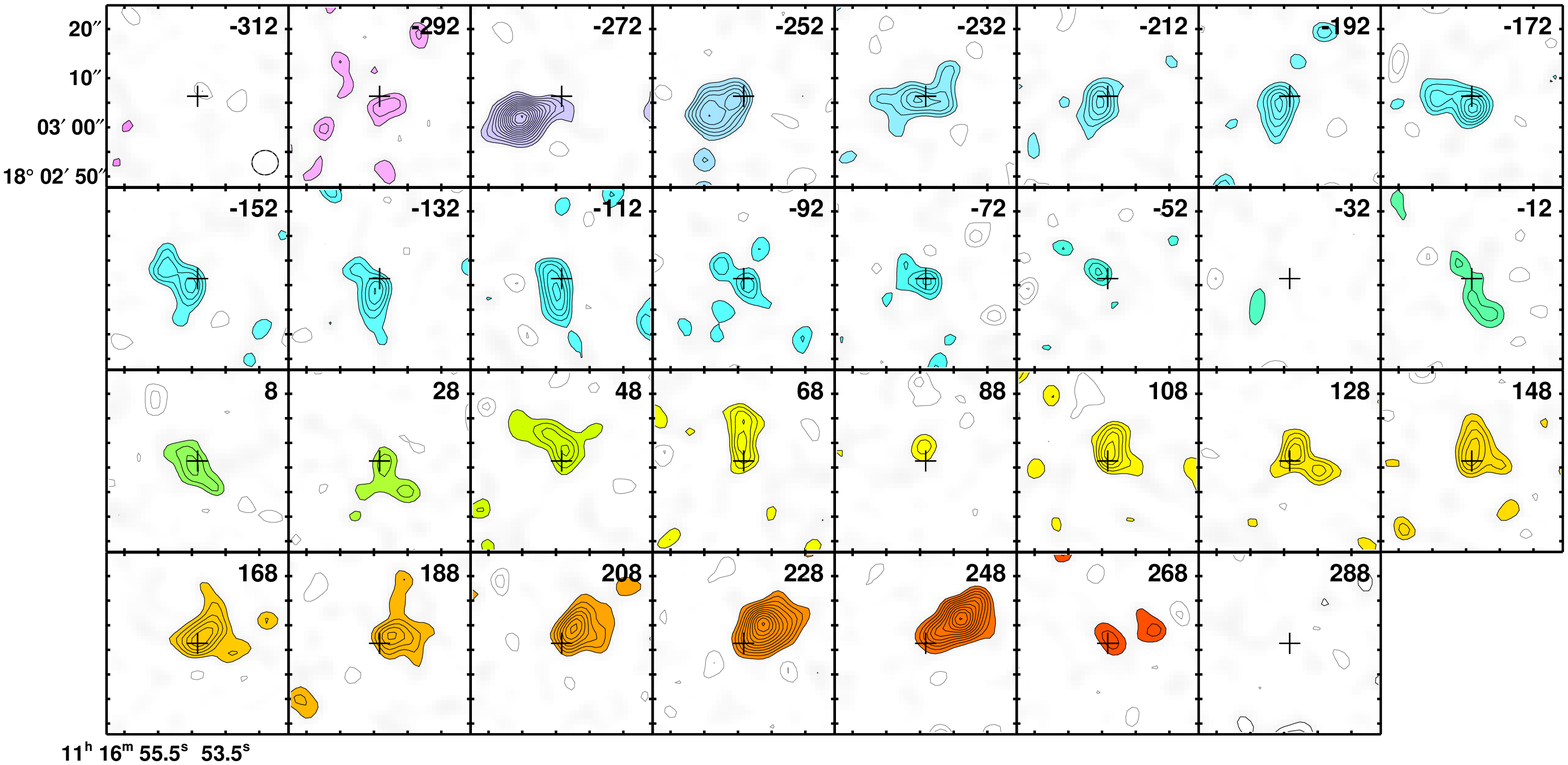}}
\caption{{\bf NGC~3607} is a field regular rotator ($M_K$ = -24.74) with normal stellar morphology.  It contains a dust disc.  The moment0 peak is 17 Jy beam$^{-1}$ \kms.  The moment1 contours are placed at 40\kms\ intervals and the PVD contours are placed at $3\sigma$ intervals.}
\end{figure*}

\clearpage
\begin{figure*}
\centering
\subfloat{\includegraphics[height=2.2in,clip,trim=2.1cm 3.2cm 0cm 2.8cm]{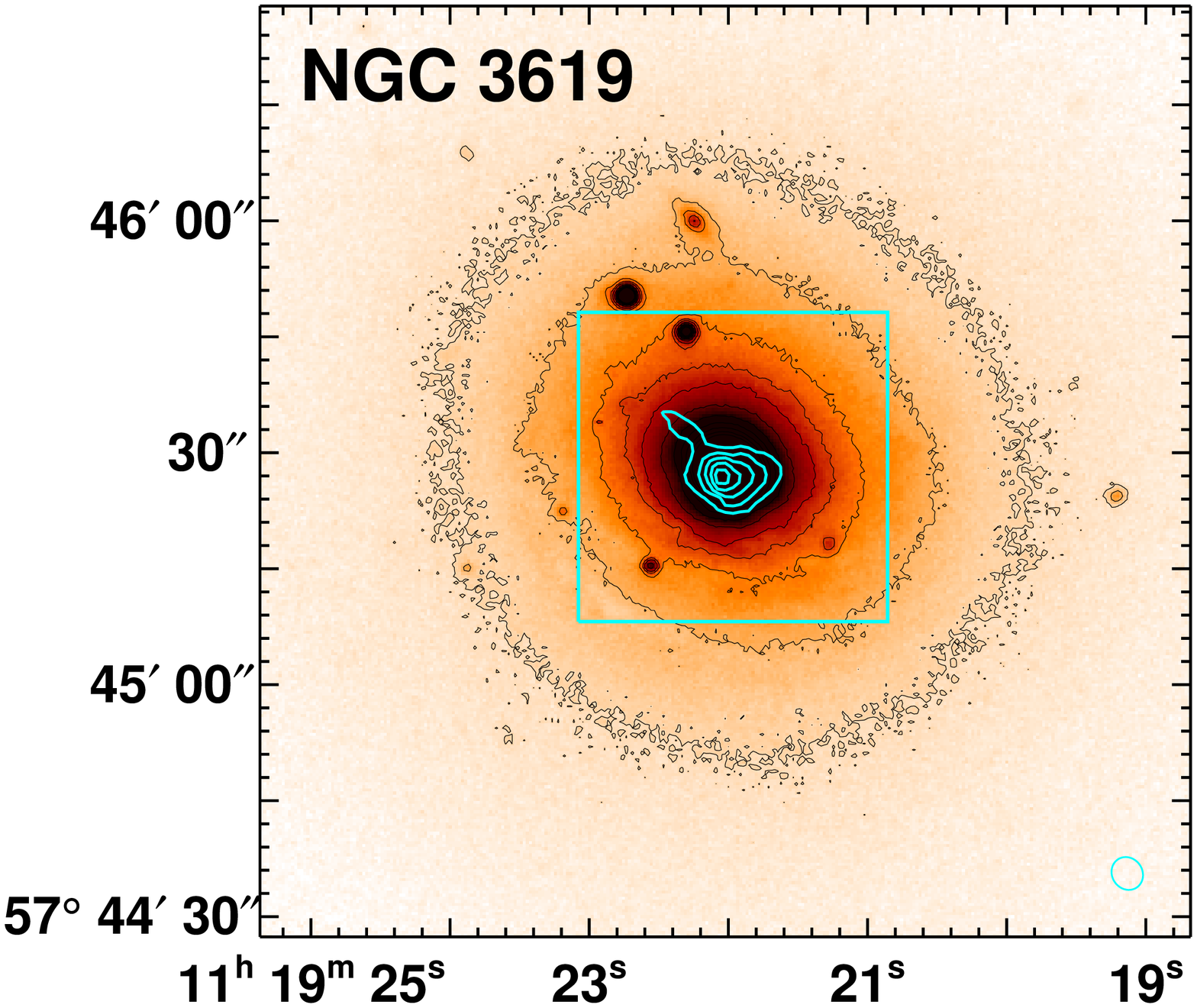}}
\subfloat{\includegraphics[height=2.2in,clip,trim=0cm 0.6cm 0.8cm 0.3cm]{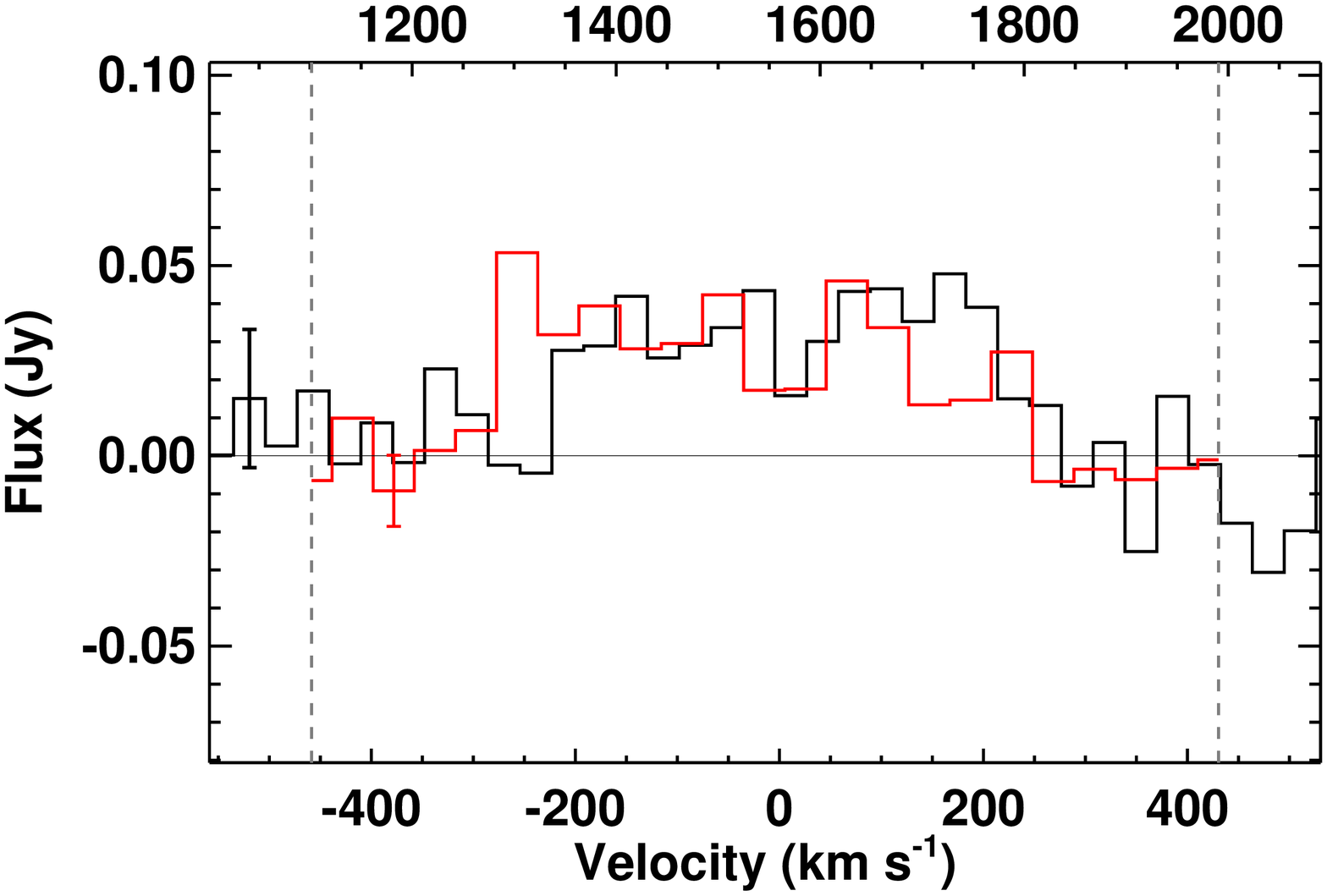}}
\end{figure*}
\begin{figure*}
\subfloat{\includegraphics[height=1.6in,clip,trim=0.1cm 1.4cm 0cm 2.4cm]{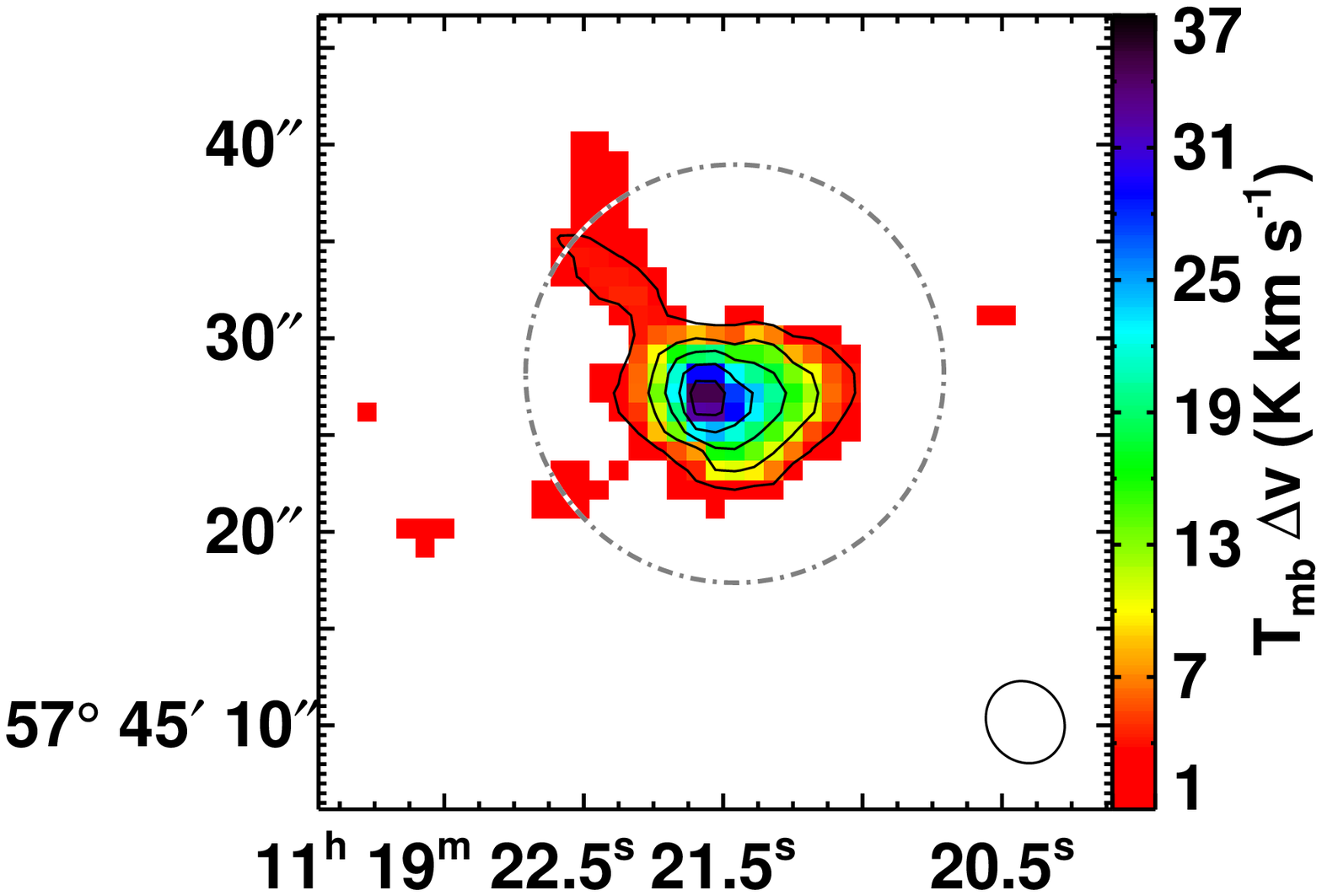}}
\subfloat{\includegraphics[height=1.6in,clip,trim=0cm 1.4cm 0cm 2.4cm]{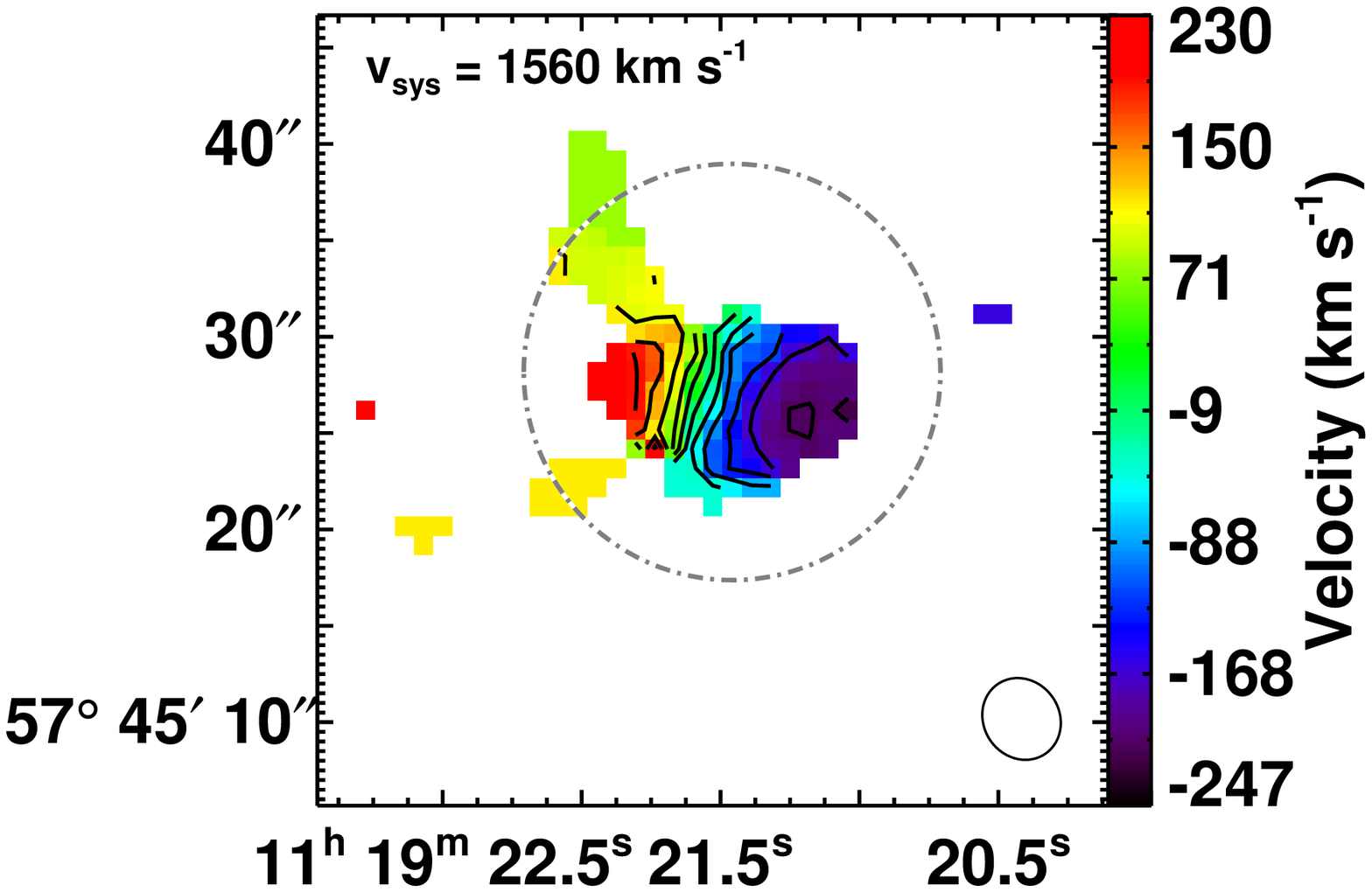}}
\subfloat{\includegraphics[height=1.6in,clip,trim=0cm 1.4cm 0cm 0.9cm]{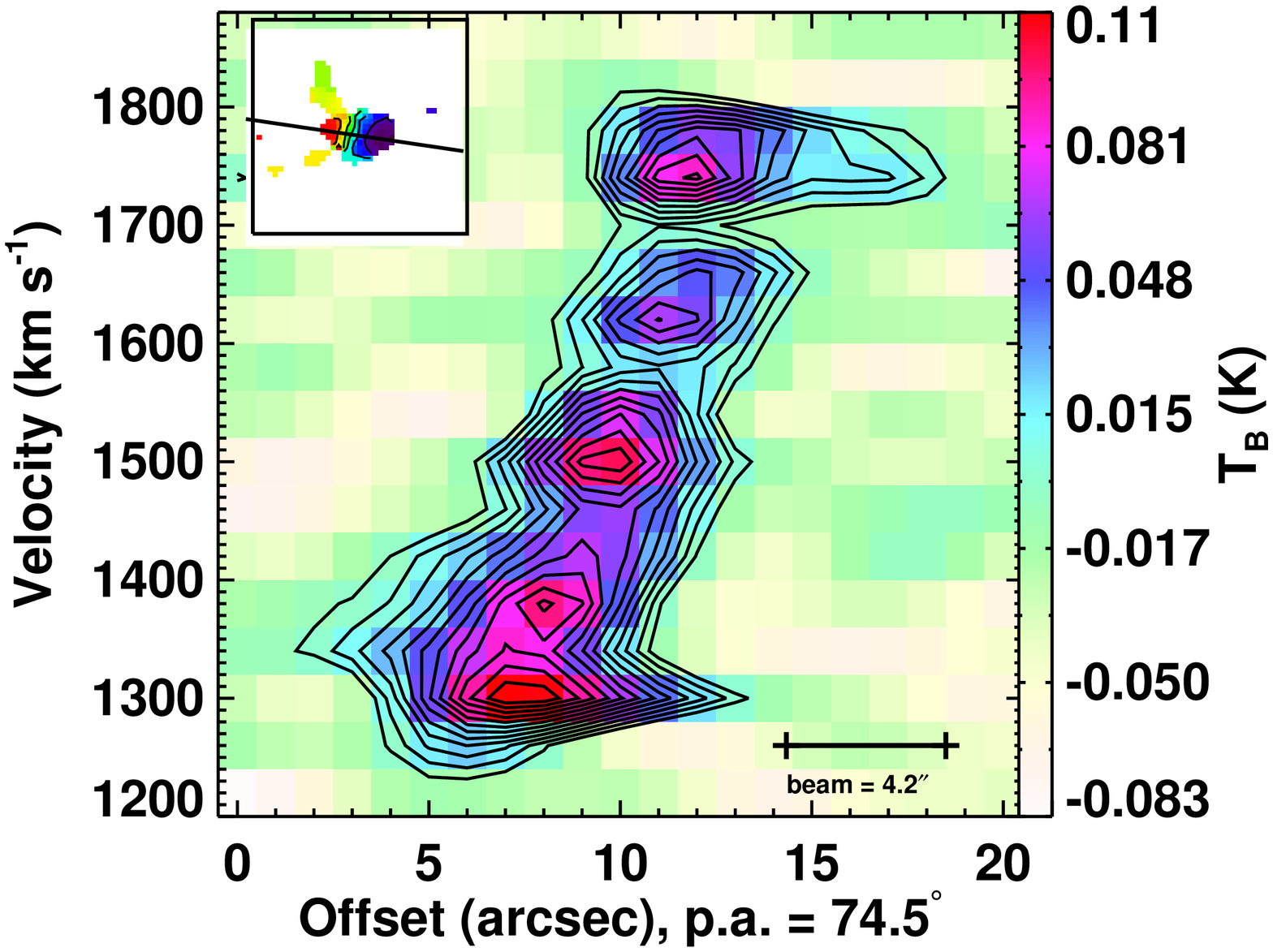}}
\end{figure*}
\begin{figure*}
\subfloat{\includegraphics[width=7in,clip,trim=1.6cm 11cm 7.5cm 0.6cm]{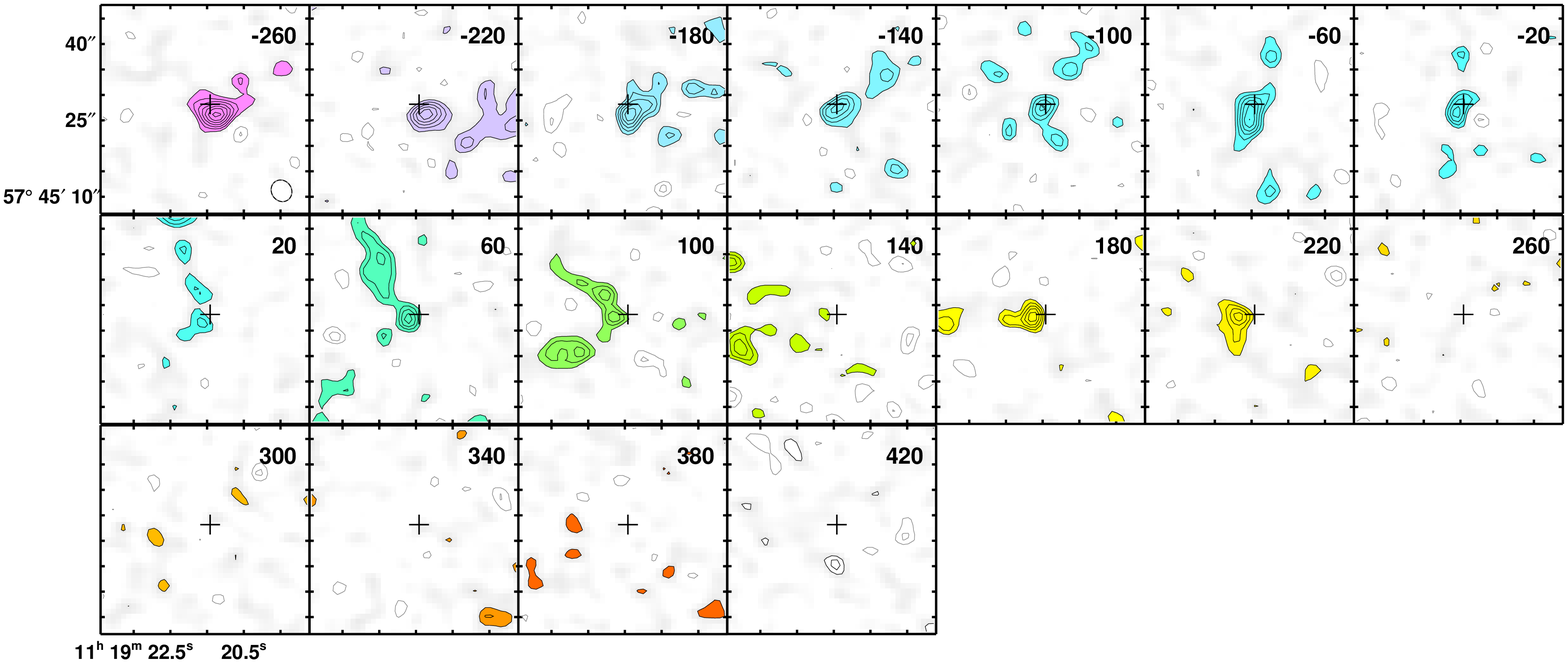}}
\caption{{\bf NGC~3619} is a field regular rotator ($M_K$ = -23.57) with a shell stellar morphology.  It contains dust filaments, bars and rings.  It is very likely that CARMA has resolved out what is possibly an extended gas disc.  Observations that are sensitive to larger size scales is required to confirm this.  The moment0 peak is 6.9 Jy beam$^{-1}$ \kms.  The moment1 contours are placed at 40\kms\ intervals.}
\end{figure*}

\clearpage
\begin{figure*}
\centering
\subfloat{\includegraphics[height=2.2in,clip,trim=3.8cm 3.2cm 0cm 2.8cm]{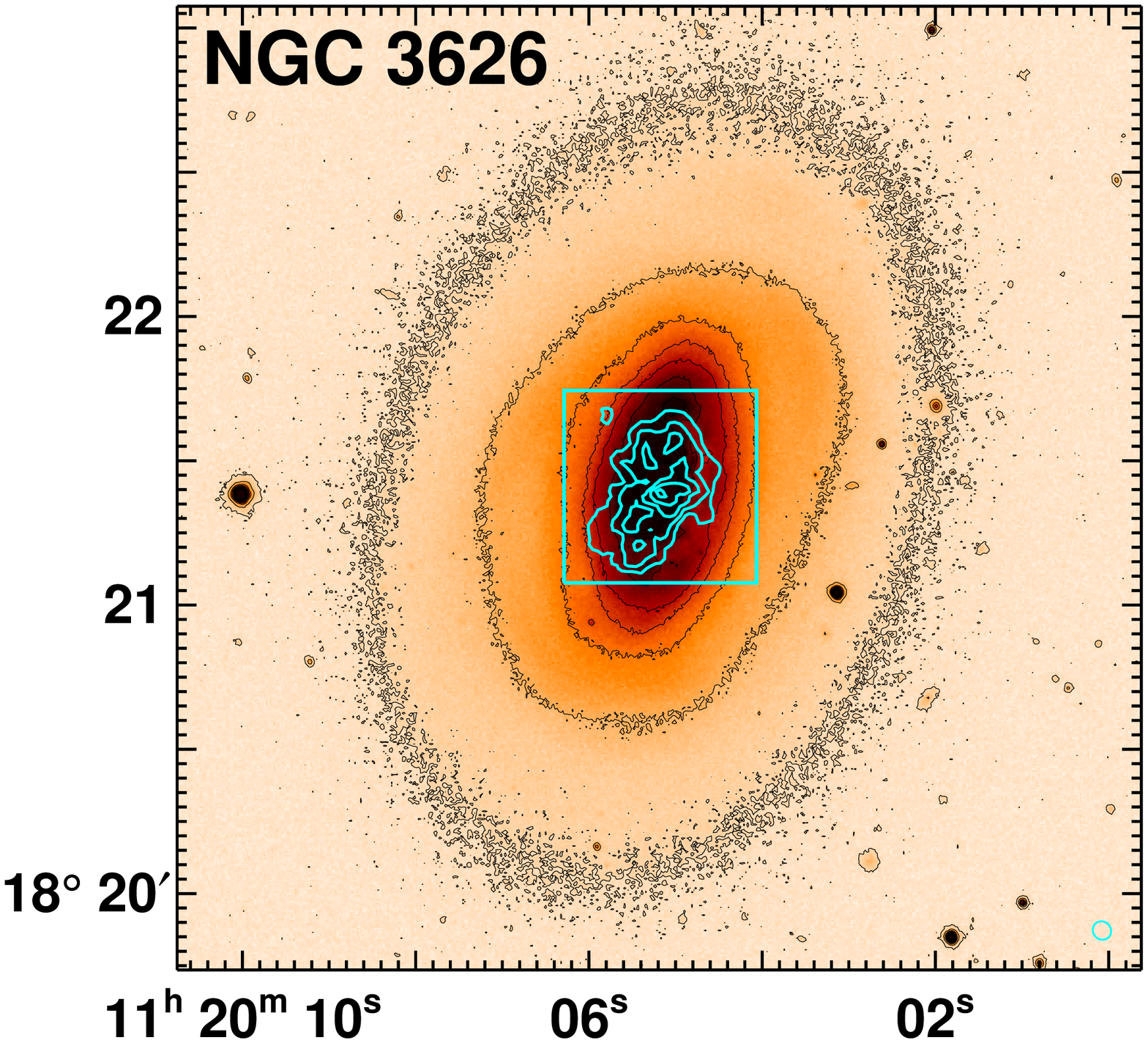}}
\subfloat{\includegraphics[height=2.2in,clip,trim=0cm 0.5cm 0.4cm 0.2cm]{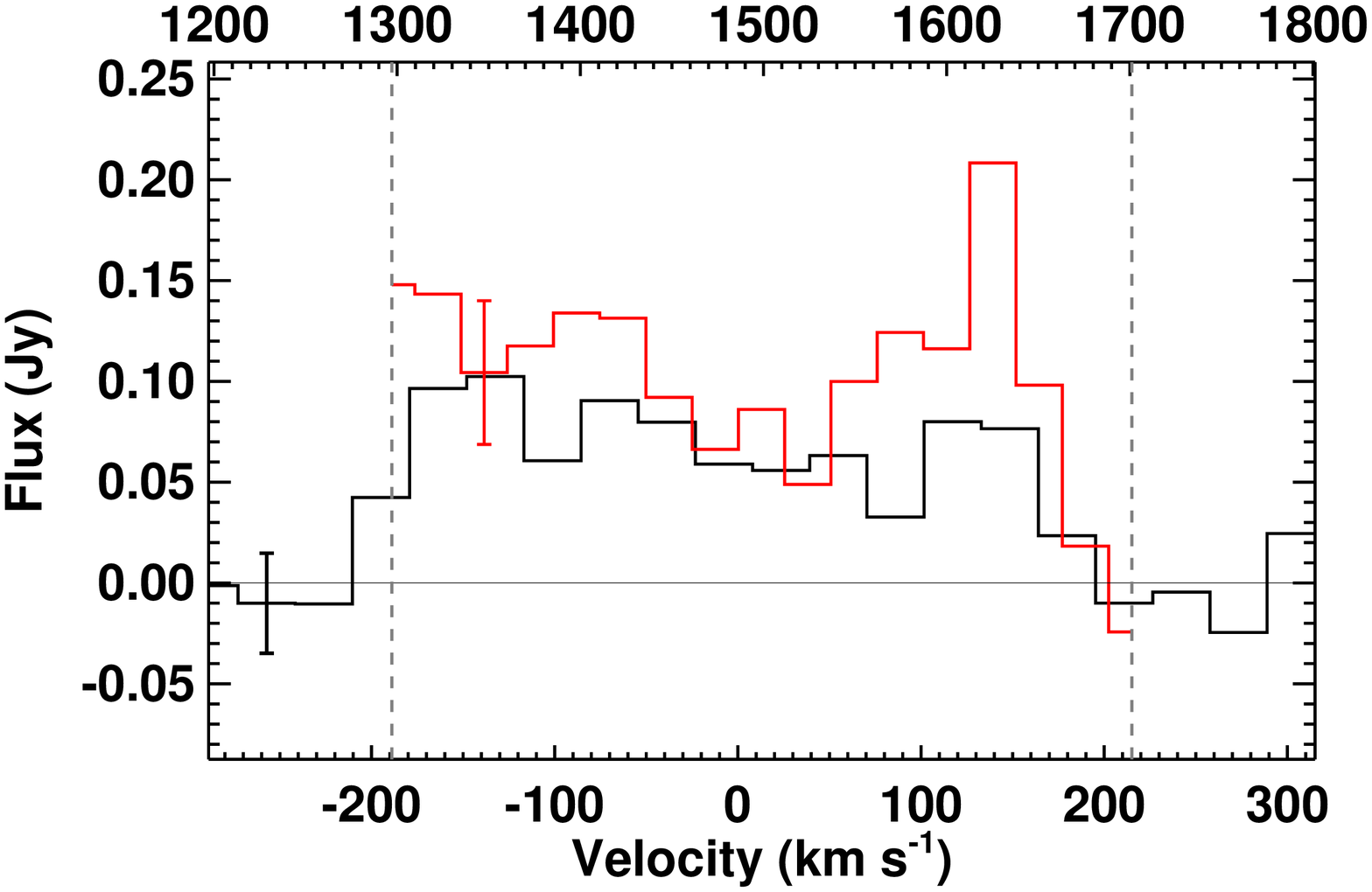}}
\end{figure*}
\begin{figure*}
\subfloat{\includegraphics[height=1.6in,clip,trim=0.1cm 1.4cm 0.4cm 2.4cm]{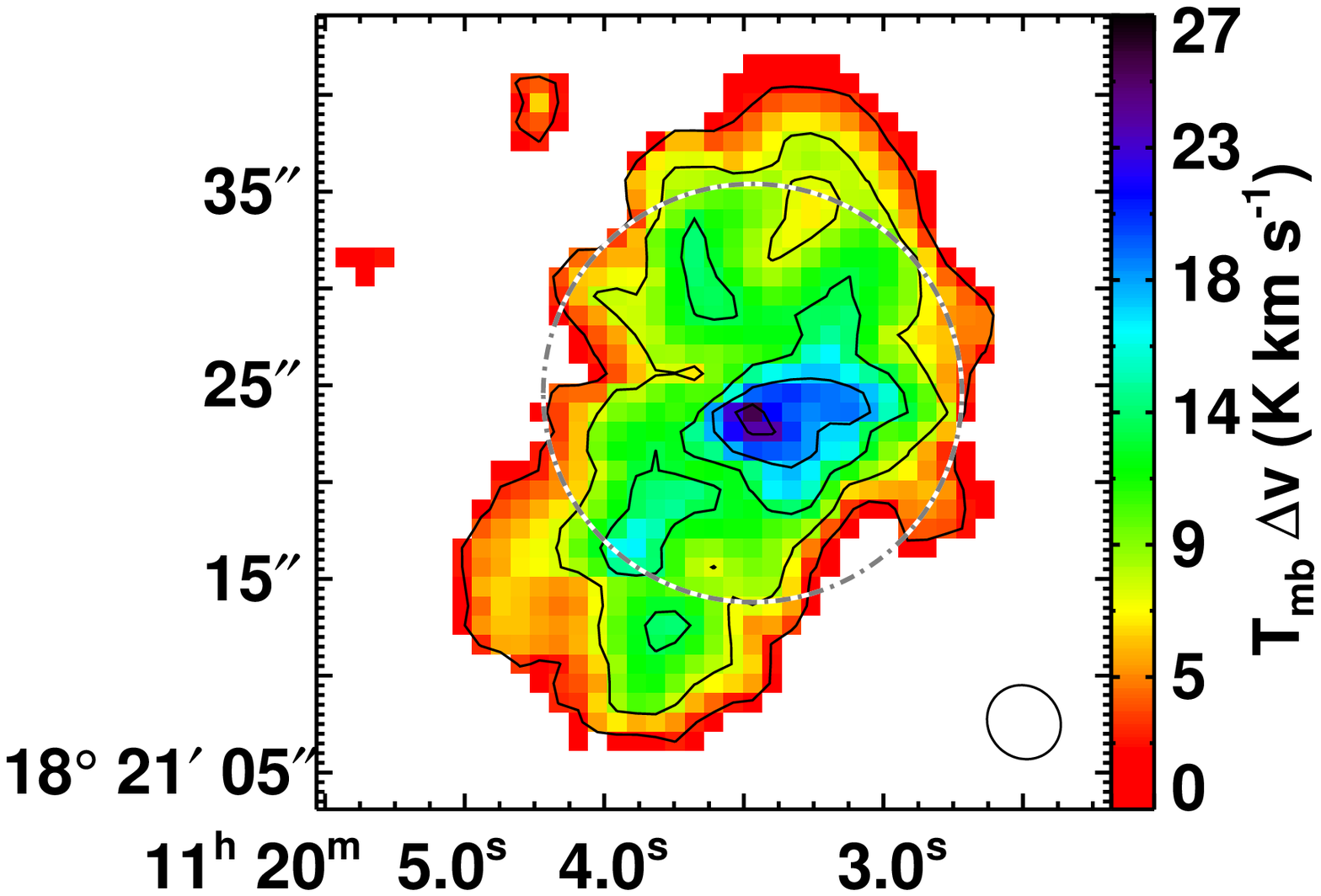}}
\subfloat{\includegraphics[height=1.6in,clip,trim=0.3cm 1.4cm 0cm 2.4cm]{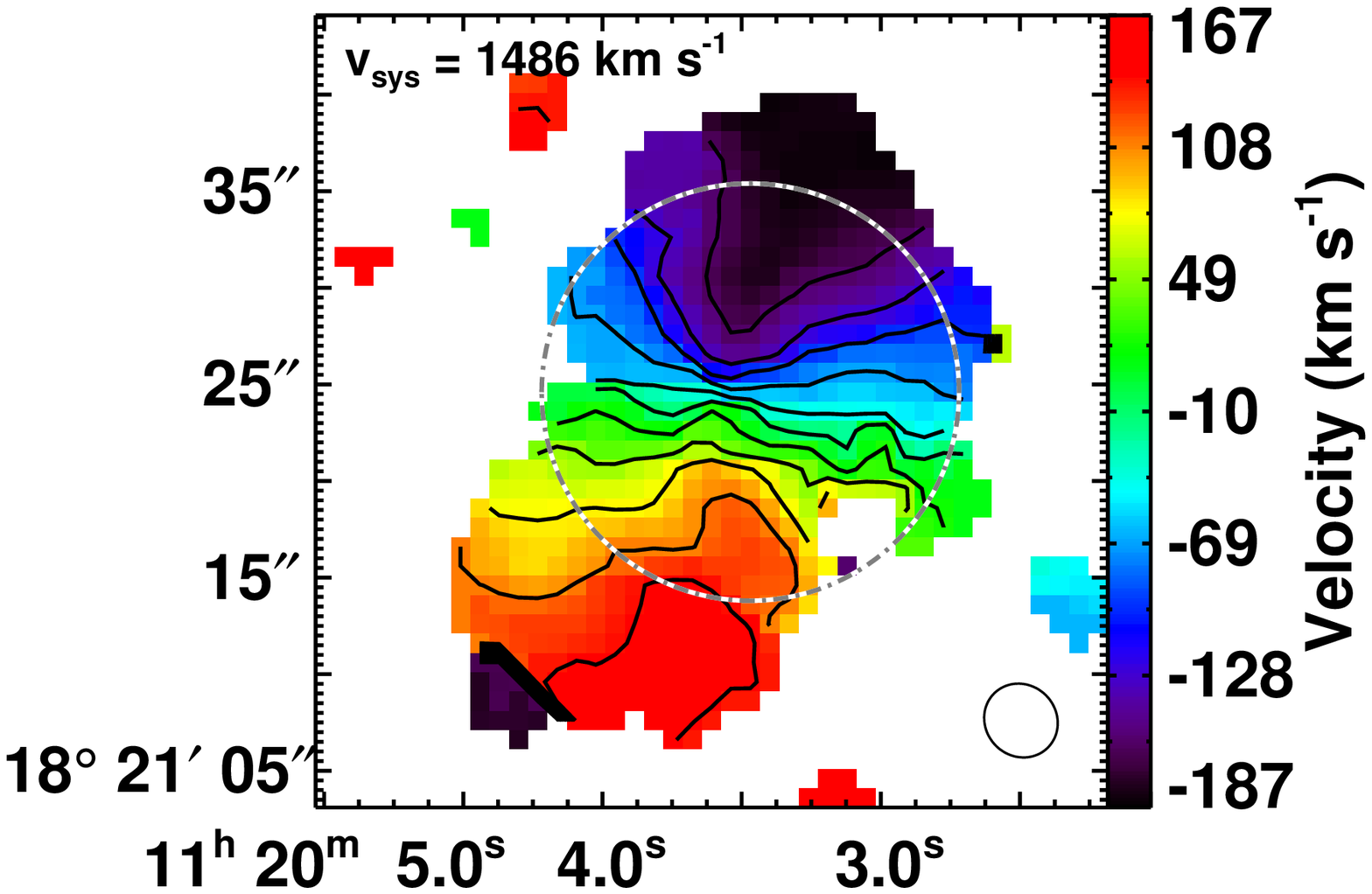}}
\subfloat{\includegraphics[height=1.6in,clip,trim=0cm 1.4cm 0cm 0.9cm]{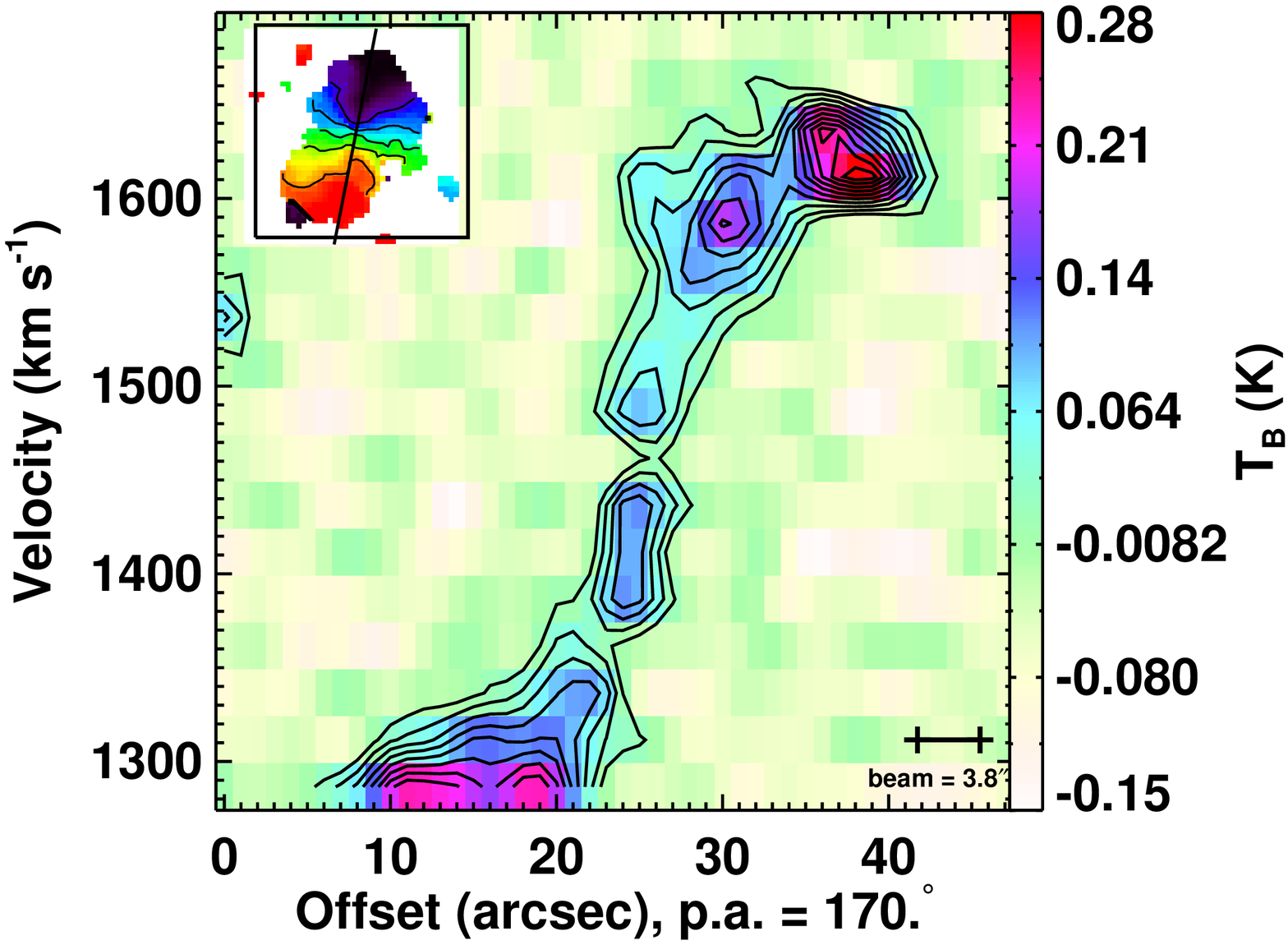}}
\end{figure*}
\begin{figure*}
\subfloat{\includegraphics[width=7in,clip,trim=1.6cm 7cm 8cm 2.6cm]{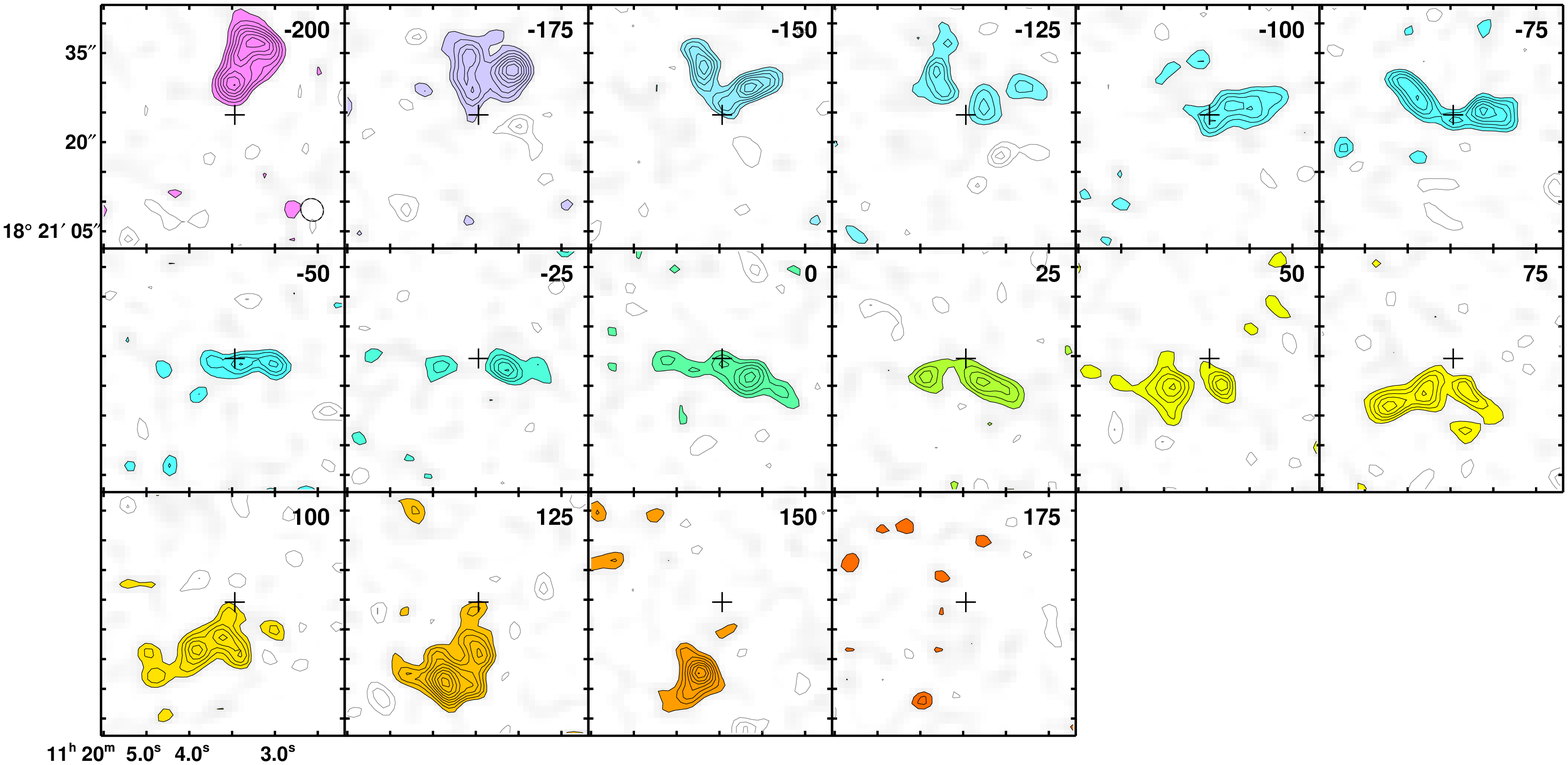}}
\caption{{\bf NGC~3626} is a group regular rotator ($M_K$ = -23.30) that includes a double maximum velocity feature, with ring stellar morphology.  It contains a dust disc. Unfortunately the central velocity of the observations were offset from $v_{\rm sys}$, and due to its large linewidth, a small amount of the blueshifted spectrum was not within the 420 \kms\ CARMA window. The moment0 peak is 4.3 Jy beam$^{-1}$ \kms. The moment1 contours are placed at 30\kms\ intervals and PVD contours are placed at $1.5\sigma$ intervals.}
\end{figure*}

\clearpage
\begin{figure*}
\centering
\subfloat{\includegraphics[height=2.2in,clip,trim=2.2cm 3.2cm 0cm 2.7cm]{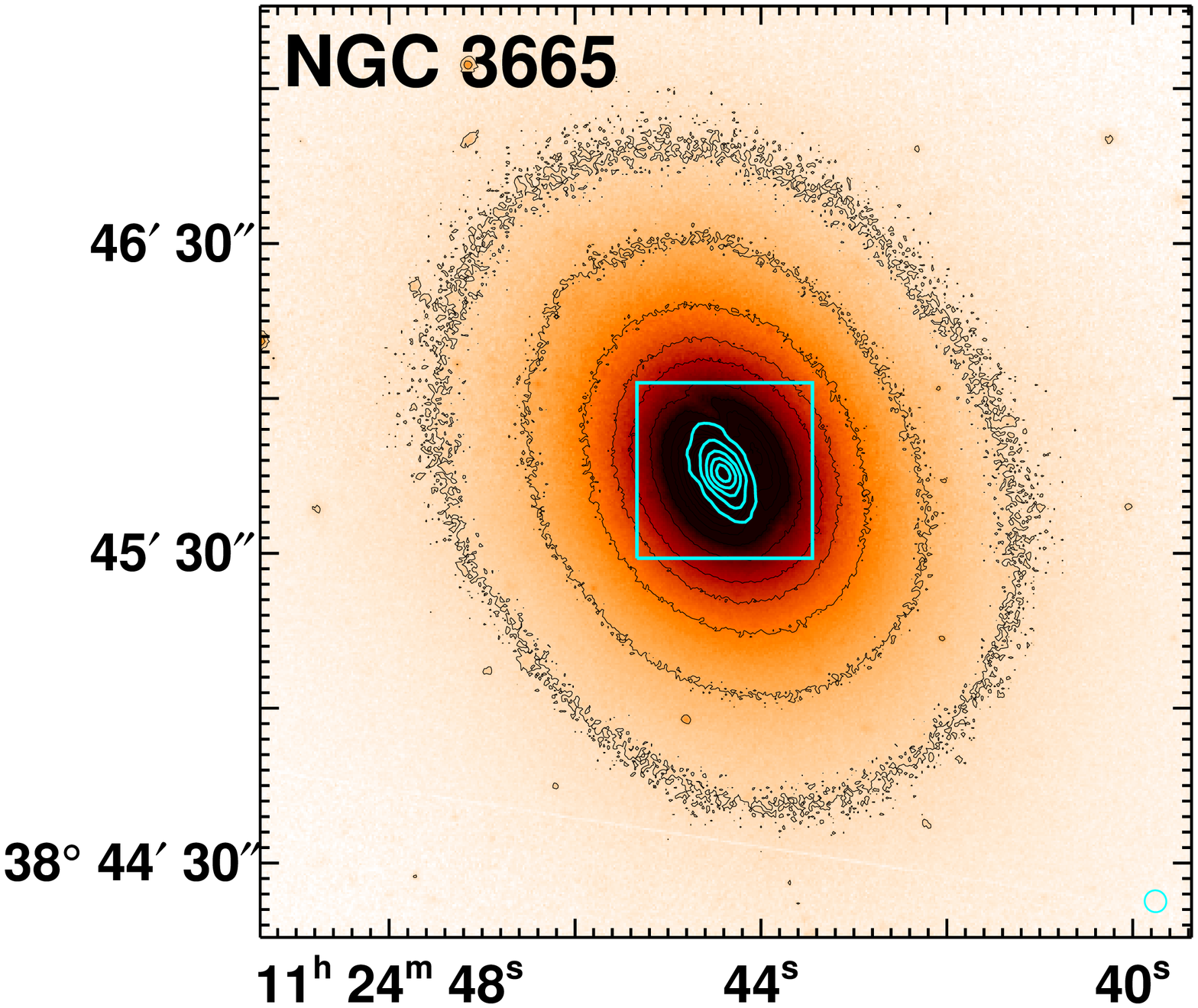}}
\subfloat{\includegraphics[height=2.2in,clip,trim=0cm 0.6cm 0.6cm 0.4cm]{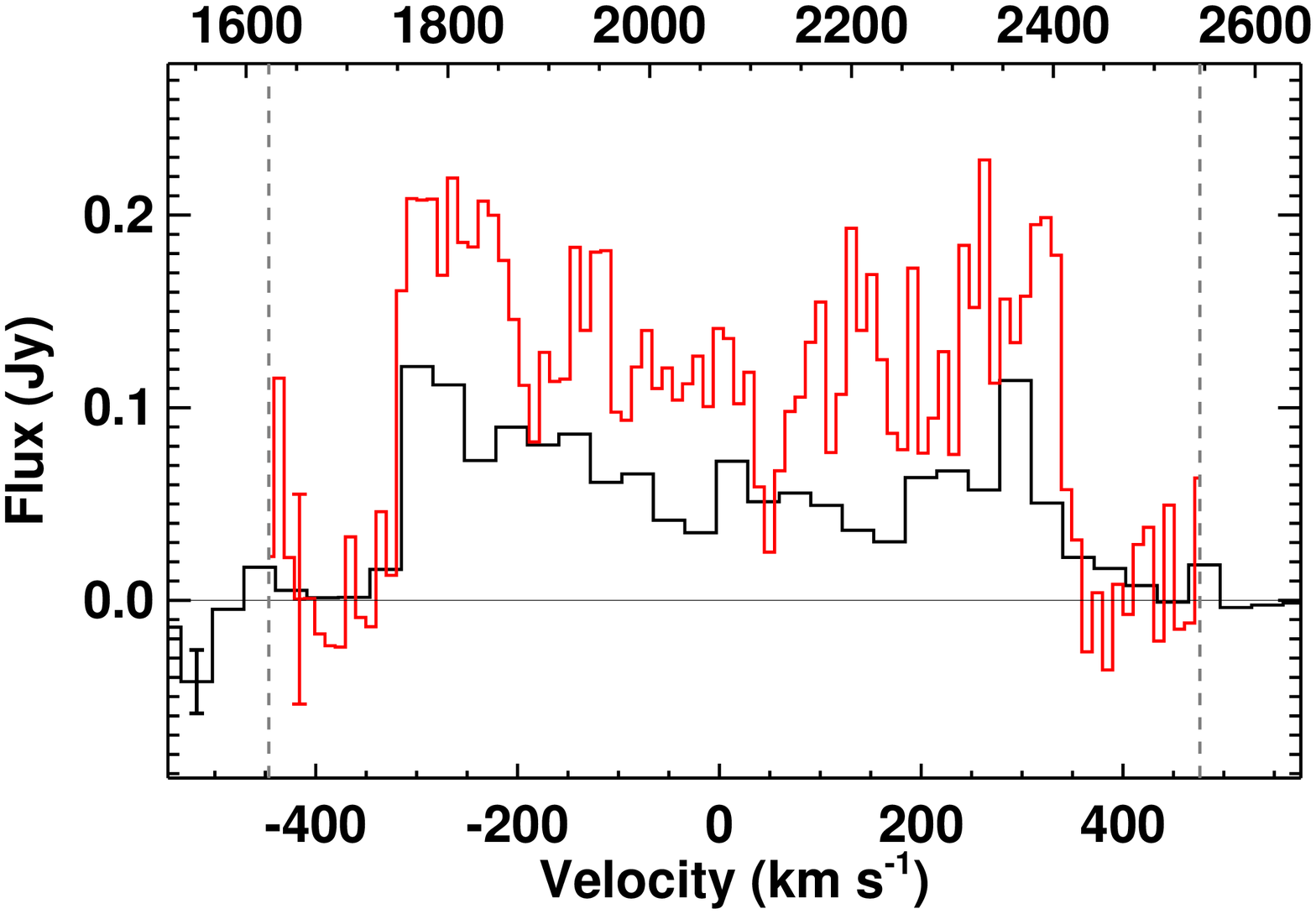}}
\end{figure*}
\begin{figure*}
\subfloat{\includegraphics[height=1.6in,clip,trim=0.1cm 1.4cm 0.4cm 2.5cm]{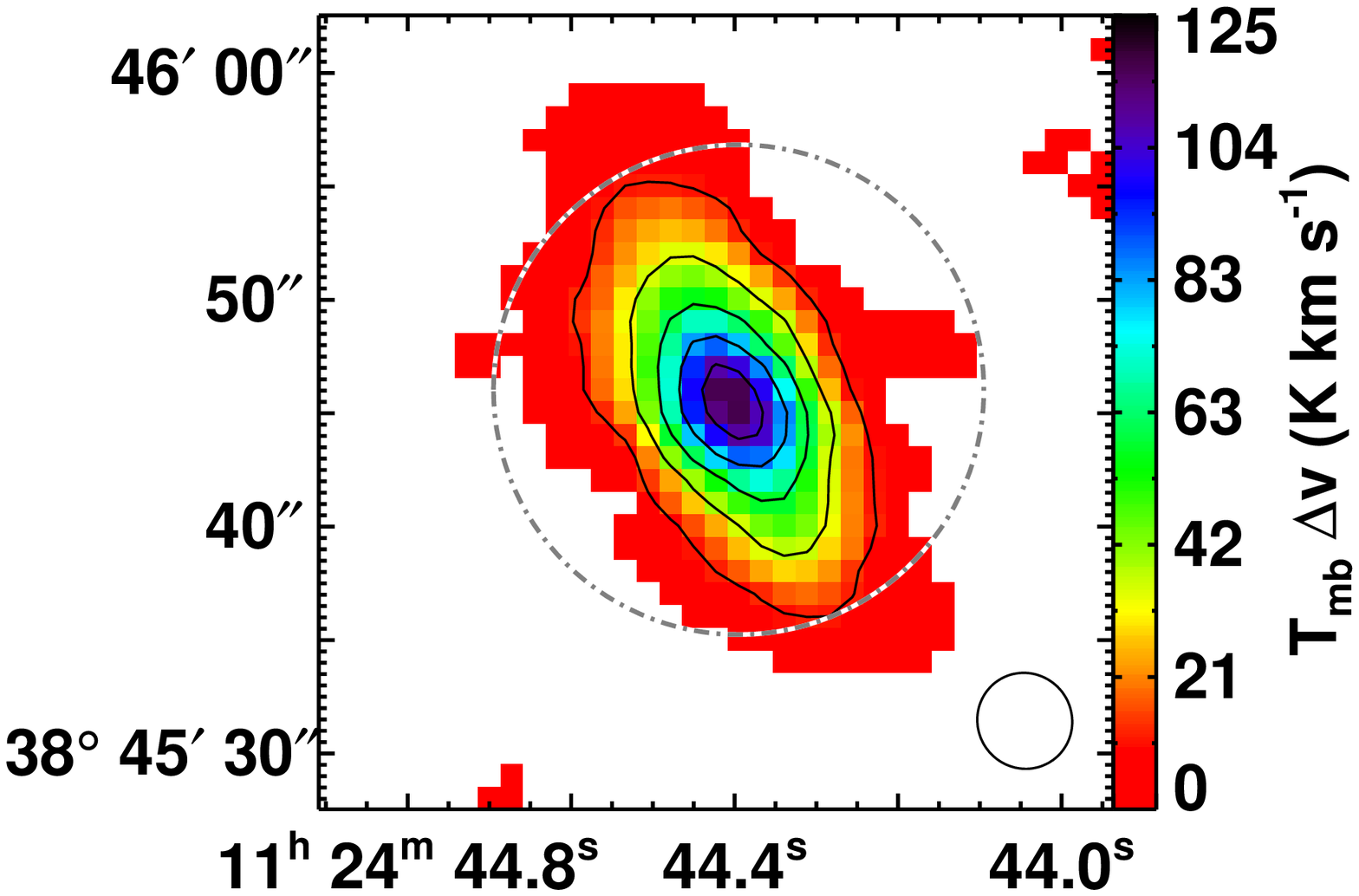}}
\subfloat{\includegraphics[height=1.6in,clip,trim=0.1cm 1.4cm 0.4cm 2.5cm]{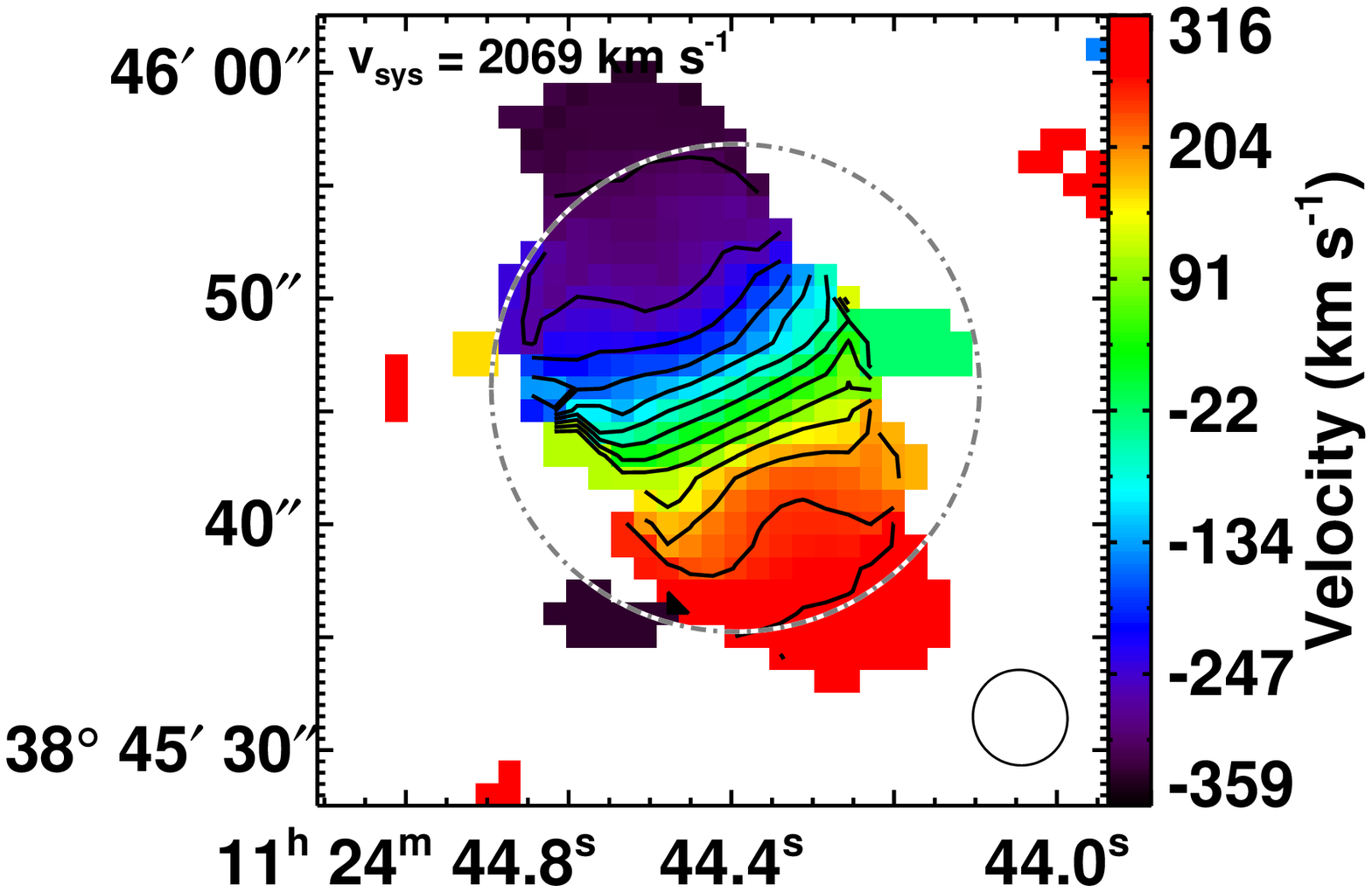}}
\subfloat{\includegraphics[height=1.6in,clip,trim=0cm 1.4cm 0cm 0.9cm]{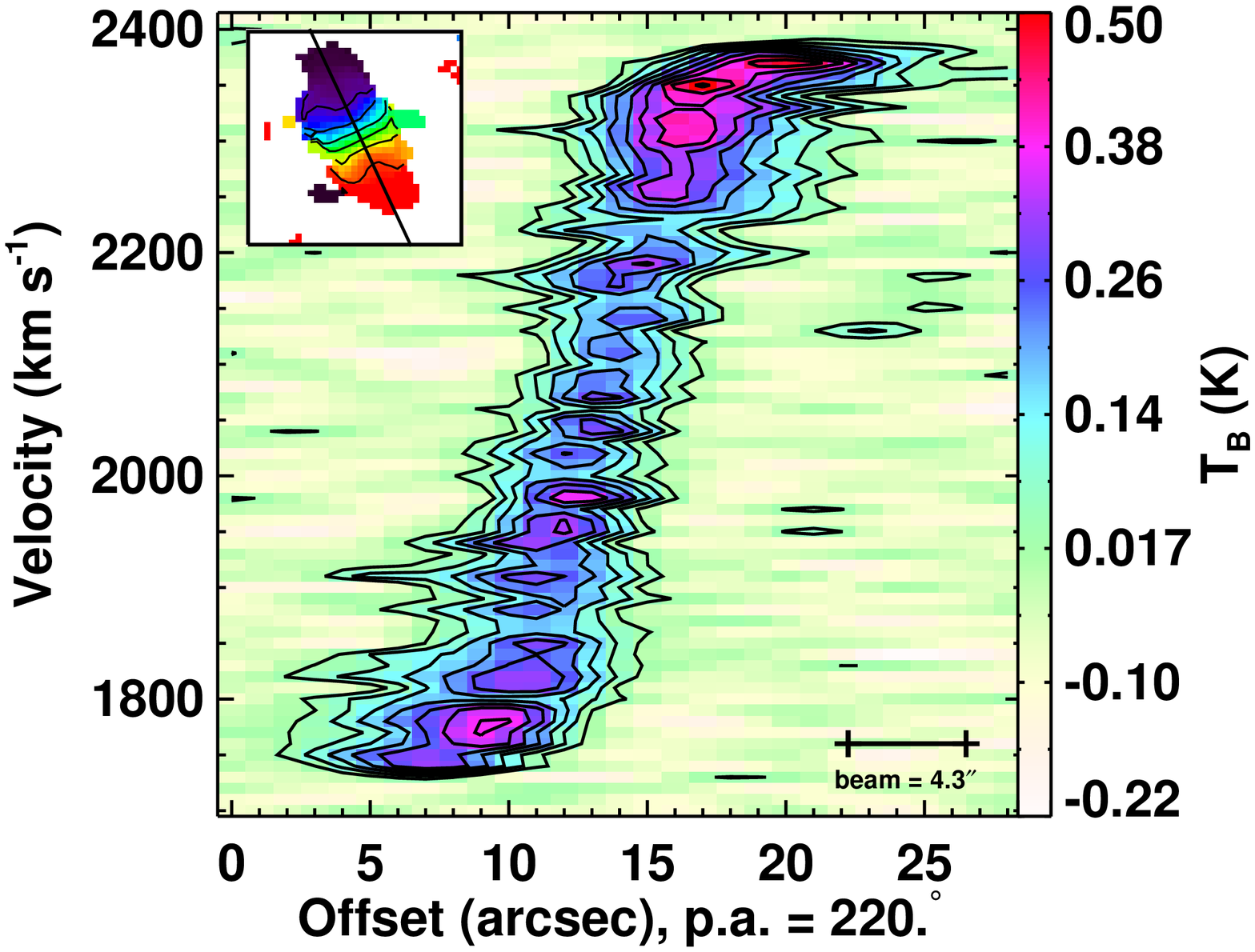}}
\end{figure*}
\begin{figure*}
\subfloat{\includegraphics[width=7in,clip,trim=1.6cm 4.1cm 8cm 4.7cm]{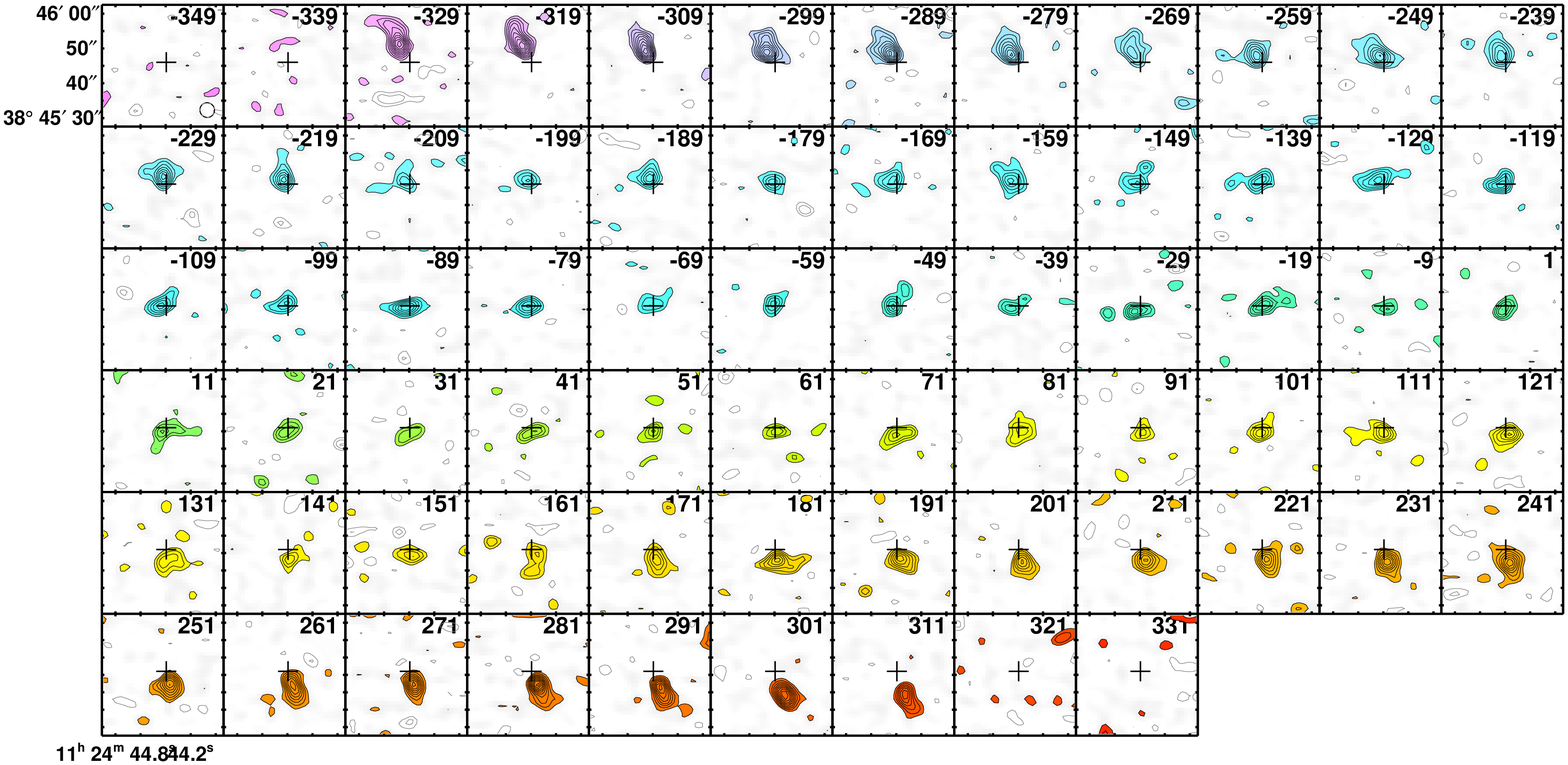}}
\caption{{\bf NGC~3665} is a group regular rotator ($M_K$ = -24.92) with normal stellar morphology.  It contains a dust disc.  The moment0 peak is 24 Jy beam$^{-1}$ \kms.  The moment1 contours are placed at 50 \kms\ intervals and the PVD contours are placed at $2\sigma$ intervals.}
\end{figure*}

\clearpage
\begin{figure*}
\centering
\subfloat{\includegraphics[height=2.2in,clip,trim=2.4cm 3.2cm 0cm 2.7cm]{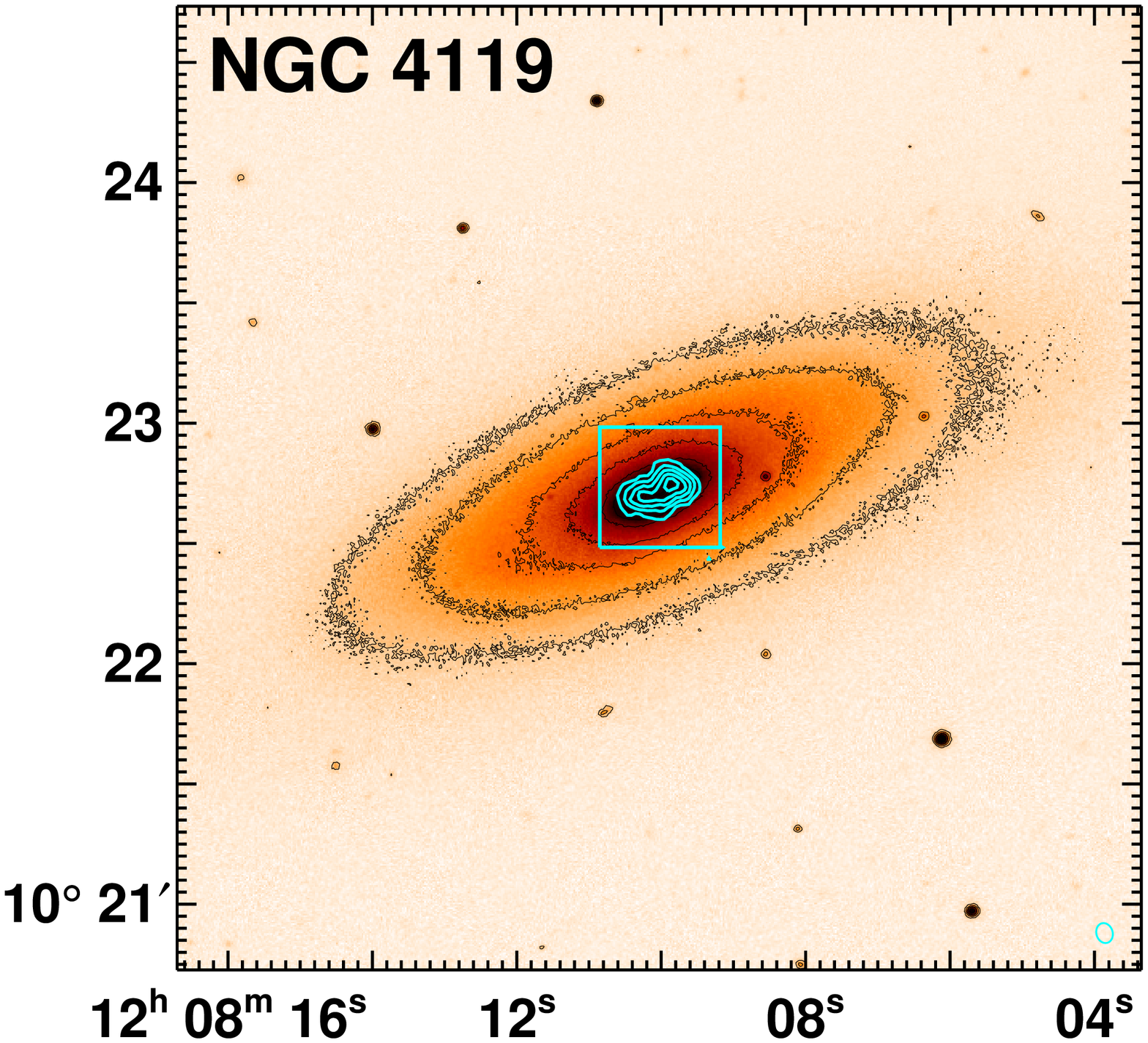}}
\subfloat{\includegraphics[height=2.2in,clip,trim=0cm 0.6cm 0.4cm 0.4cm]{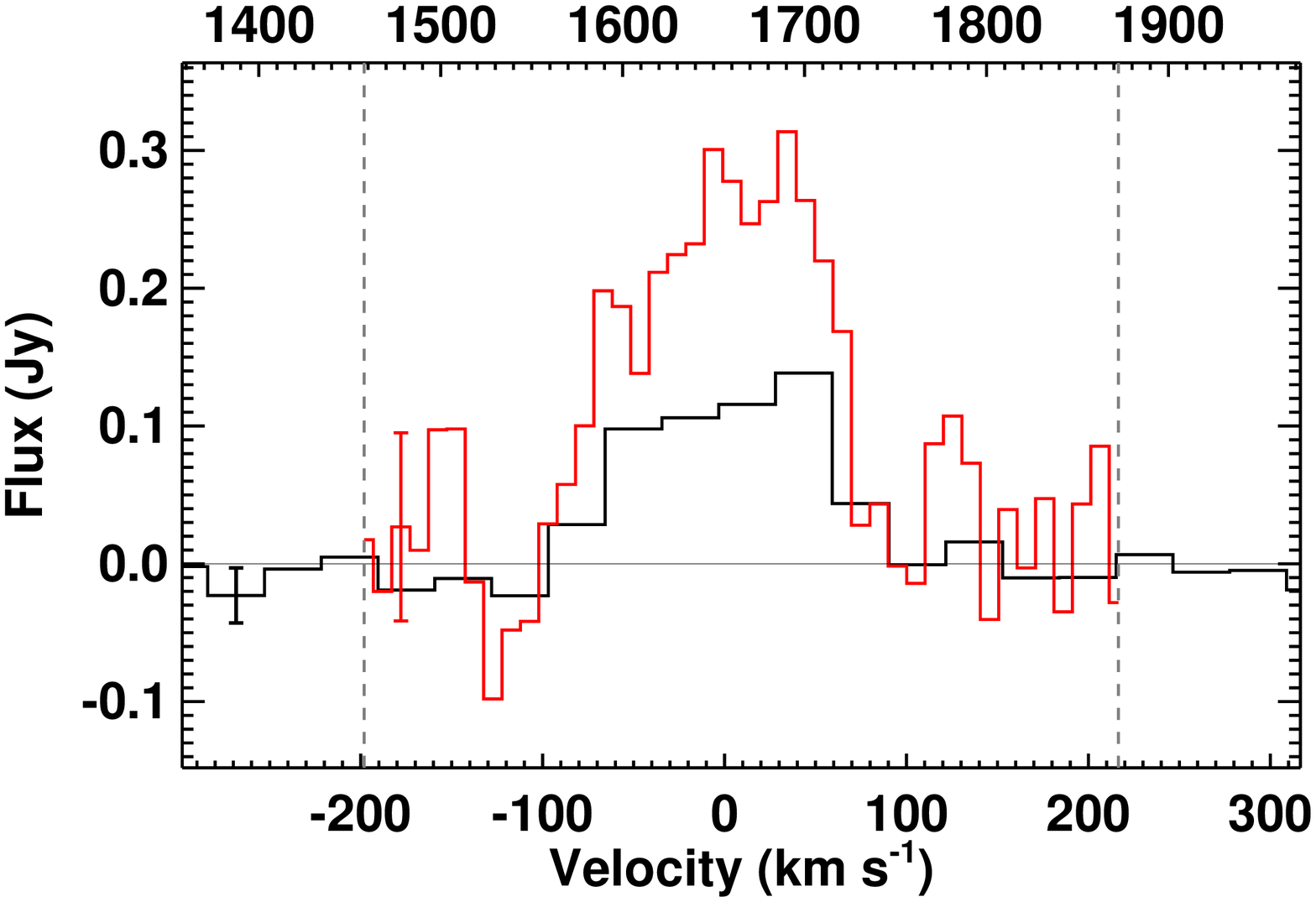}}
\end{figure*}
\begin{figure*}
\subfloat{\includegraphics[height=1.6in,clip,trim=0.1cm 1.4cm 0.4cm 2.4cm]{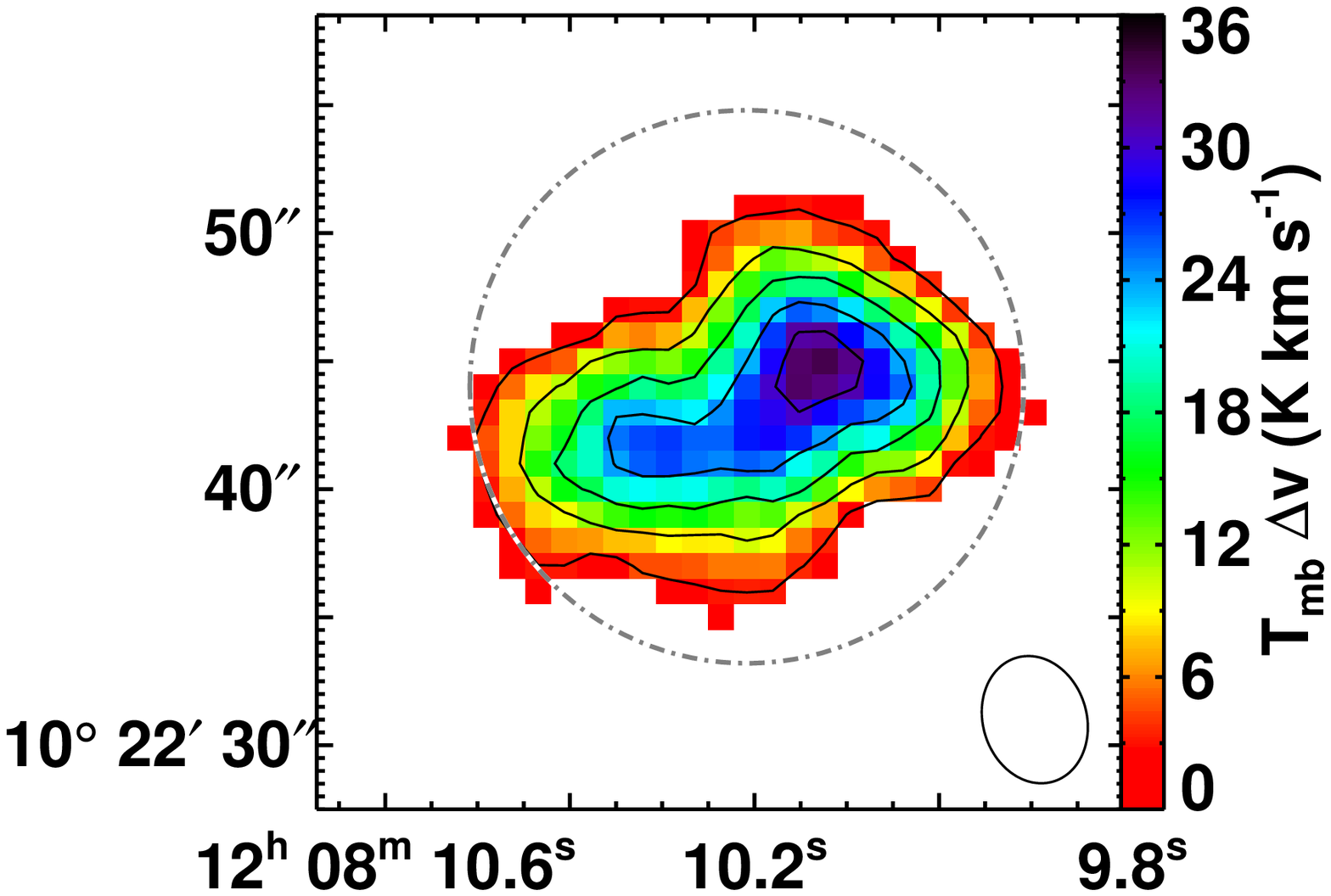}}
\subfloat{\includegraphics[height=1.6in,clip,trim=0.1cm 1.4cm 0.4cm 2.4cm]{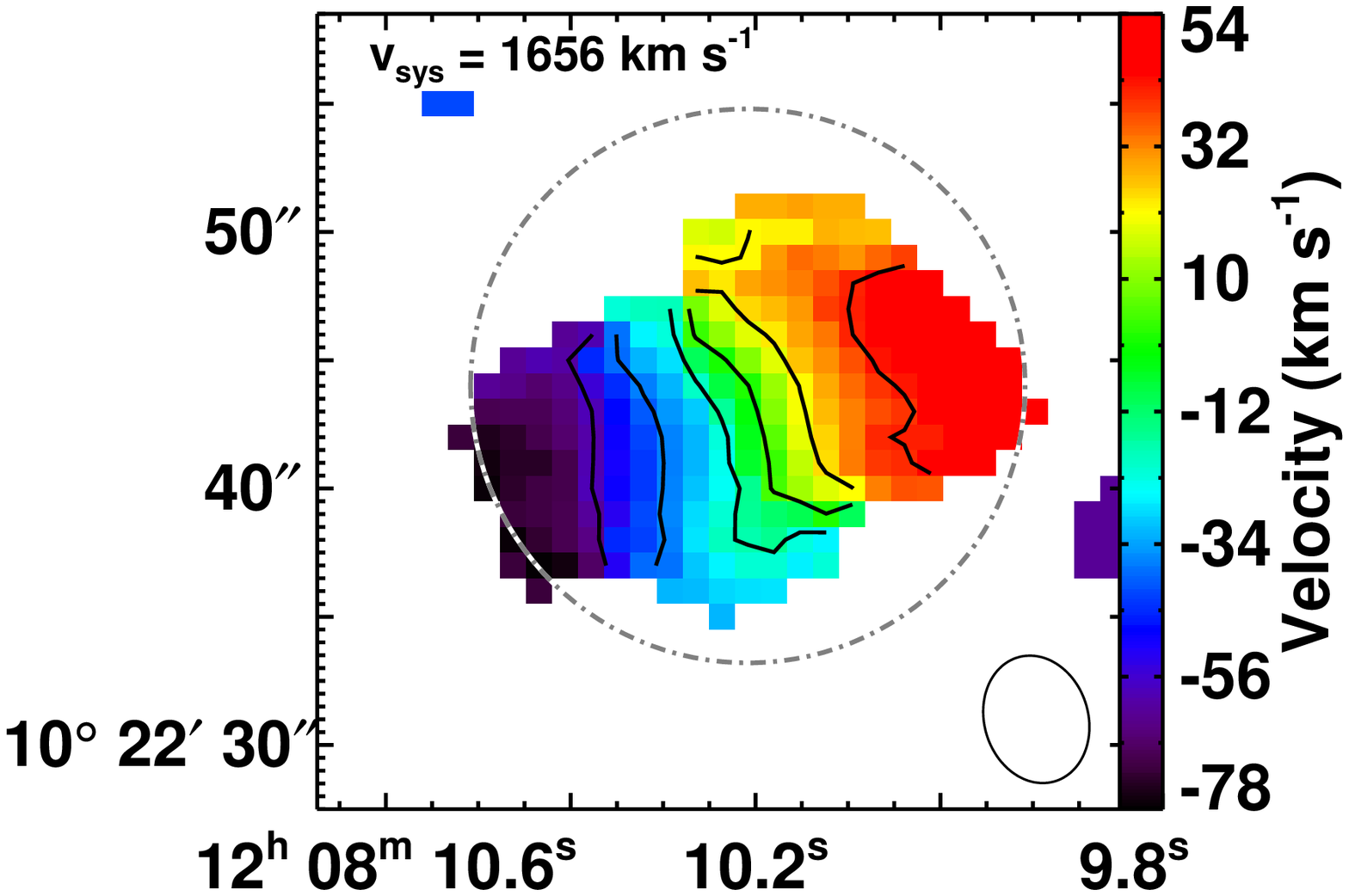}}
\subfloat{\includegraphics[height=1.6in,clip,trim=0cm 1.4cm 0cm 0.9cm]{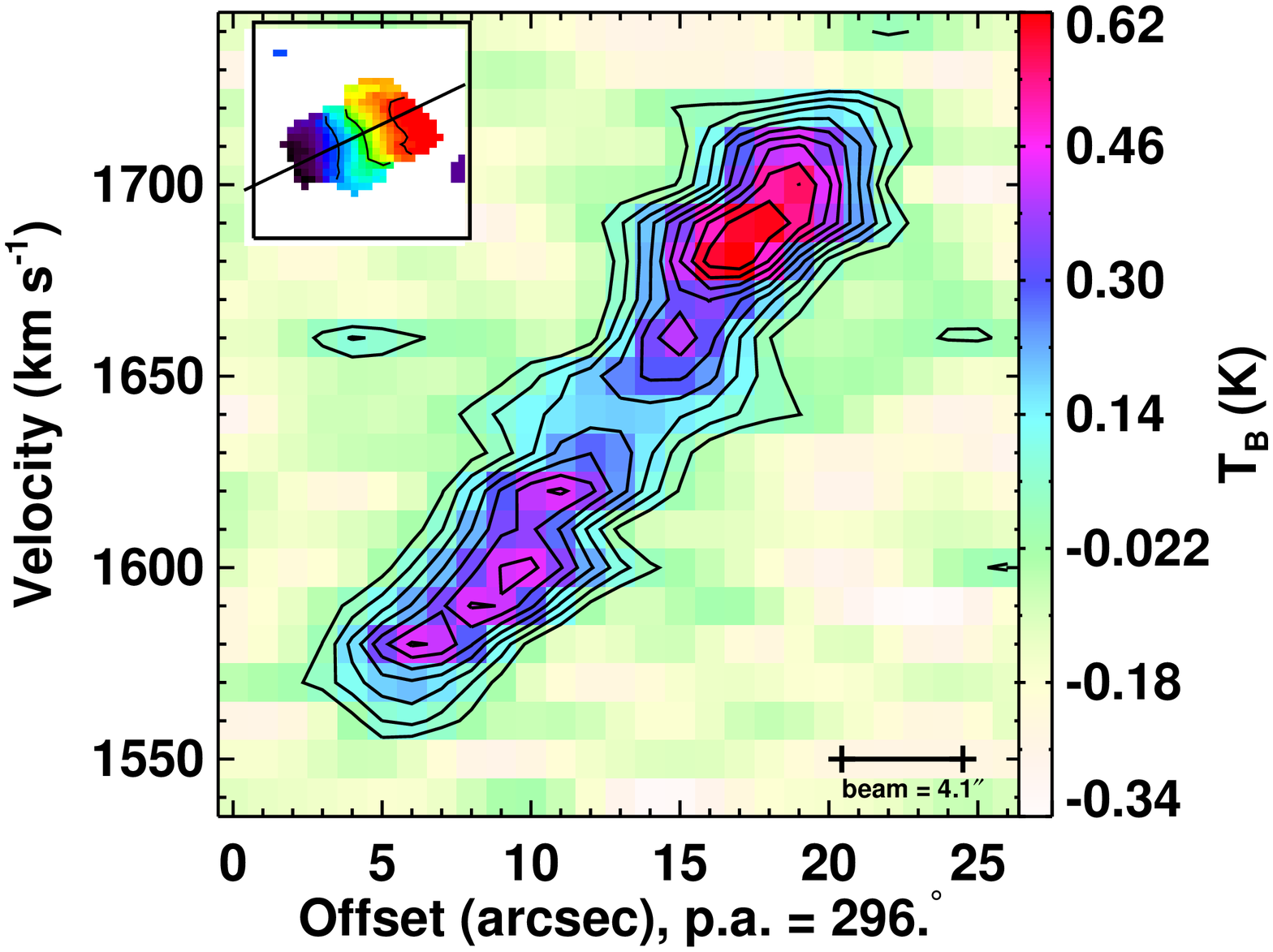}}
\end{figure*}
\begin{figure*}
\subfloat{\includegraphics[width=7in,clip,trim=1.6cm 6.2cm 8cm 5.7cm]{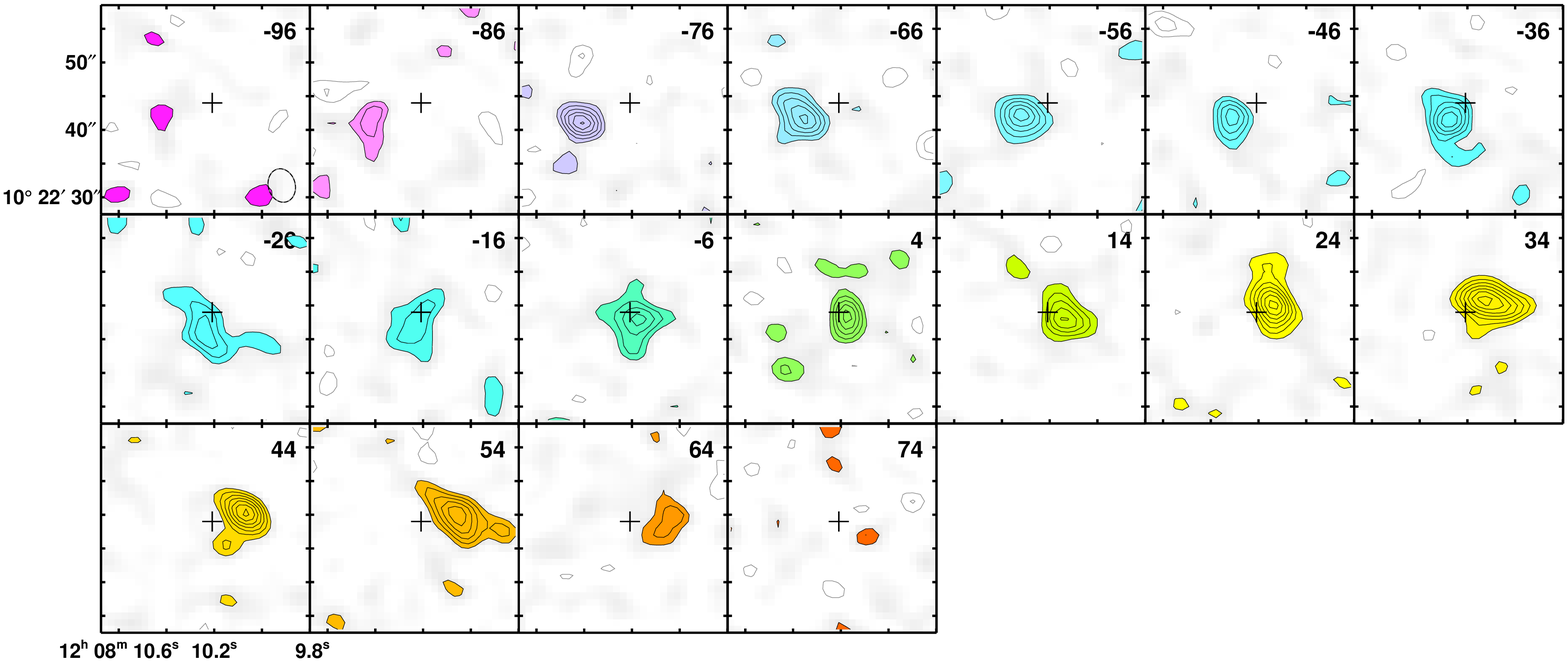}}
\caption{{\bf NGC~4119} is a Virgo regular rotator ($M_K$ = -22.60) with normal stellar morphology.  It contains a dust disc.  The moment0 peak is 8.1 Jy beam$^{-1}$ \kms.  The PVD contours are placed at $1.5\sigma$ intervals.}
\end{figure*}

\clearpage
\begin{figure*}
\centering
\subfloat{\includegraphics[height=2.2in,clip,trim=2.2cm 3.2cm 0cm 2.7cm]{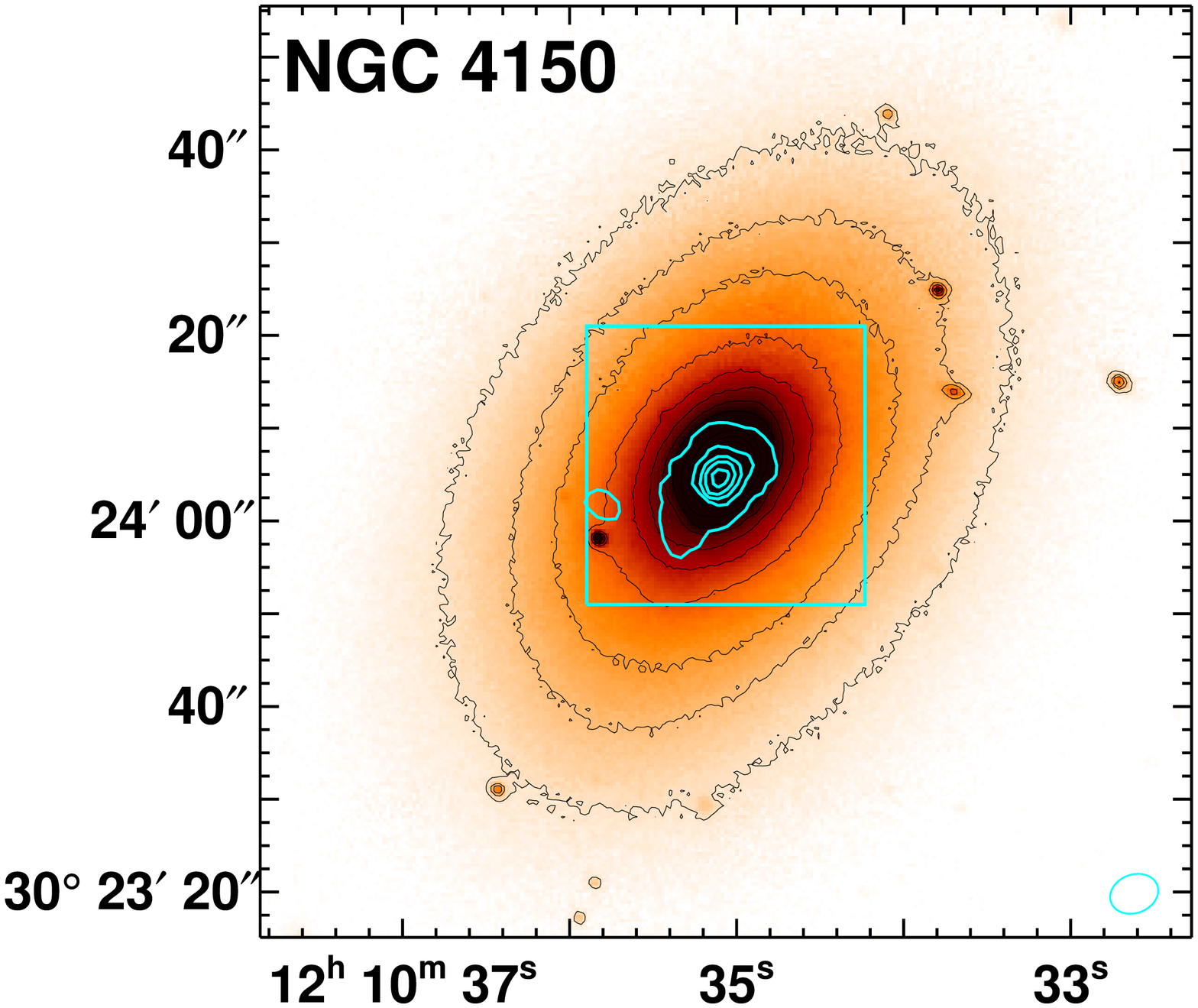}}
\subfloat{\includegraphics[height=2.2in,clip,trim=0cm 0.6cm 0.4cm 0.4cm]{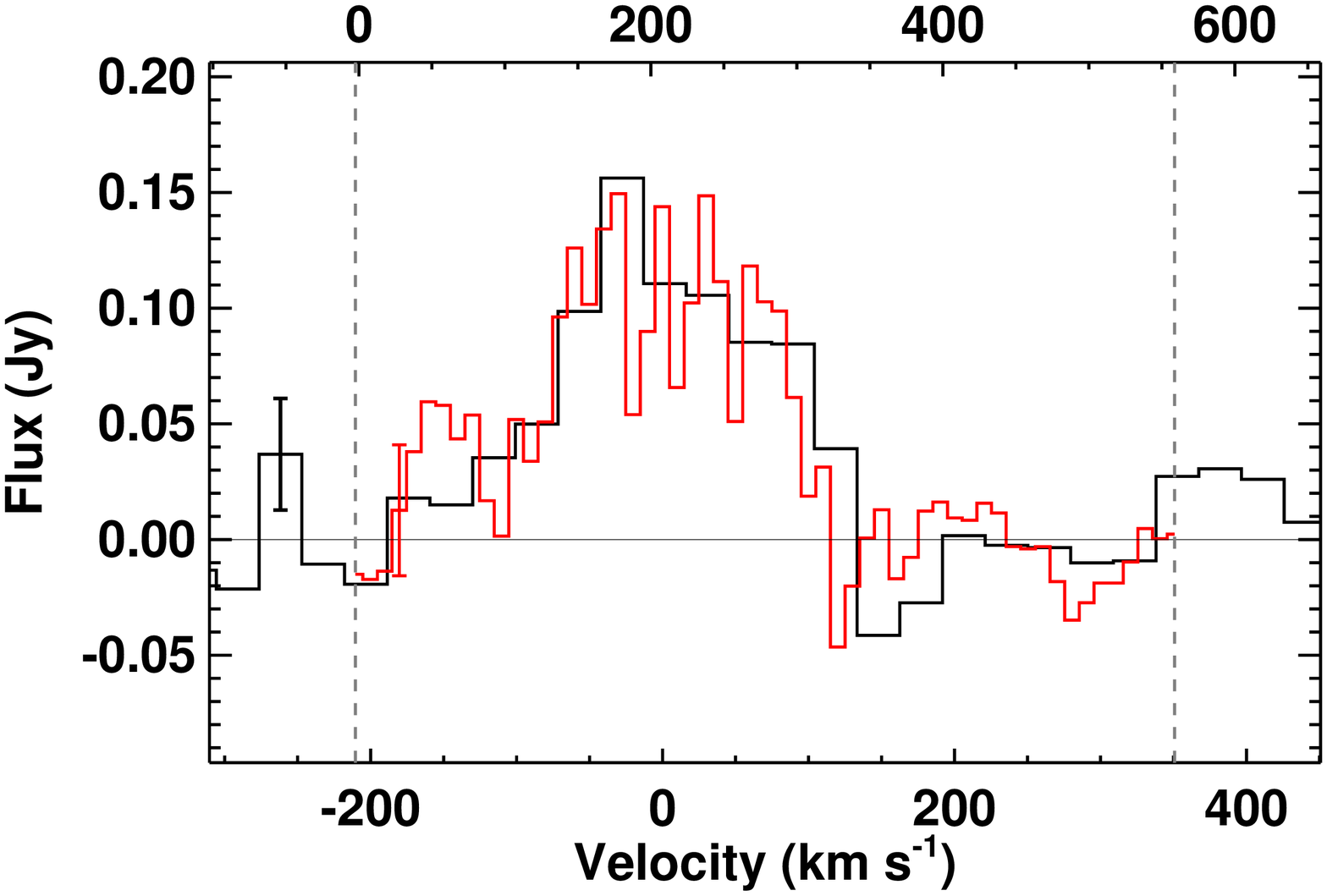}}
\end{figure*}
\begin{figure*}
\subfloat{\includegraphics[height=1.6in,clip,trim=0cm 1.4cm 0cm 2.4cm]{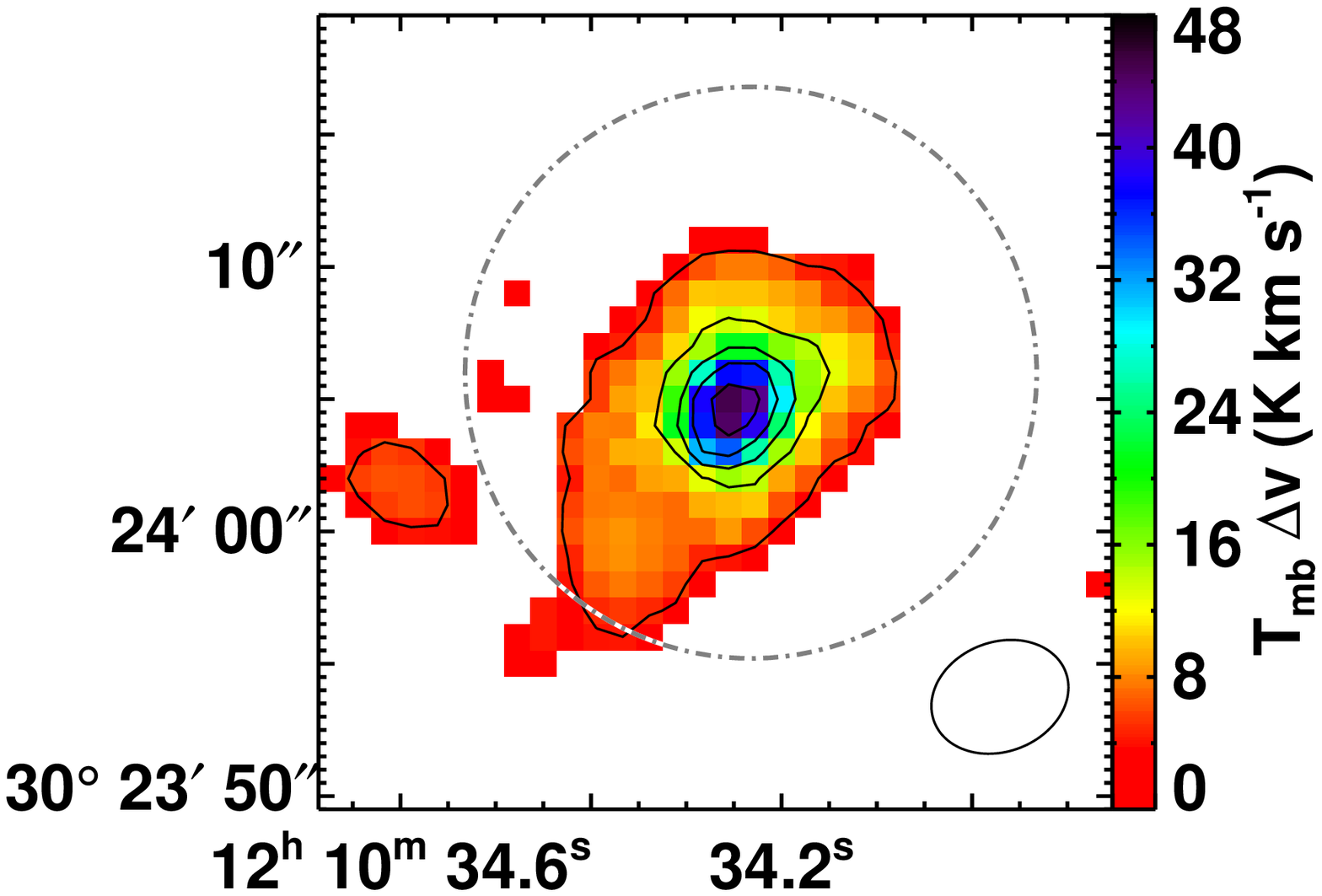}}
\subfloat{\includegraphics[height=1.6in,clip,trim=0cm 1.4cm 0cm 2.4cm]{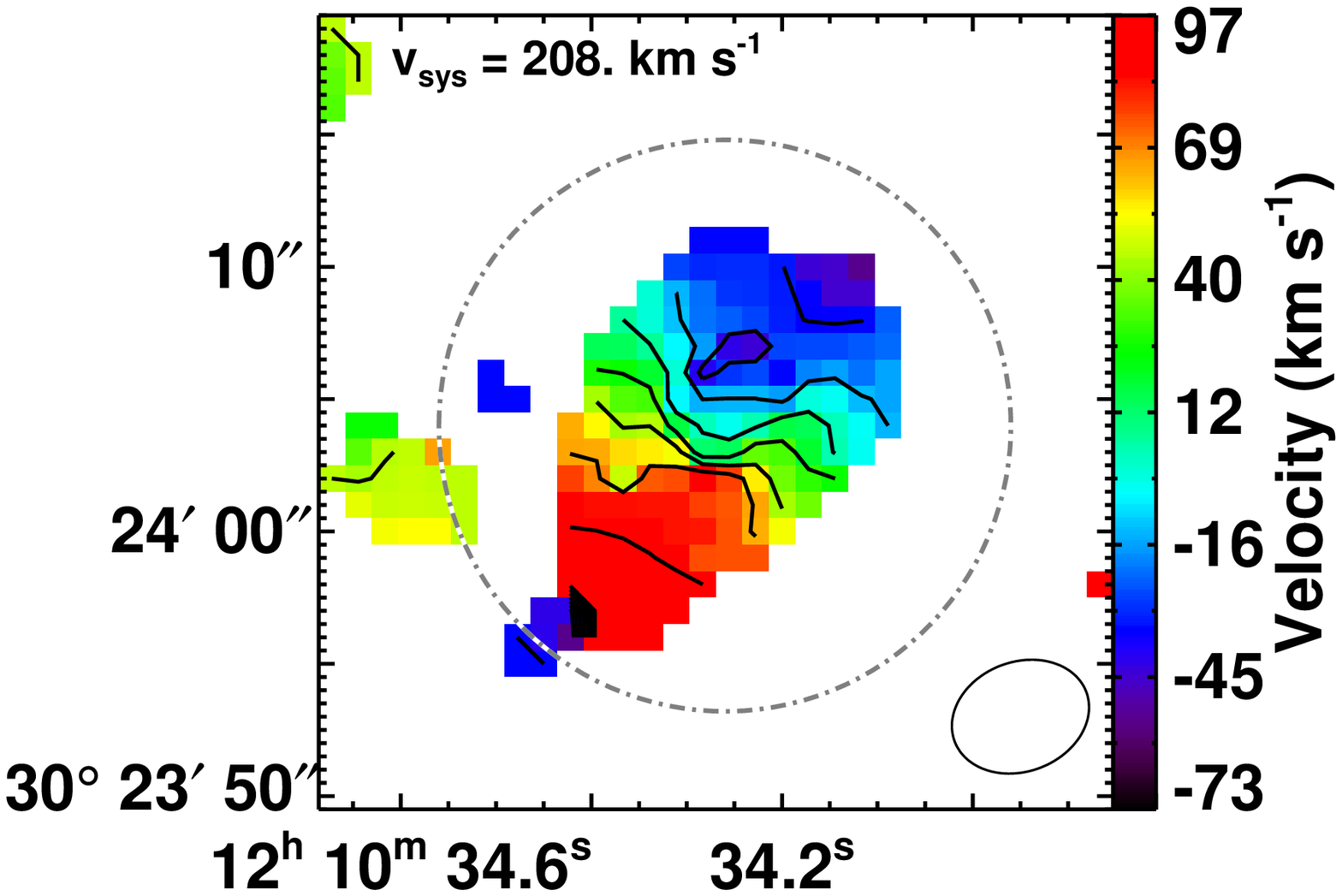}}
\subfloat{\includegraphics[height=1.6in,clip,trim=0cm 1.4cm 0cm 0.9cm]{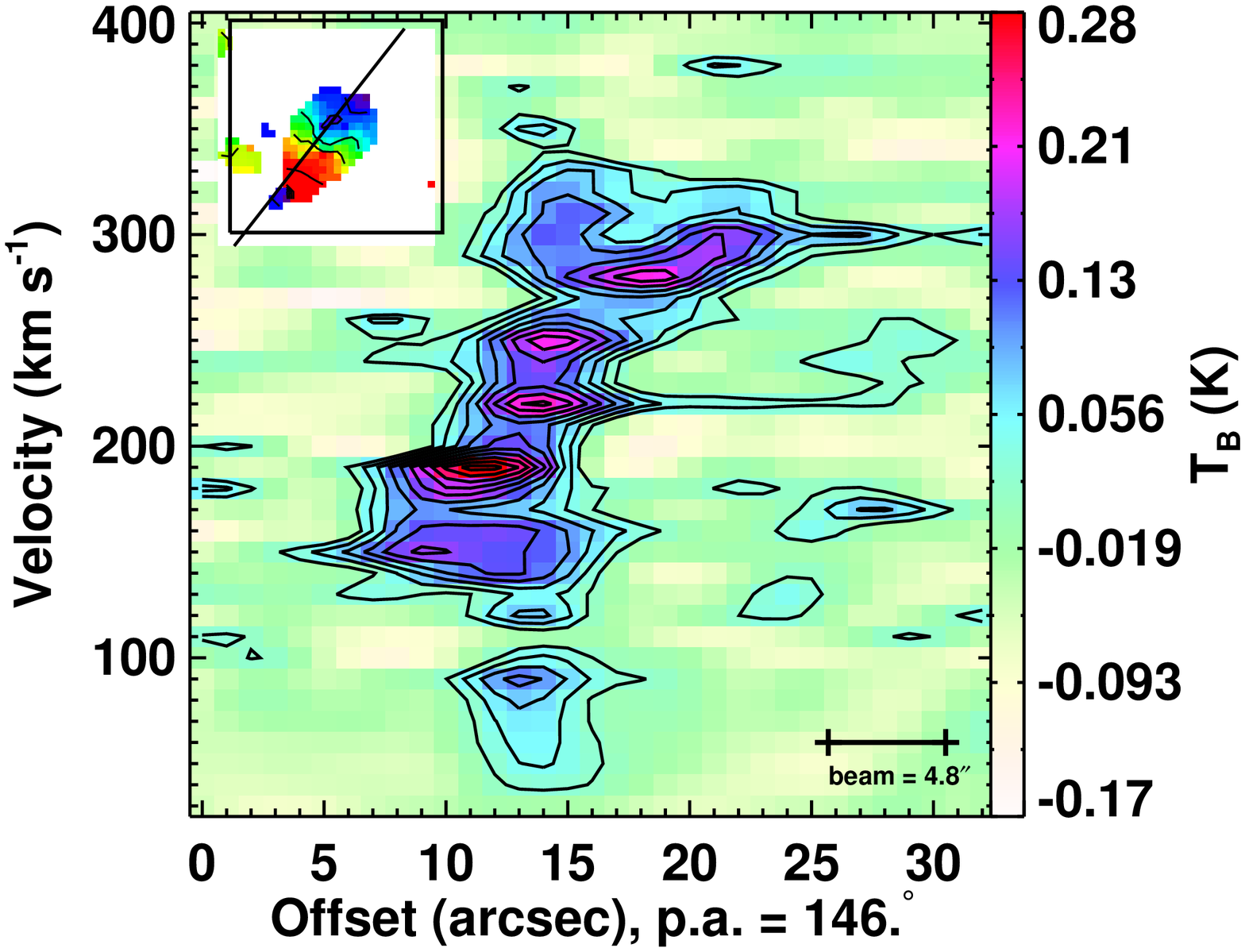}}
\end{figure*}
\begin{figure*}
\subfloat{\includegraphics[width=7in,clip,trim=1.2cm 1.9cm 2.5cm 0cm]{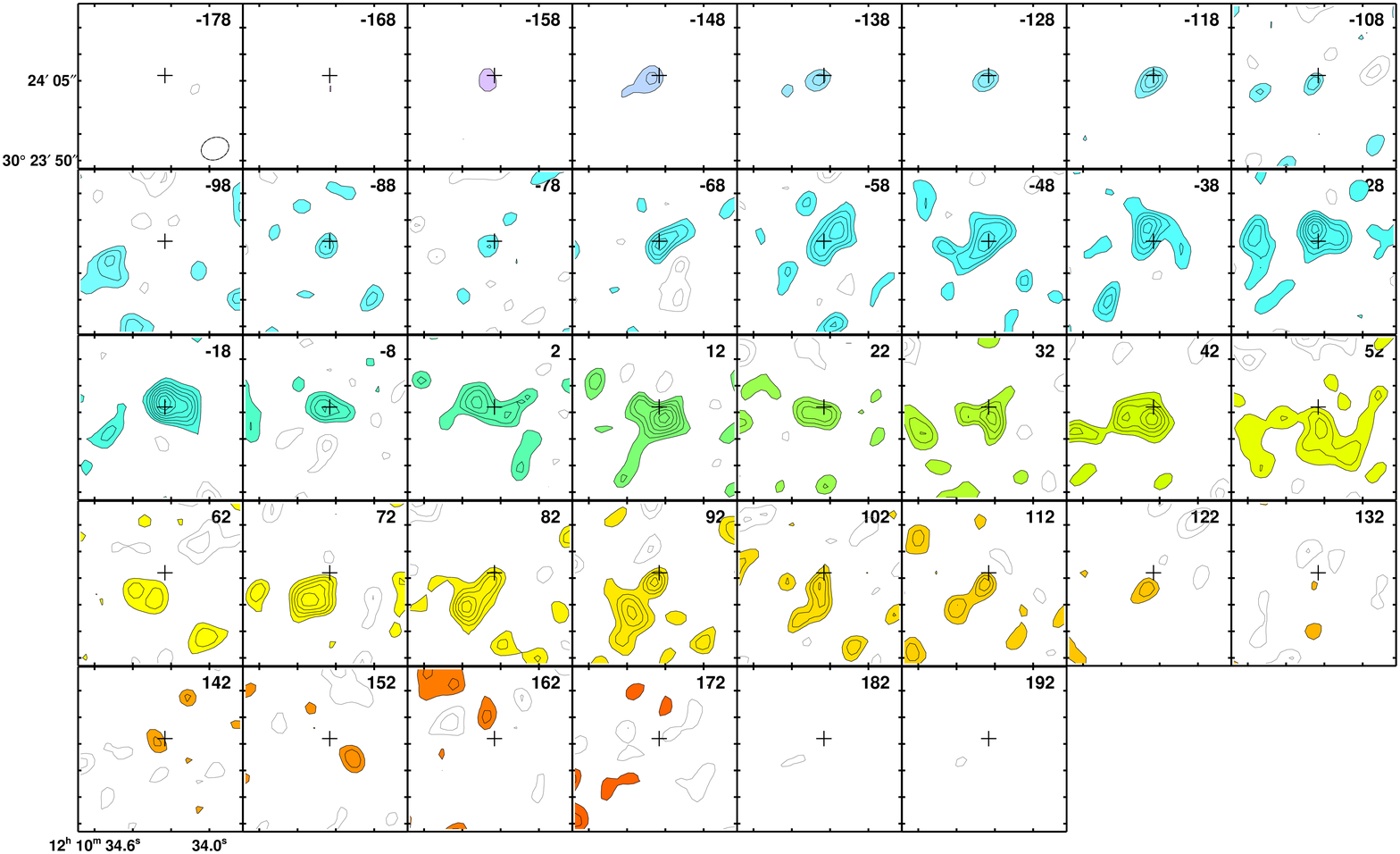}}
\caption{{\bf NGC~4150} is a Virgo regular rotator ($M_K$ = -21.65) with normal stellar morphology.  The moment0 peak is 9.9 Jy beam$^{-1}$ \kms.  The PVD contours are placed at $1.5\sigma$ intervals.}
\end{figure*}

\clearpage
\begin{figure*}
\centering
\subfloat{\includegraphics[height=2.2in,clip,trim=2.2cm 3.2cm 0cm 2.7cm]{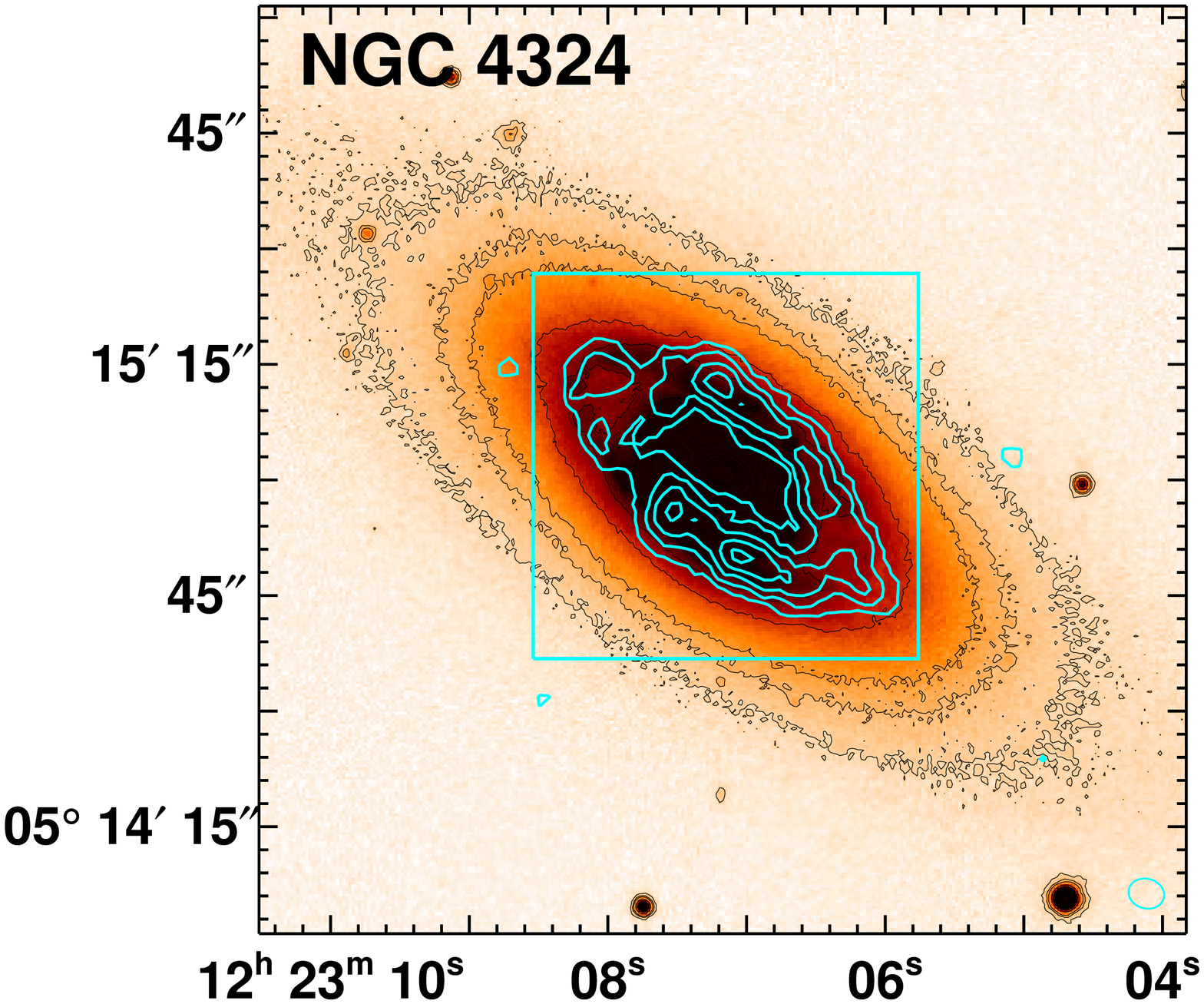}}
\subfloat{\includegraphics[height=2.2in,clip,trim=0cm 0.6cm 0.4cm 0.4cm]{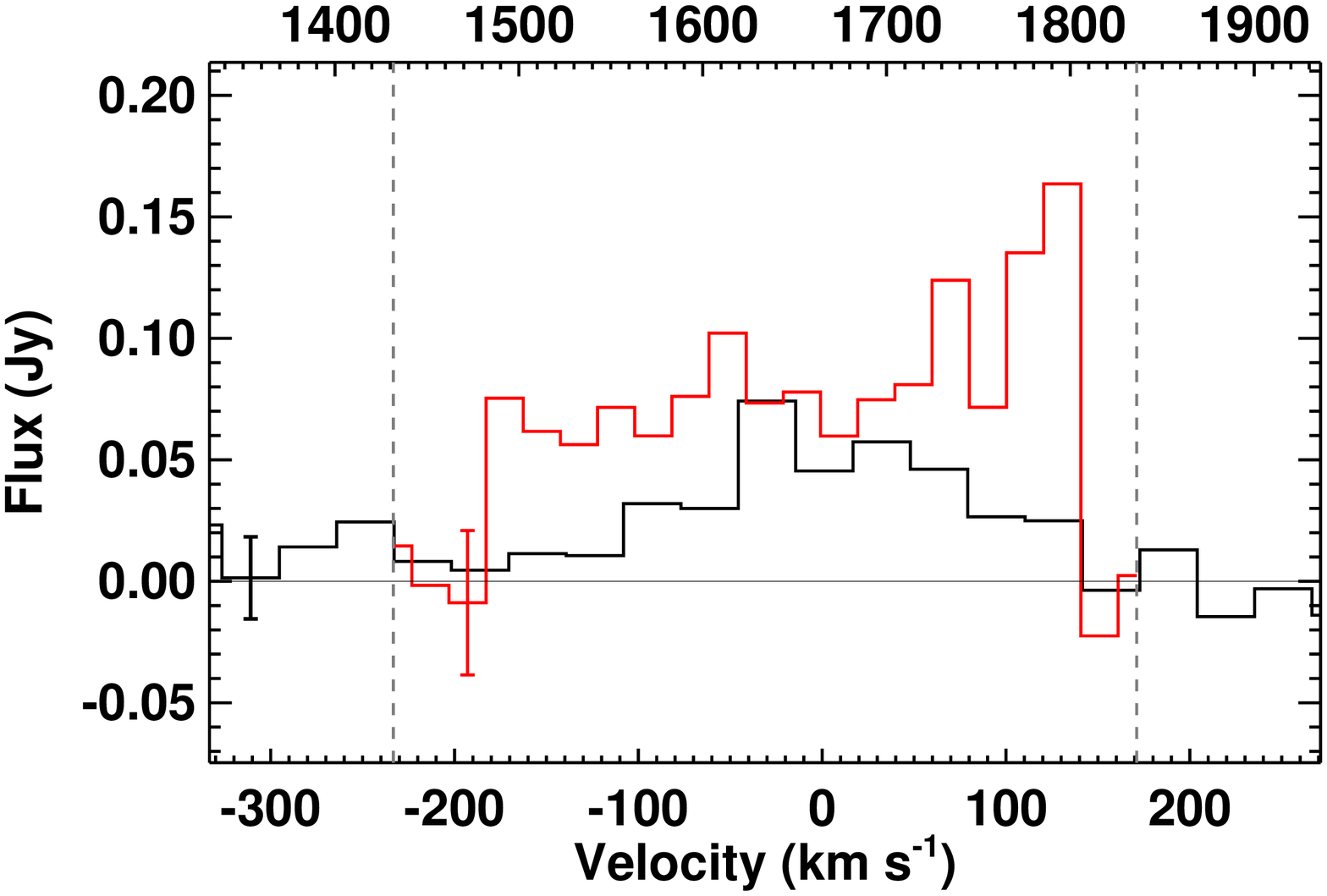}}
\end{figure*}
\begin{figure*}
\subfloat{\includegraphics[height=1.6in,clip,trim=0.1cm 1.4cm 0cm 2.4cm]{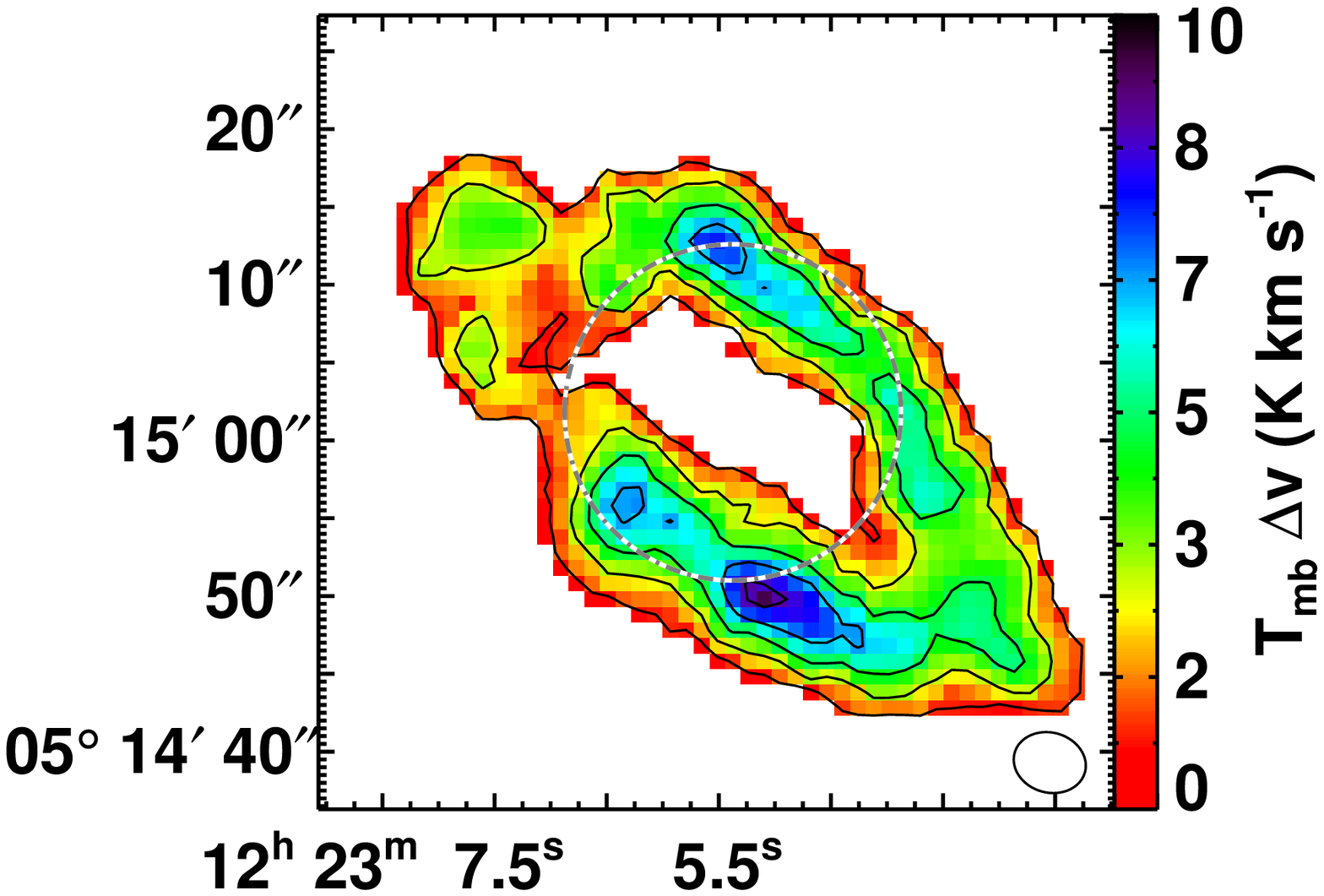}}
\subfloat{\includegraphics[height=1.6in,clip,trim=0cm 1.4cm 0cm 2.4cm]{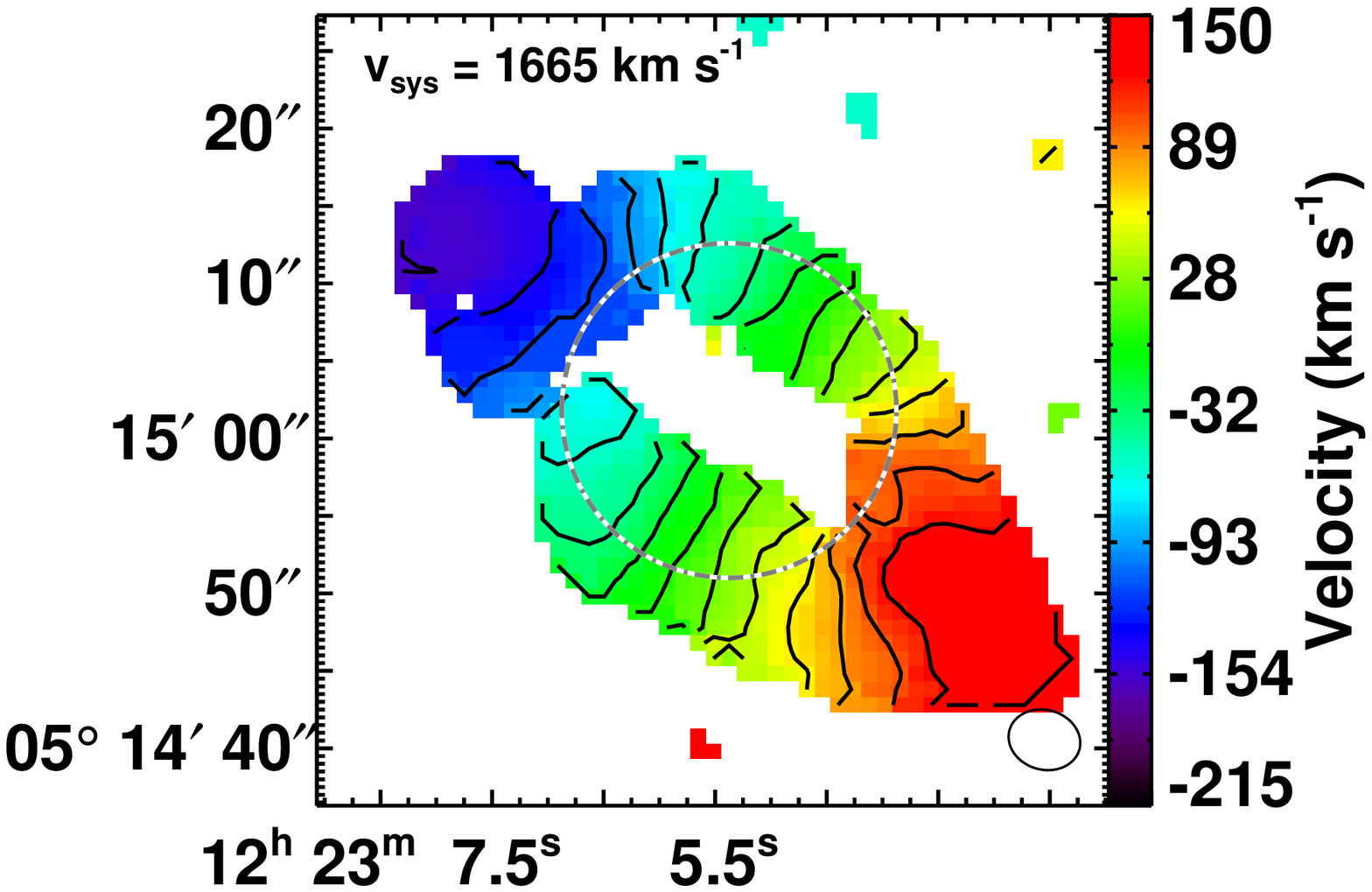}}
\subfloat{\includegraphics[height=1.6in,clip,trim=0cm 1.4cm 0cm 0.9cm]{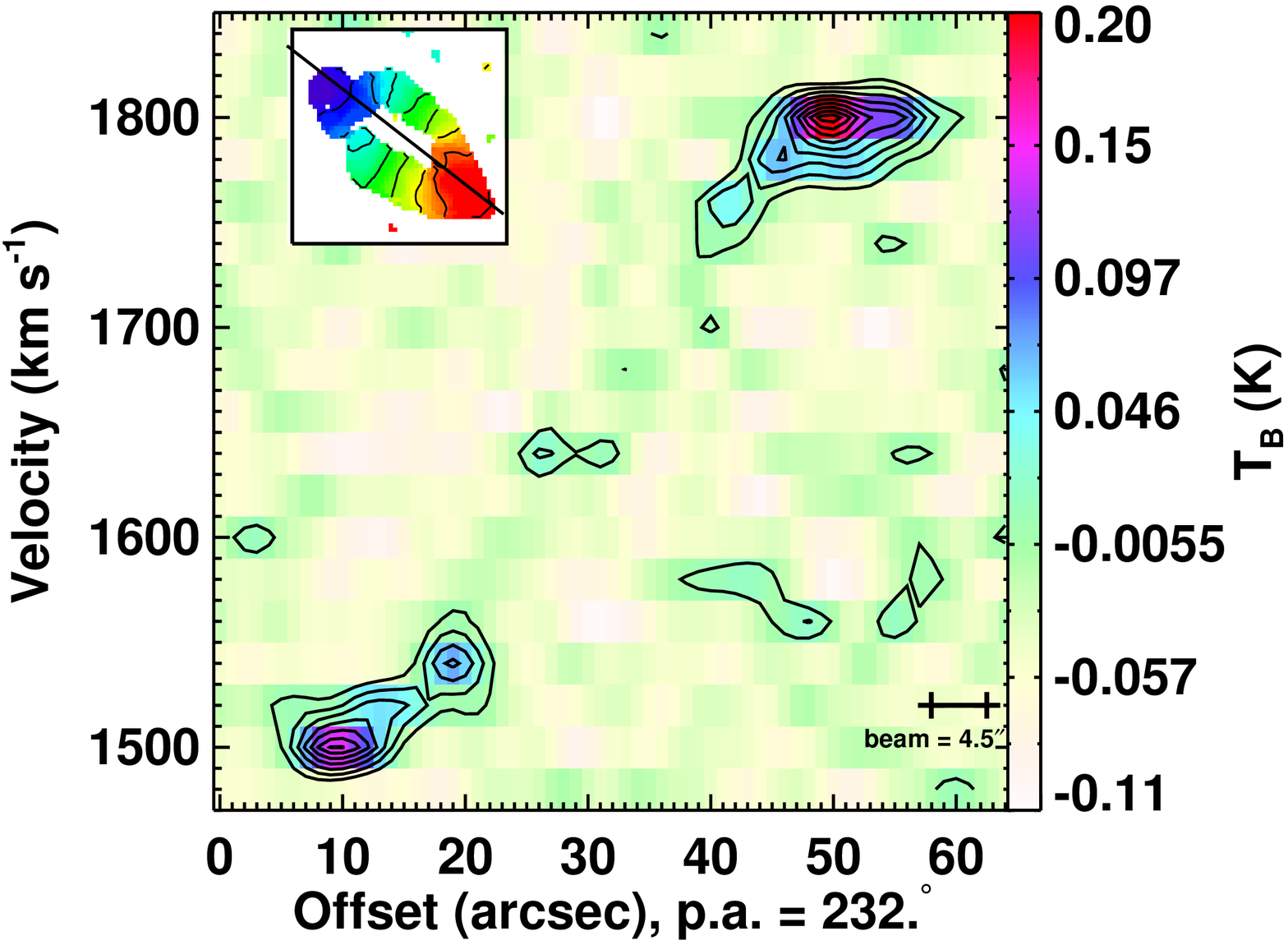}}
\end{figure*}
\begin{figure*}
\subfloat{\includegraphics[width=7in,clip,trim=1.6cm 3.8cm 2.5cm 0cm]{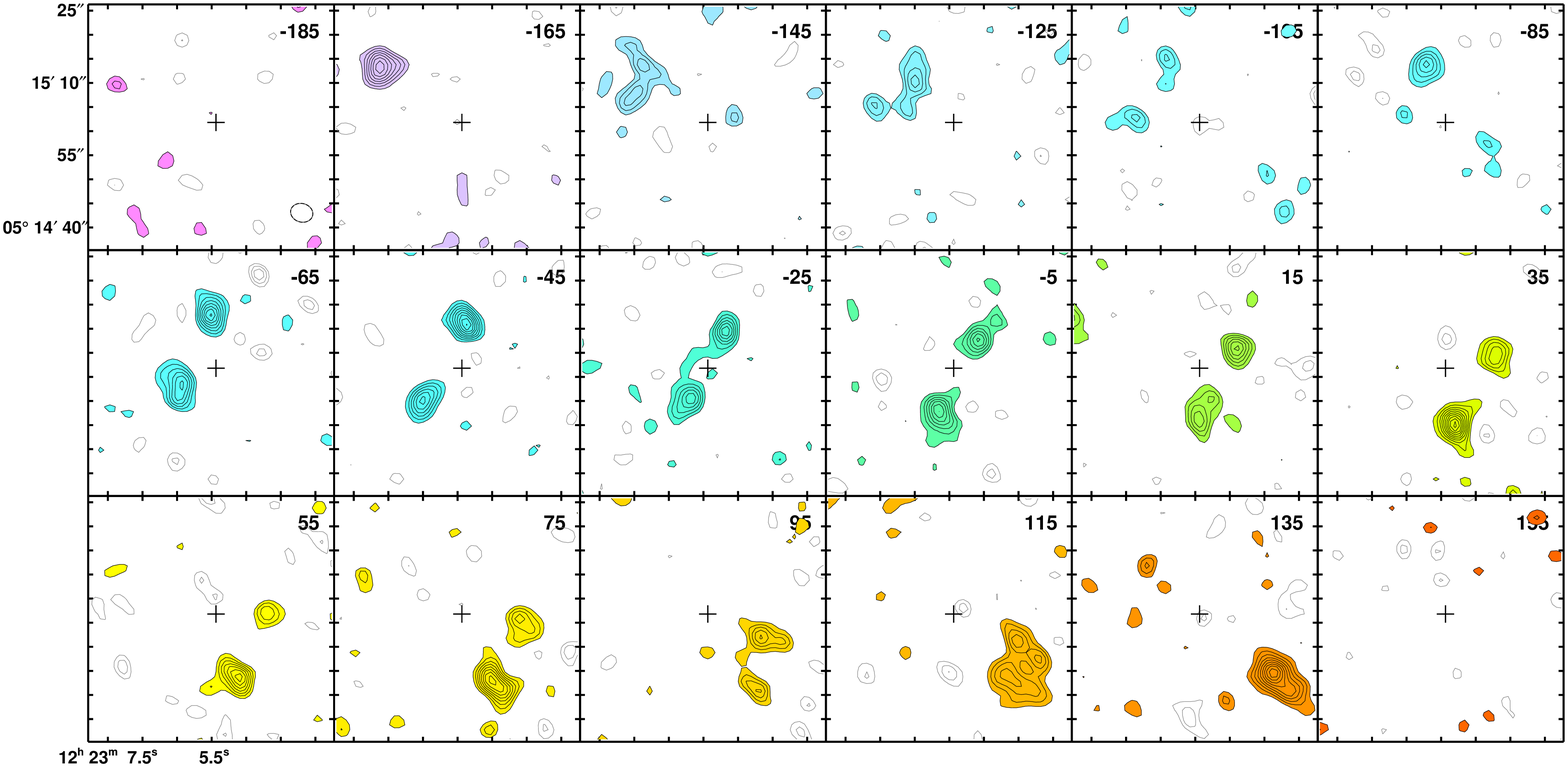}}
\caption{{\bf NGC~4324} is a field regular rotator ($M_K$ = -21.61) that includes two velocity maxima.  It contains a ring stellar morphology, as well as a dust disc, bar, and ring.  The moment0 peak is 2.0 Jy beam$^{-1}$ \kms.  The PVD contours are placed at $2\sigma$ intervals.}
\end{figure*}

\clearpage
\begin{figure*}
\centering
\subfloat{\includegraphics[height=2.2in,clip,trim=2.2cm 3.2cm 0cm 2.7cm]{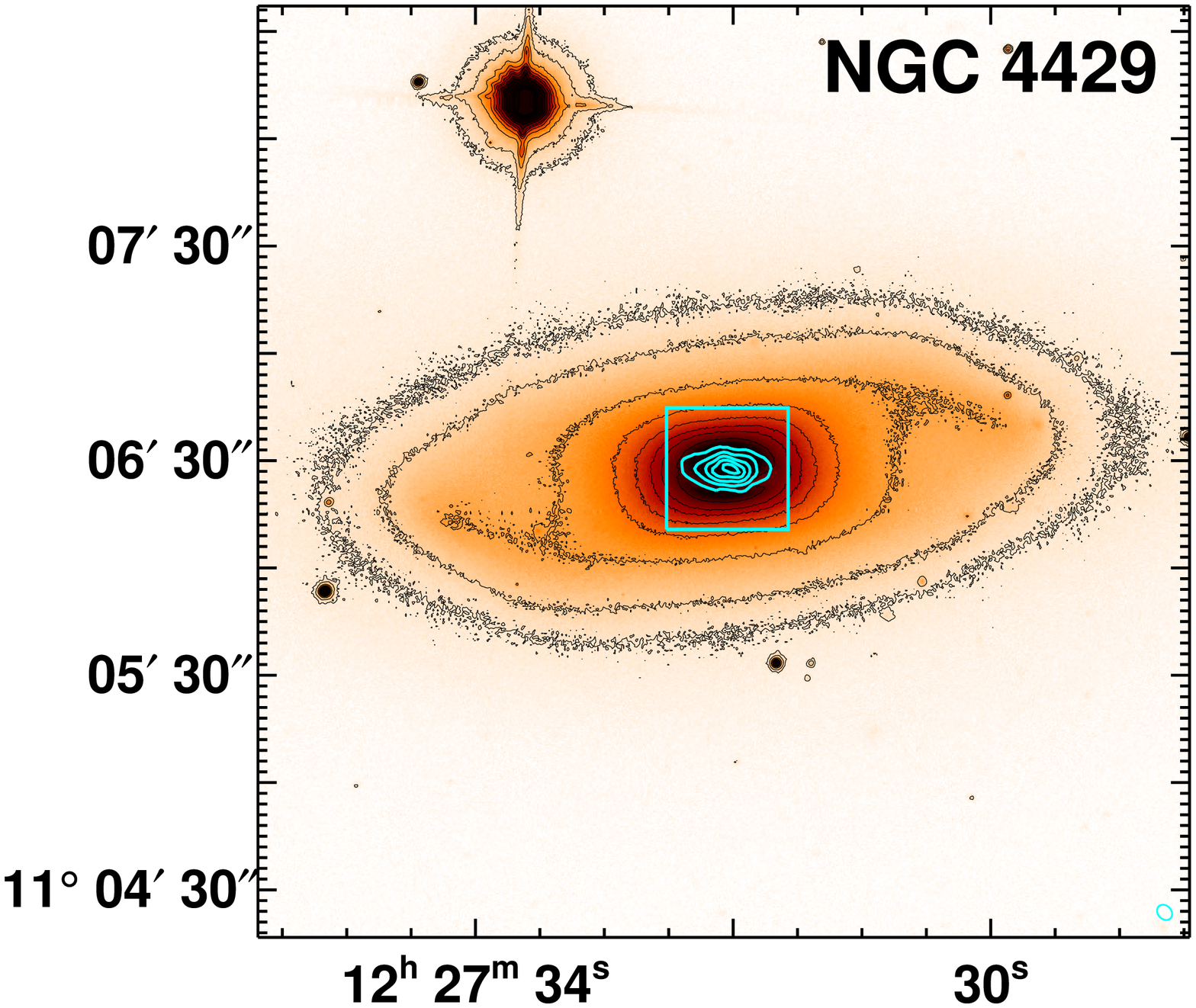}}
\subfloat{\includegraphics[height=2.2in,clip,trim=0cm 0.6cm 0.4cm 0.4cm]{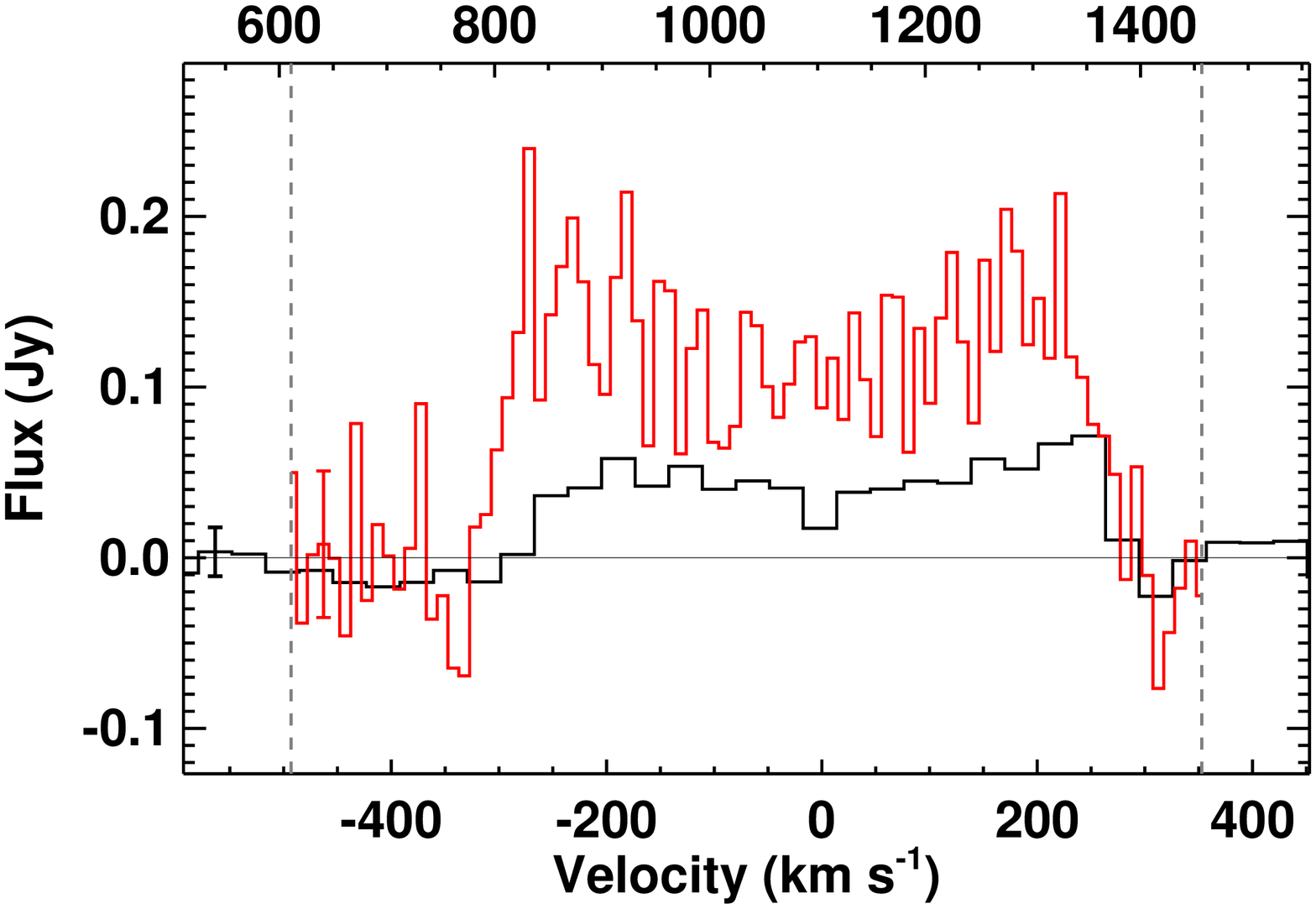}}
\end{figure*}
\begin{figure*}
\subfloat{\includegraphics[height=1.6in,clip,trim=0.1cm 1.4cm 0.6cm 2.5cm]{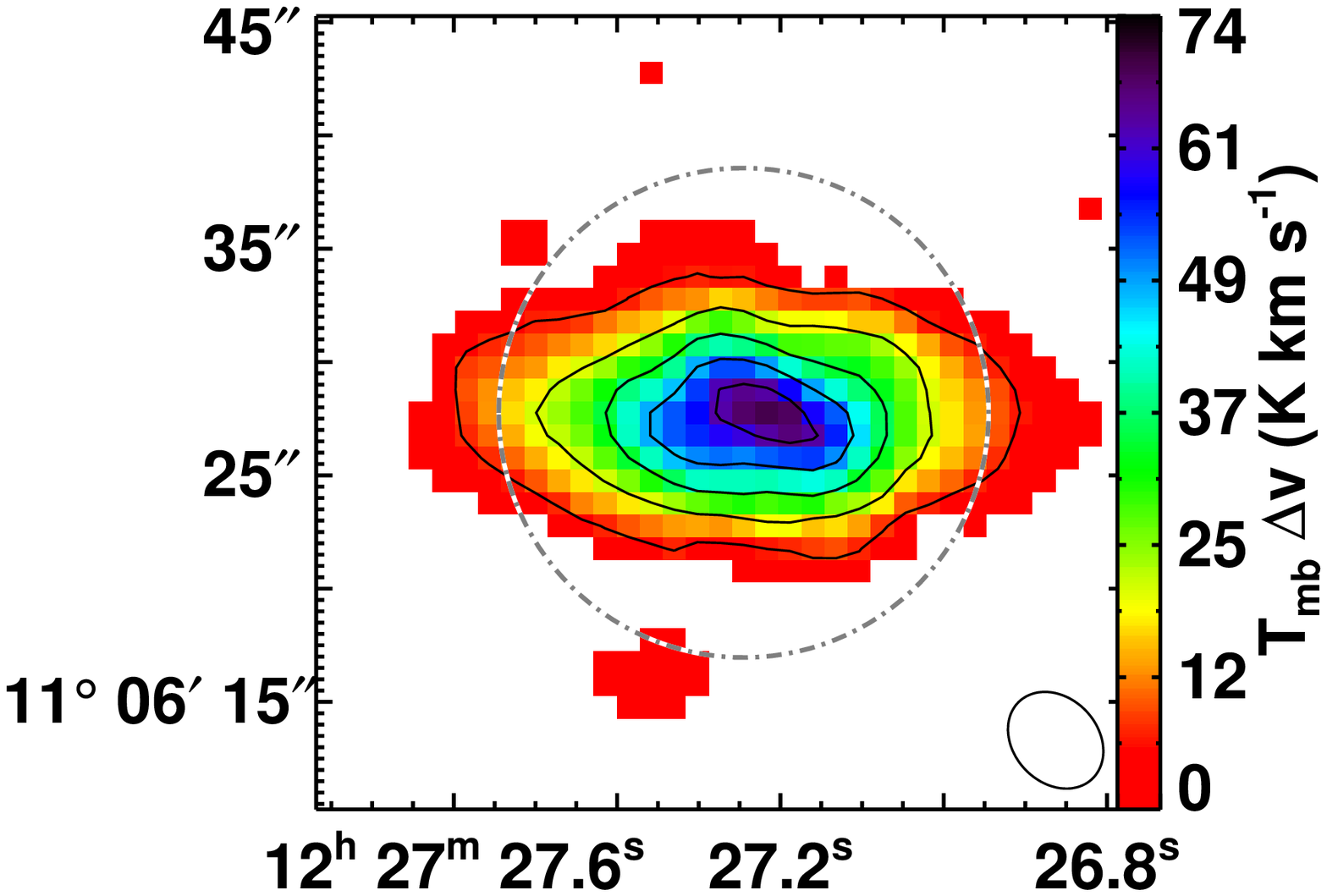}}
\subfloat{\includegraphics[height=1.6in,clip,trim=0.1cm 1.4cm 0cm 2.4cm]{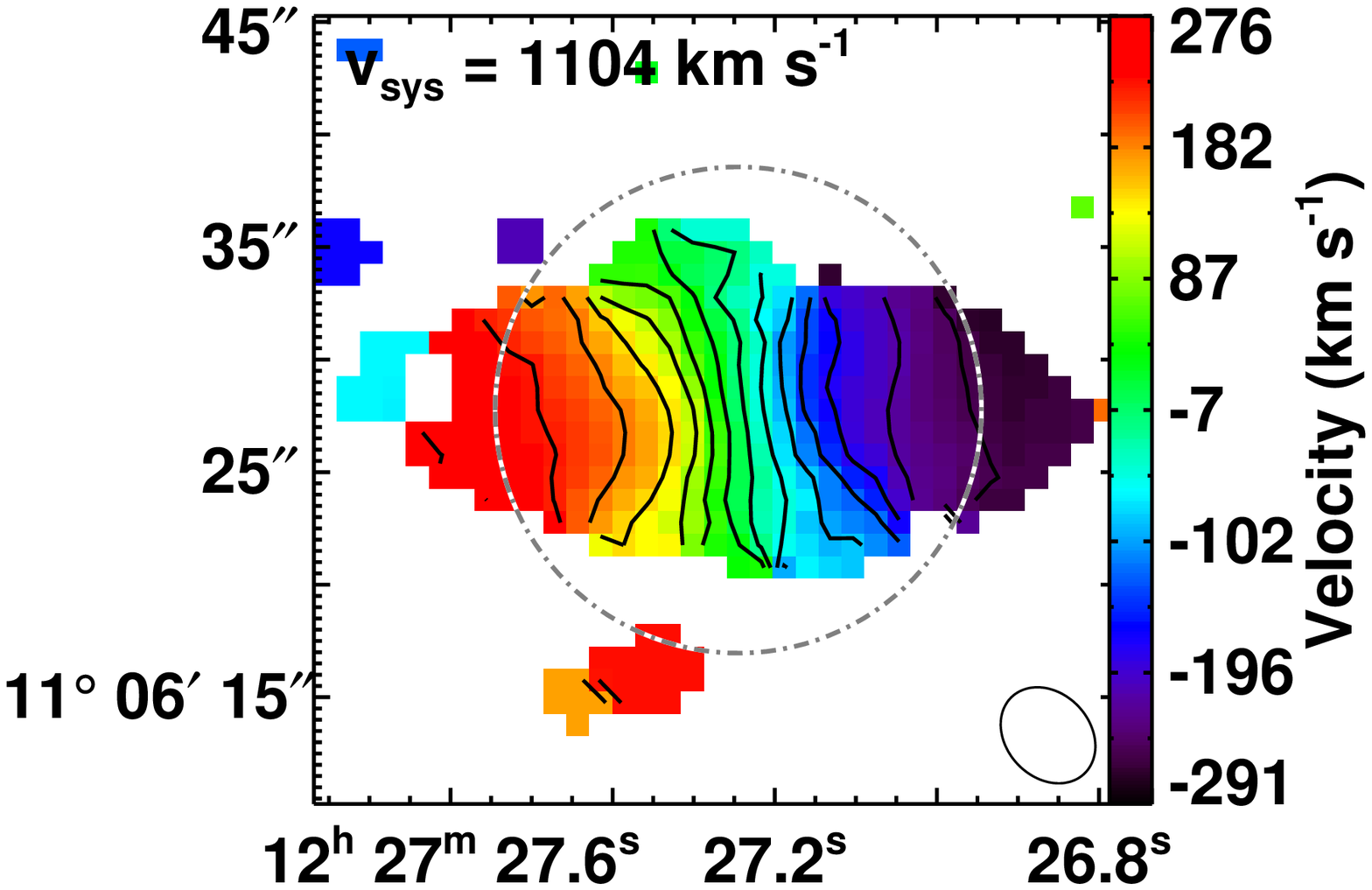}}
\subfloat{\includegraphics[height=1.6in,clip,trim=0cm 1.4cm 0cm 0.9cm]{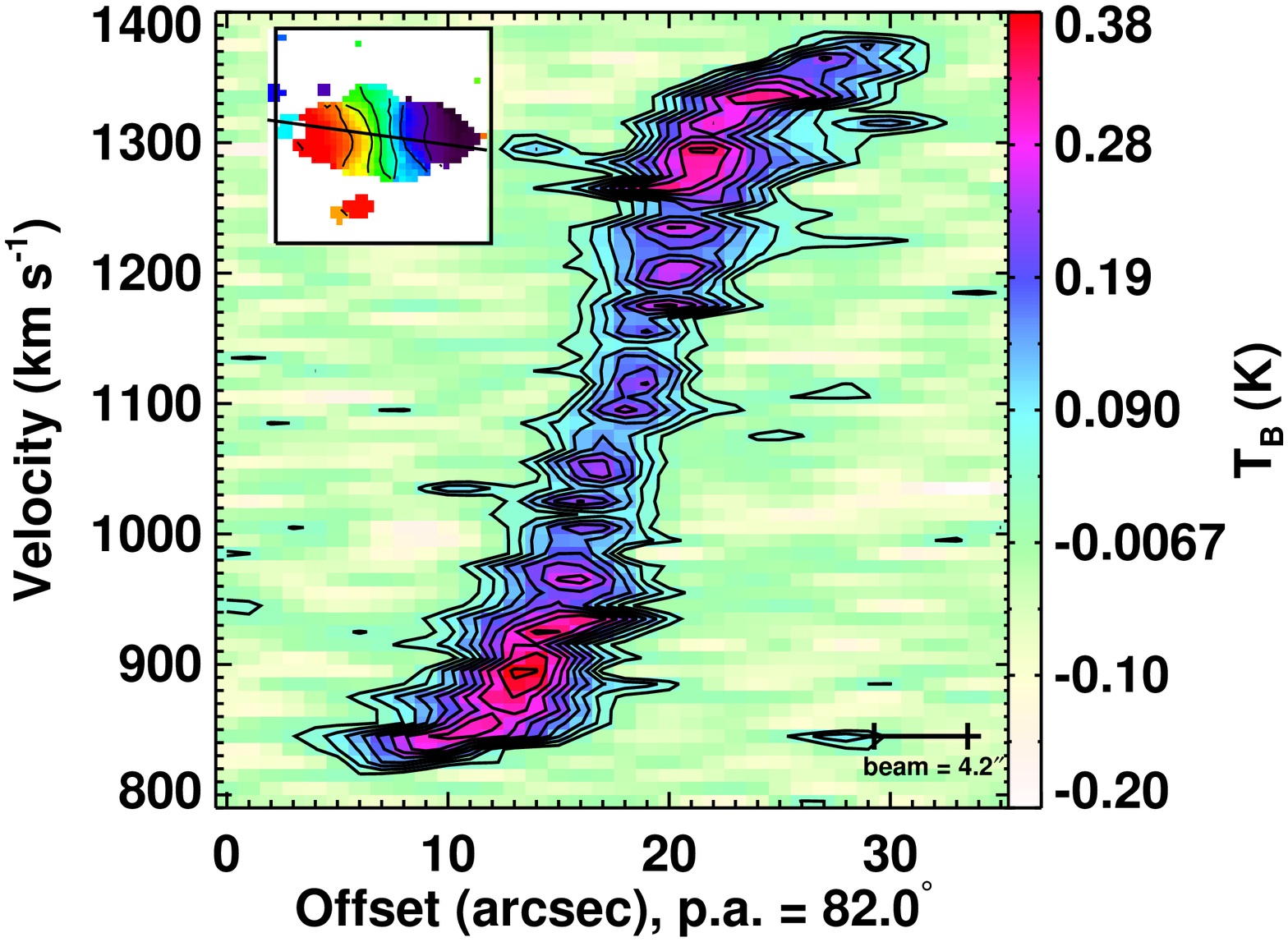}}
\end{figure*}
\begin{figure*}
\subfloat{\includegraphics[width=7in,clip,trim=1.6cm 2.3cm 2.6cm 3cm]{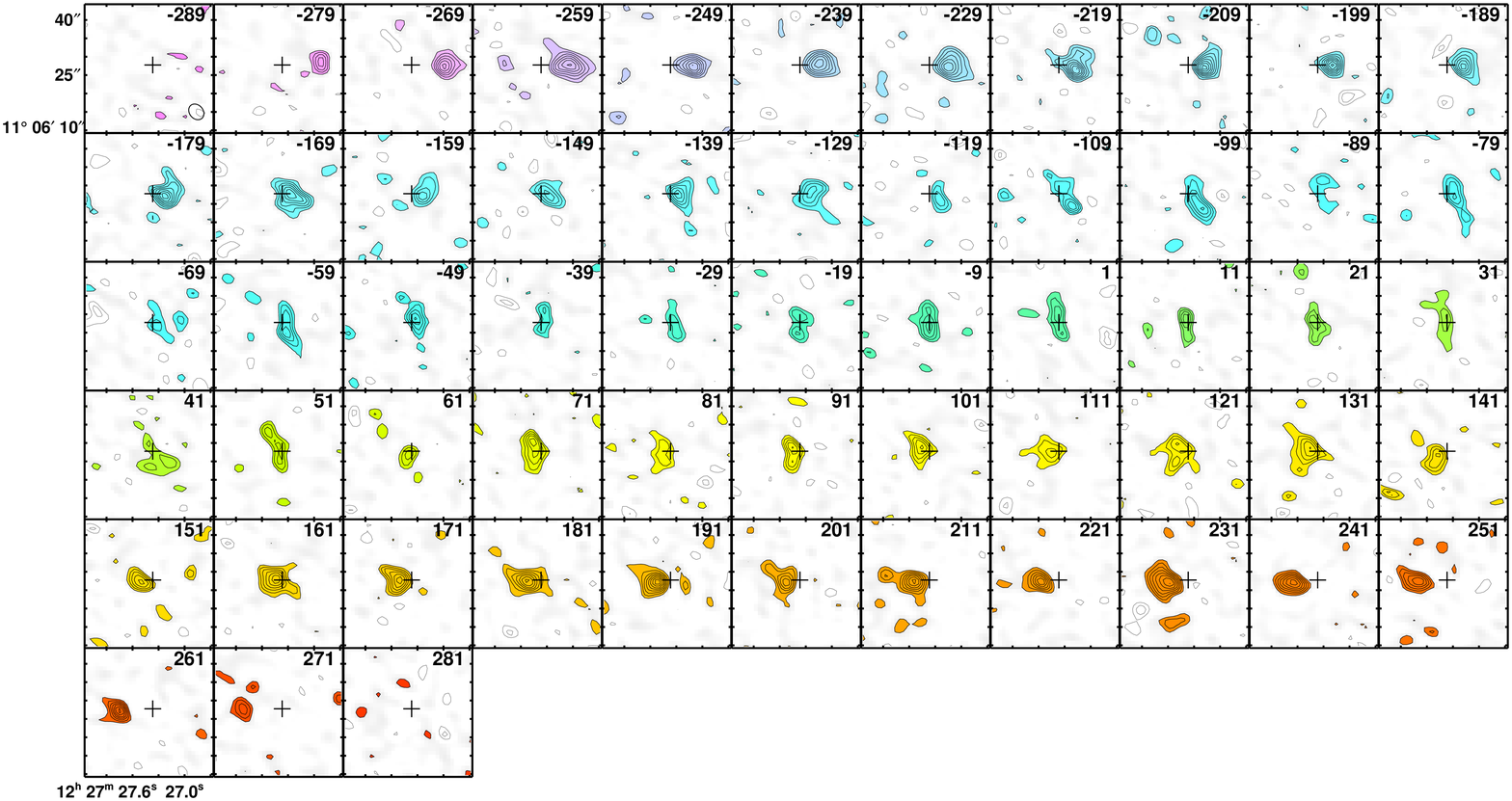}}
\caption{{\bf NGC~4429} is a Virgo regular rotator ($M_K$ = -24.32) that includes two velocity maxima, with a bar and ring stellar morphology.  It contains a dust disc.  The moment0 peak is 14 Jy beam$^{-1}$ \kms.  The moment1 contours are at 40 \kms\ intervals and the PVD contours are placed at $1.5\sigma$ intervals.}
\end{figure*}

\clearpage
\begin{figure*}
\centering
\subfloat{\includegraphics[height=2.2in,clip,trim=2.2cm 3.2cm 0cm 2.7cm]{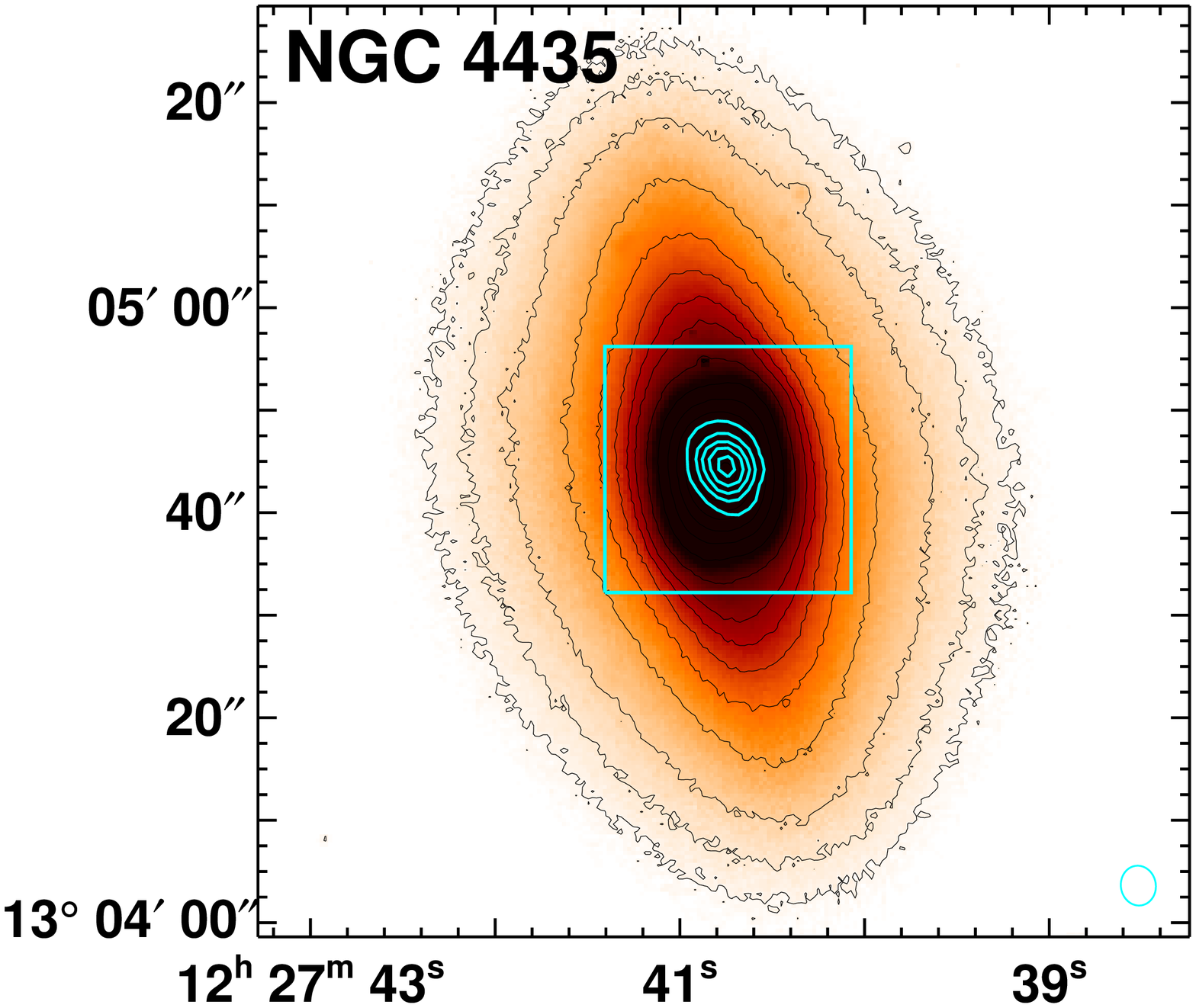}}
\subfloat{\includegraphics[height=2.2in,clip,trim=0cm 0.6cm 0.4cm 0.4cm]{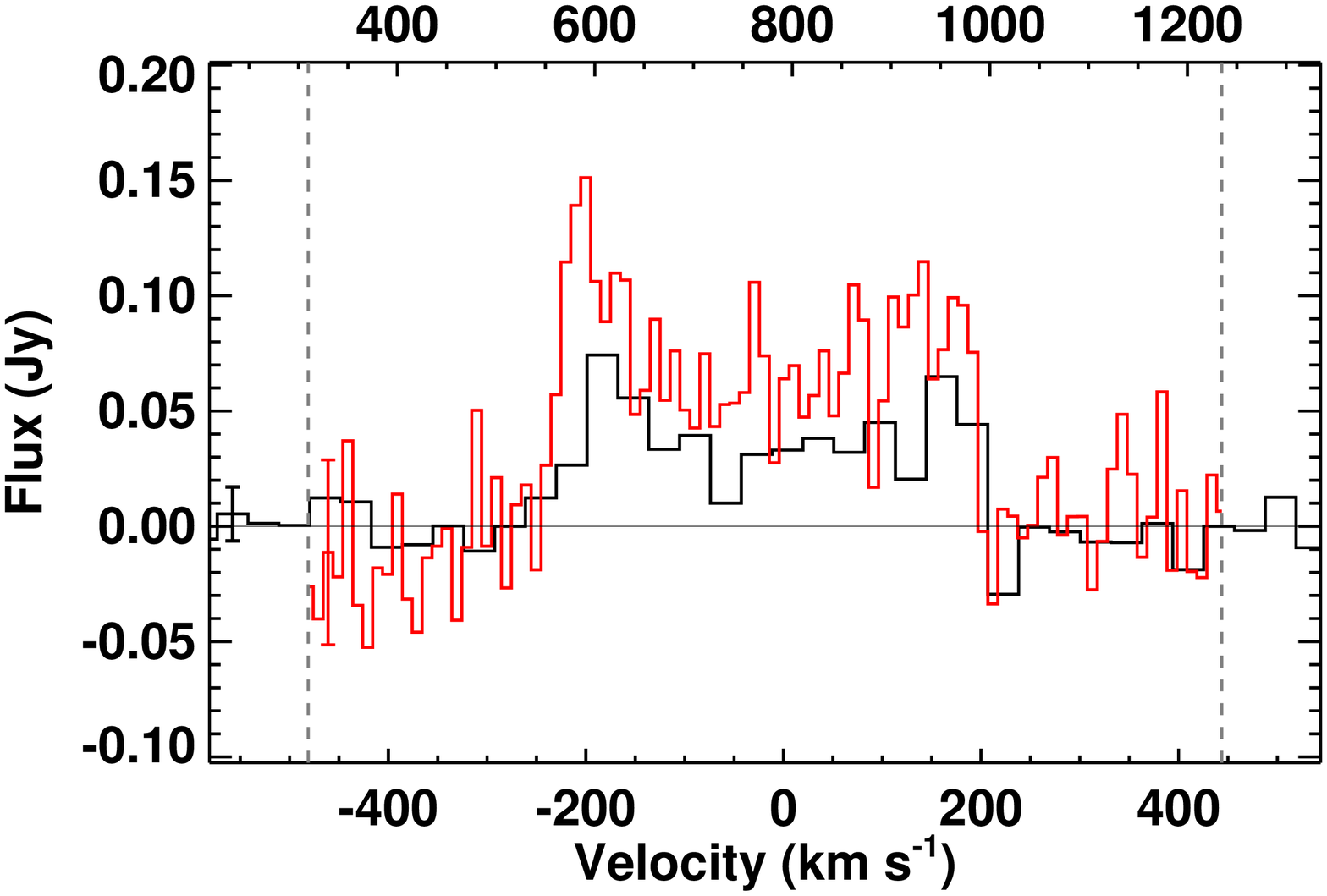}}
\end{figure*}
\begin{figure*}
\subfloat{\includegraphics[height=1.6in,clip,trim=0.1cm 1.4cm 0.4cm 2.4cm]{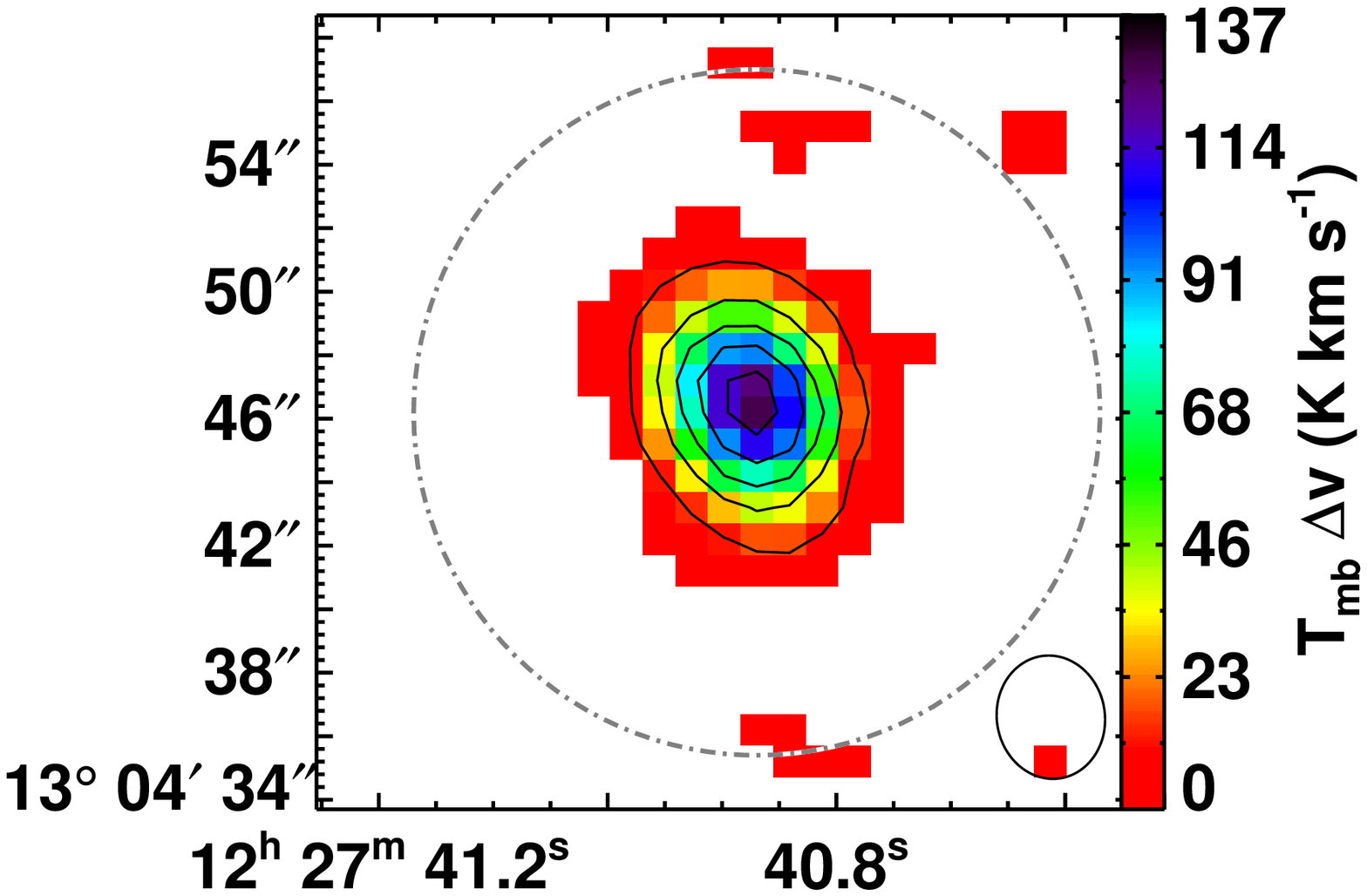}}
\subfloat{\includegraphics[height=1.6in,clip,trim=0.1cm 1.4cm 0.4cm 2.4cm]{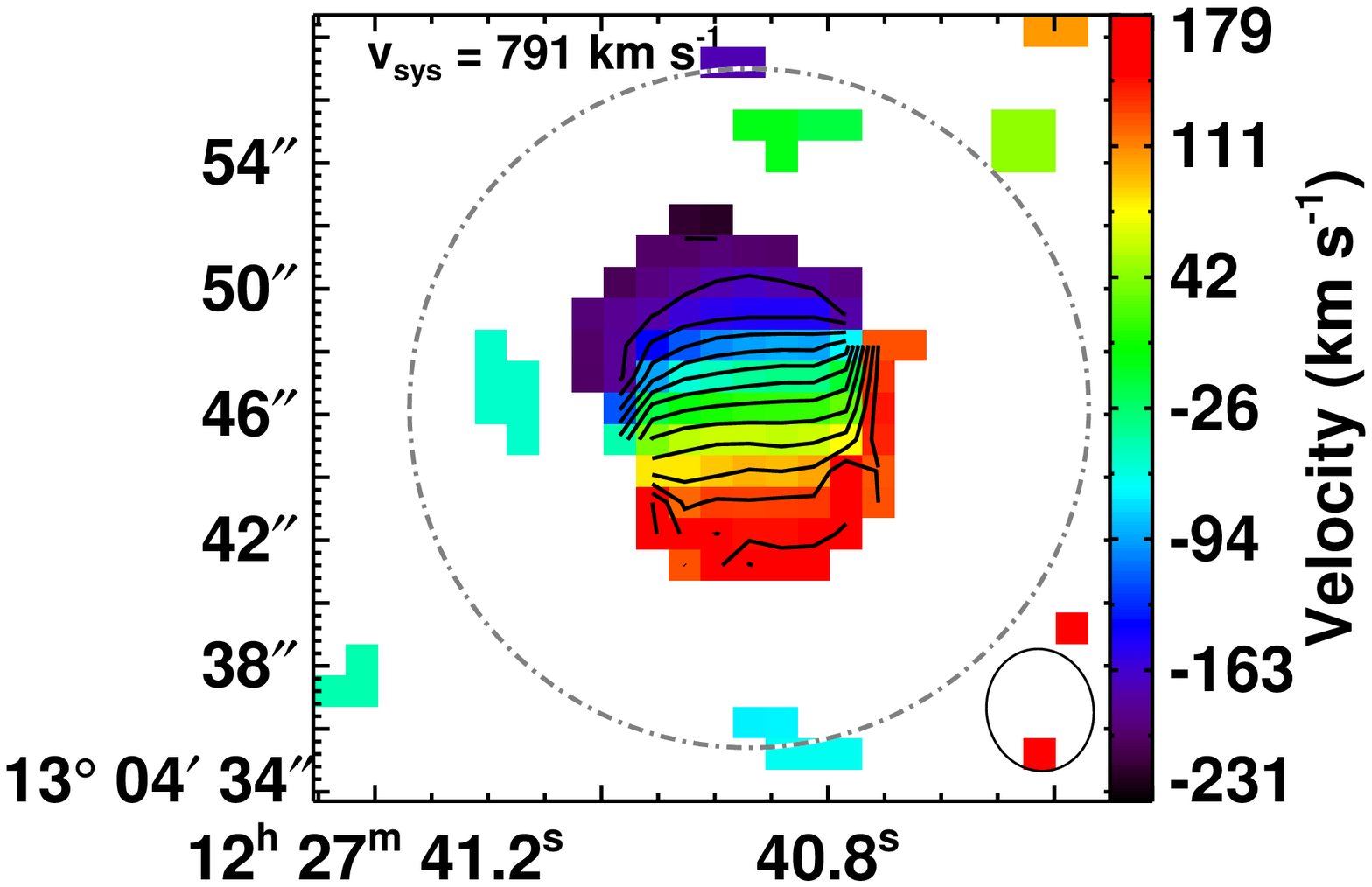}}
\subfloat{\includegraphics[height=1.6in,clip,trim=0cm 1.4cm 0cm 0.9cm]{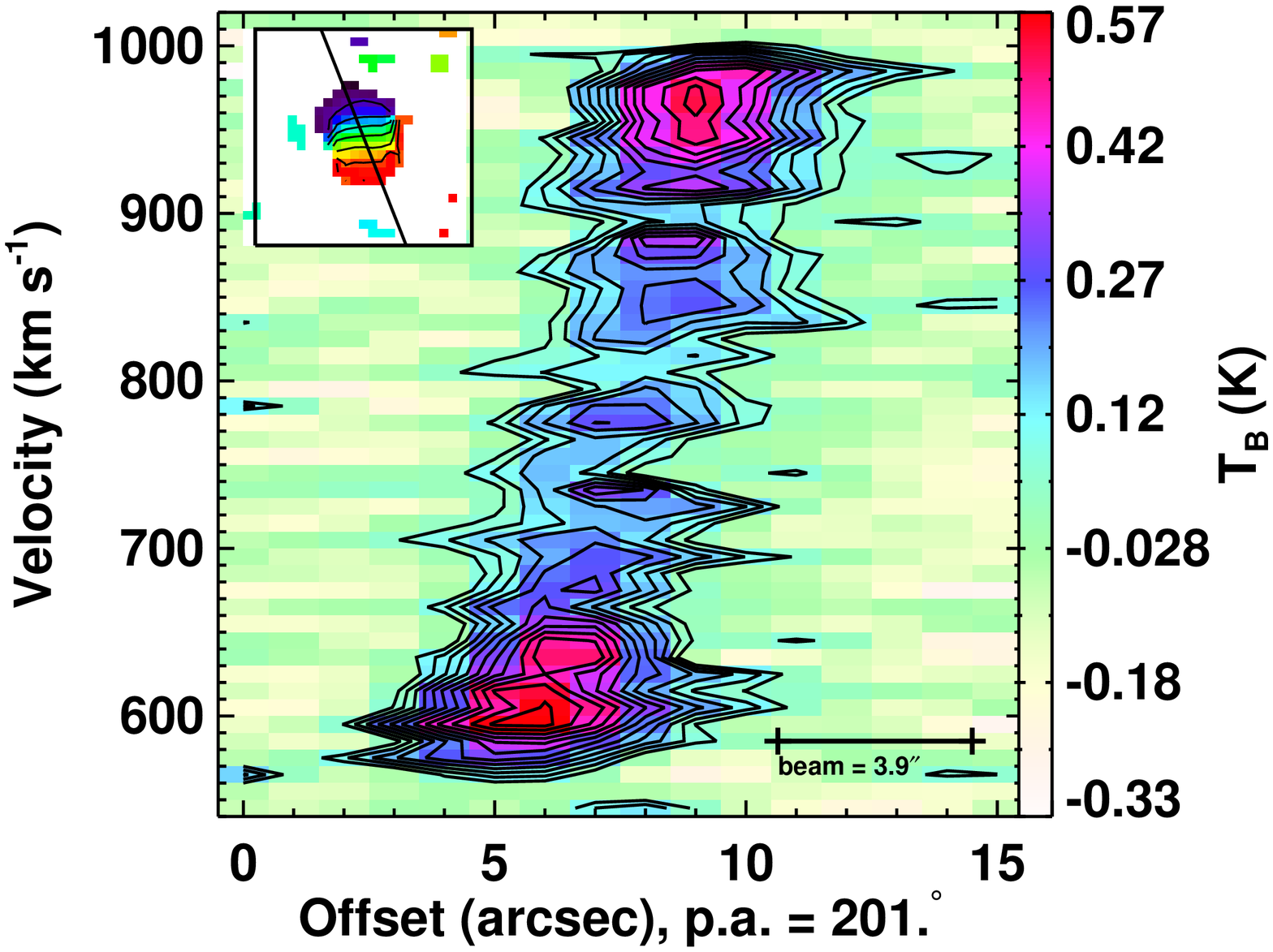}}
\end{figure*}
\begin{figure*}
\subfloat{\includegraphics[width=7in,clip,trim=1.6cm 4cm 3.1cm 3.2cm]{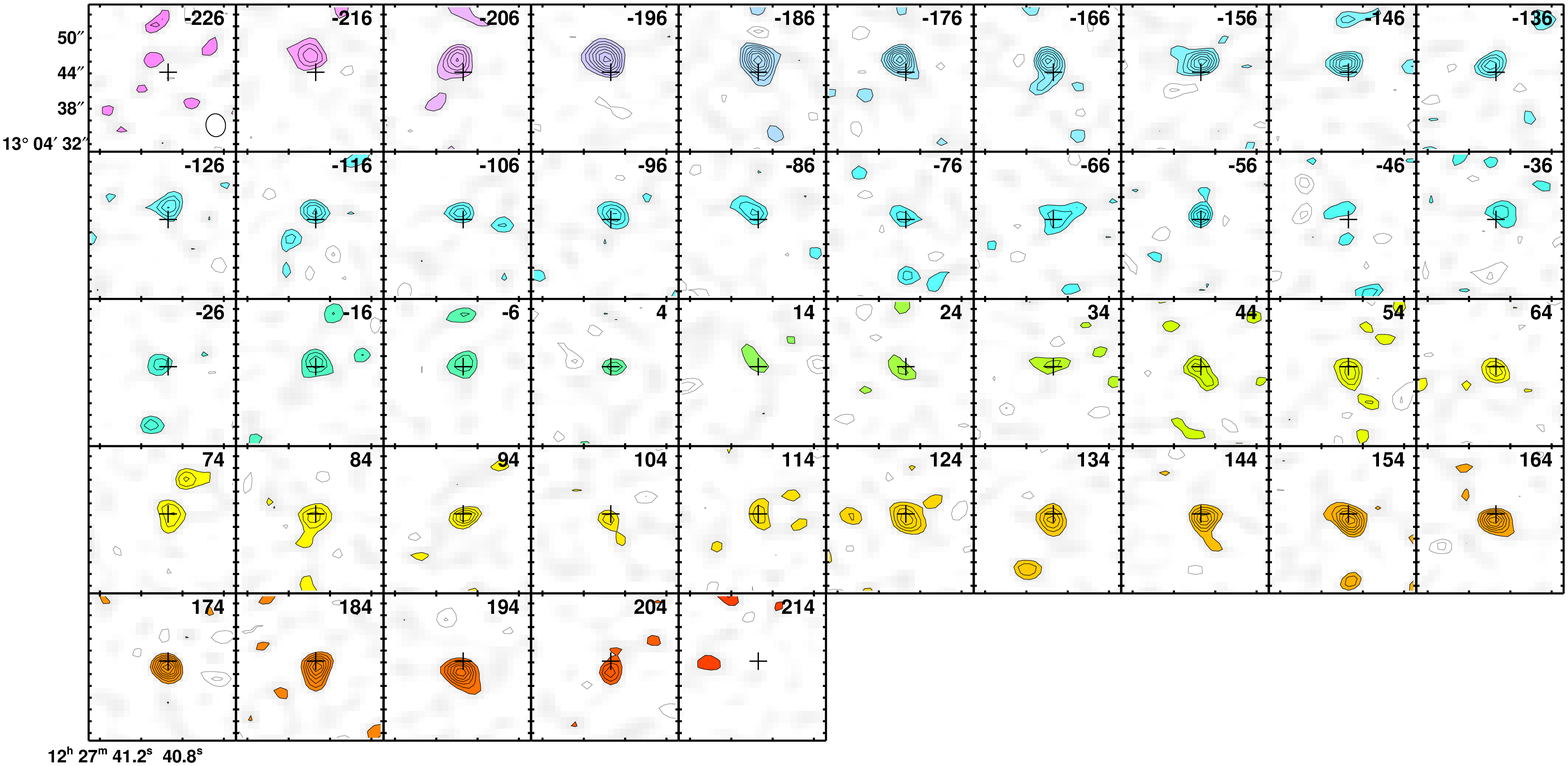}}
\caption{{\bf NGC~4435} is a Virgo regular rotator ($M_K$ = -23.82) that includes two velocity maxima, with a normal stellar morphology.  It contains a dust disc.  The moment0 peak is 20 Jy beam$^{-1}$ \kms.  The moment1 contours are at 30 \kms\ intervals.}
\end{figure*}

\clearpage
\begin{figure*}
\centering
\subfloat{\includegraphics[height=2.2in,clip,trim=2.2cm 3.2cm 0cm 2.7cm]{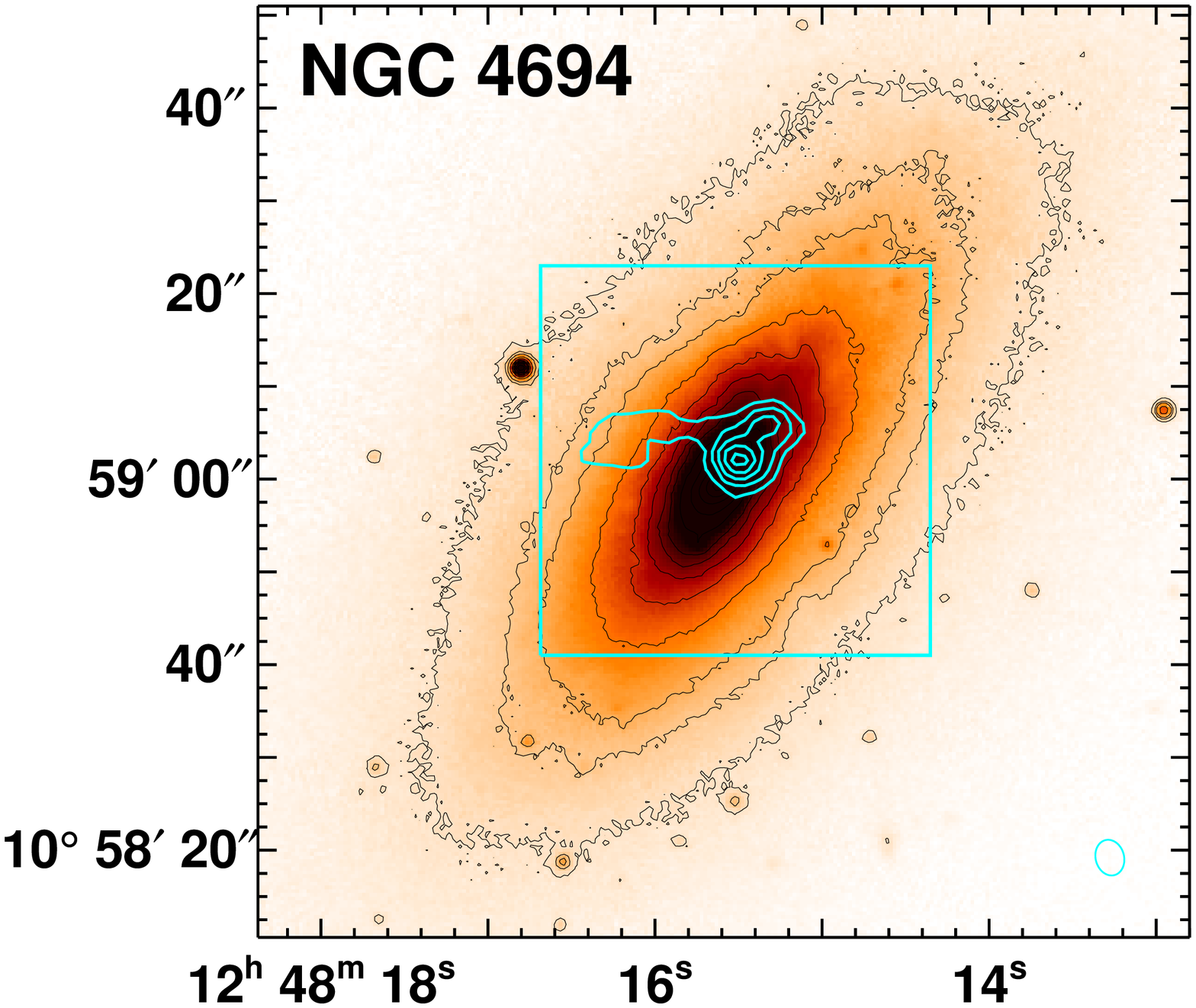}}
\subfloat{\includegraphics[height=2.2in,clip,trim=0cm 0.6cm 0.4cm 0.4cm]{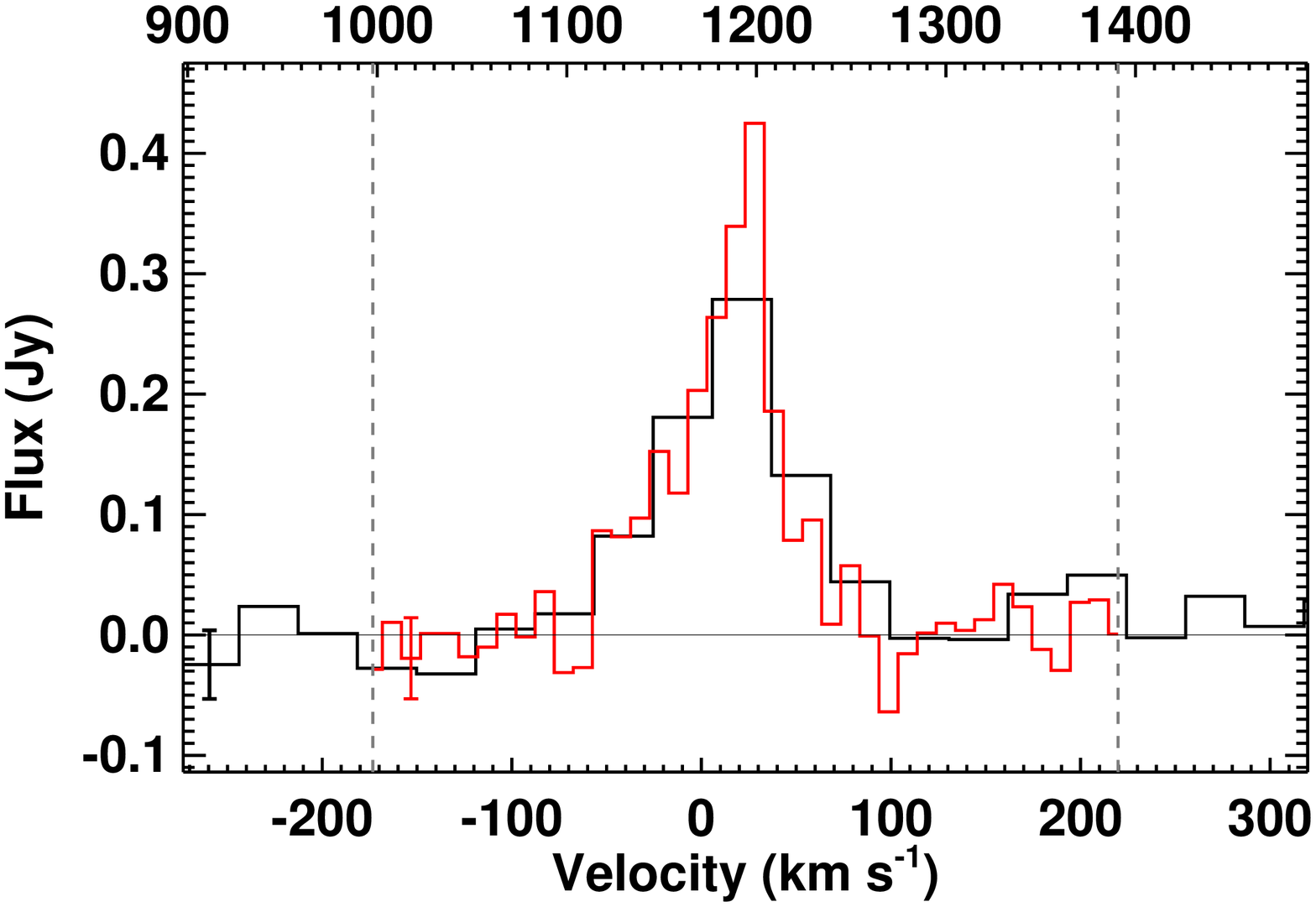}}
\end{figure*}
\begin{figure*}
\subfloat{\includegraphics[height=1.6in,clip,trim=0.1cm 1.4cm 0.6cm 2.4cm]{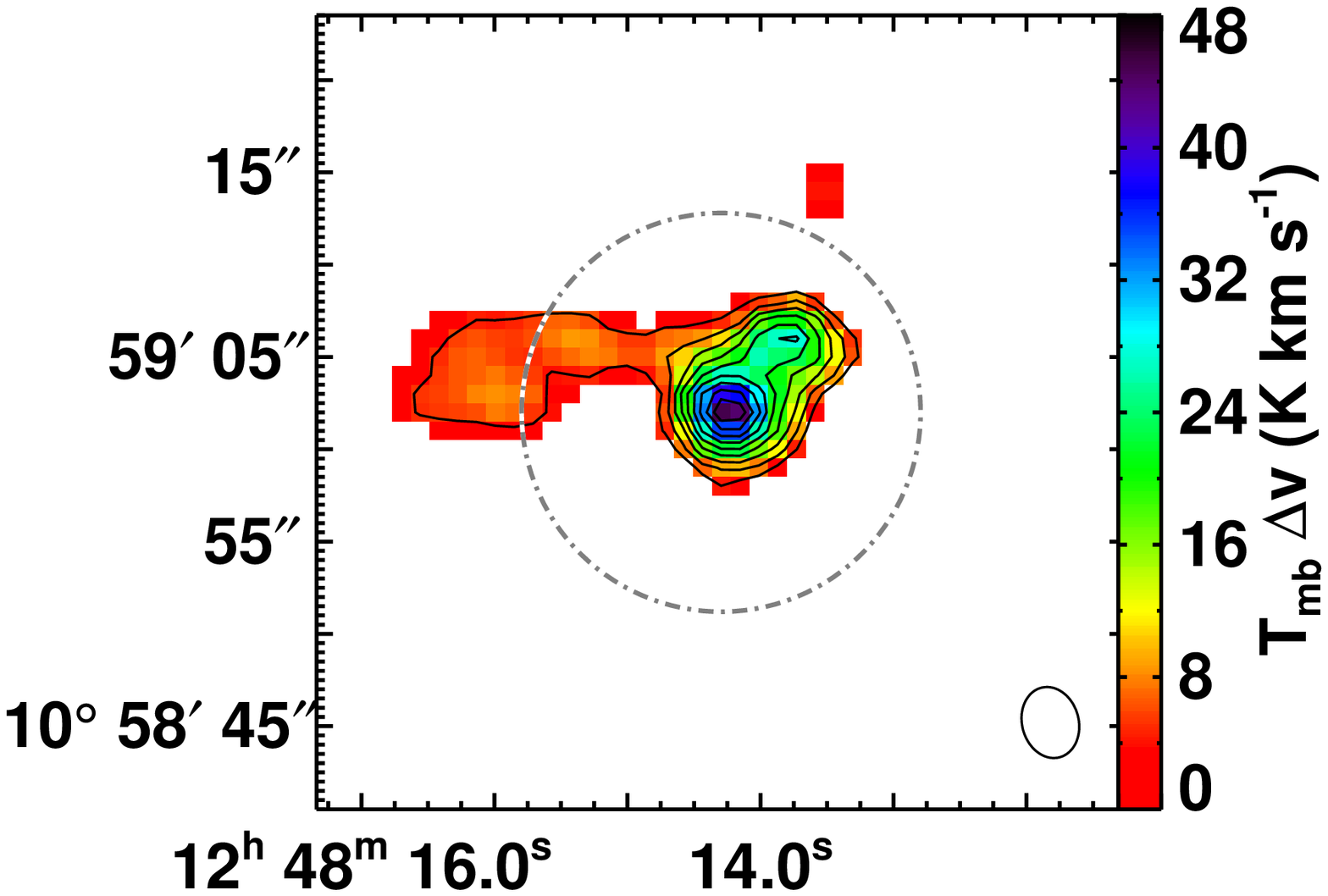}}
\subfloat{\includegraphics[height=1.6in,clip,trim=0.1cm 1.4cm 0cm 2.4cm]{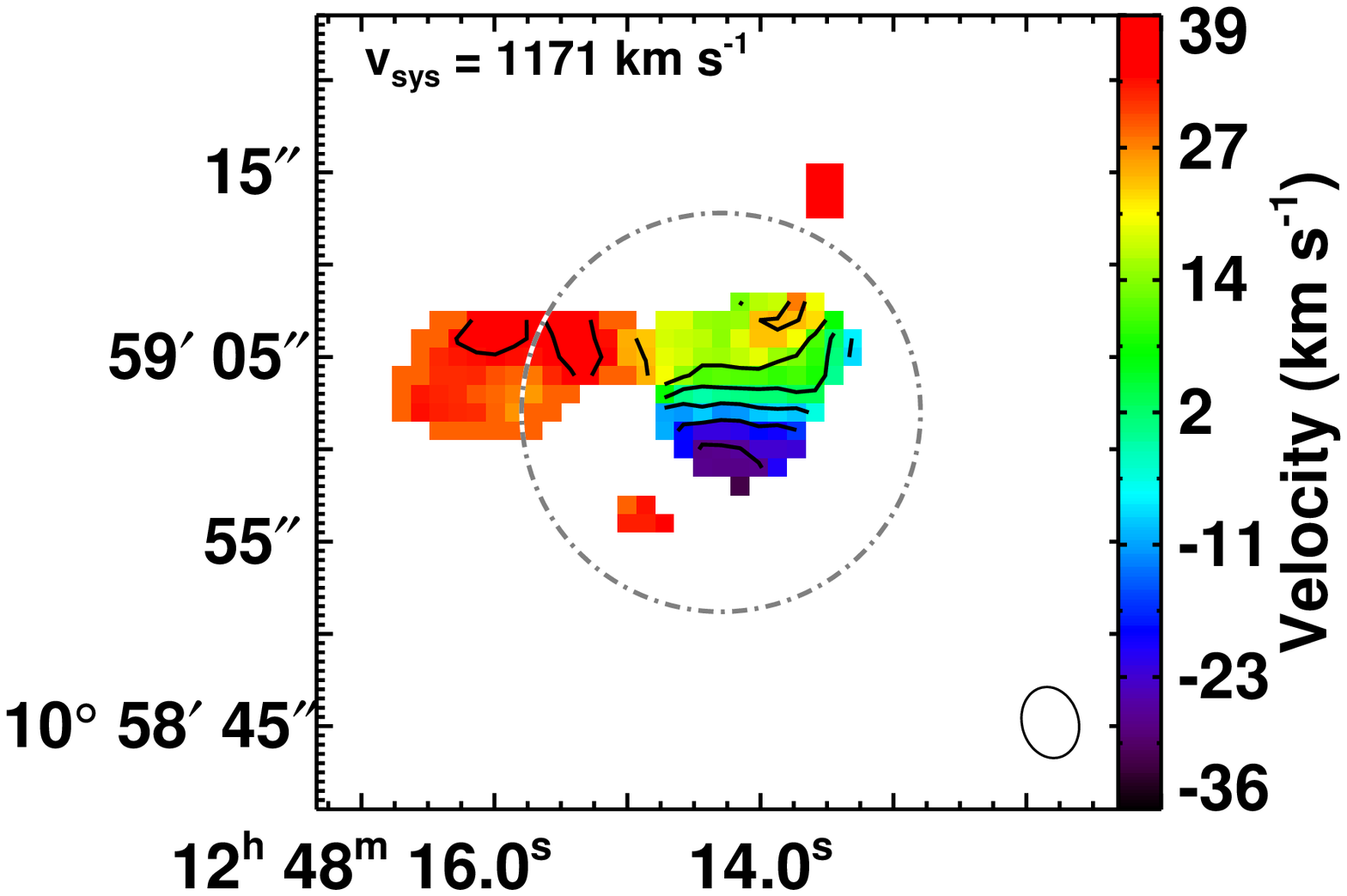}}
\subfloat{\includegraphics[height=1.6in,clip,trim=0cm 1.4cm 0cm 0.9cm]{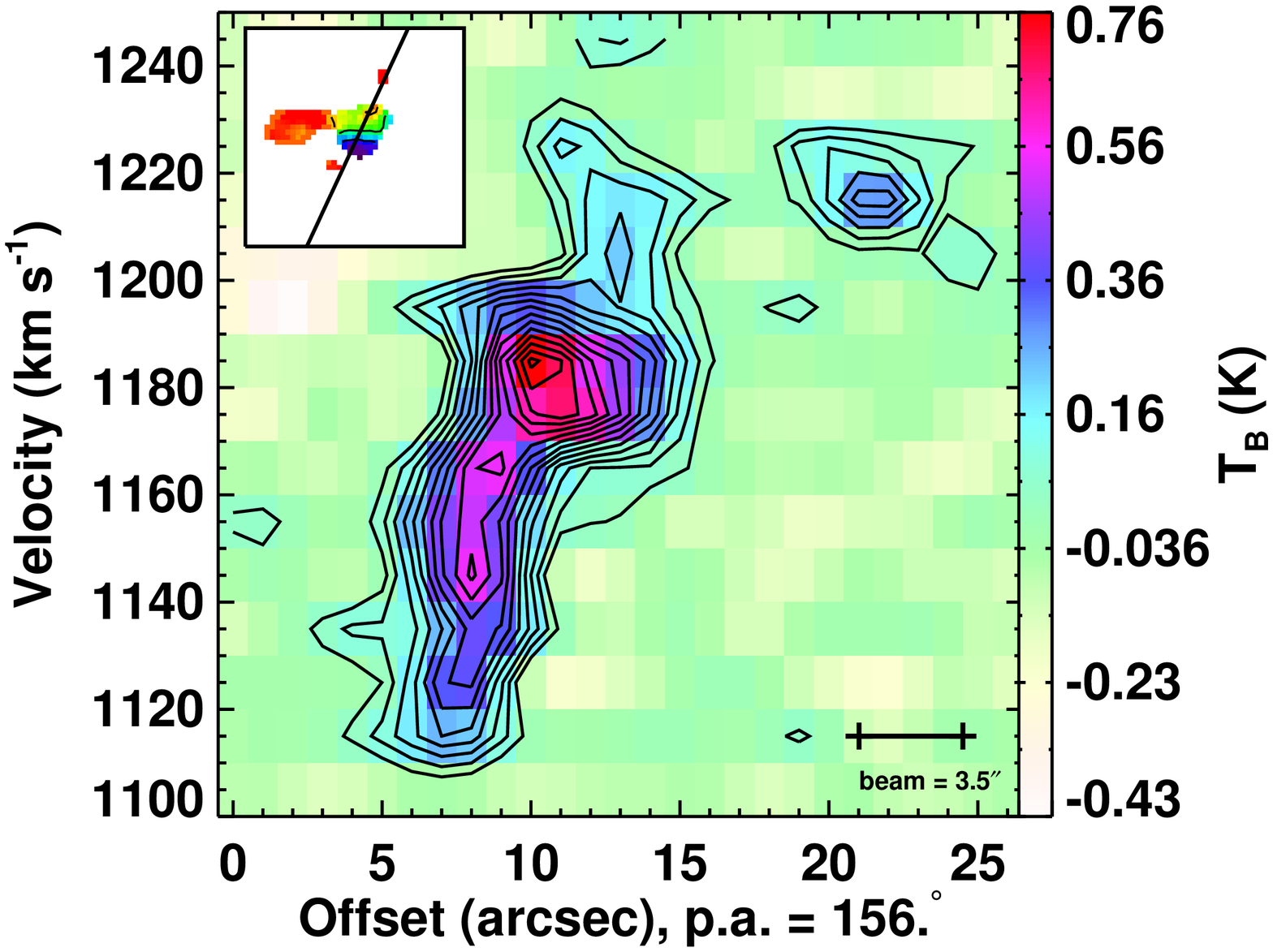}}
\end{figure*}
\begin{figure*}
\subfloat{\includegraphics[width=7in,clip,trim=0cm 1.3cm 2.6cm 1.8cm]{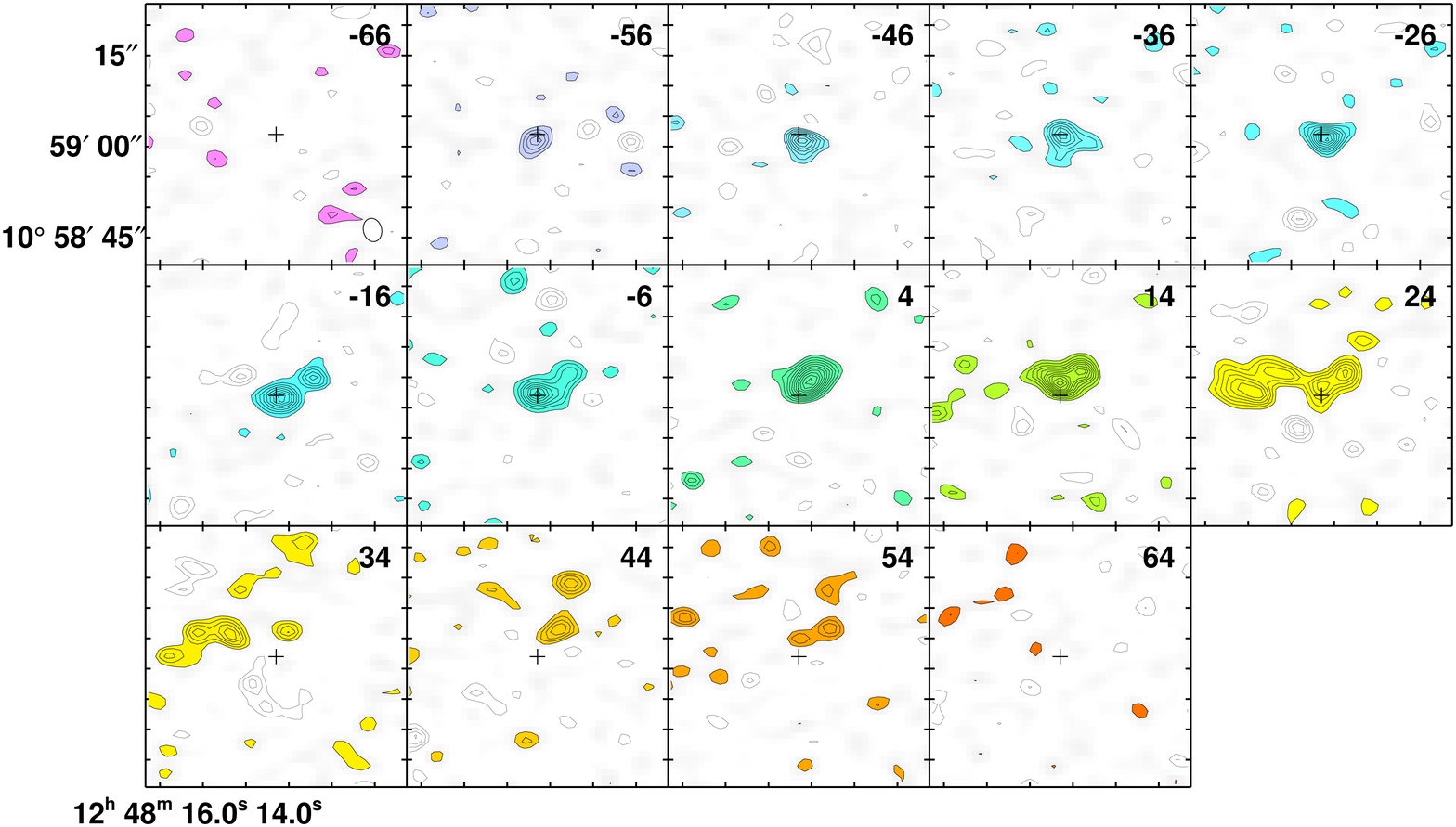}}
\caption{{\bf NGC~4694} is a Virgo regular rotator ($M_K$ = -22.15) with normal stellar morphology.  It appears to be on its first approach into Virgo.  It contains a dust filaments and bars.  The moment0 peak is 6.0 Jy beam$^{-1}$ \kms.  The moment1 contours are placed at 10\kms\ intervals and the PVD contours are placed at $1.5\sigma$ intervals.}
\end{figure*}

\clearpage
\begin{figure*}
\centering
\subfloat{\includegraphics[height=2.2in,clip,trim=2.4cm 3.2cm 0cm 2.7cm]{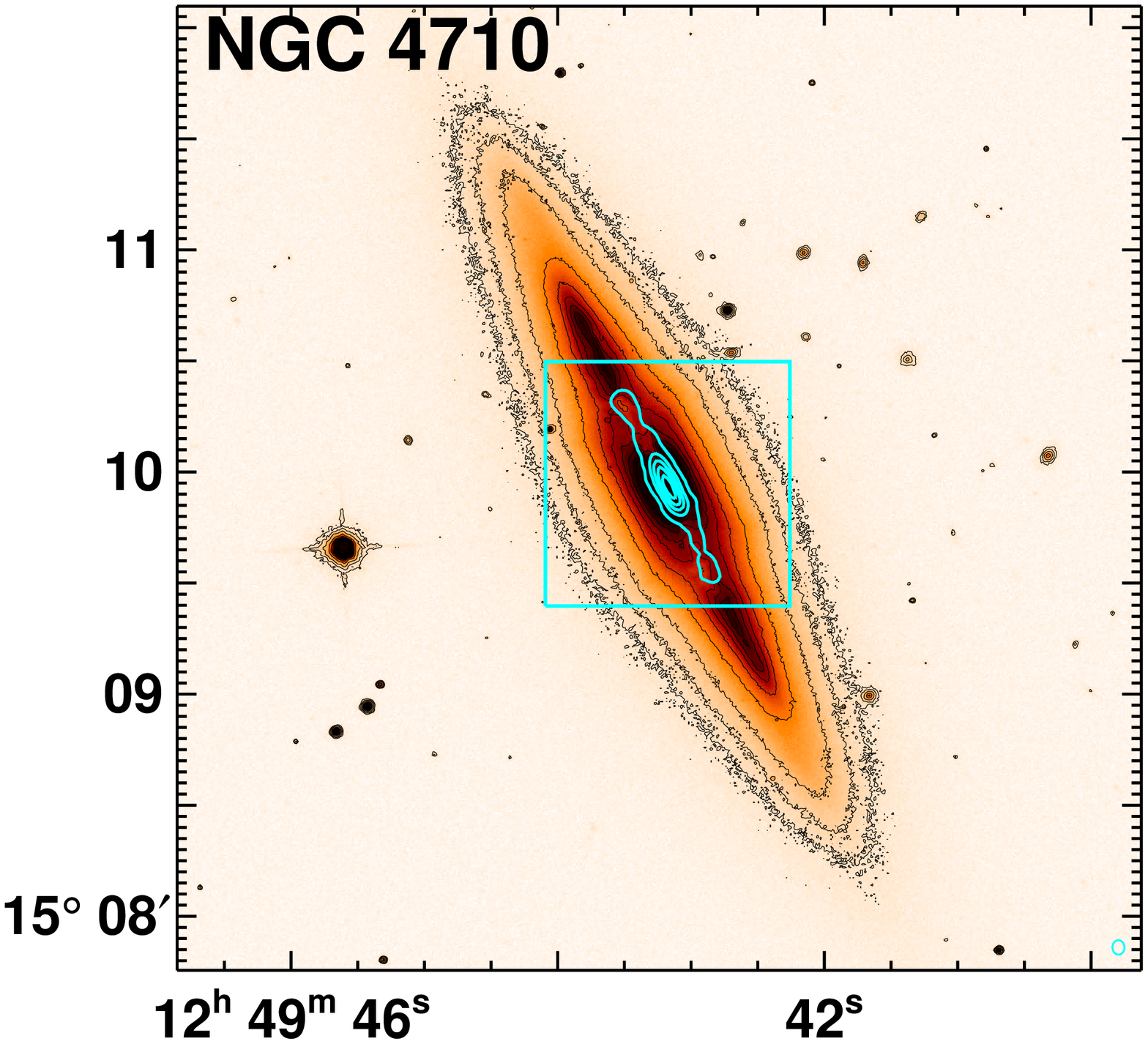}}
\subfloat{\includegraphics[height=2.2in,clip,trim=0cm 0.6cm 0.4cm 0.4cm]{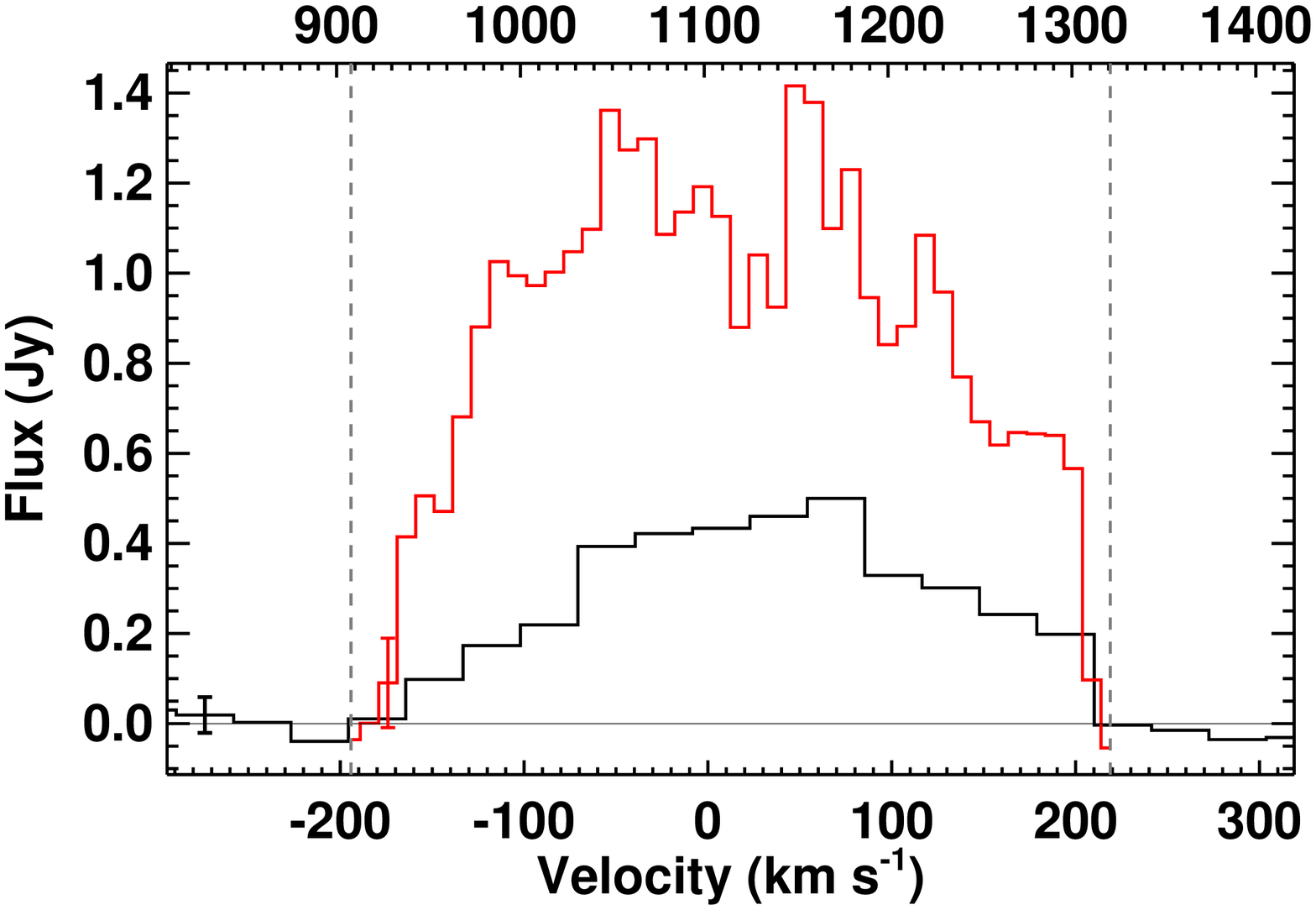}}
\end{figure*}
\begin{figure*}
\subfloat{\includegraphics[height=1.6in,clip,trim=0cm 1.4cm 0cm 2cm]{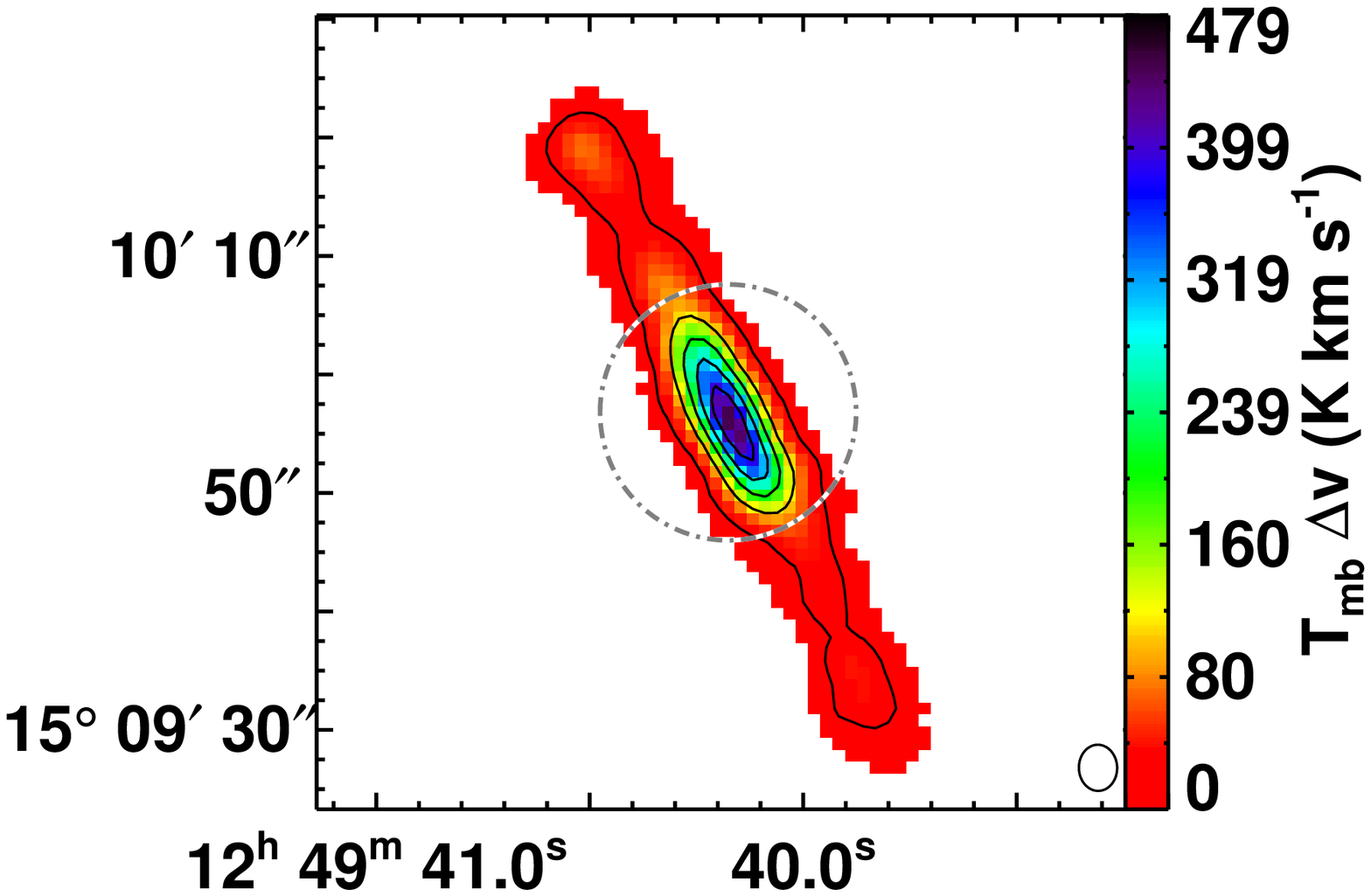}}
\subfloat{\includegraphics[height=1.6in,clip,trim=0cm 1.4cm 0cm 2cm]{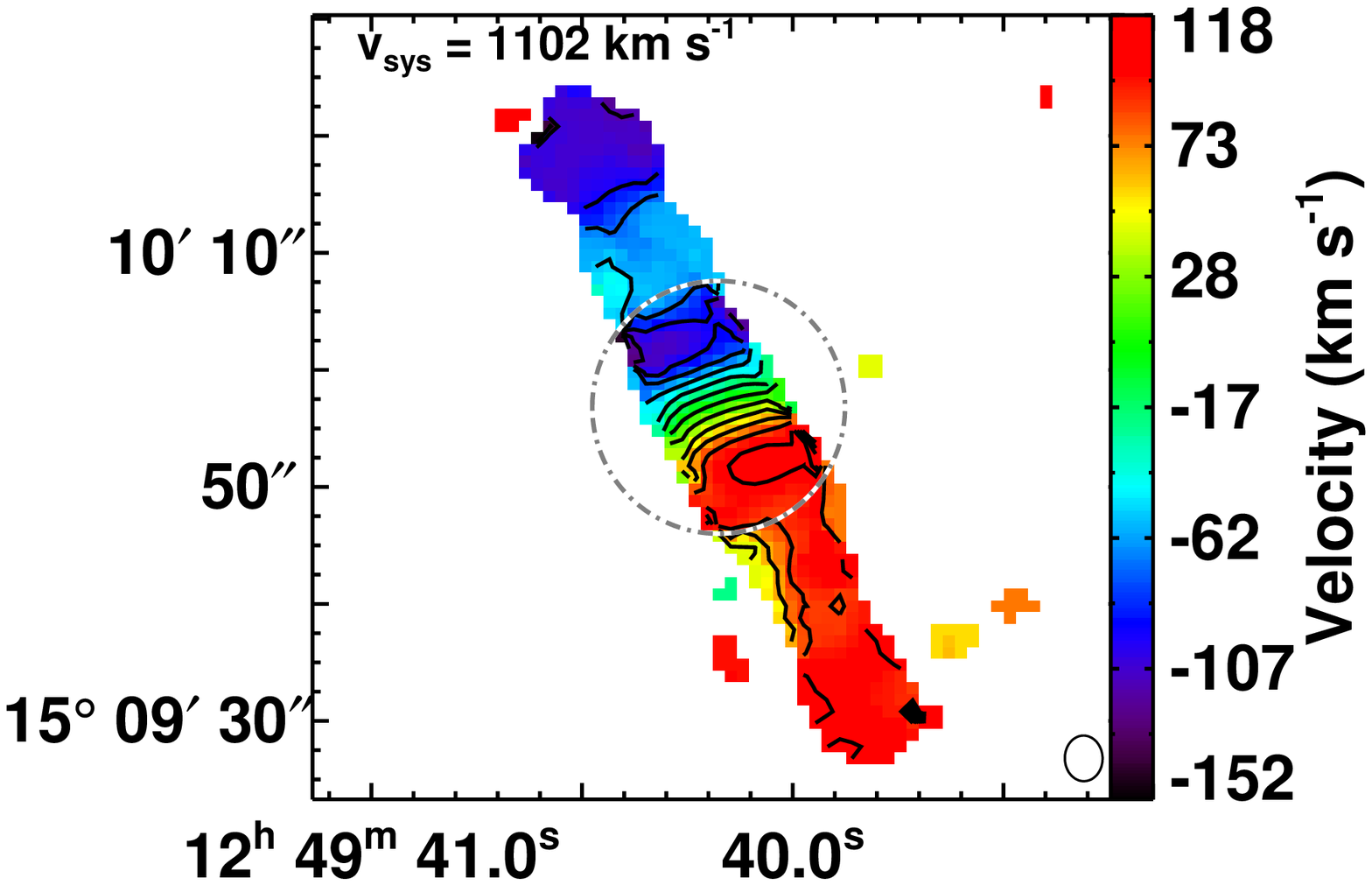}}
\subfloat{\includegraphics[height=1.6in,clip,trim=0cm 1.4cm 0.6cm 0.8cm]{figures/ngc4710_pv.eps}}
\end{figure*}
\begin{figure*}
\subfloat{\includegraphics[width=7.0in,clip,trim=1.1cm 1.5cm 0.3cm 1.5cm]{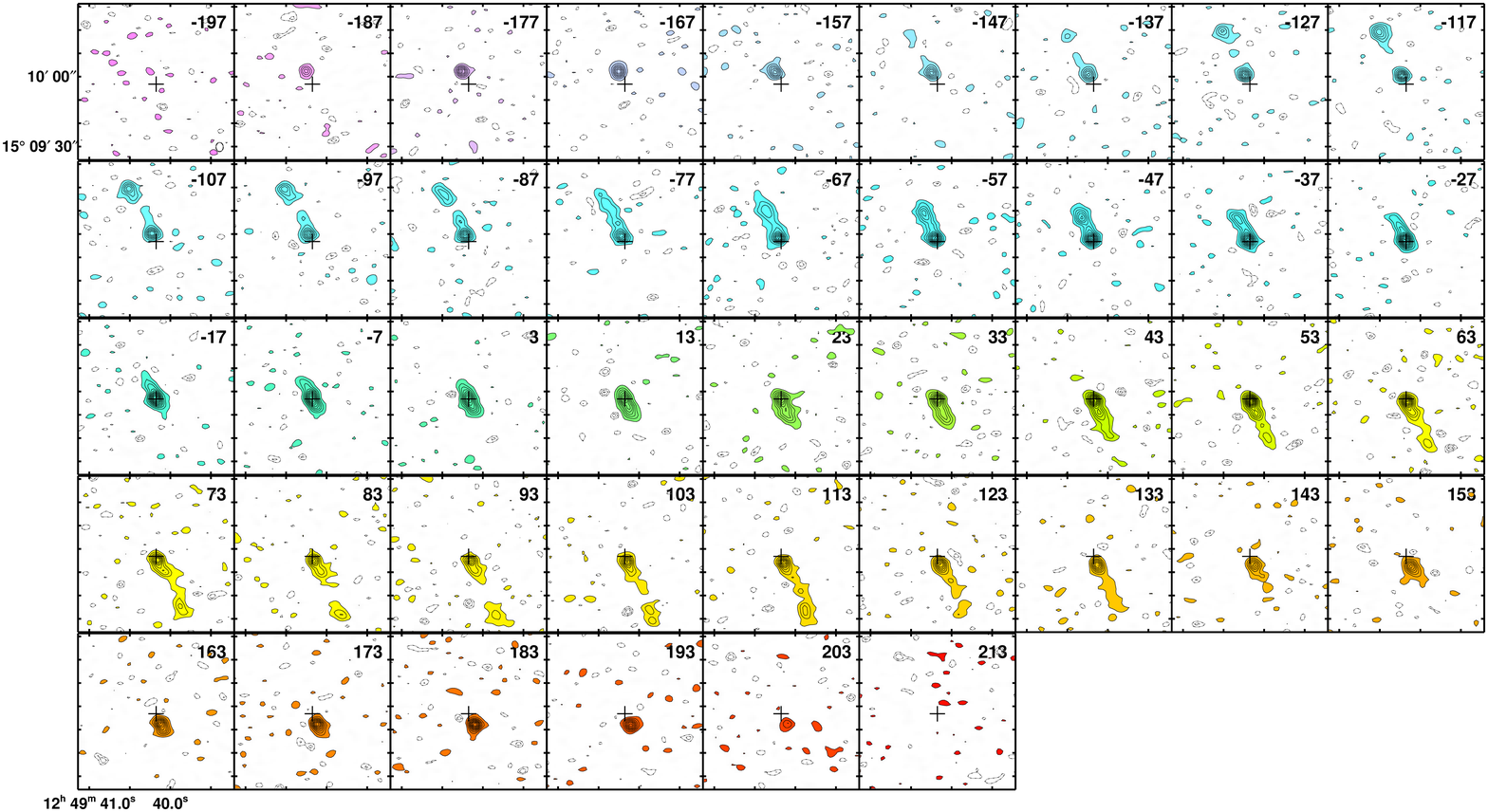}}
\caption{{\bf NGC~4710} is a Virgo regular rotator ($M_K$ = -23.52) with normal stellar morphology.  It appears to be on the outskirts of Virgo and is likely to be on its first approach.  It contains a dust disc, and is one of the nearest to edge-on systems in the sample.  The moment0 peak is 62 Jy beam$^{-1}$ \kms.  Channel map and PVD contours are placed at 3$\sigma$ intervals.}
\end{figure*}

\clearpage
\begin{figure*}
\centering
\subfloat{\includegraphics[height=2.2in,clip,trim=2.3cm 3.2cm 0cm 2.7cm]{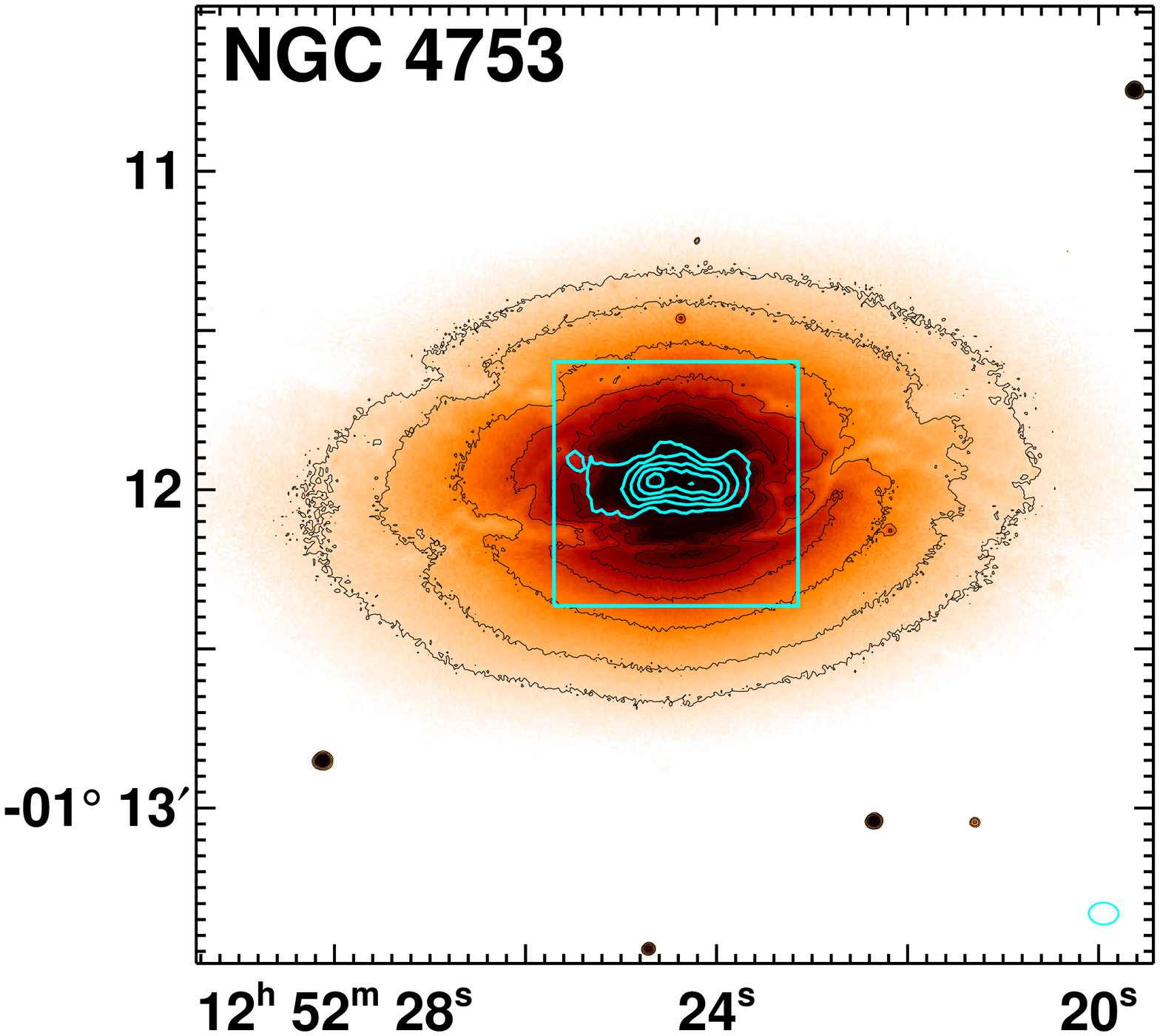}}
\subfloat{\includegraphics[height=2.2in,clip,trim=0cm 0.6cm 0.4cm 0.4cm]{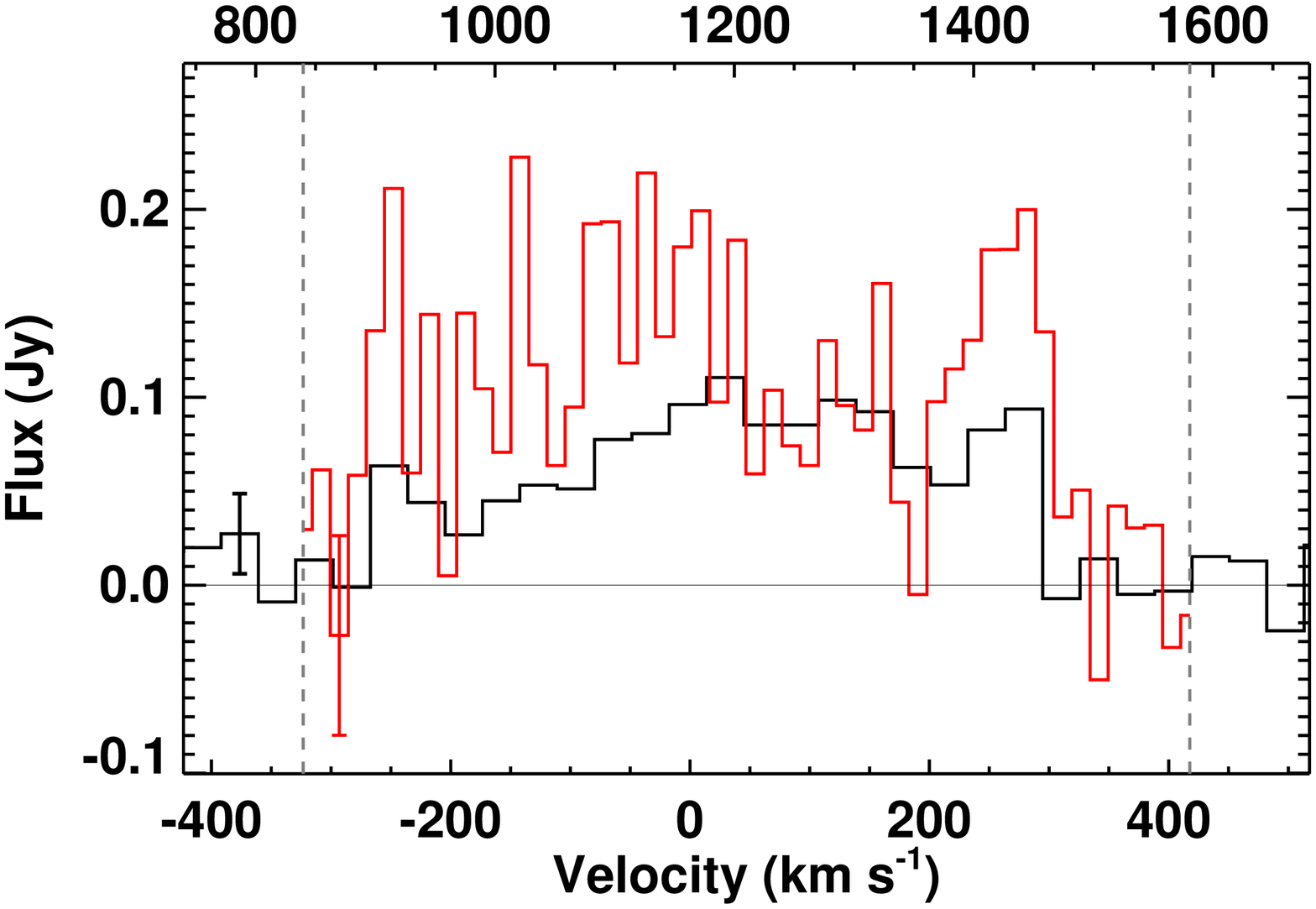}}
\end{figure*}
\begin{figure*}
\subfloat{\includegraphics[height=1.6in,clip,trim=0.1cm 1.4cm 0.8cm 2.4cm]{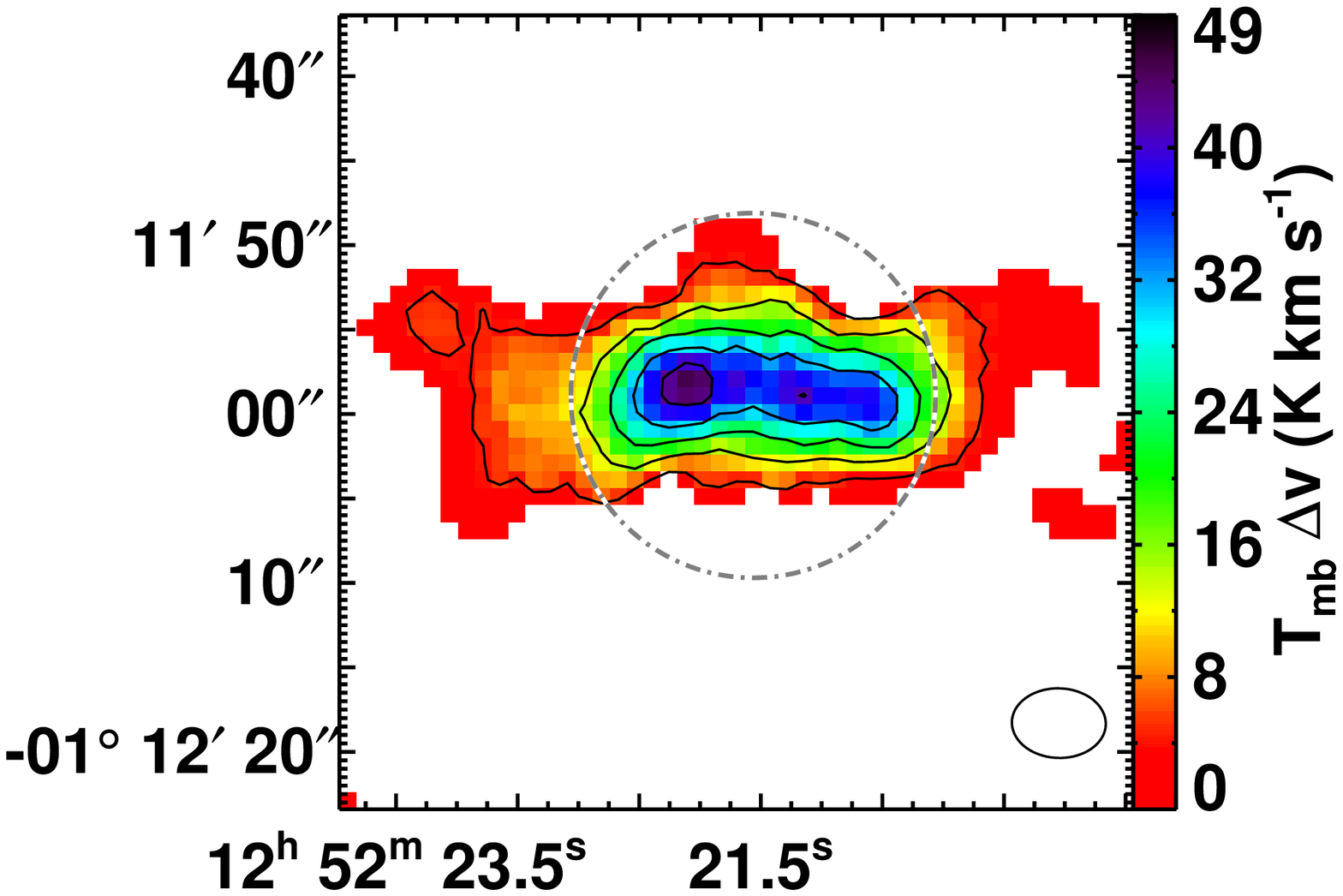}}
\subfloat{\includegraphics[height=1.6in,clip,trim=0.1cm 1.4cm 0cm 2.4cm]{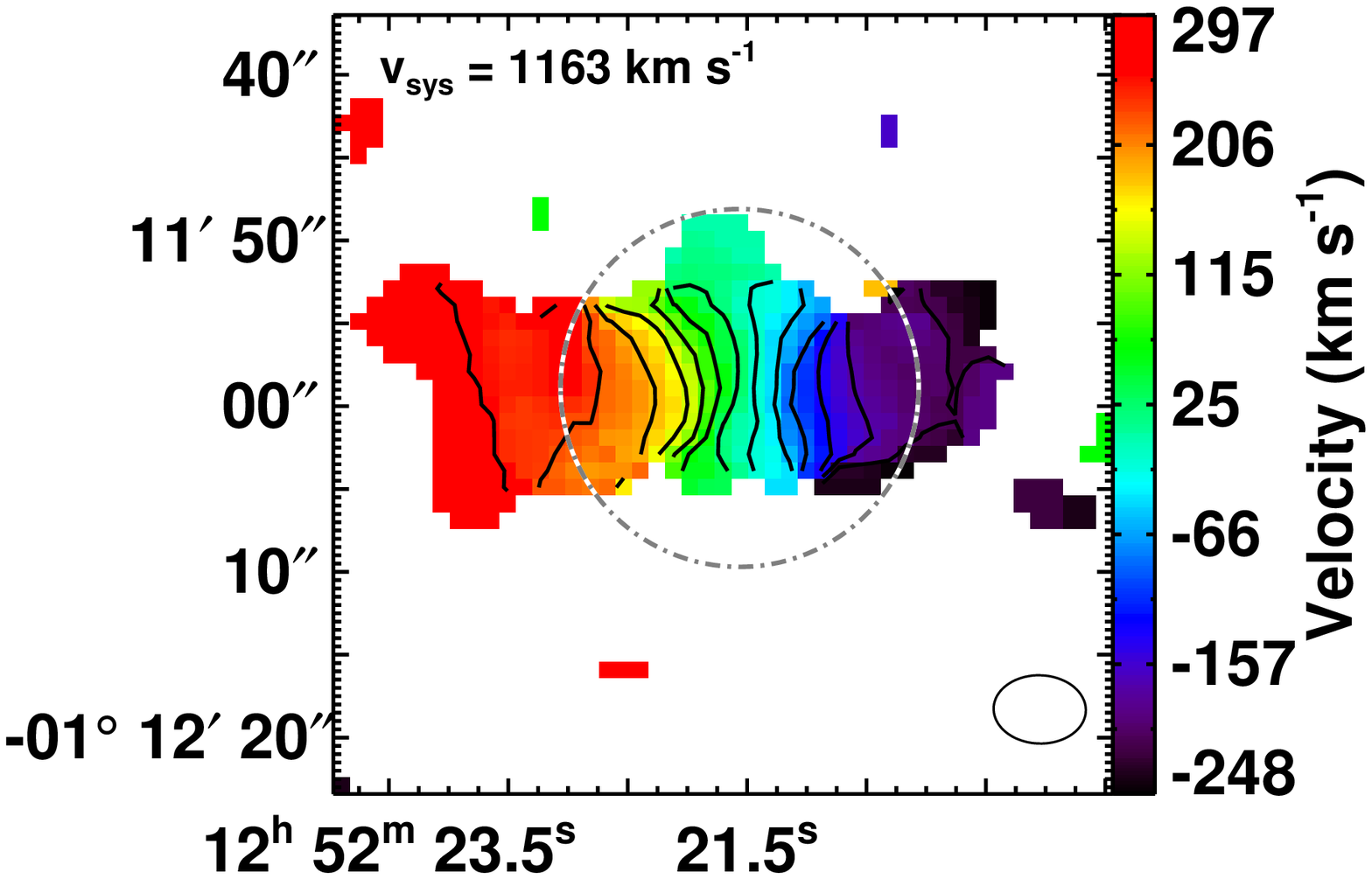}}
\subfloat{\includegraphics[height=1.6in,clip,trim=0cm 1.4cm 0cm 0.9cm]{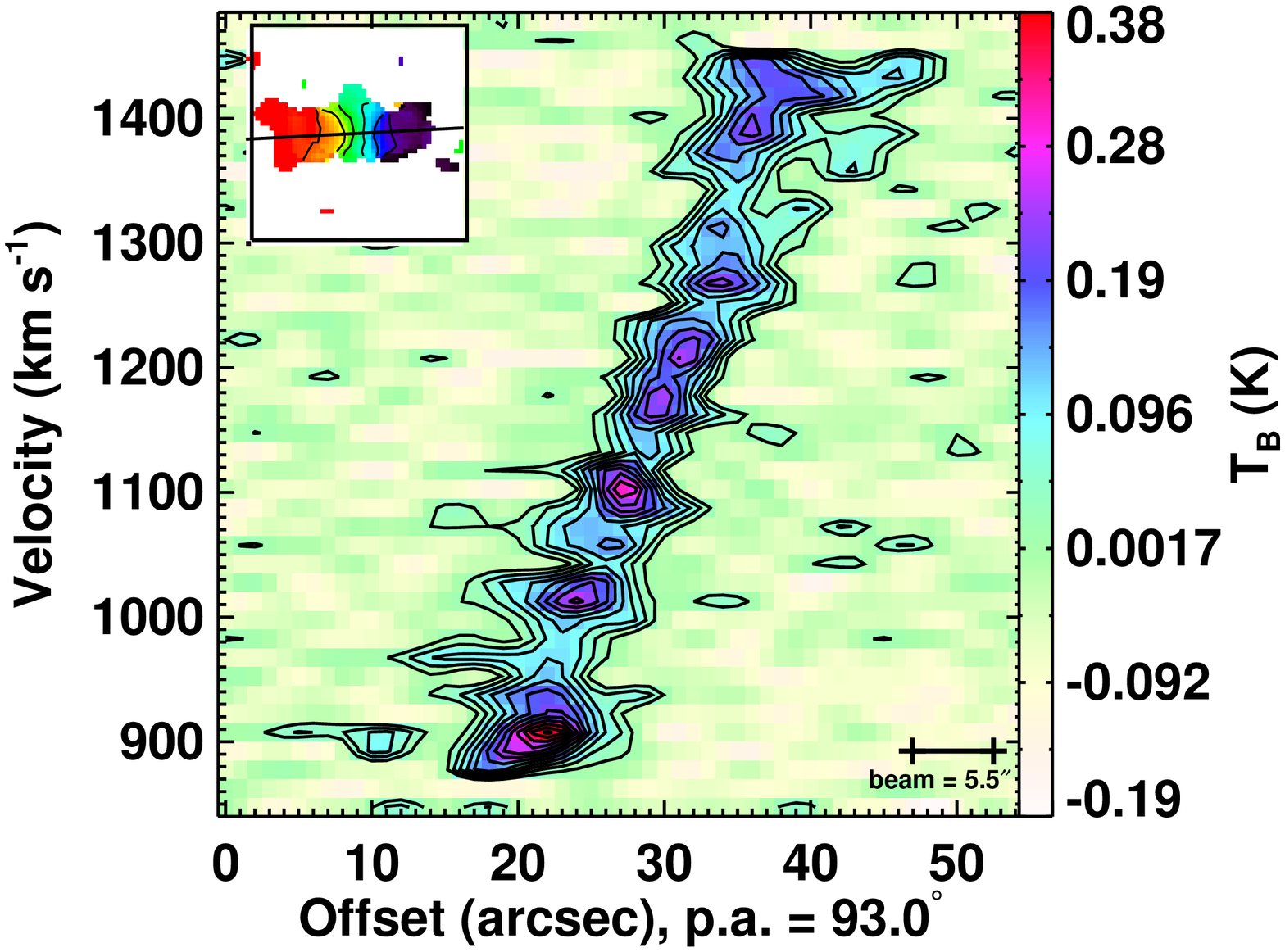}}
\end{figure*}
\begin{figure*}
\subfloat{\includegraphics[width=7in,clip,trim=1cm 1.0cm 0.3cm 4.5cm]{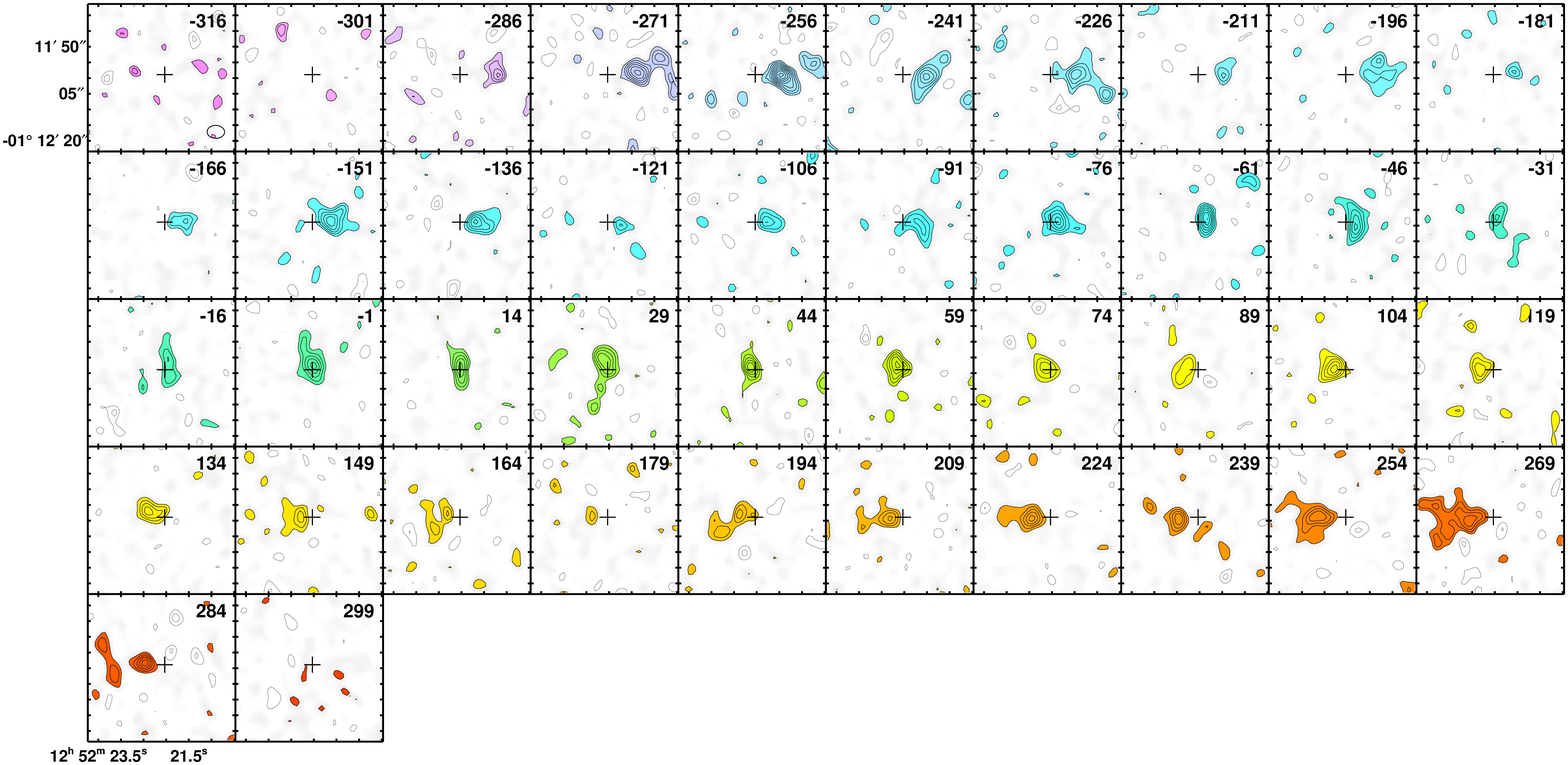}}
\caption{{\bf NGC~4753} is a field regular rotator ($M_K$ = -25.09) that includes two velocity maxima, with stellar morphology consistent with interaction.  It contains a dust filament.  It is also the most massive galaxy in the CARMA \atlas\ survey.  The moment0 peak is 12 Jy beam$^{-1}$ \kms.  Moment1 contours are placed at 40\kms\ intervals and PVD contours are placed at $1.5\sigma$ intervals.}
\end{figure*}

\clearpage
\begin{figure*}
\centering
\subfloat{\includegraphics[height=2.2in,clip,trim=2.2cm 3.2cm 0cm 2.7cm]{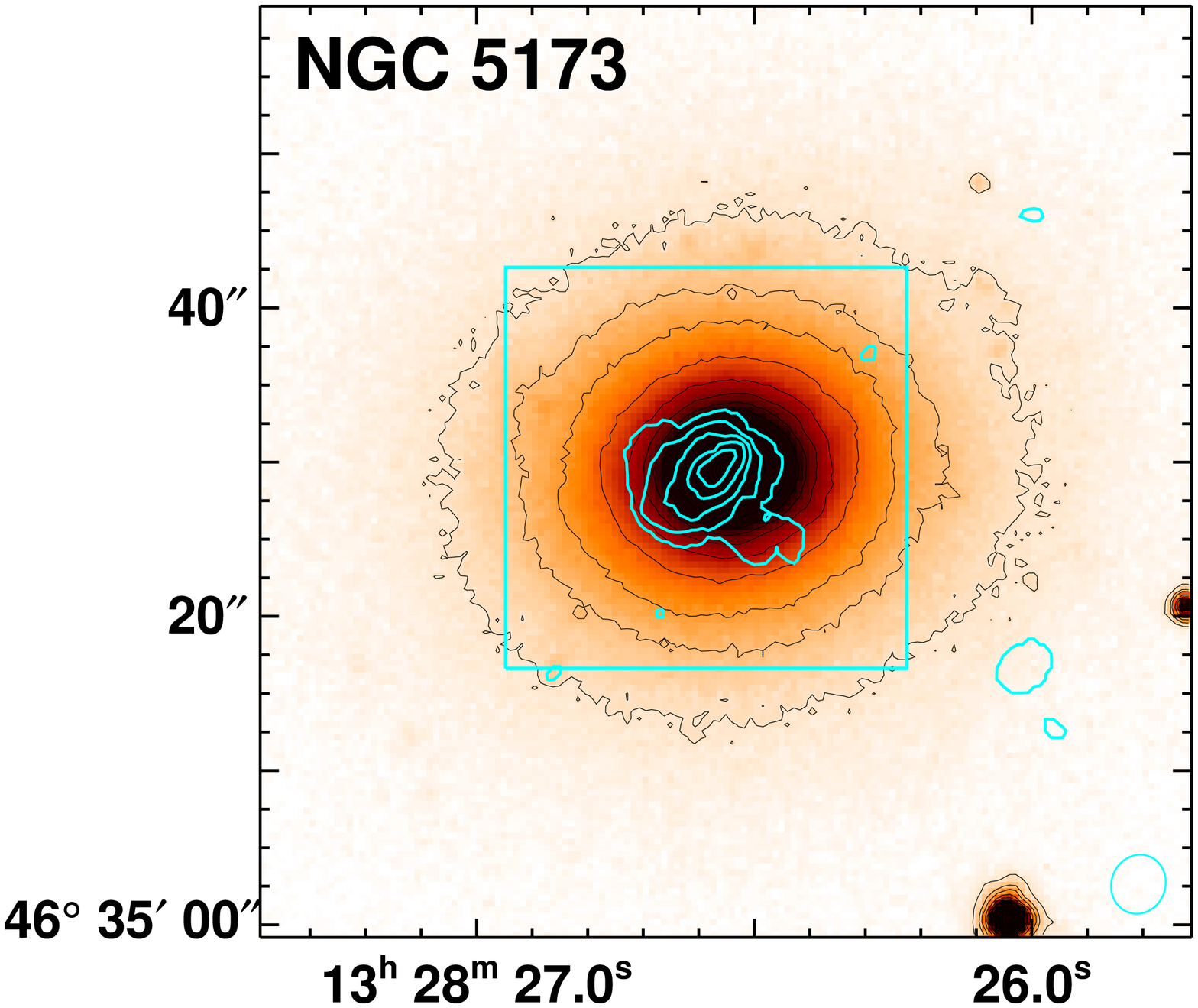}}
\subfloat{\includegraphics[height=2.2in,clip,trim=0cm 0.6cm 0.4cm 0.4cm]{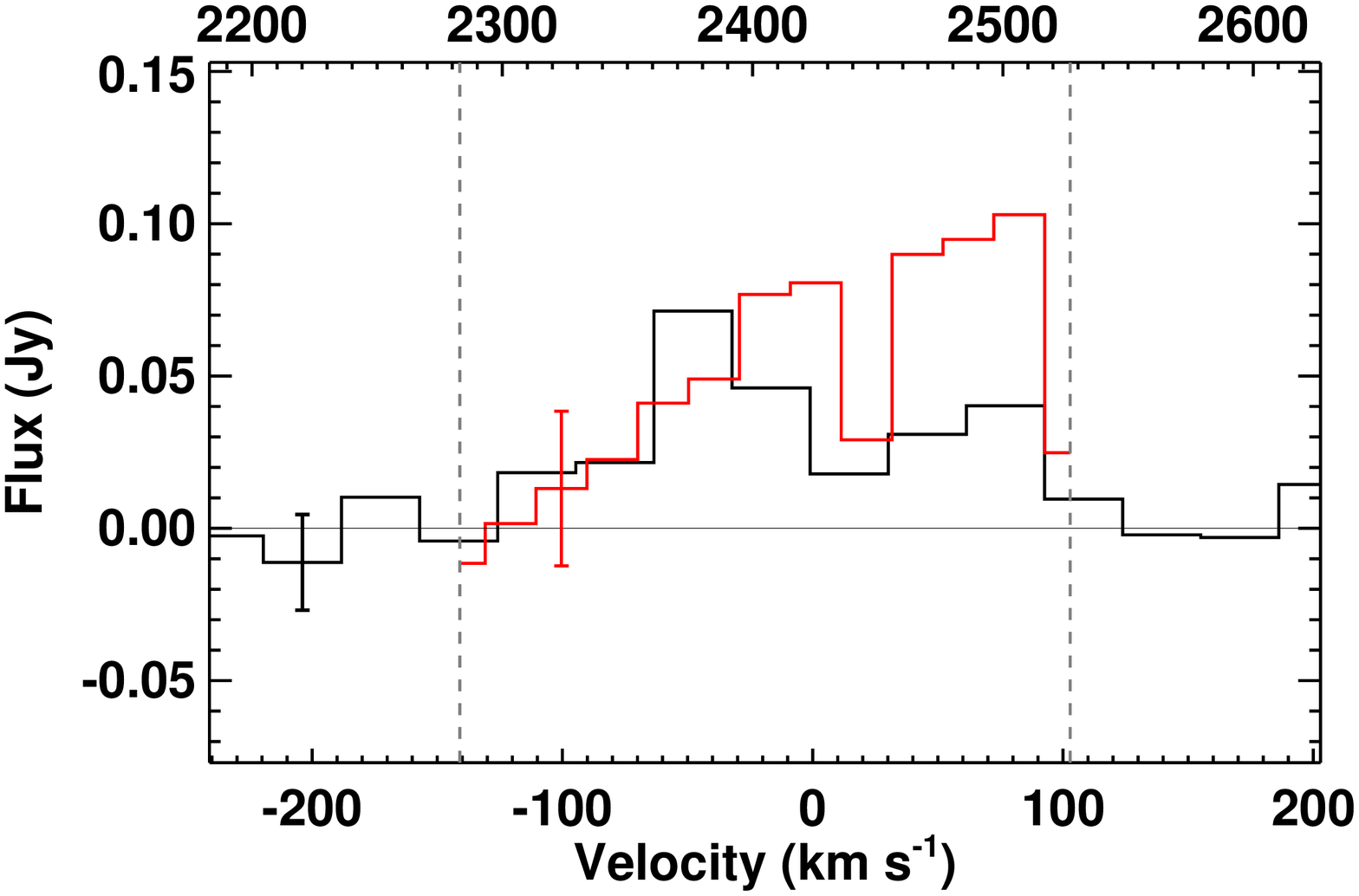}}
\end{figure*}
\begin{figure*}
\subfloat{\includegraphics[height=1.6in,clip,trim=0.1cm 1.4cm 0.6cm 2.4cm]{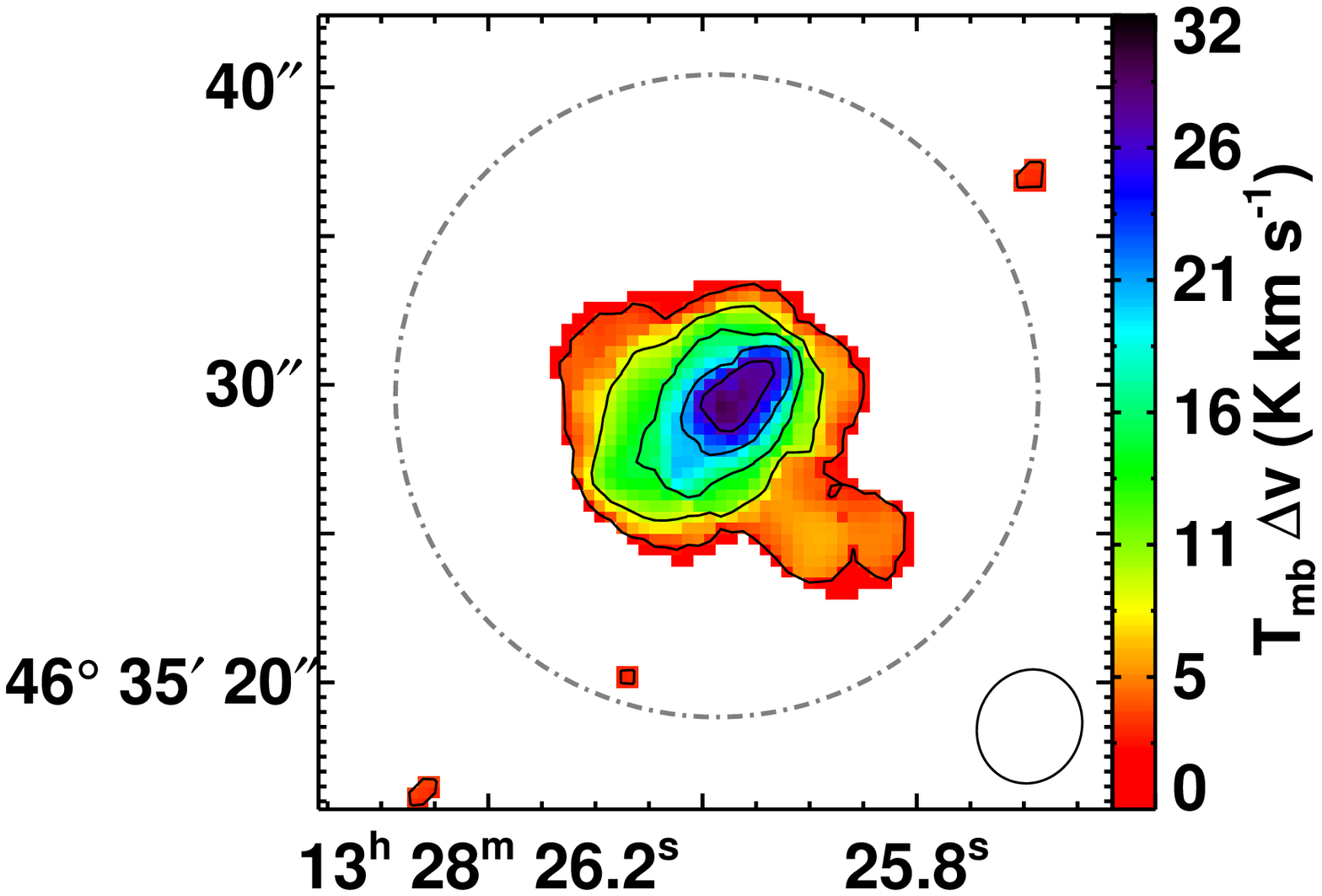}}
\subfloat{\includegraphics[height=1.6in,clip,trim=0.1cm 1.4cm 0cm 2.4cm]{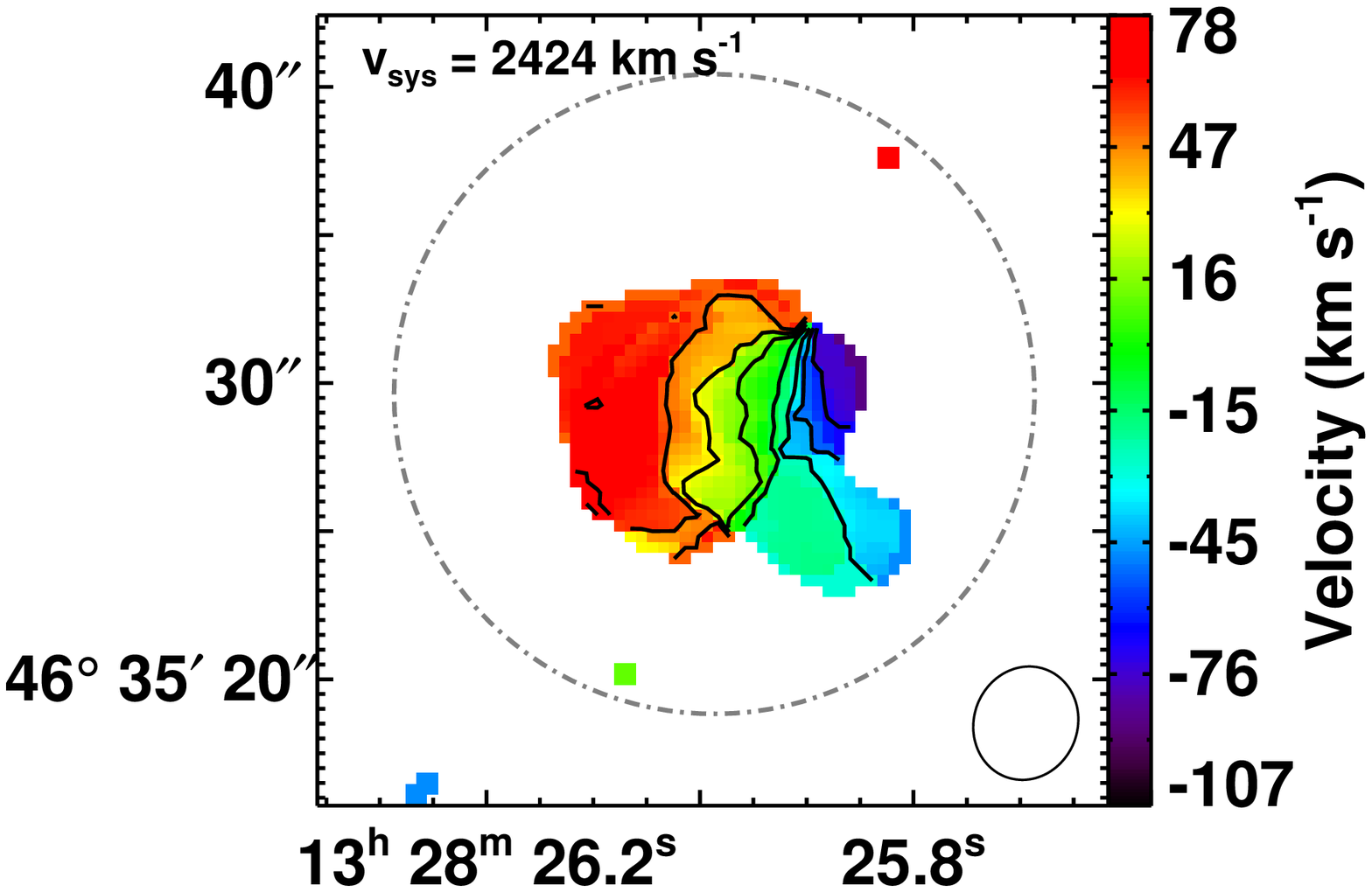}}
\subfloat{\includegraphics[height=1.6in,clip,trim=0cm 1.4cm 0cm 0.9cm]{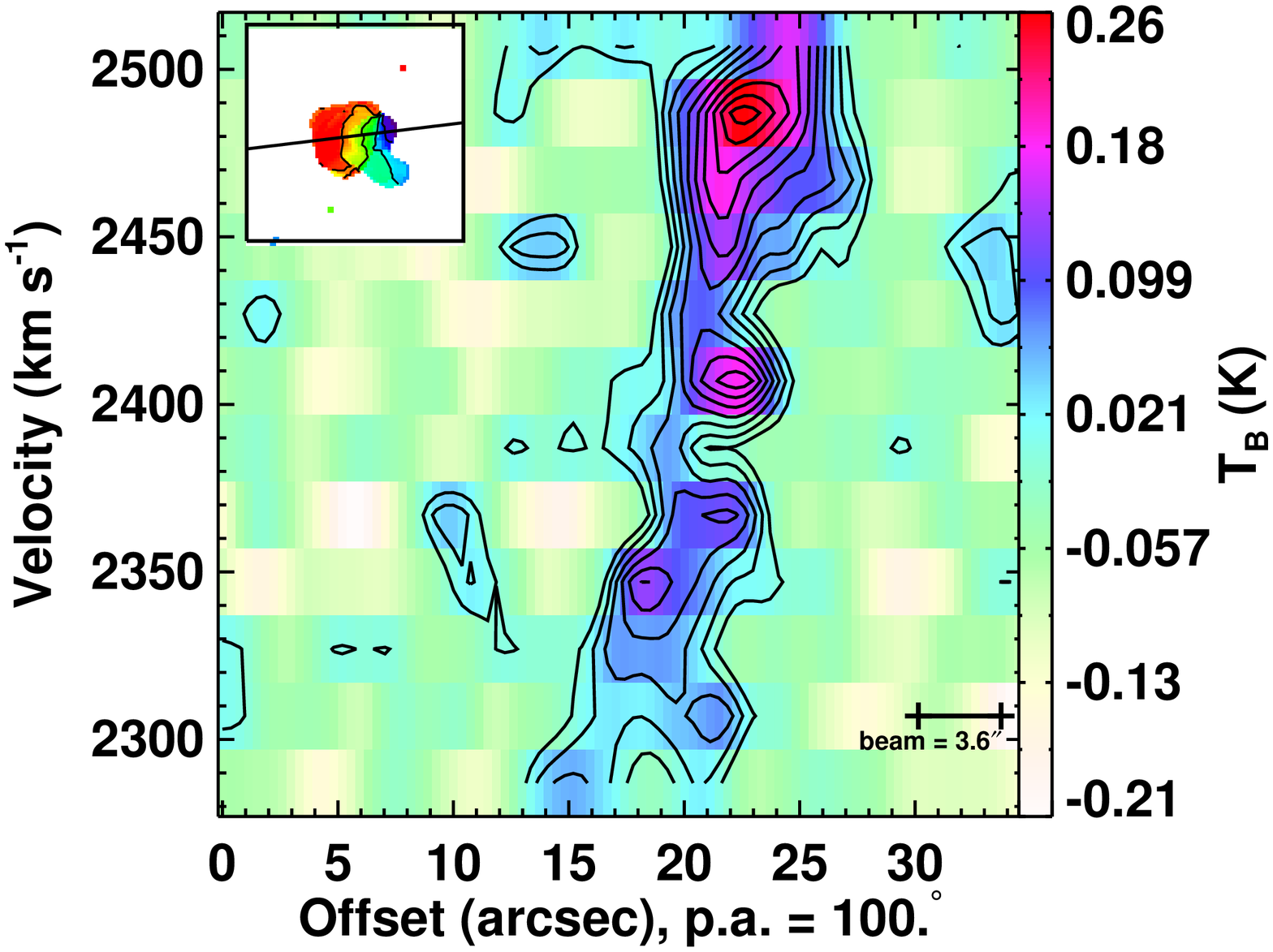}}
\end{figure*}
\begin{figure*}
\subfloat{\includegraphics[width=7in,clip,trim=0.3cm 3.7cm 6.7cm 1.5cm]{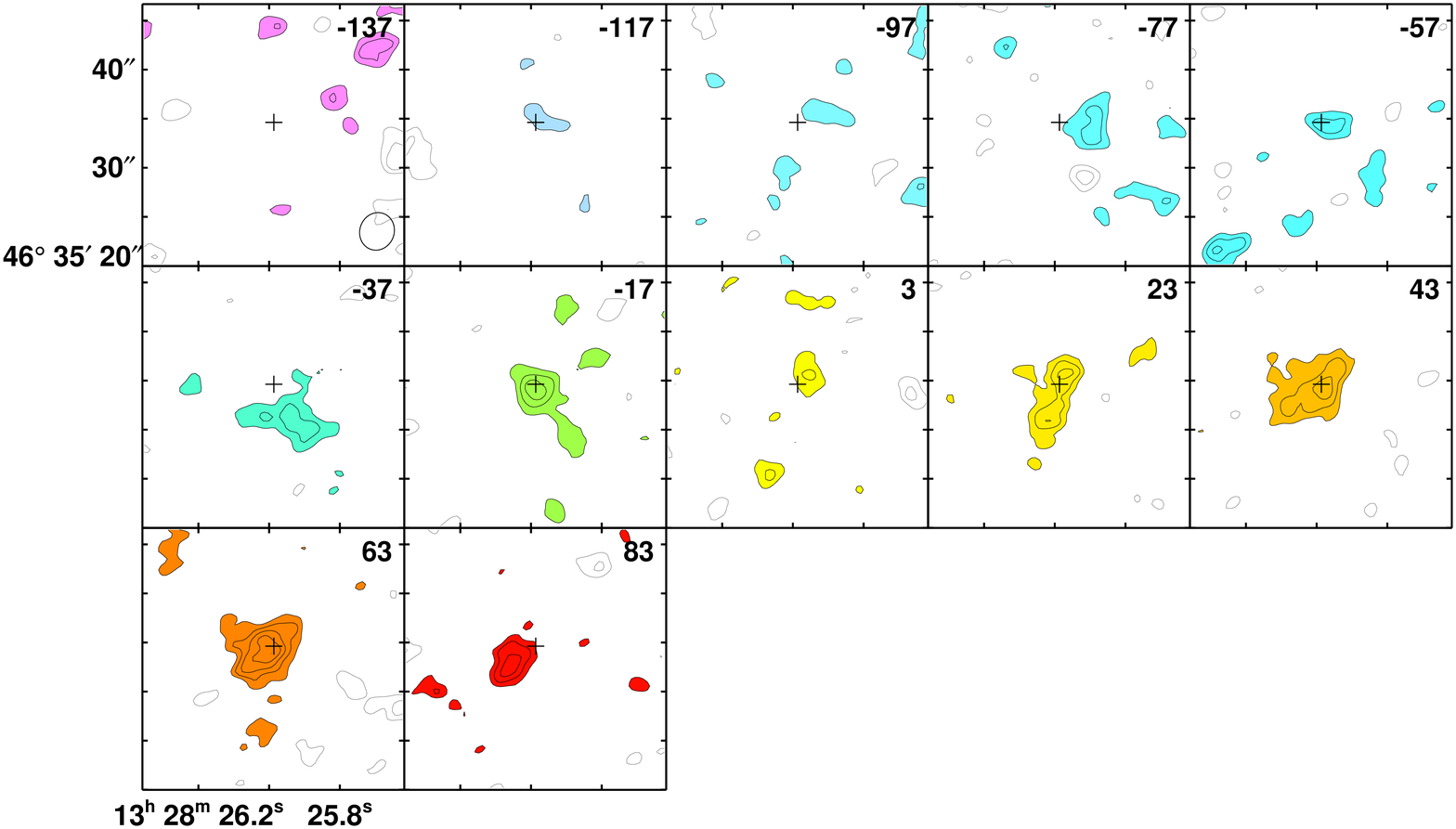}}
\caption{{\bf NGC~5173} is a group regular rotator ($M_K$ = -22.88) with normal stellar morphology.  It contains a dust bar.  NGC~5173 was observed in CARMA D- and C- arrays, and the data were shared between this survey, and the work of \citealt{wei+10b}, and thus uses the smaller pixel size of $0.4''$, to reflect the superior resolution of the observations.  The moment0 peak is 4.6 Jy beam$^{-1}$ \kms.  PVD contours are placed at $1.5\sigma$ intervals.}
\end{figure*}

\clearpage
\begin{figure*}
\centering
\subfloat{\includegraphics[height=2.2in,clip,trim=2.2cm 3.2cm 0cm 2.7cm]{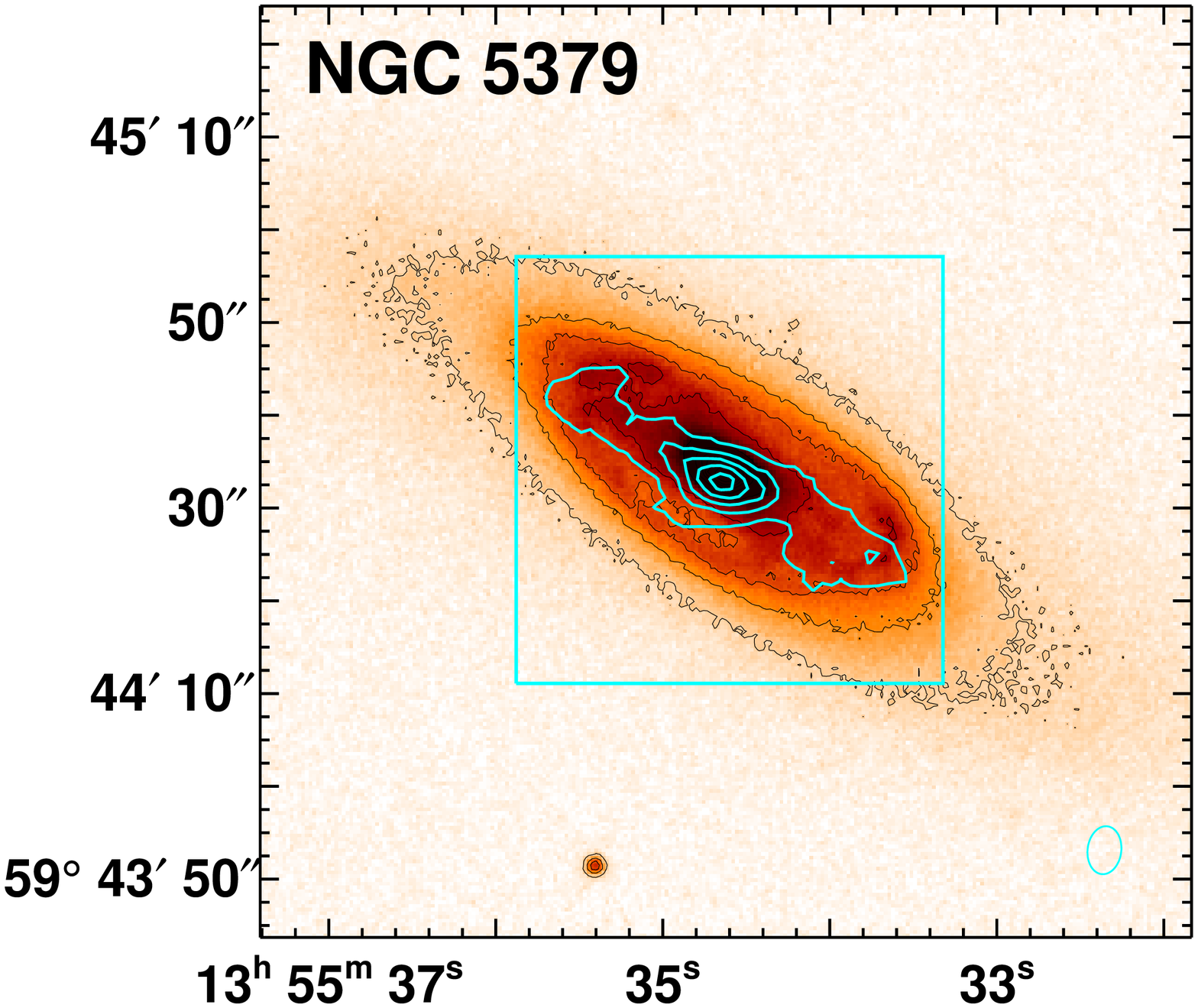}}
\subfloat{\includegraphics[height=2.2in,clip,trim=0cm 0.6cm 0.4cm 0.4cm]{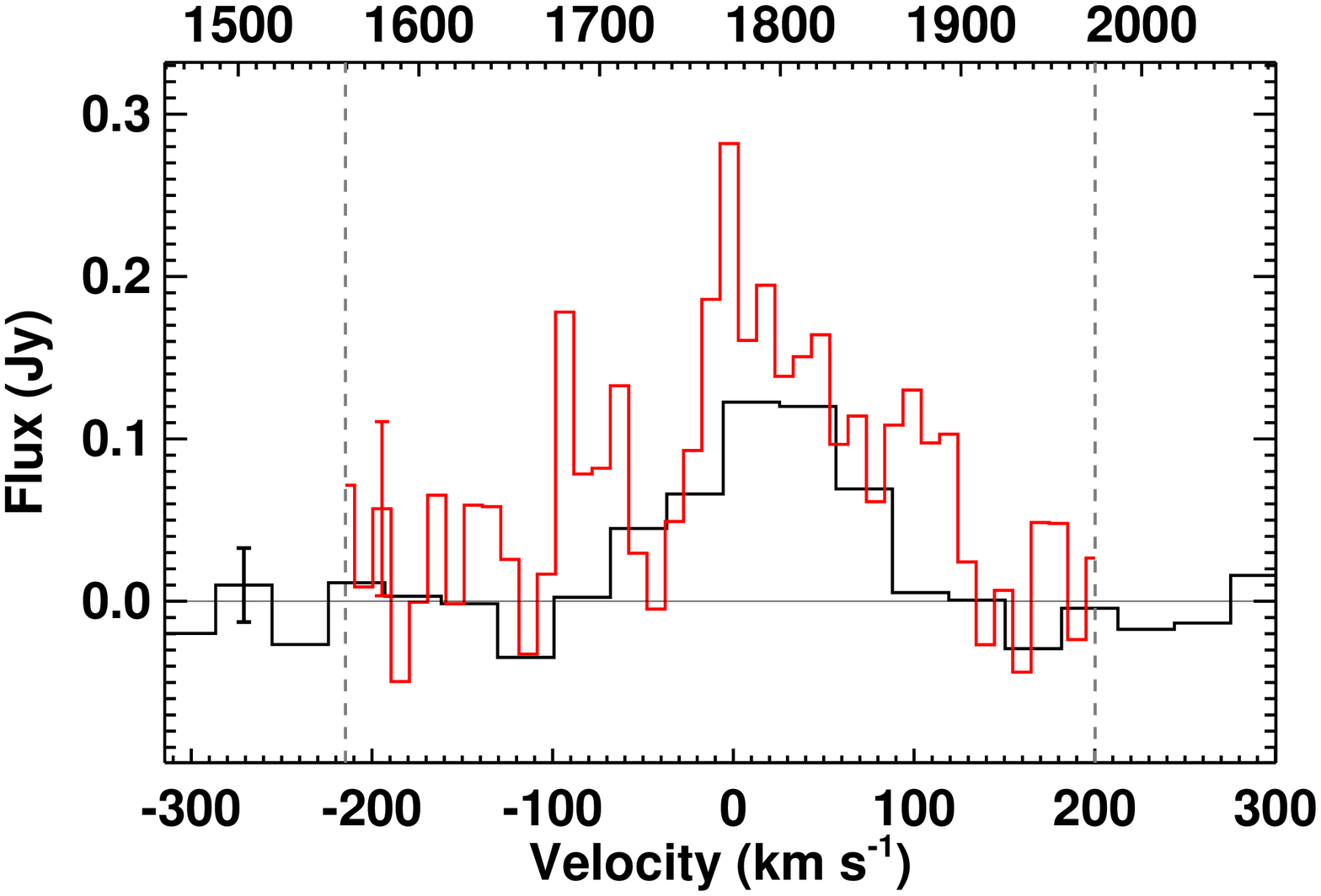}}
\end{figure*}
\begin{figure*}
\subfloat{\includegraphics[height=1.6in,clip,trim=0.1cm 1.4cm 0.4cm 2.4cm]{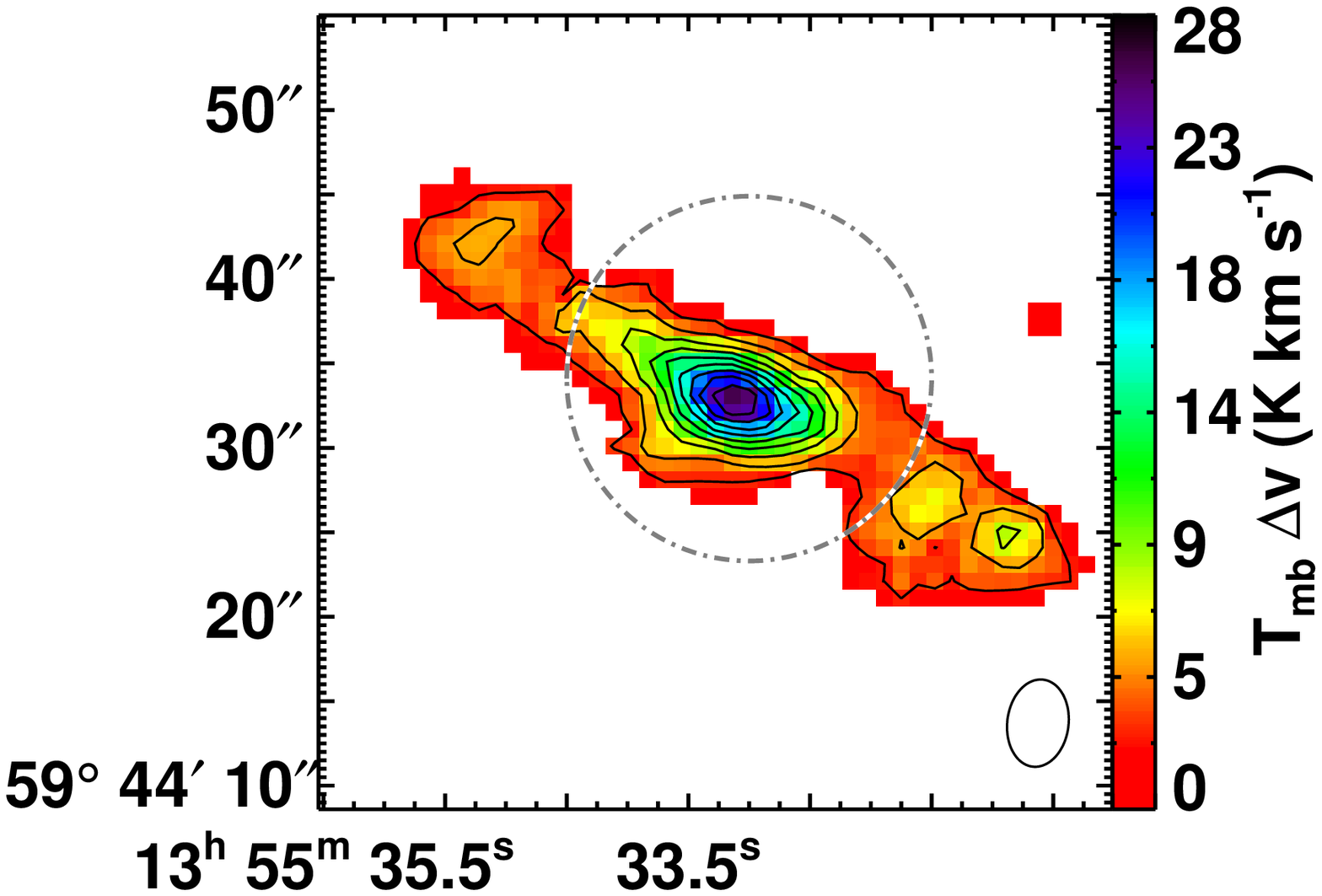}}
\subfloat{\includegraphics[height=1.6in,clip,trim=0.1cm 1.4cm 0cm 2.4cm]{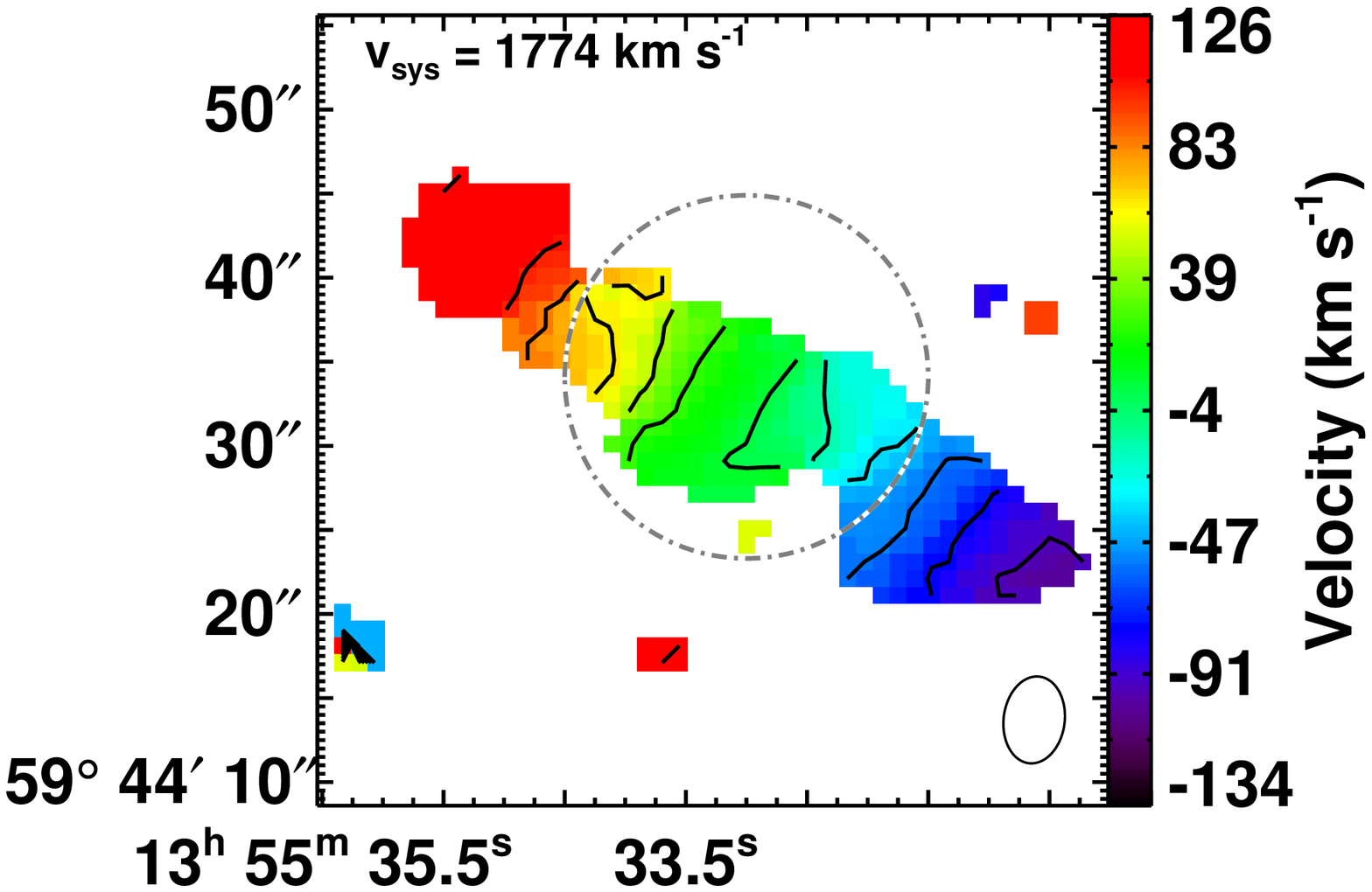}}
\subfloat{\includegraphics[height=1.6in,clip,trim=0cm 1.4cm 0cm 0.9cm]{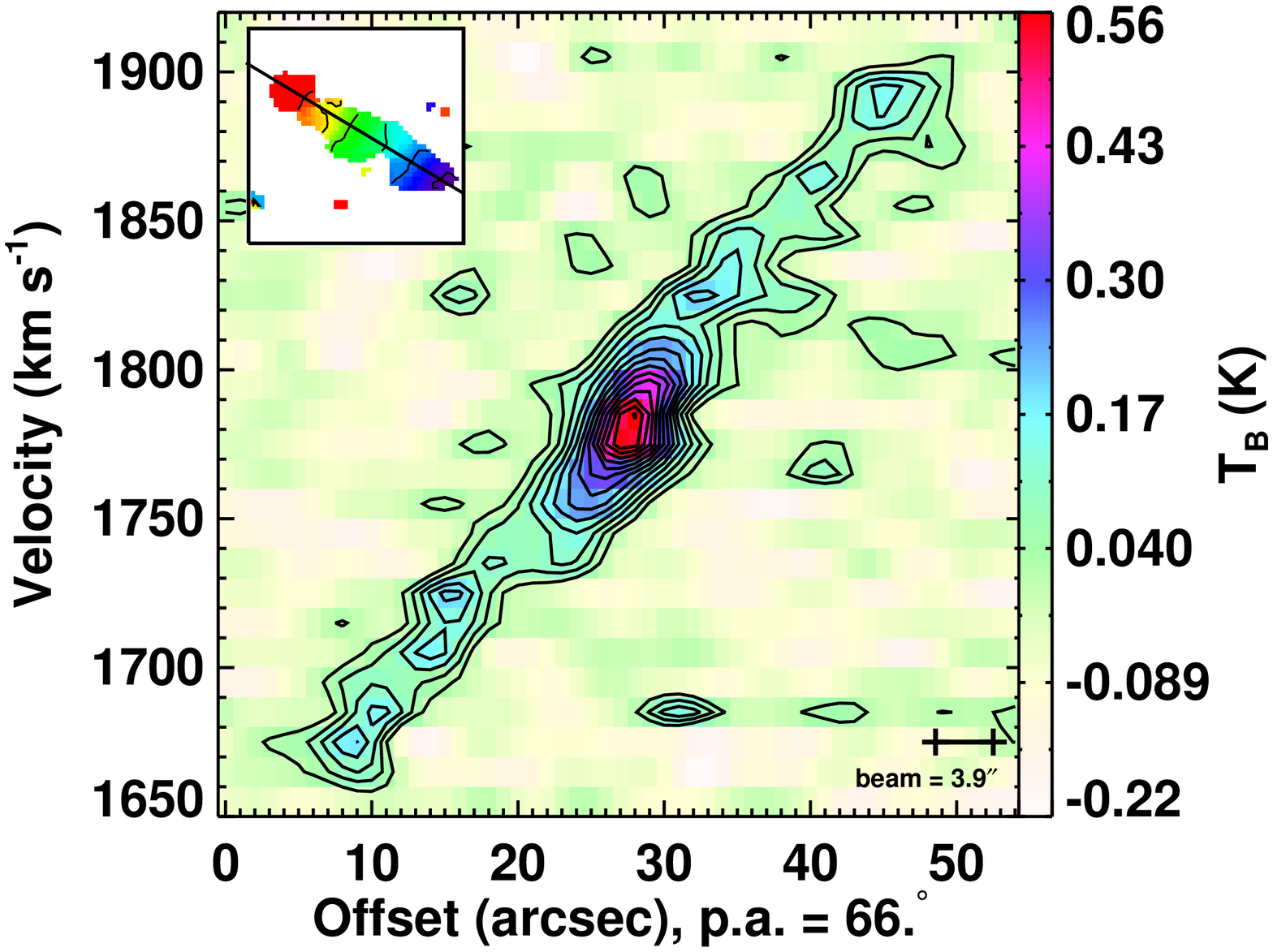}}
\end{figure*}
\begin{figure*}
\subfloat{\includegraphics[width=7in,clip,trim=0.6cm 1.3cm 0.3cm 1.3cm]{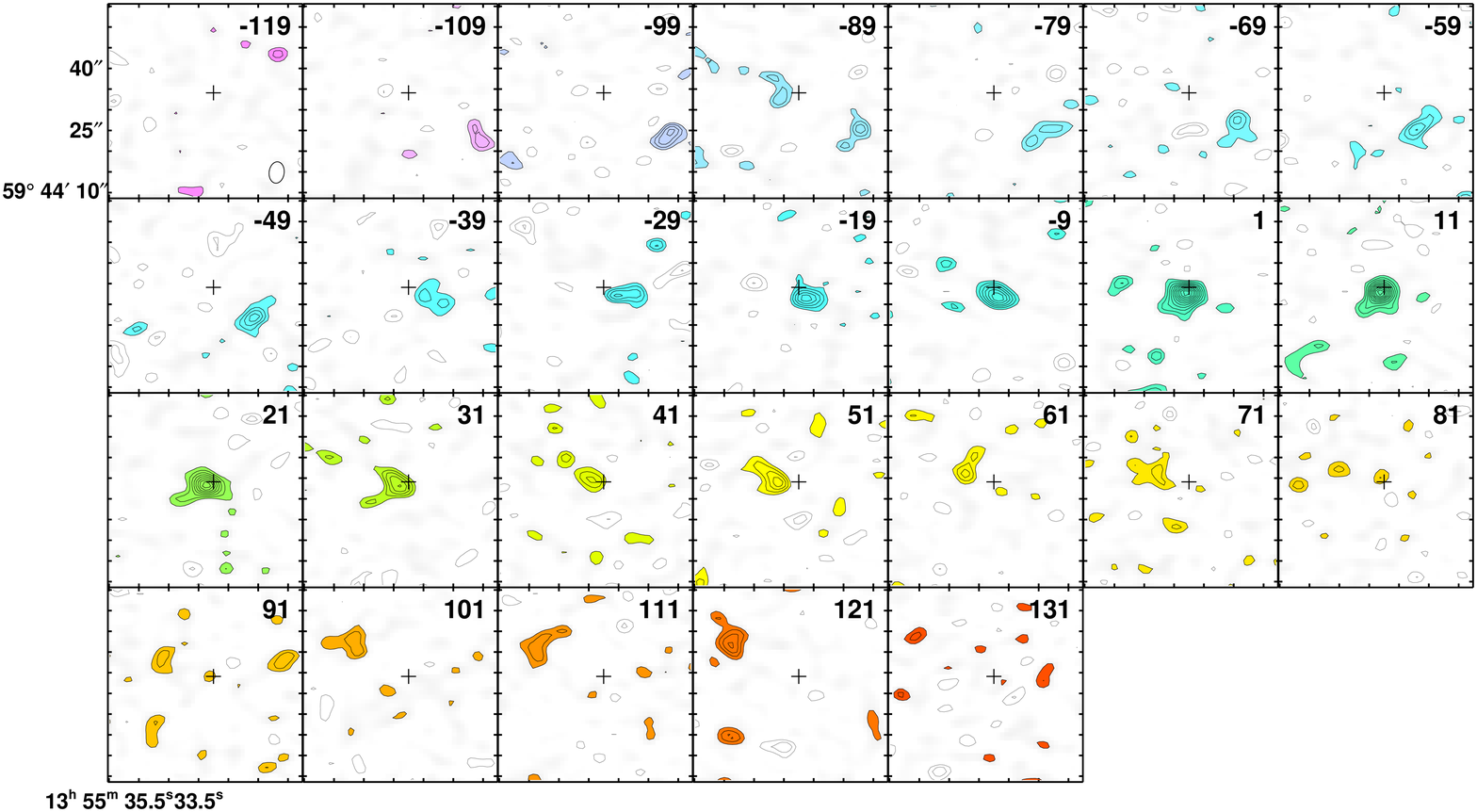}}
\caption{{\bf NGC~5379} is a group regular rotator ($M_K$ = -22.08) with ring stellar morphology.  It contains a dust bar, ring, and filaments.  The moment0 peak is 5.7 Jy beam$^{-1}$ \kms.  PVD contours are placed at $1.5\sigma$ intervals.}
\end{figure*}

\clearpage
\begin{figure*}
\centering
\subfloat{\includegraphics[height=2.2in,clip,trim=2.5cm 3.2cm 0cm 2.7cm]{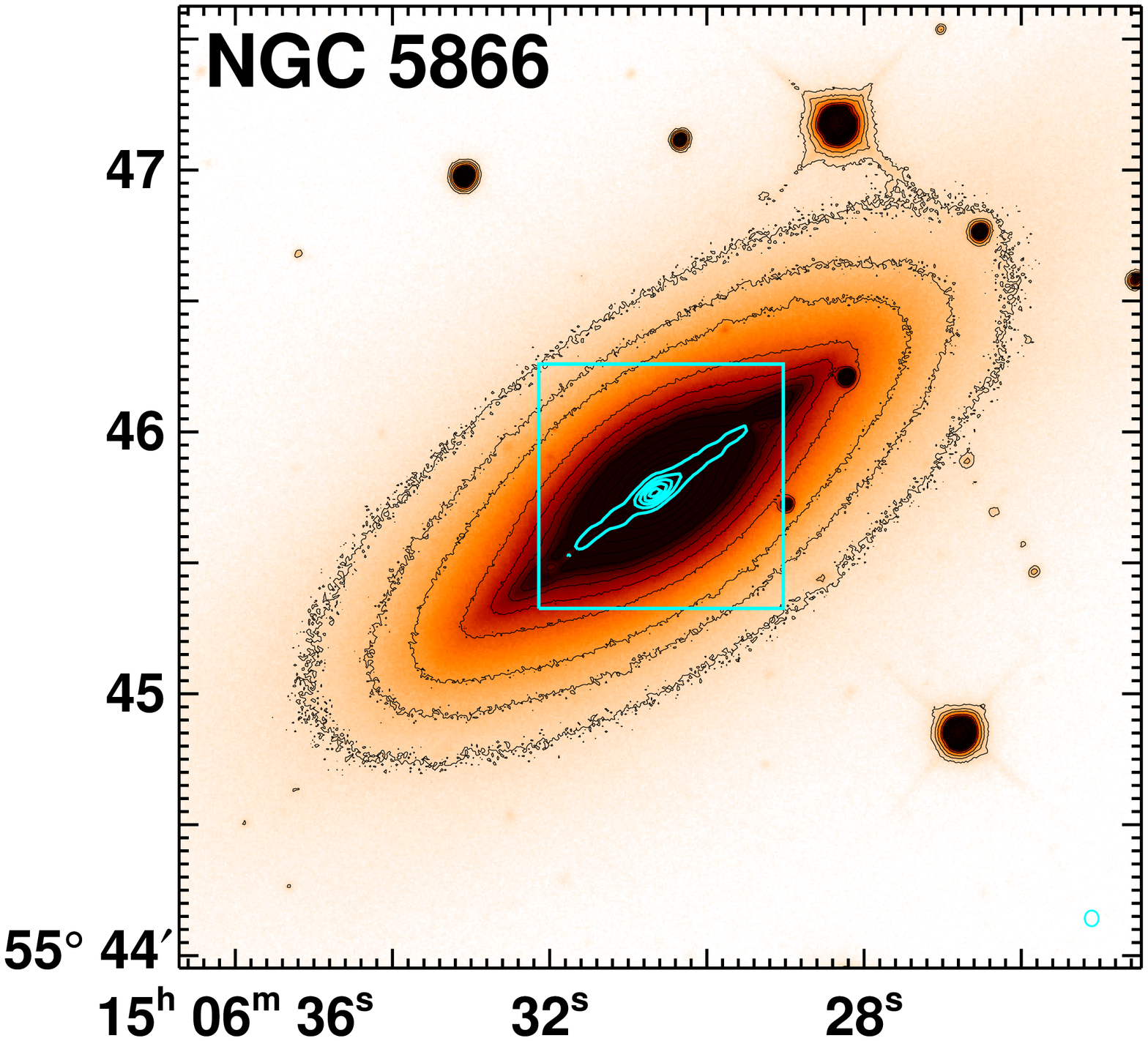}}
\subfloat{\includegraphics[height=2.2in,clip,trim=0cm 0.6cm 0.4cm 0.4cm]{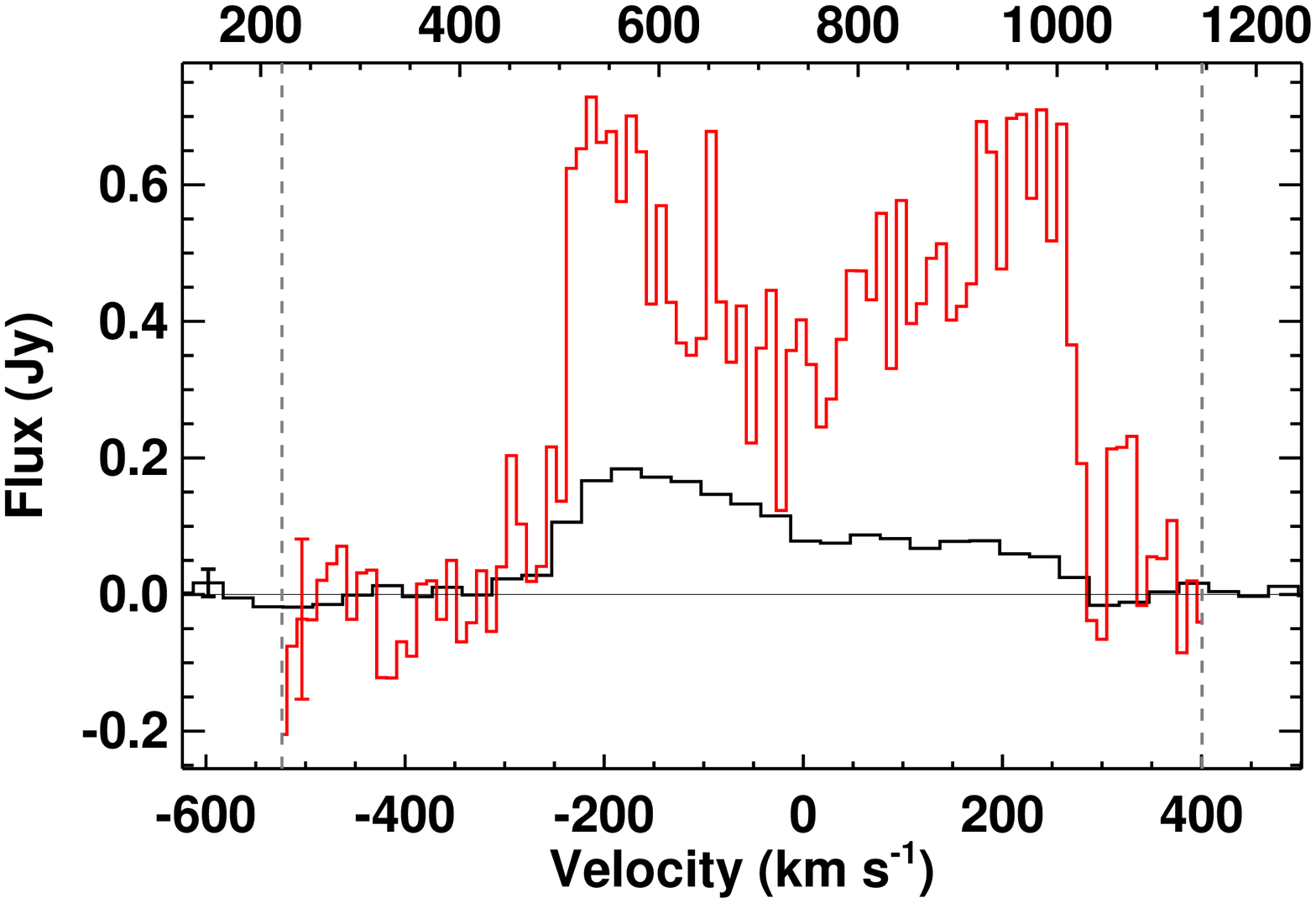}}
\end{figure*}
\begin{figure*}
\subfloat{\includegraphics[height=1.6in,clip,trim=0.1cm 1.4cm 0.7cm 2.5cm]{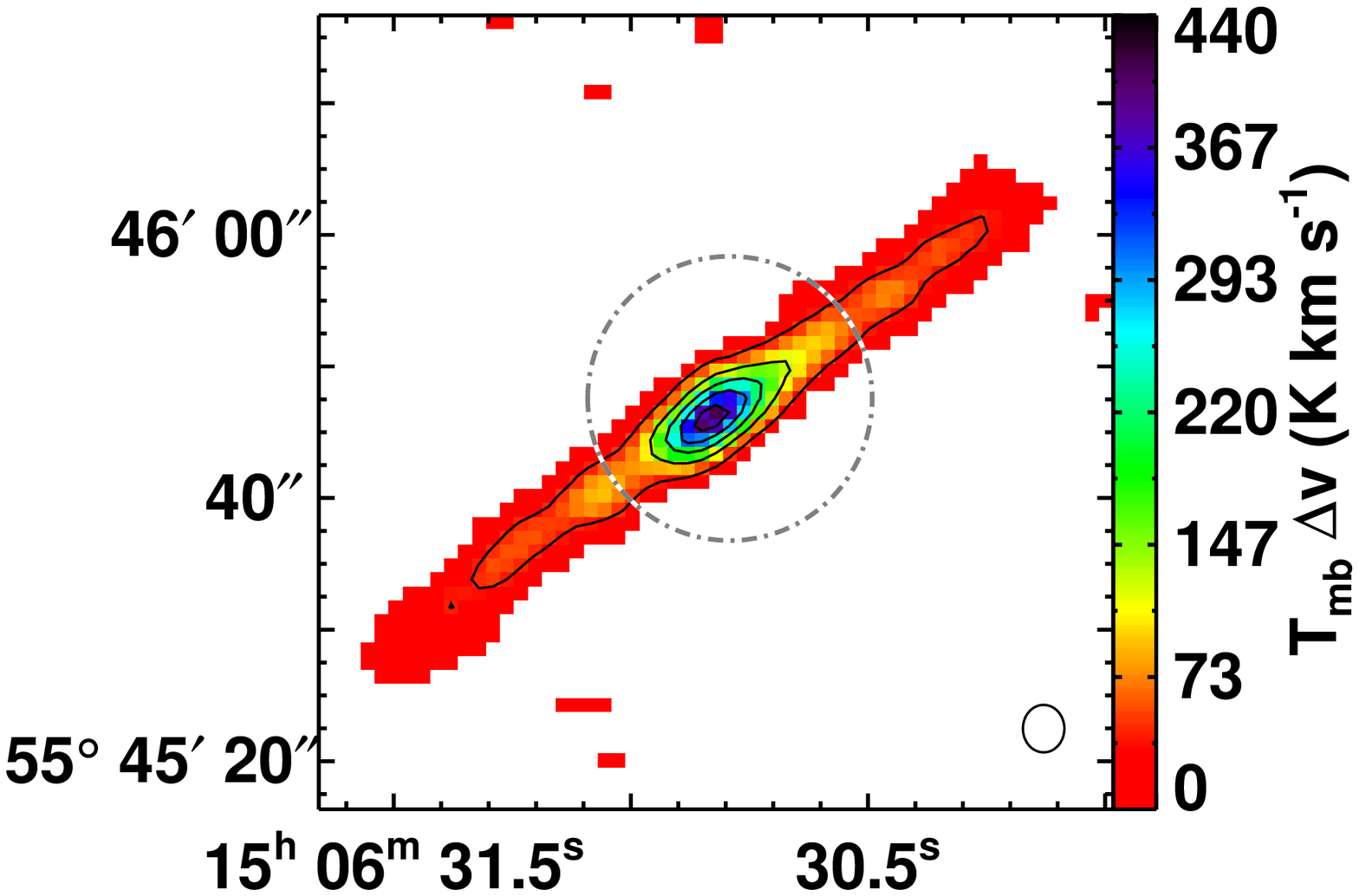}}
\subfloat{\includegraphics[height=1.6in,clip,trim=0cm 1.4cm 0cm 2.4cm]{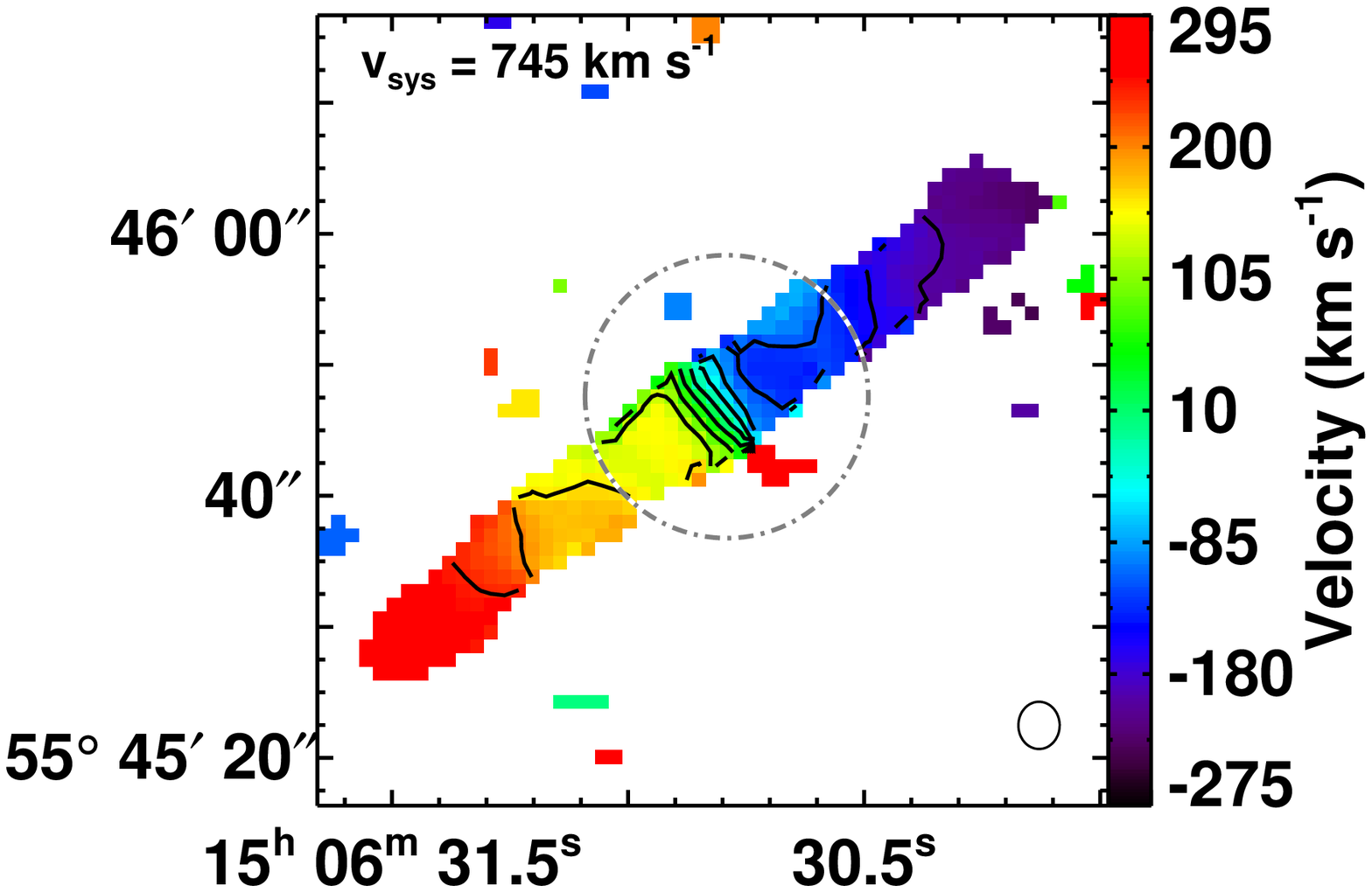}}
\subfloat{\includegraphics[height=1.6in,clip,trim=0cm 1.4cm 0cm 0.9cm]{figures/ngc5866_pv.eps}}
\end{figure*}
\begin{figure*}
\subfloat{\includegraphics[width=7in,clip,trim=1.1cm 0.8cm 0.4cm 0.7cm]{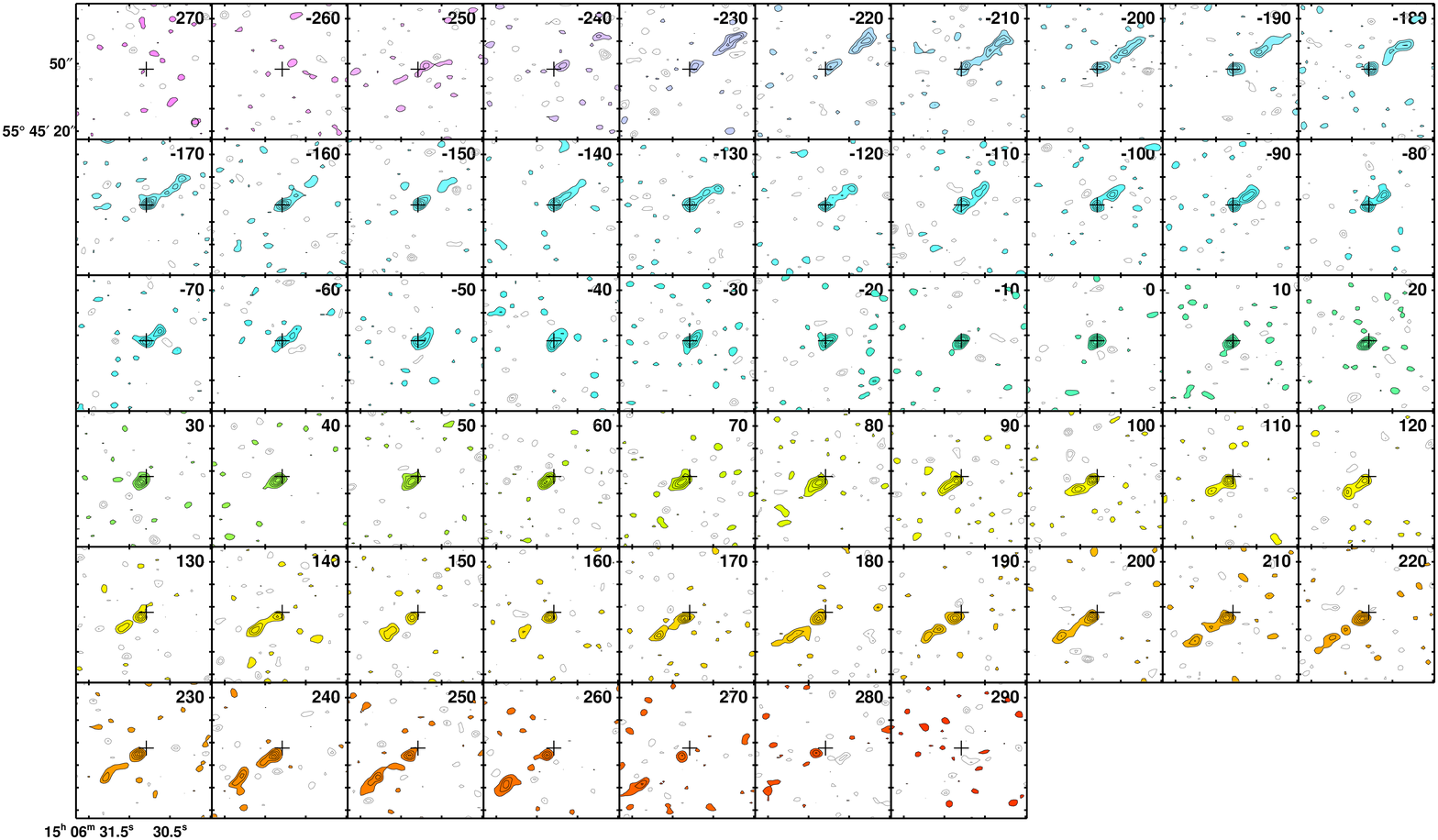}}
\caption{{\bf NGC~5866} is a field regular rotator ($M_K$ = -24.00) with normal stellar morphology.  It contains a dust disc.  The molecular gas also strongly suggests that this galaxy contains a bar.  The moment0 peak is 54 Jy beam$^{-1}$ \kms.  Moment1 contours are placed at 40\kms\ intervals.  Channel map contours are placed at $2\sigma$ intervals and PVD contours are placed at $2\sigma$ intervals.}
\end{figure*}

\clearpage
\begin{figure*}
\centering
\subfloat{\includegraphics[height=2.2in,clip,trim=2.2cm 3.2cm 0cm 2.7cm]{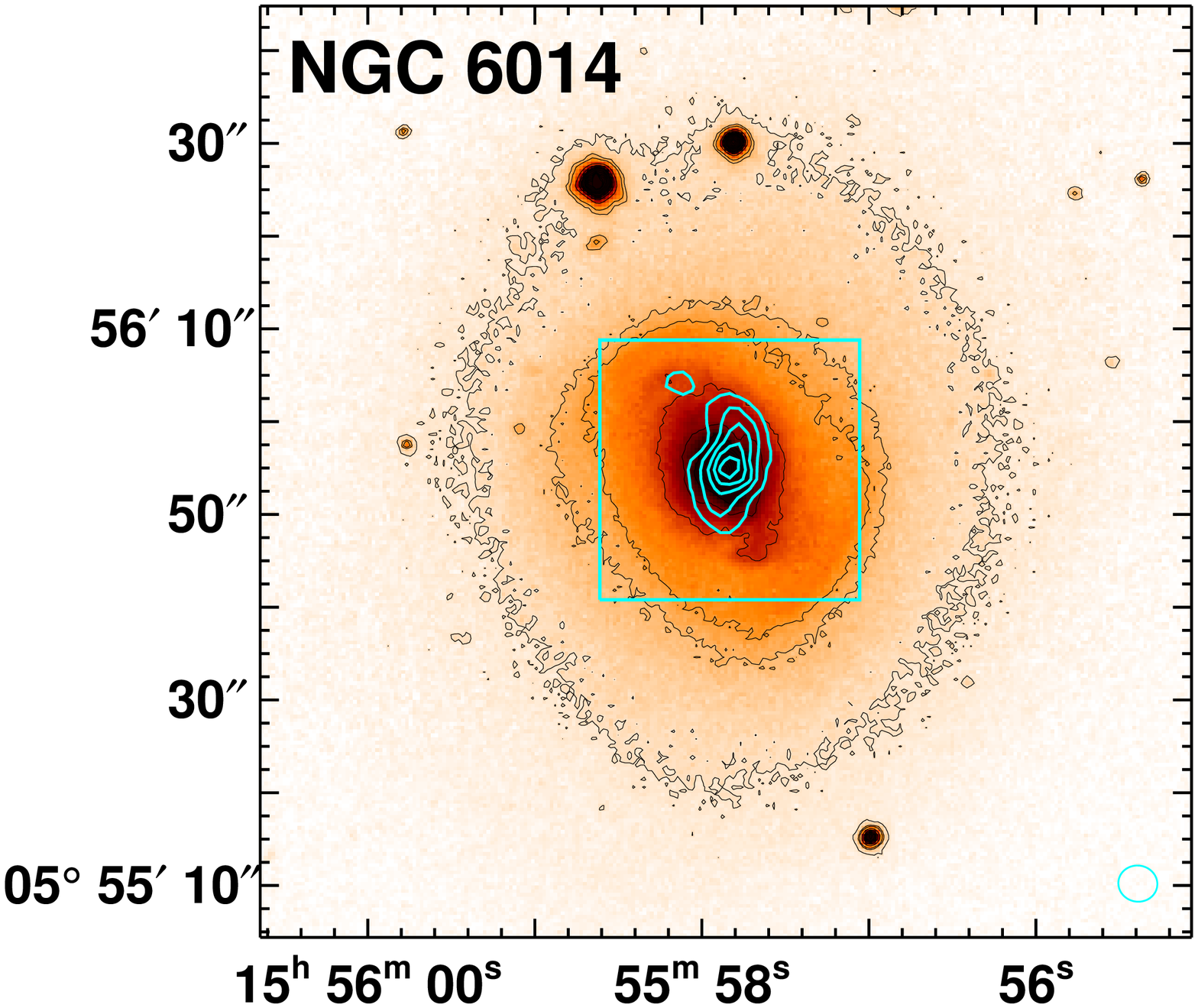}}
\subfloat{\includegraphics[height=2.2in,clip,trim=0cm 0.6cm 0.4cm 0.4cm]{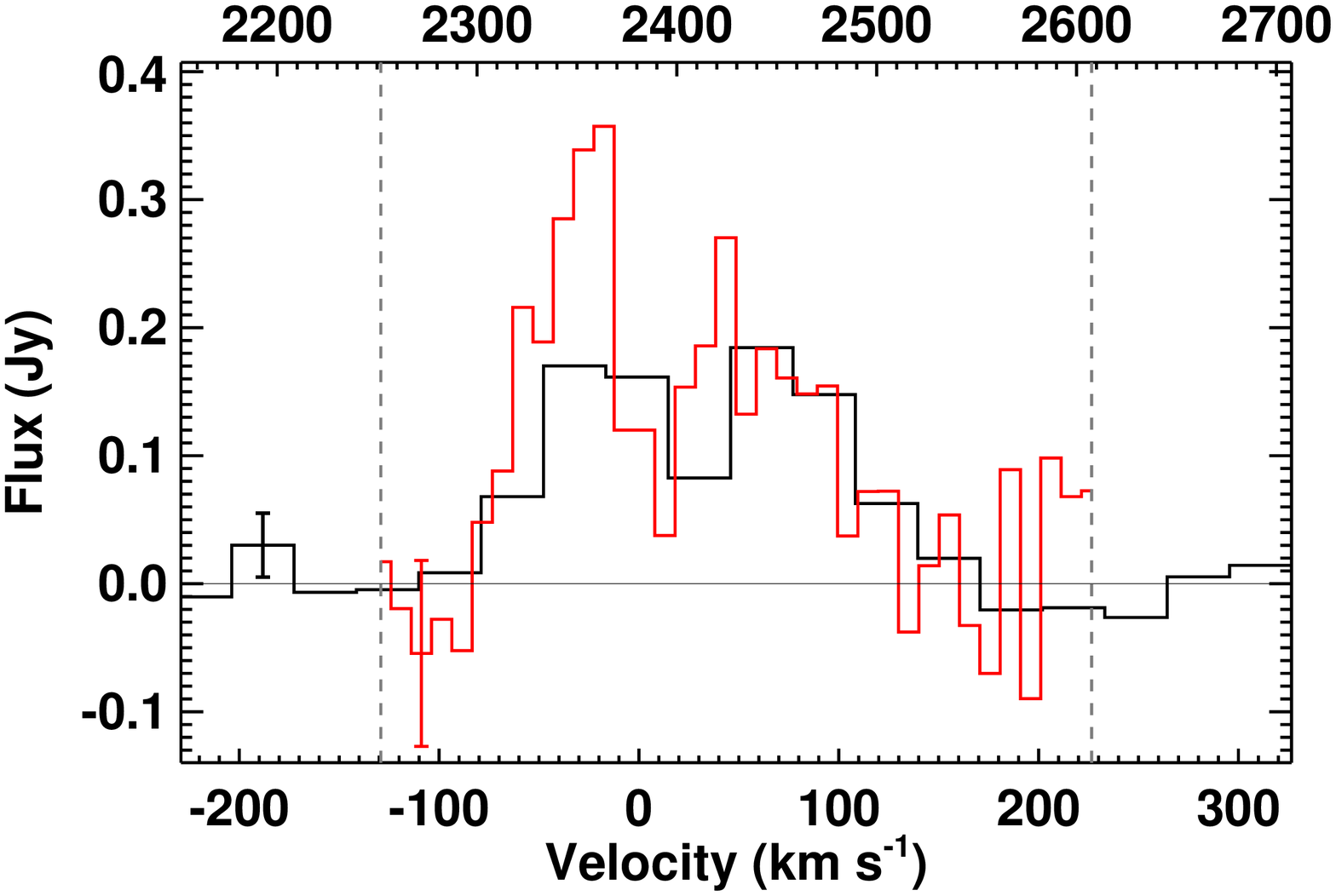}}
\end{figure*}
\begin{figure*}
\subfloat{\includegraphics[height=1.6in,clip,trim=0.1cm 1.4cm 0.6cm 2.4cm]{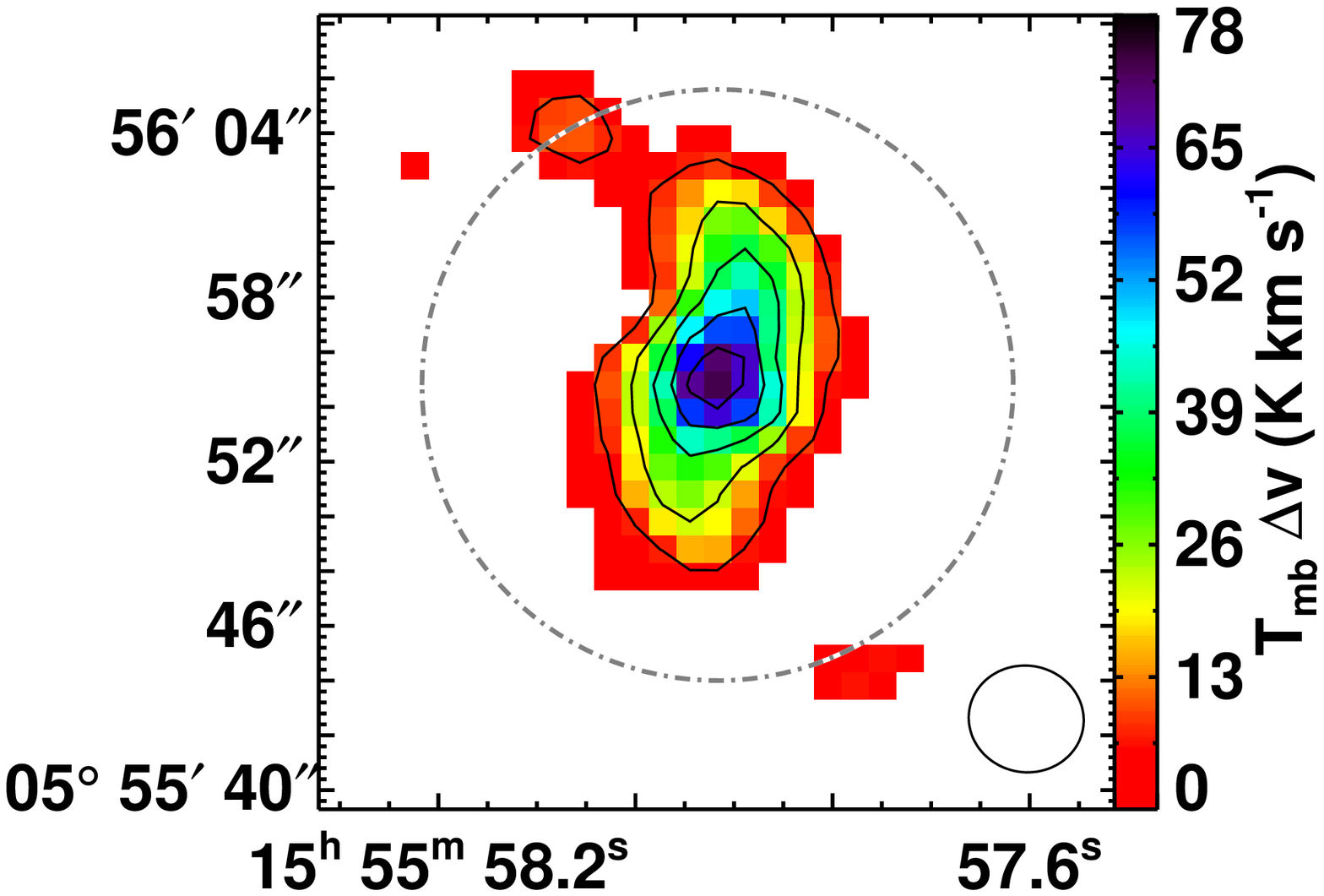}}
\subfloat{\includegraphics[height=1.6in,clip,trim=0.1cm 1.4cm 0cm 2.4cm]{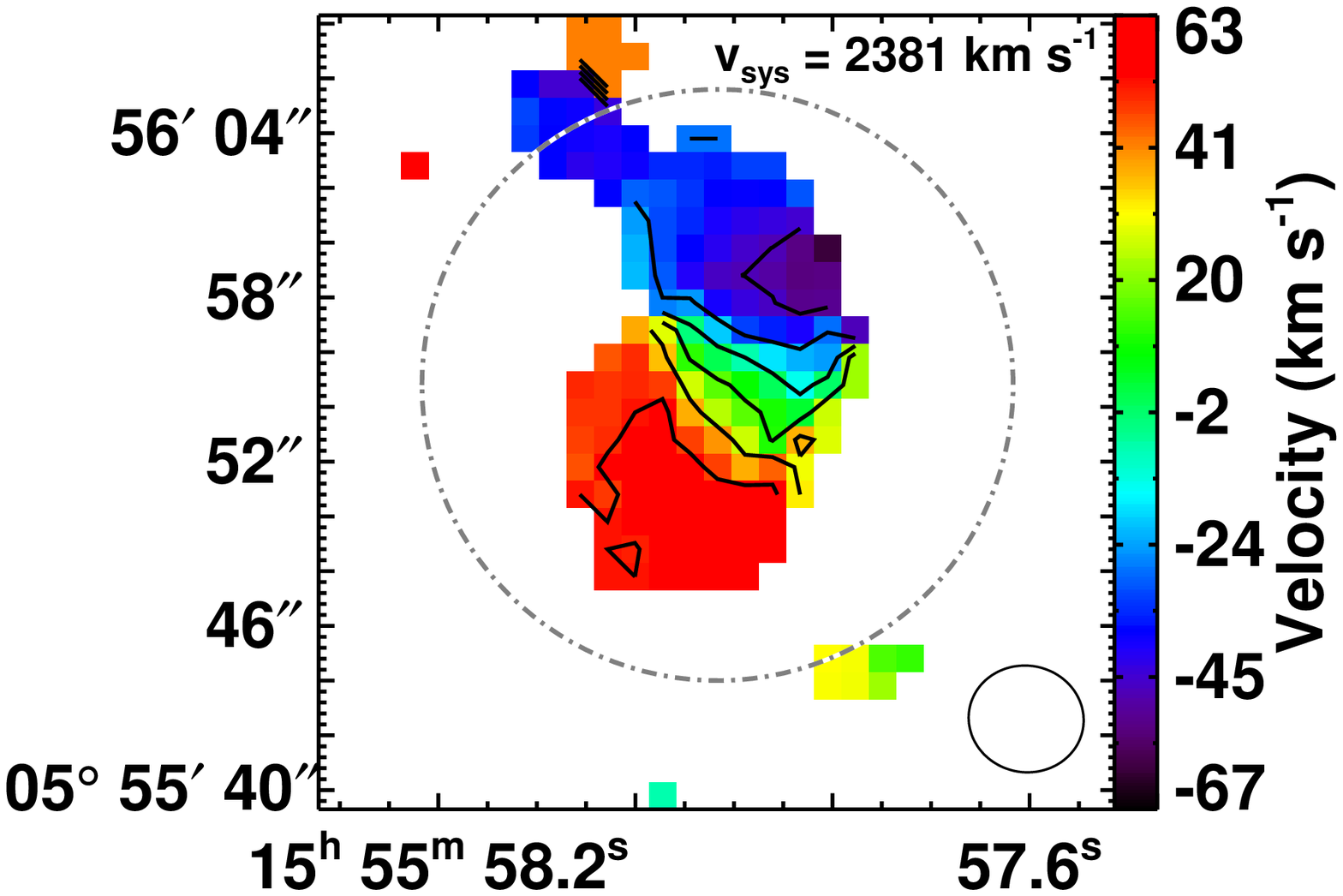}}
\subfloat{\includegraphics[height=1.6in,clip,trim=0cm 1.4cm 0cm 0.9cm]{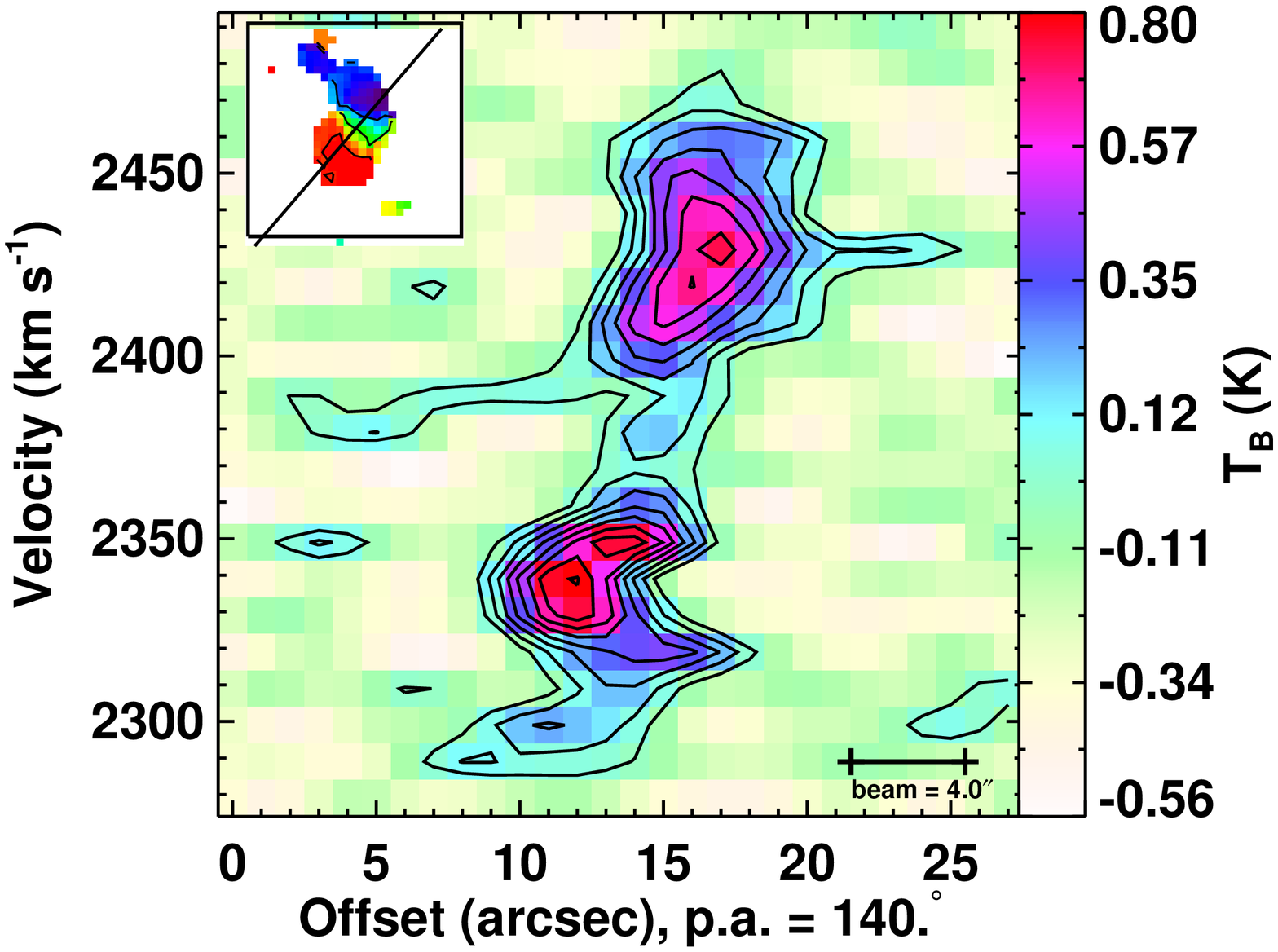}}
\end{figure*}
\begin{figure*}
\subfloat{\includegraphics[width=7in,clip,trim=1.3cm 2cm 0.4cm 0.2cm]{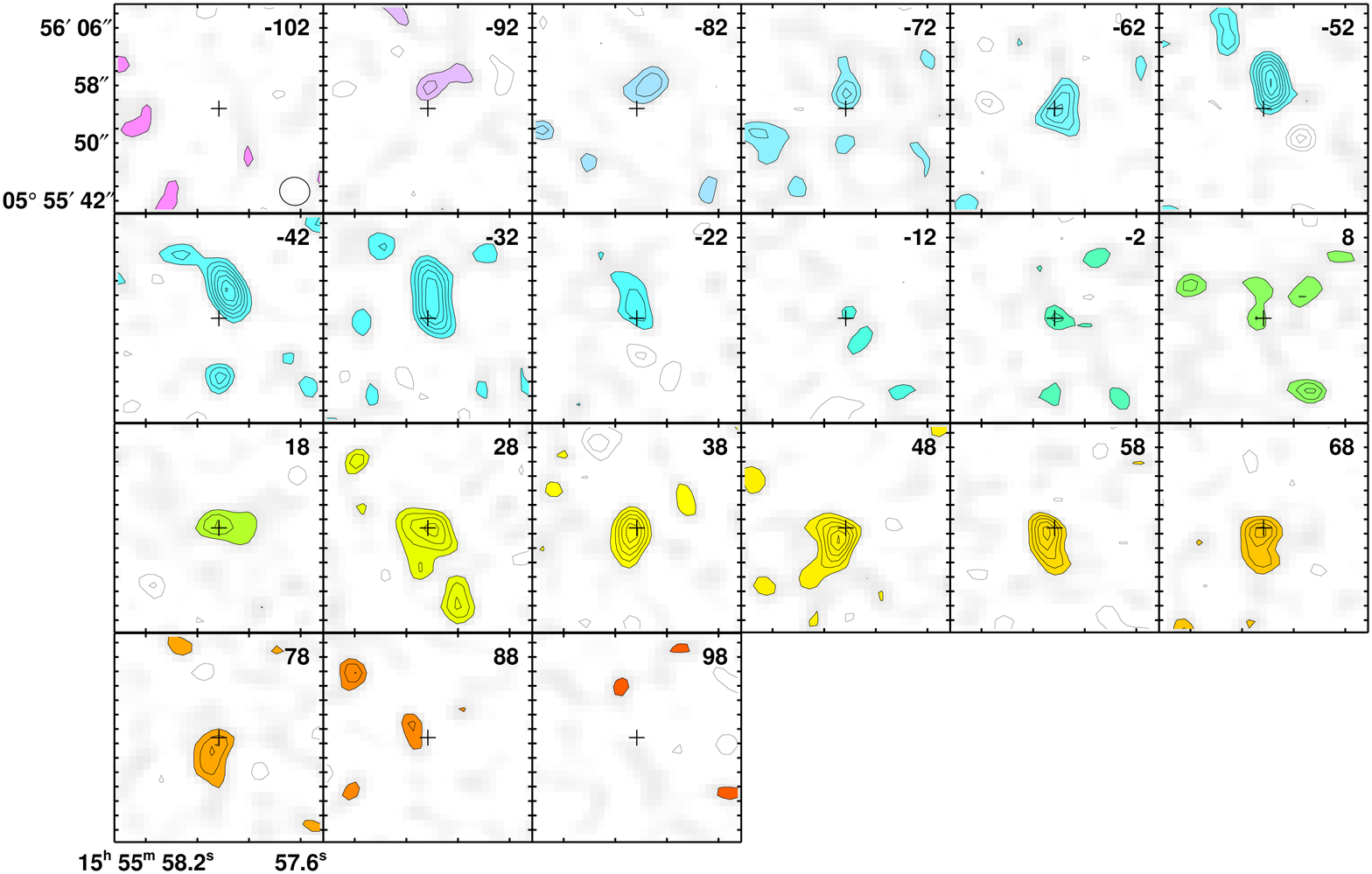}}
\caption{{\bf NGC~6014} is a field regular rotator ($M_K$ = -22.99) with normal stellar morphology.  It contains a dust bar, ring and filaments.  The moment0 peak is 14 Jy beam$^{-1}$ \kms.  PVD contours are placed at $1.5\sigma$ intervals.}
\end{figure*}

\clearpage
\begin{figure*}
\centering
\subfloat{\includegraphics[height=2.2in,clip,trim=2.2cm 3.2cm 0cm 2.7cm]{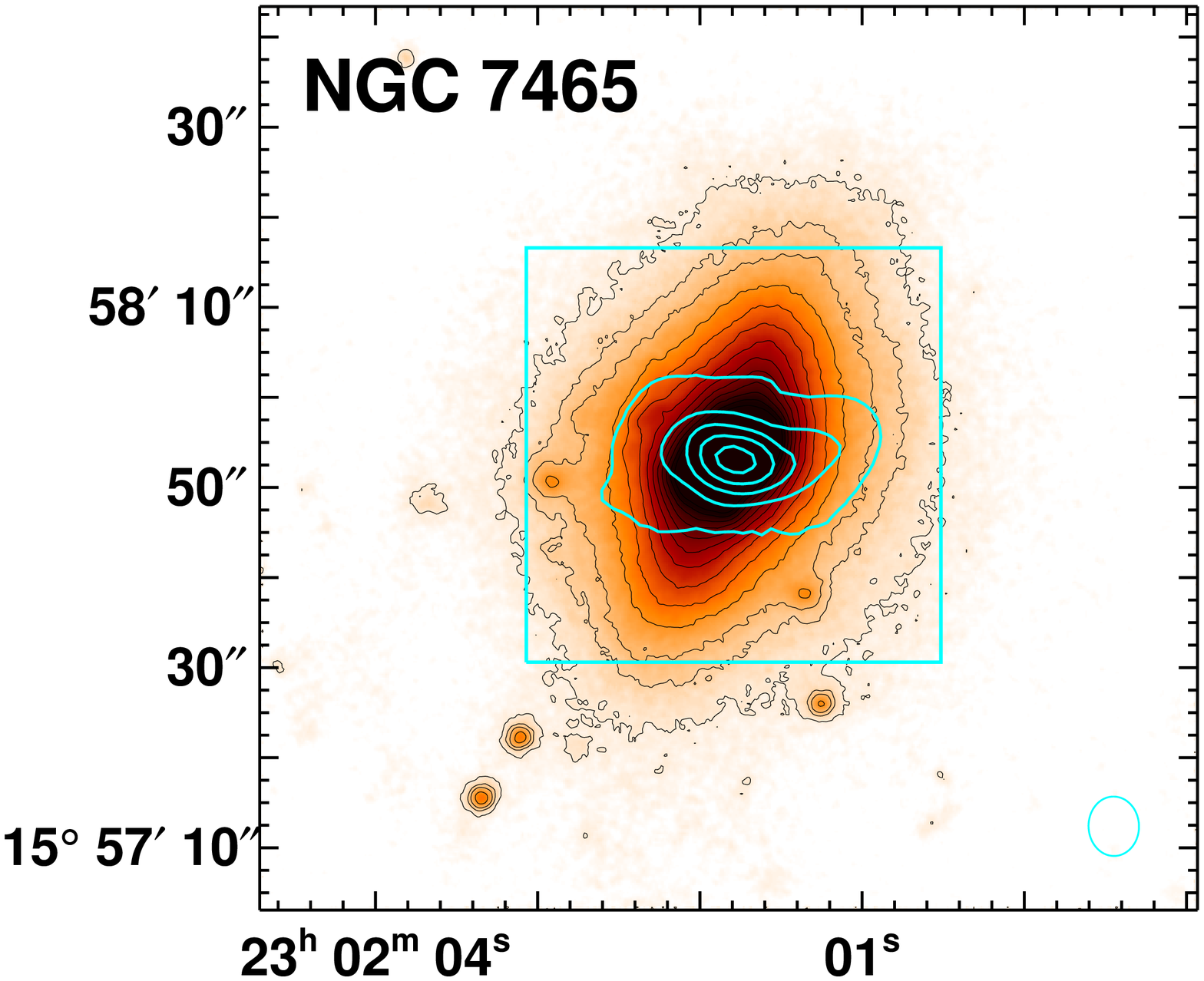}}
\subfloat{\includegraphics[height=2.2in,clip,trim=0cm 0.6cm 0.4cm 0.4cm]{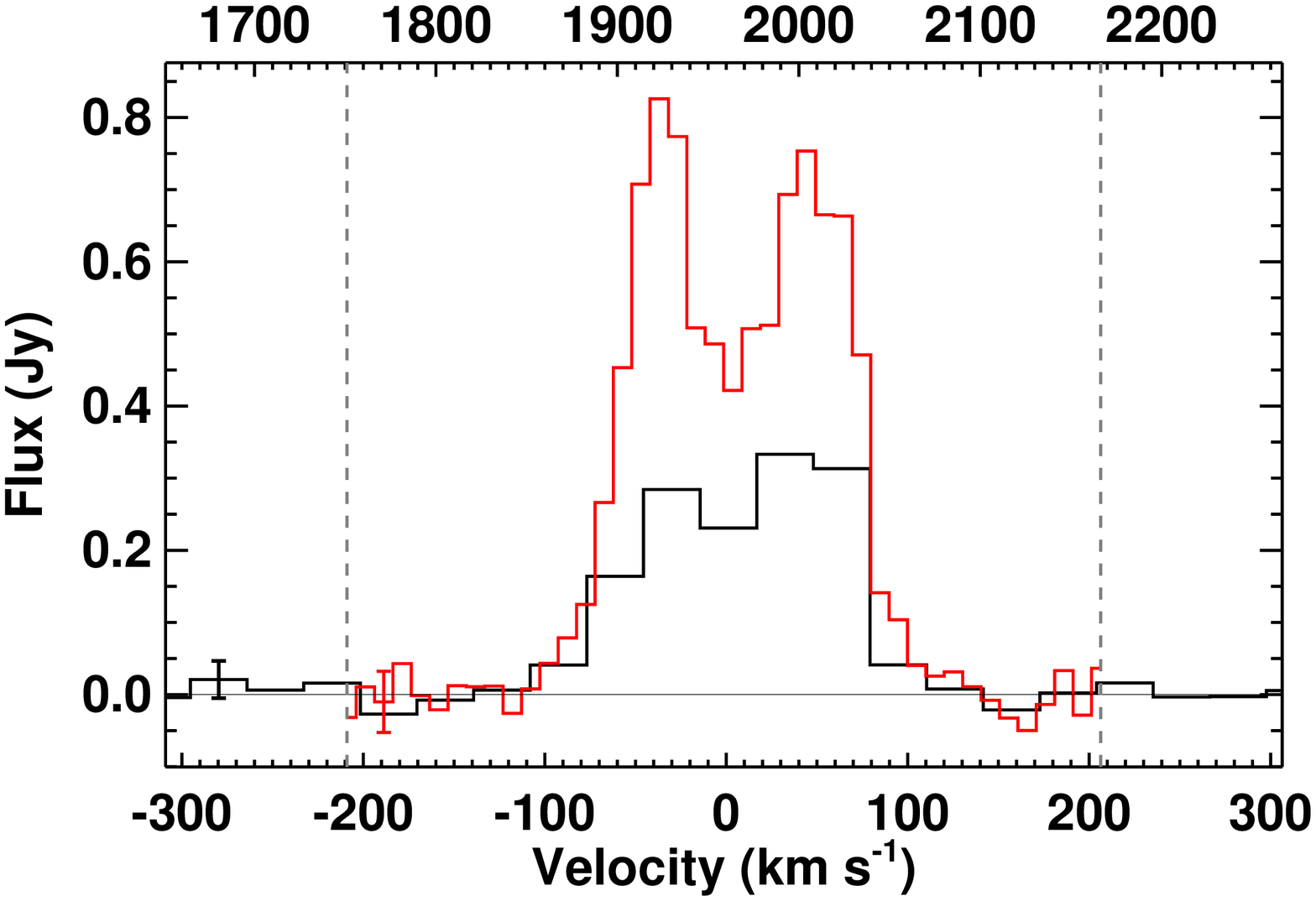}}
\end{figure*}
\begin{figure*}
\subfloat{\includegraphics[height=1.6in,clip,trim=0.1cm 1.4cm 0.6cm 2.4cm]{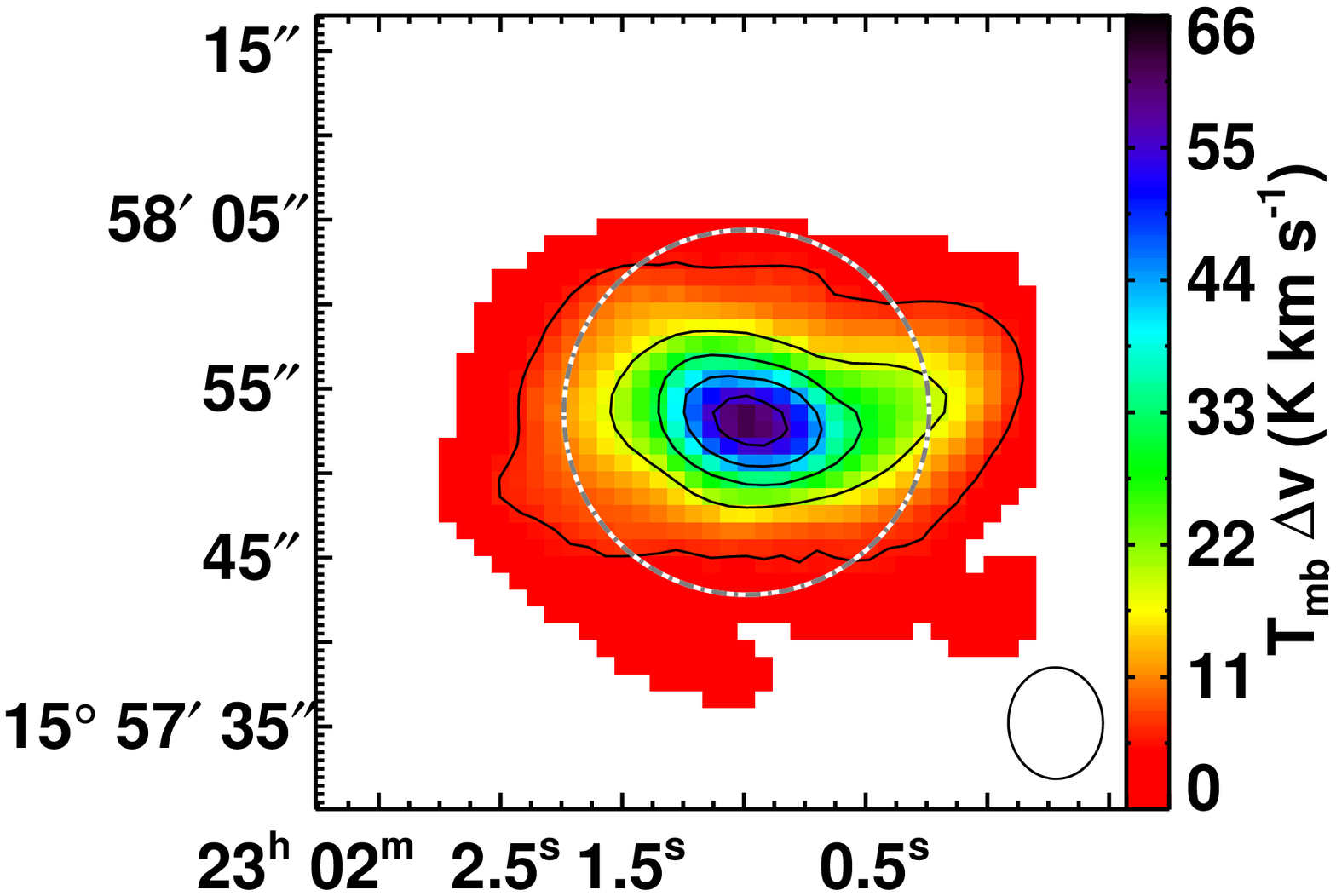}}
\subfloat{\includegraphics[height=1.6in,clip,trim=0.1cm 1.4cm 0cm 2.4cm]{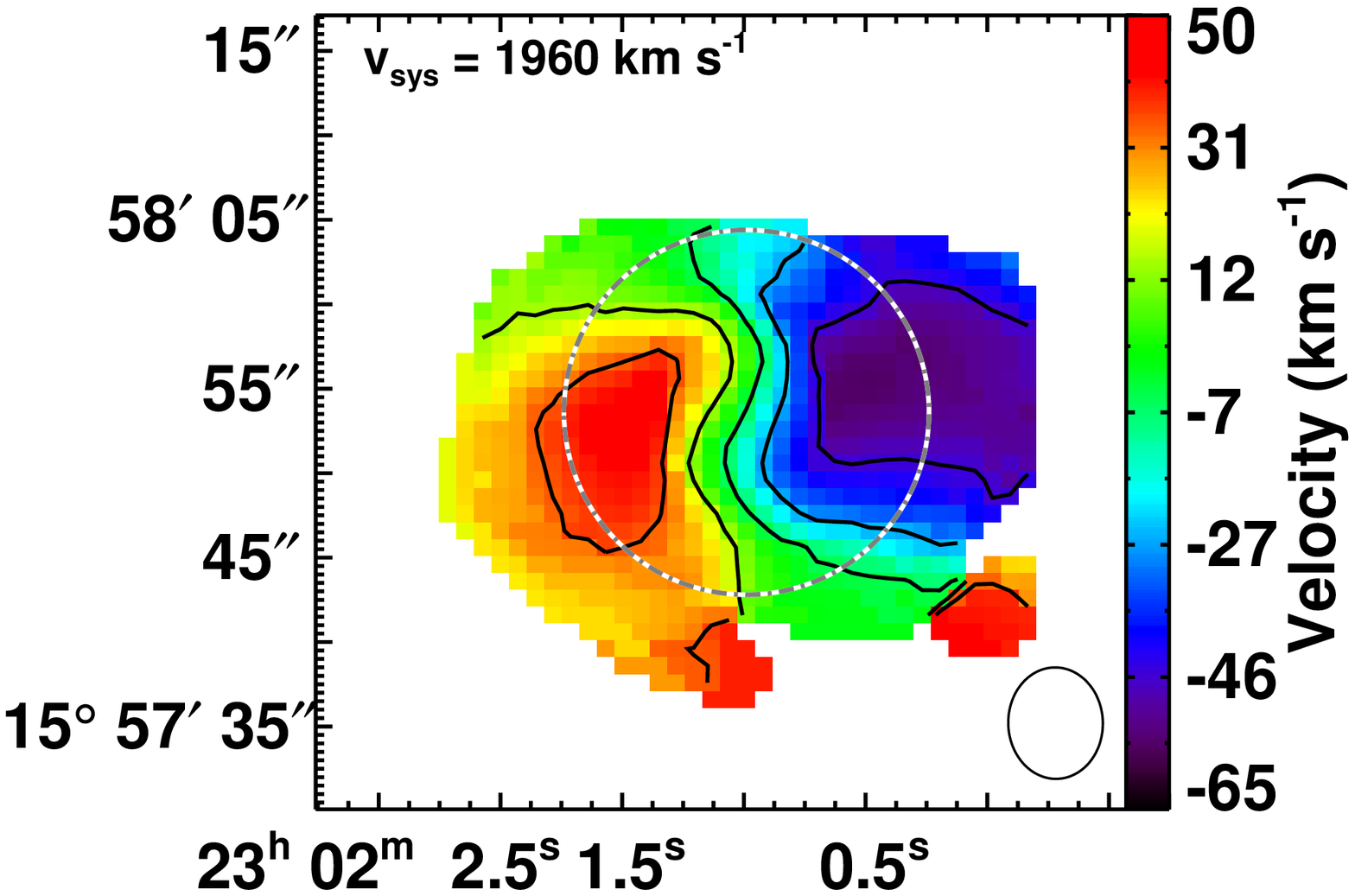}}
\subfloat{\includegraphics[height=1.6in,clip,trim=0cm 1.4cm 0cm 0.9cm]{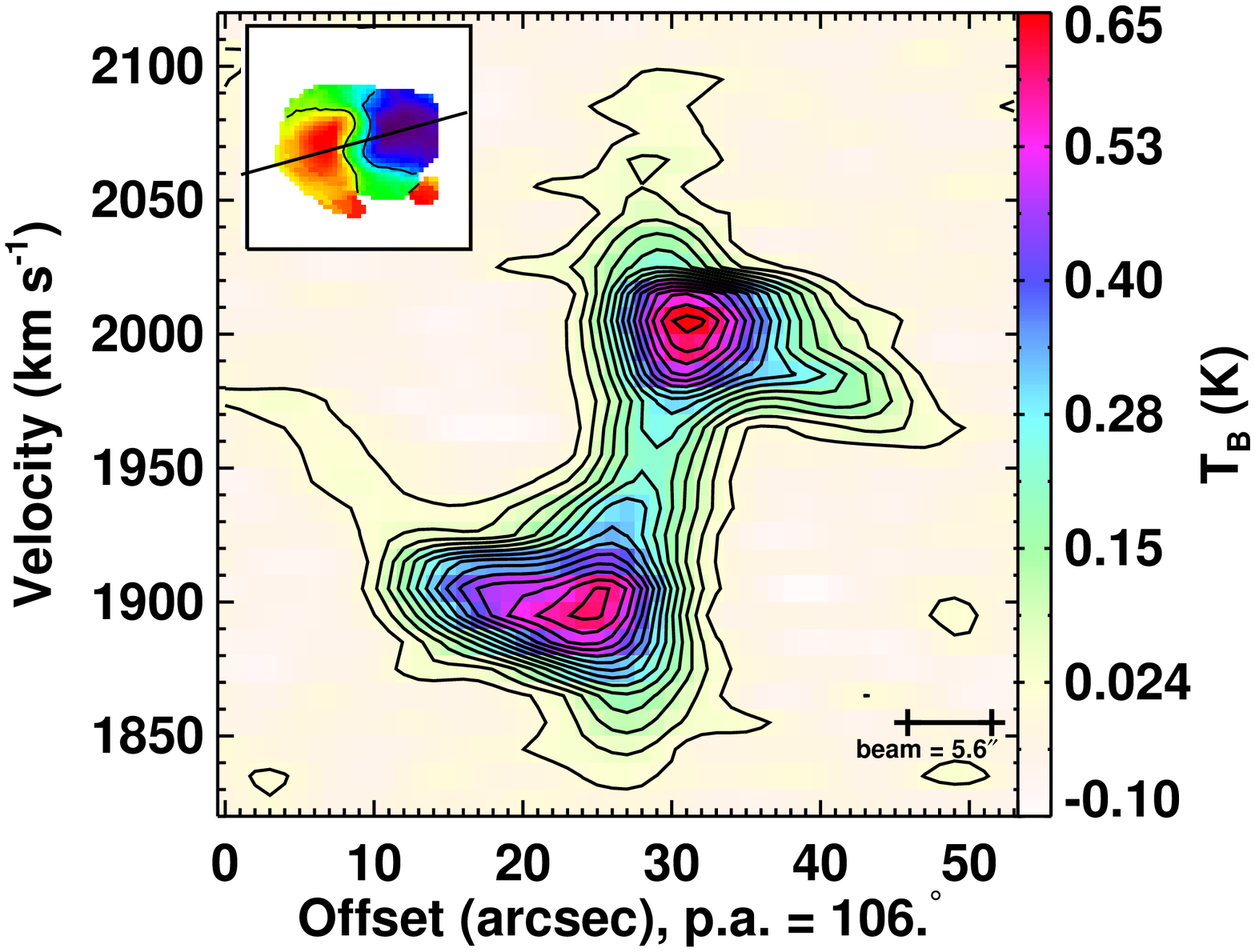}}
\end{figure*}
\begin{figure*}
\subfloat{\includegraphics[width=7in,clip,trim=0cm 0.8cm 0.4cm 1.3cm]{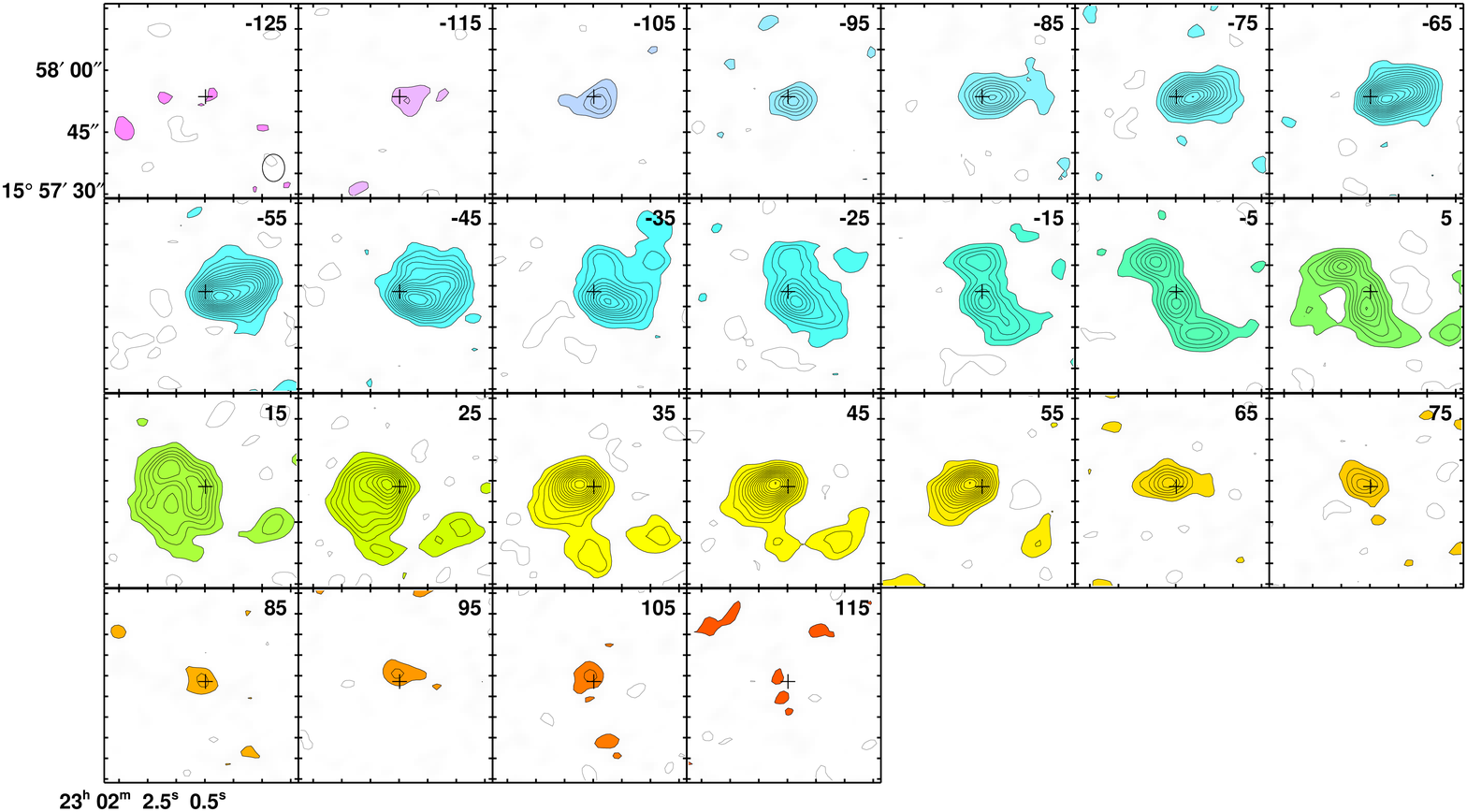}}
\caption{{\bf NGC~7465} is a field non-regular rotator ($M_K$ = -22.82) with a kinematically-decoupled core (KDC) with stellar morphology indicative of interaction.  It contains  dust filaments.  The moment0 peak is 26 Jy beam$^{-1}$ \kms.  Channel map contours are placed at $2\sigma$ intervals and PVD contours are placed at $4\sigma$ intervals.}
\end{figure*}

\clearpage
\begin{figure*}
\centering
\subfloat{\includegraphics[height=2.2in,clip,trim=2.2cm 3.2cm 0cm 2.7cm]{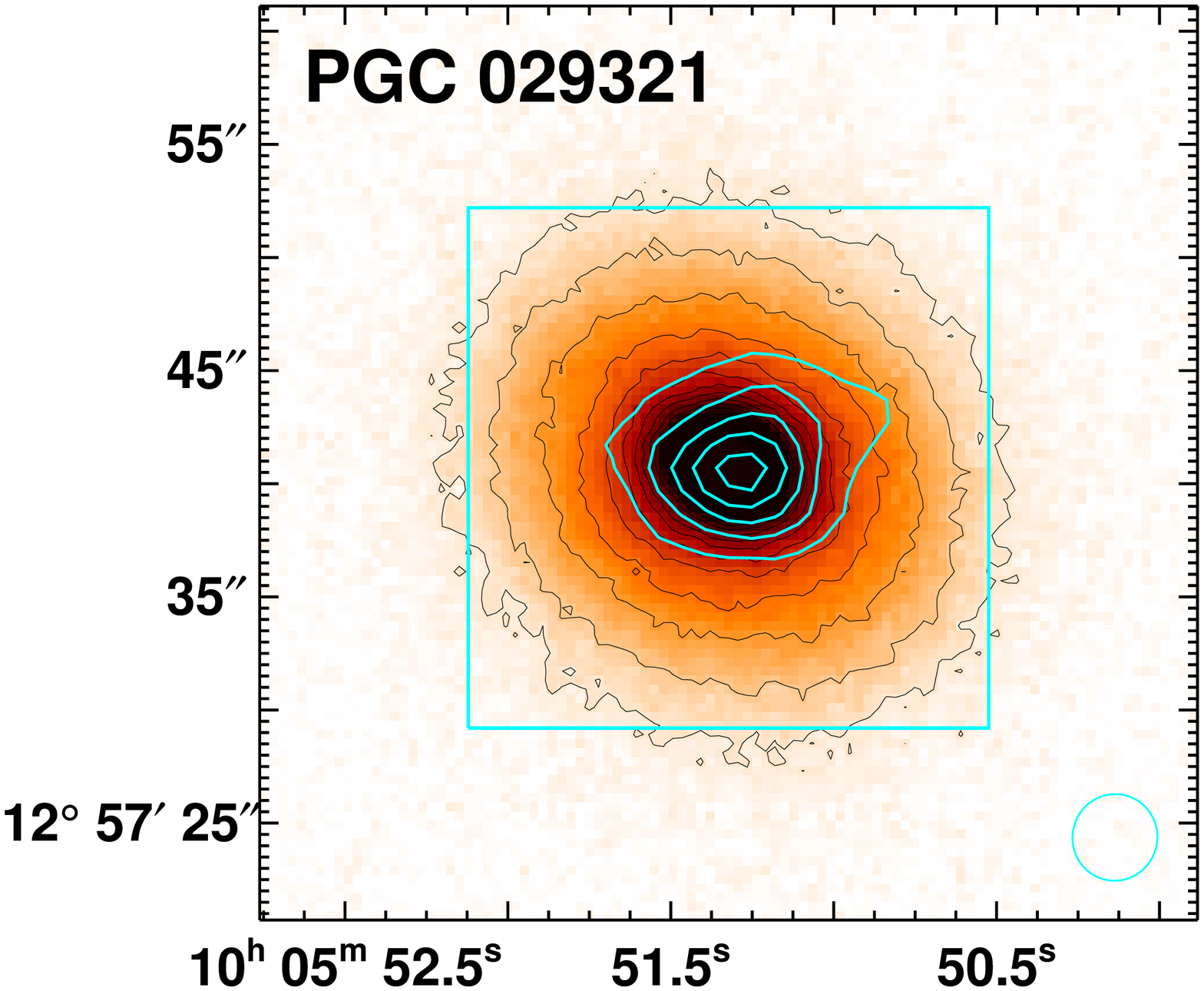}}
\subfloat{\includegraphics[height=2.2in,clip,trim=0cm 0.6cm 0.4cm 0.4cm]{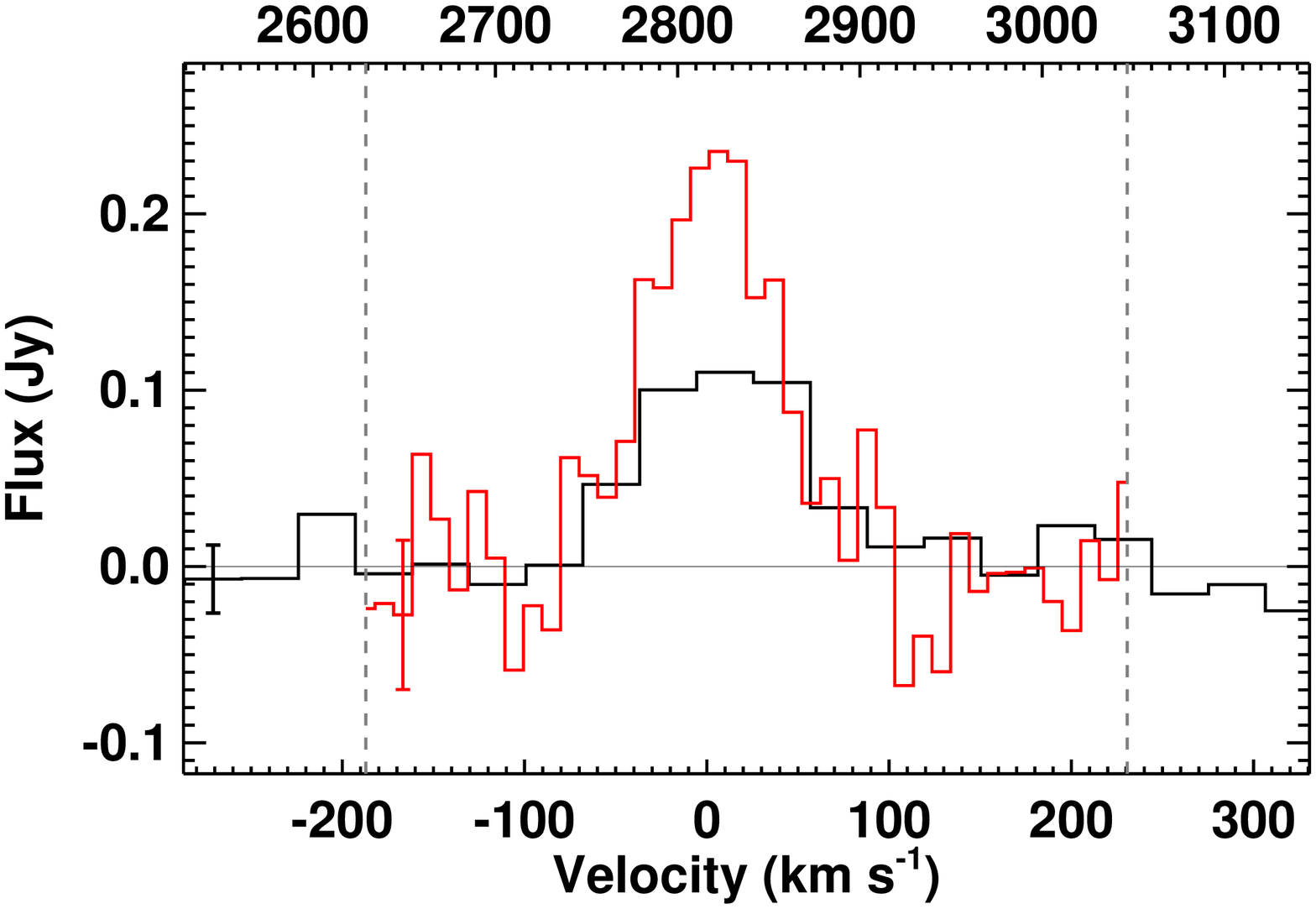}}
\end{figure*}
\begin{figure*}
\subfloat{\includegraphics[height=1.6in,clip,trim=0.1cm 1.4cm 0.6cm 2.4cm]{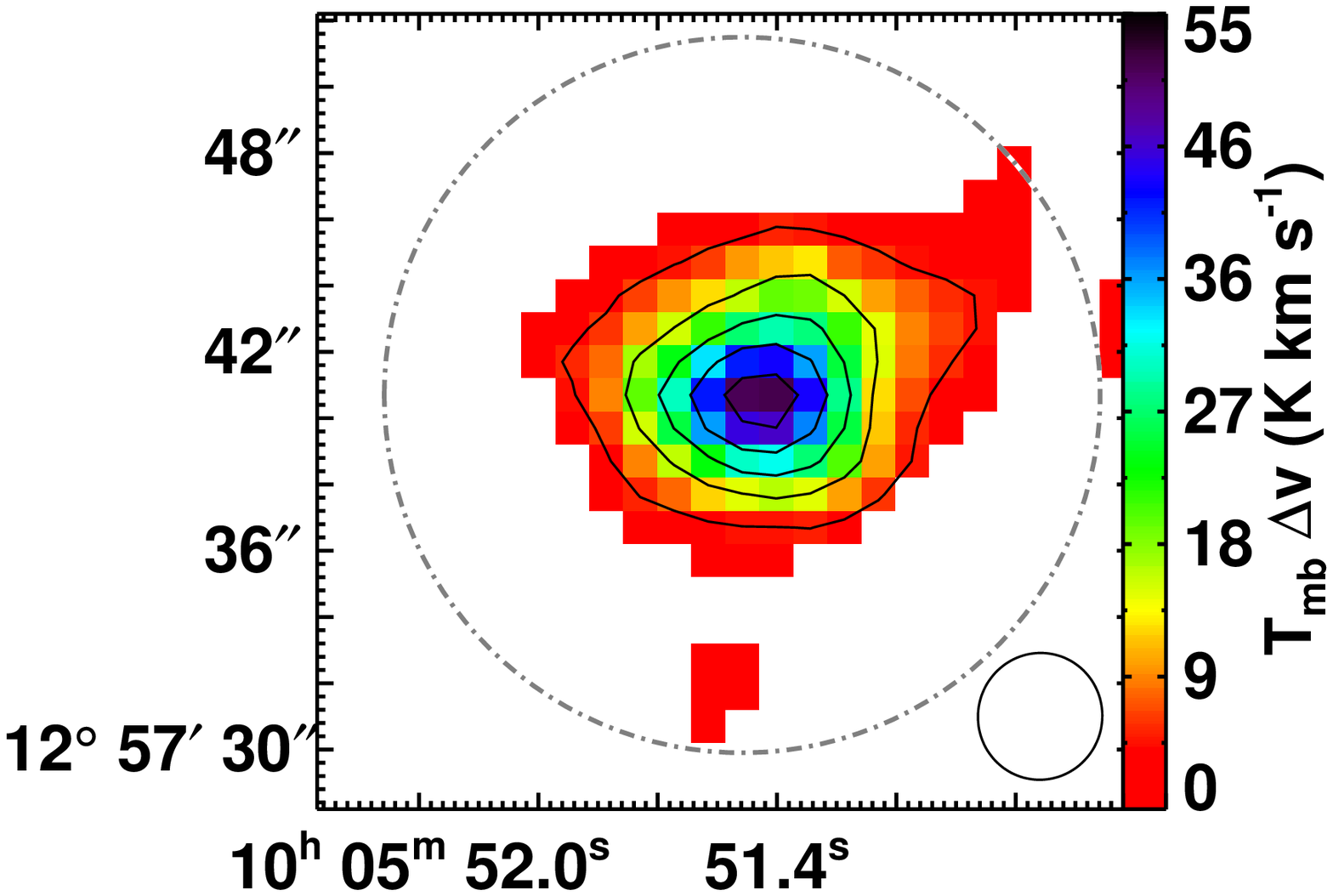}}
\subfloat{\includegraphics[height=1.6in,clip,trim=0.1cm 1.4cm 0cm 2.4cm]{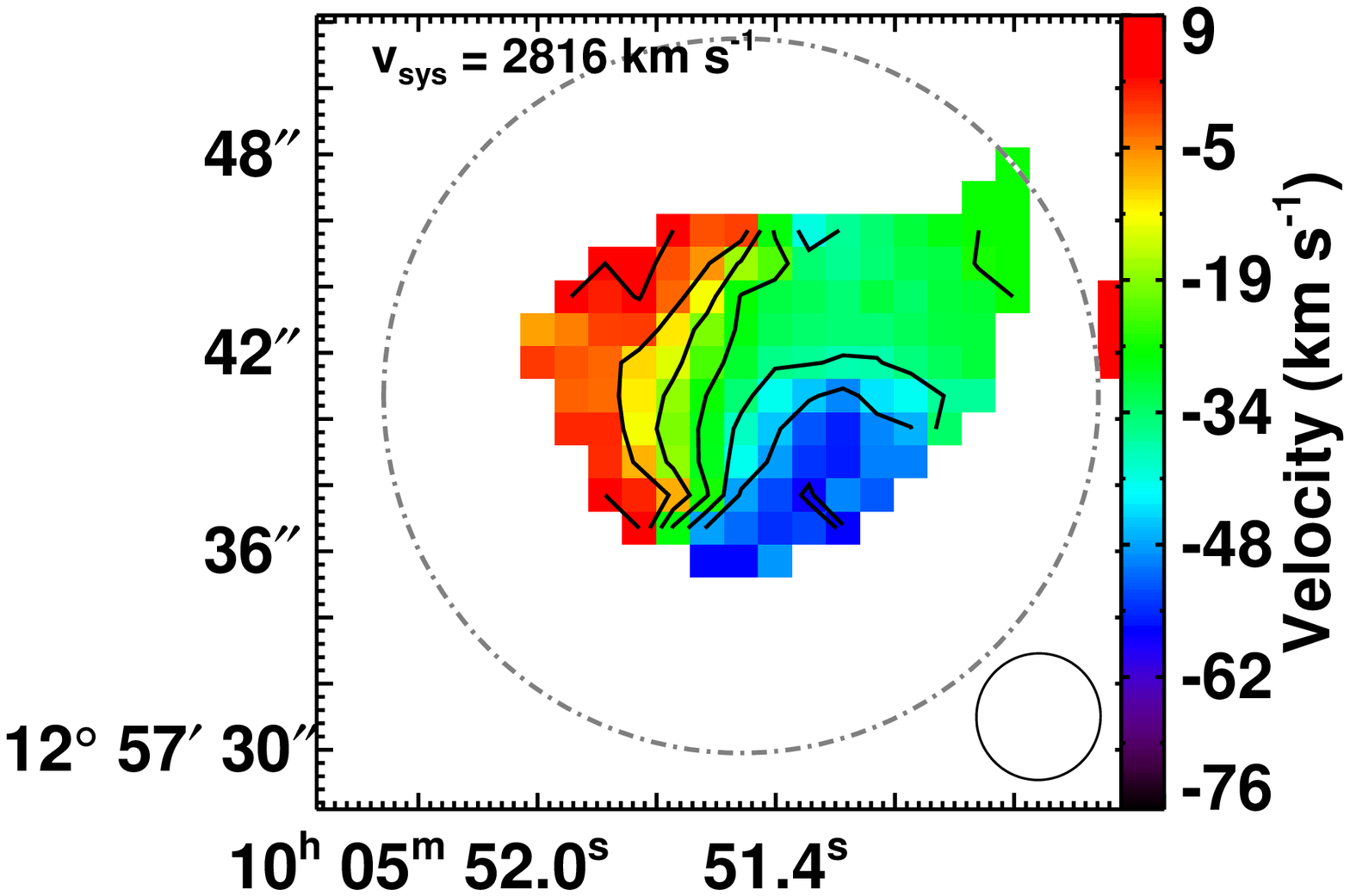}}
\subfloat{\includegraphics[height=1.6in,clip,trim=0cm 1.4cm 0cm 0.9cm]{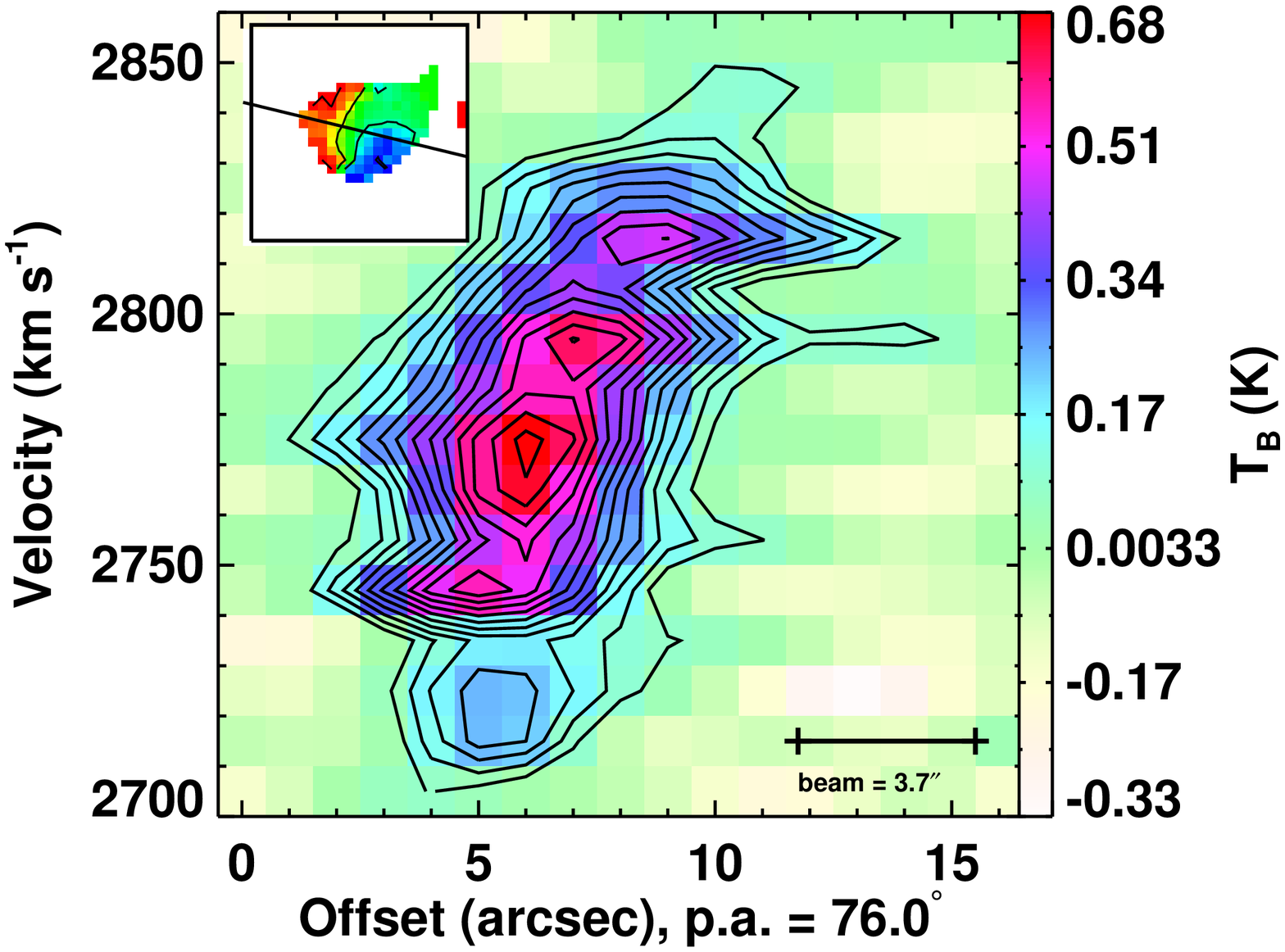}}
\end{figure*}
\begin{figure*}
\subfloat{\includegraphics[width=7in,clip,trim=0cm 1cm 0.1cm 4.1cm]{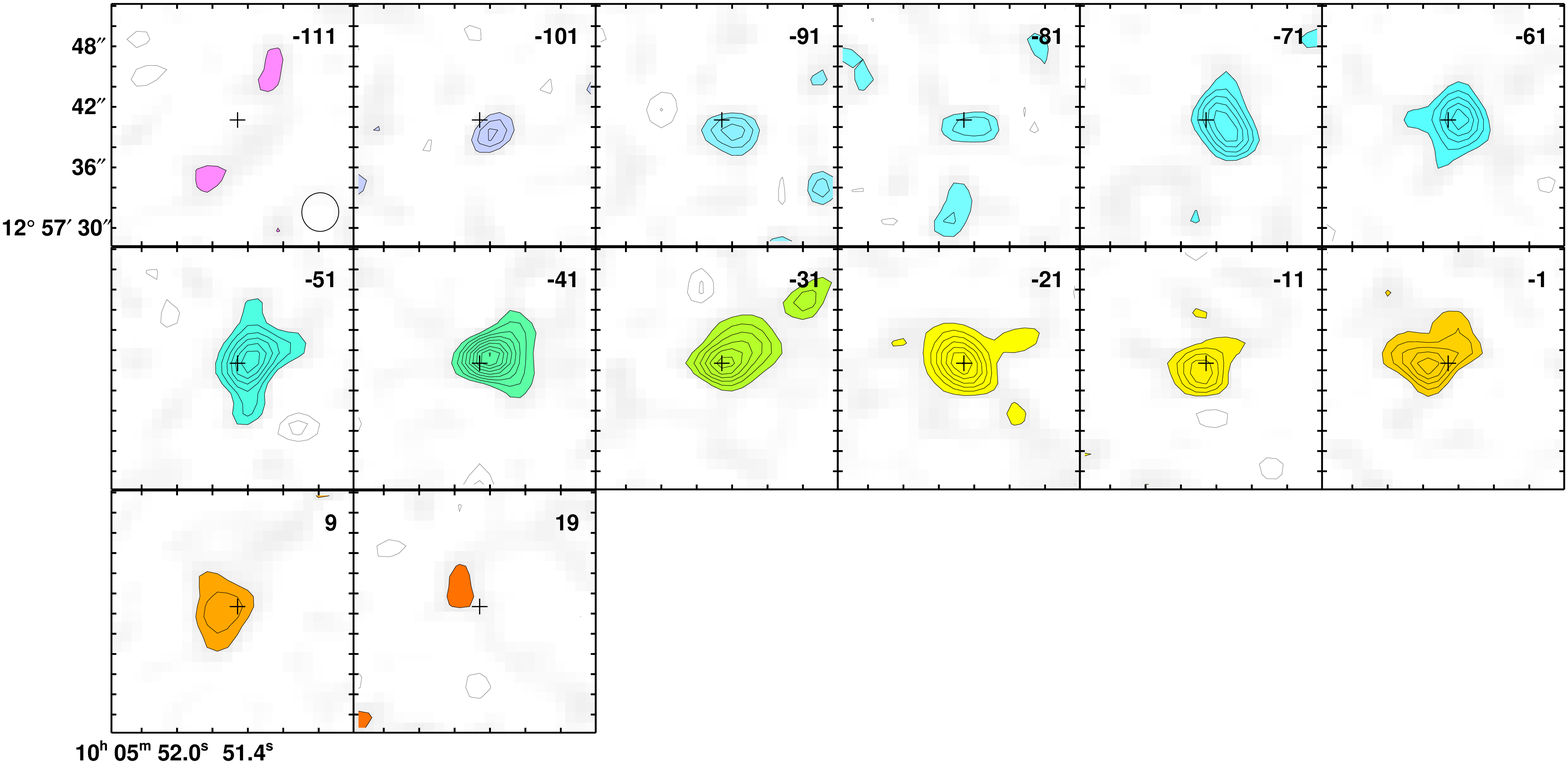}}
\caption{{\bf PGC~029321} is a field regular rotator ($M_K$ = -21.66) with normal stellar morphology.  It contains dust filaments.  The moment0 peak is 8.5 Jy beam$^{-1}$ \kms.}
\end{figure*}

\clearpage
\begin{figure*}
\centering
\subfloat{\includegraphics[height=2.2in,clip,trim=2.2cm 2.9cm 0cm 2.6cm]{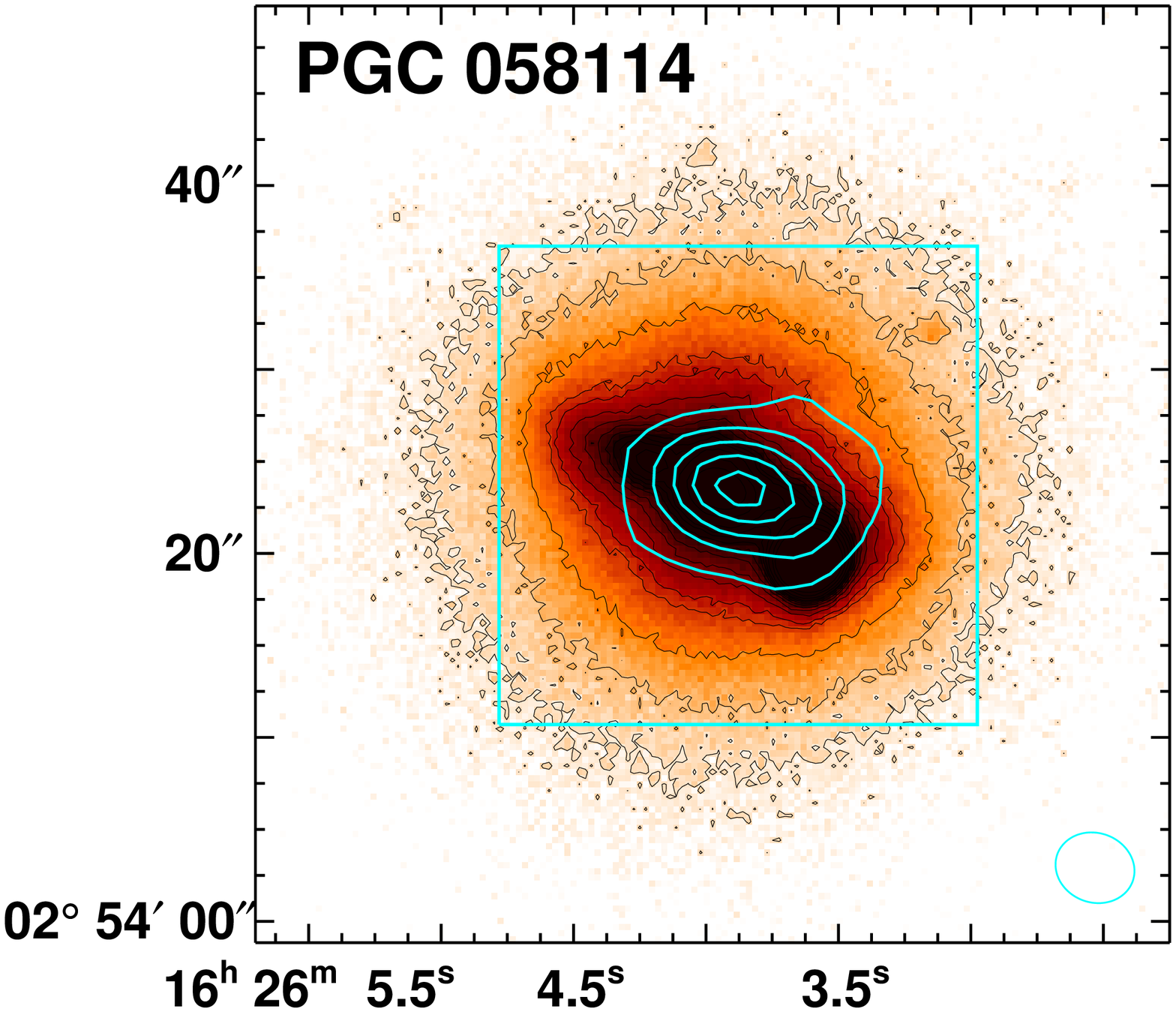}}
\subfloat{\includegraphics[height=2.2in,clip,trim=0cm 0.6cm 0.4cm 0.4cm]{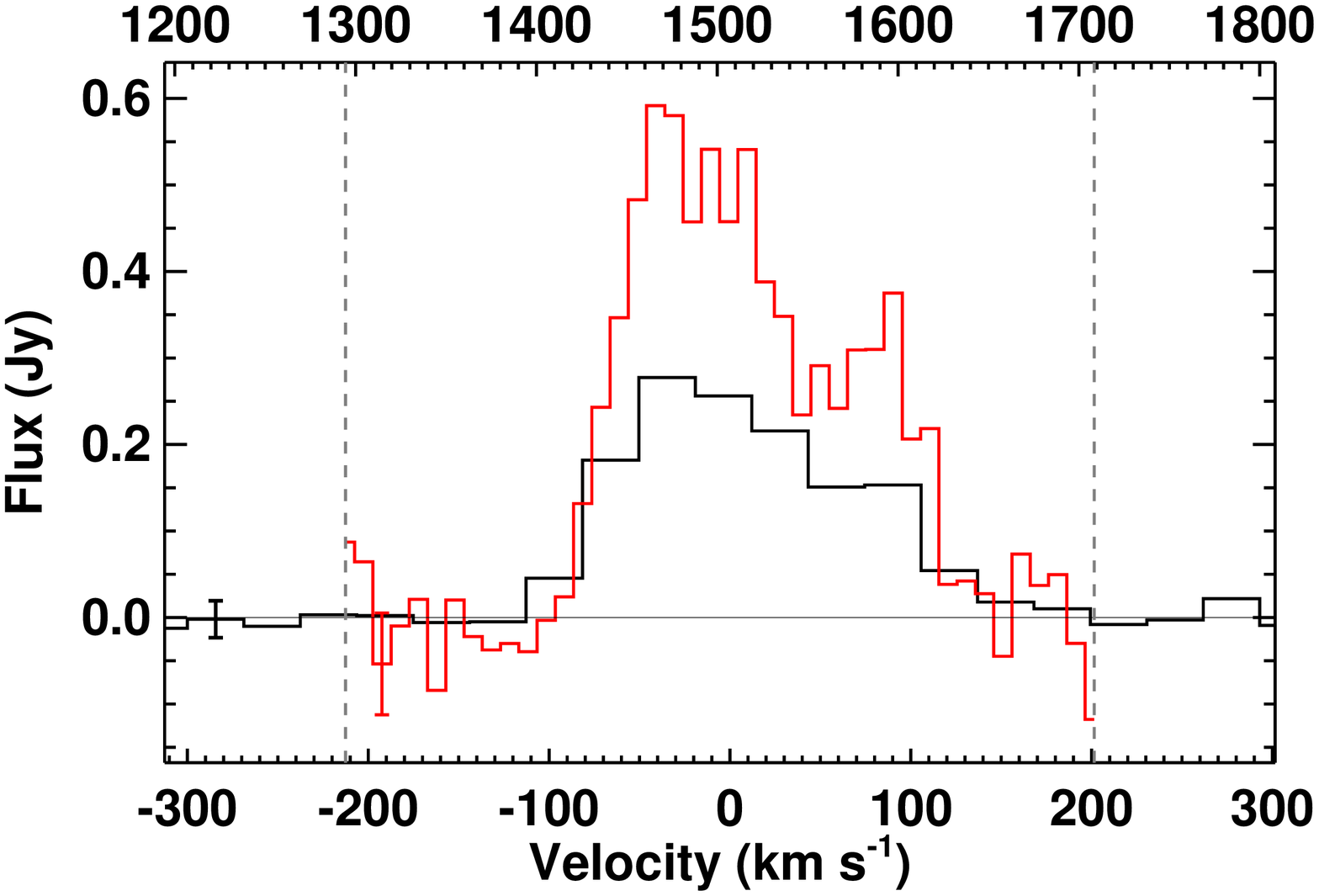}}
\end{figure*}
\begin{figure*}
\subfloat{\includegraphics[height=1.6in,clip,trim=0.1cm 1.4cm 0.6cm 2.4cm]{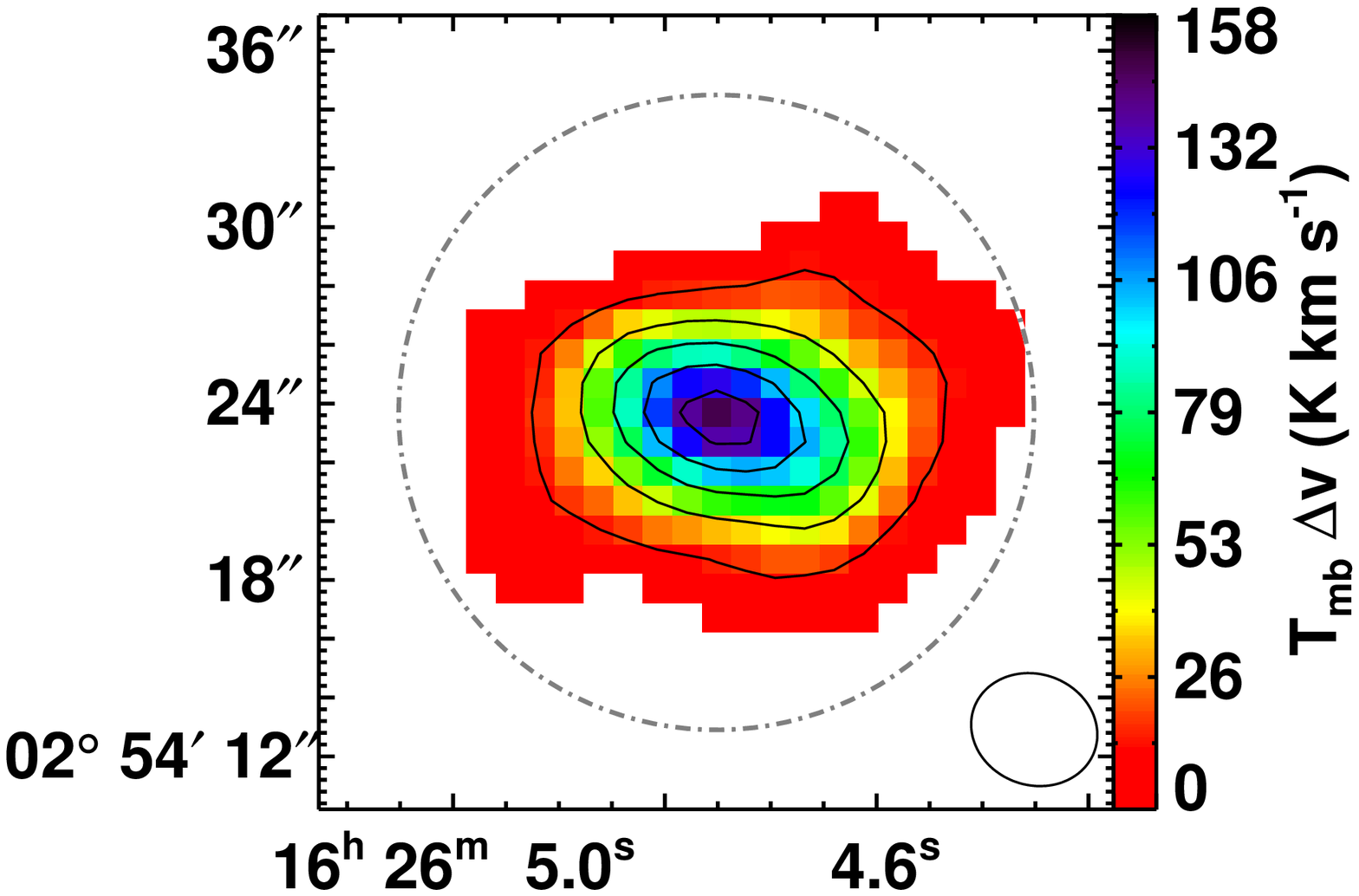}}
\subfloat{\includegraphics[height=1.6in,clip,trim=0.1cm 1.4cm 0cm 2.4cm]{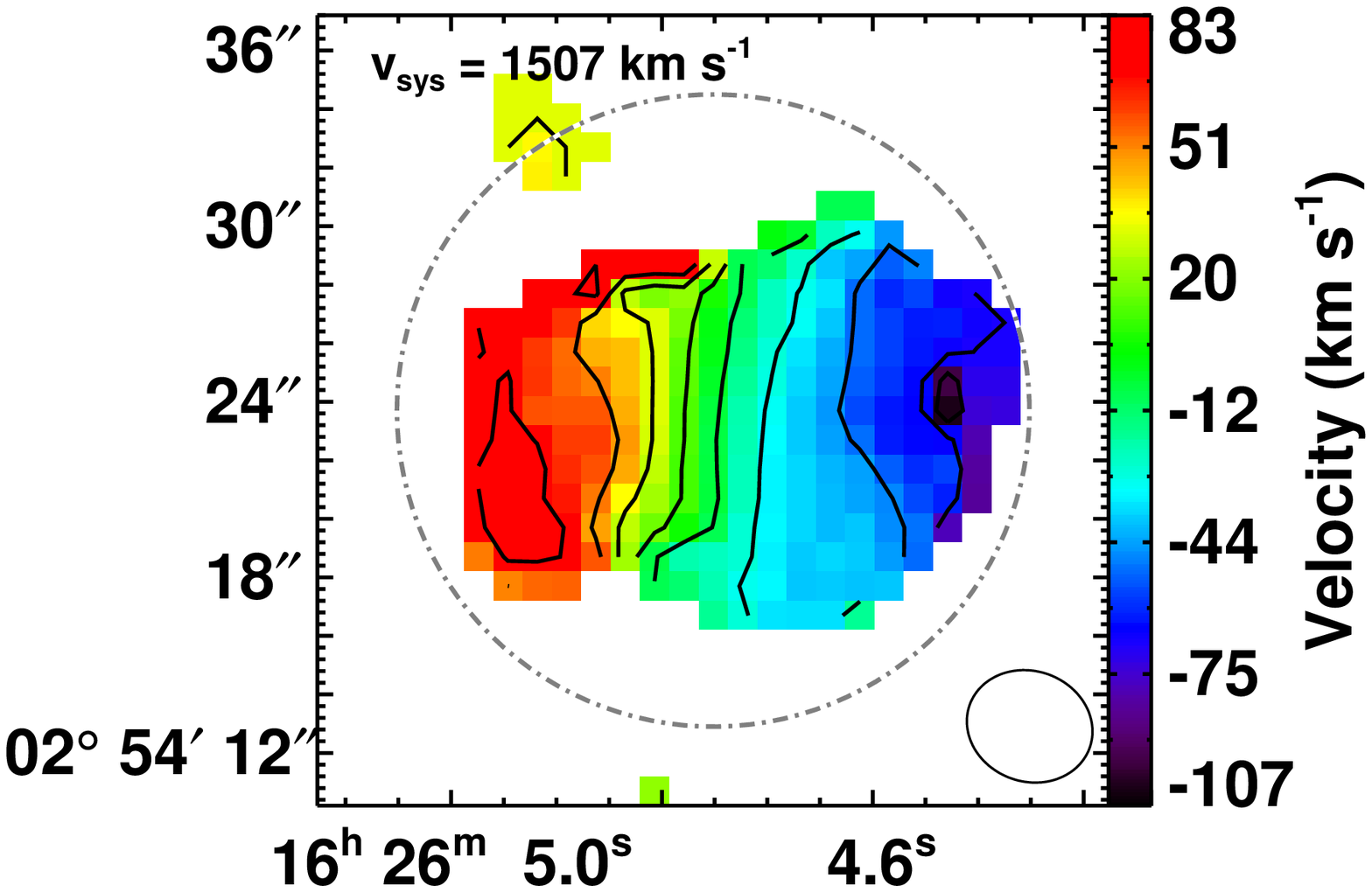}}
\subfloat{\includegraphics[height=1.6in,clip,trim=0cm 1.4cm 0cm 0.9cm]{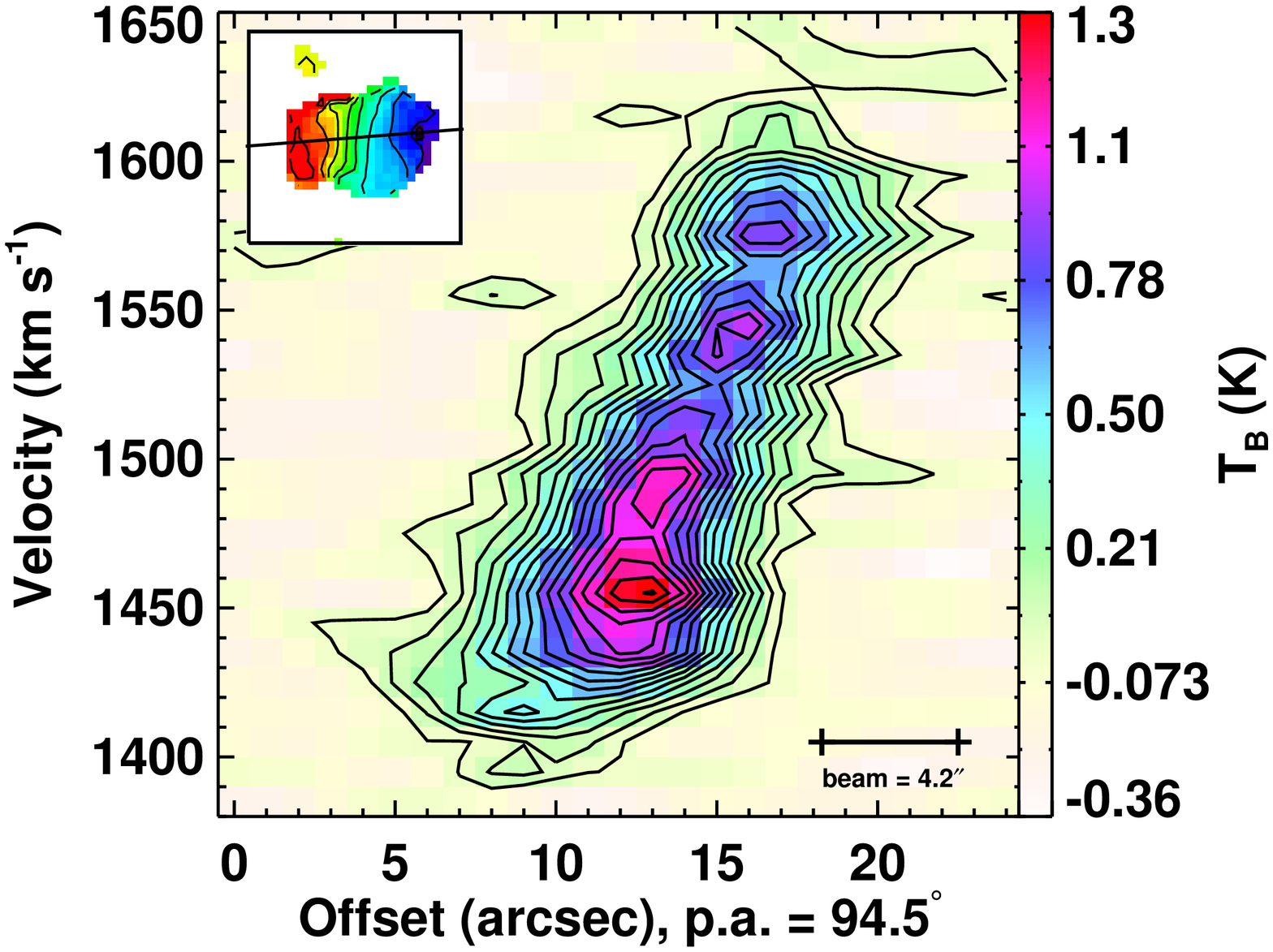}}
\end{figure*}
\begin{figure*}
\subfloat{\includegraphics[width=7in,clip,trim=0cm 0.8cm 0.1cm 1.3cm]{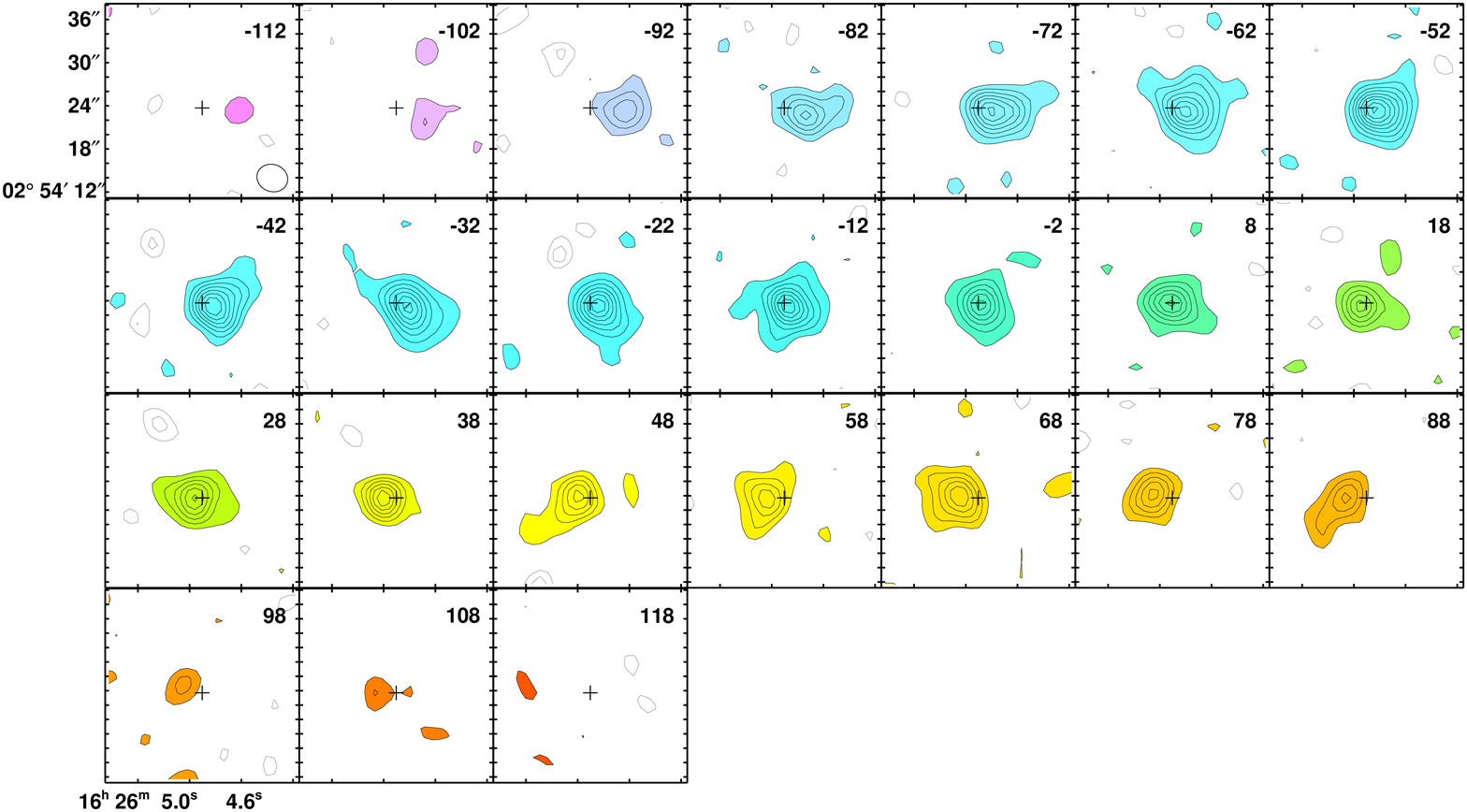}}
\caption{{\bf PGC~058114} is a field galaxy ($M_K$ = -21.57) with unknown kinematic structure or morphology.  The moment0 peak is 28 Jy beam$^{-1}$ \kms.  Channel map and PVD contours are placed at $2\sigma$ intervals.}
\end{figure*}

\clearpage
\begin{figure*}
\centering
\subfloat{\includegraphics[height=2.2in,clip,trim=2.2cm 3.2cm 0cm 2.7cm]{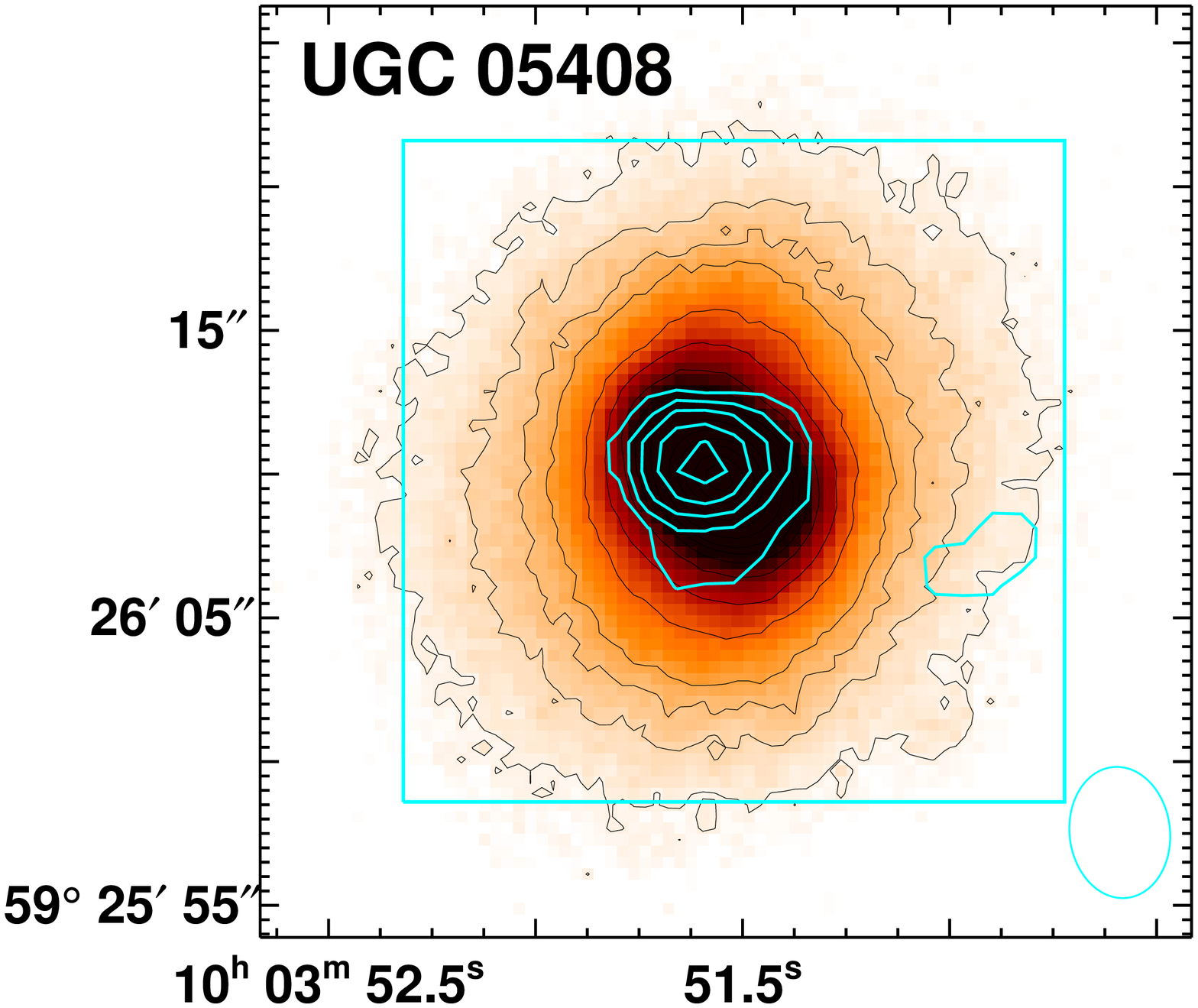}}
\subfloat{\includegraphics[height=2.2in,clip,trim=0cm 0.6cm 0.4cm 0.4cm]{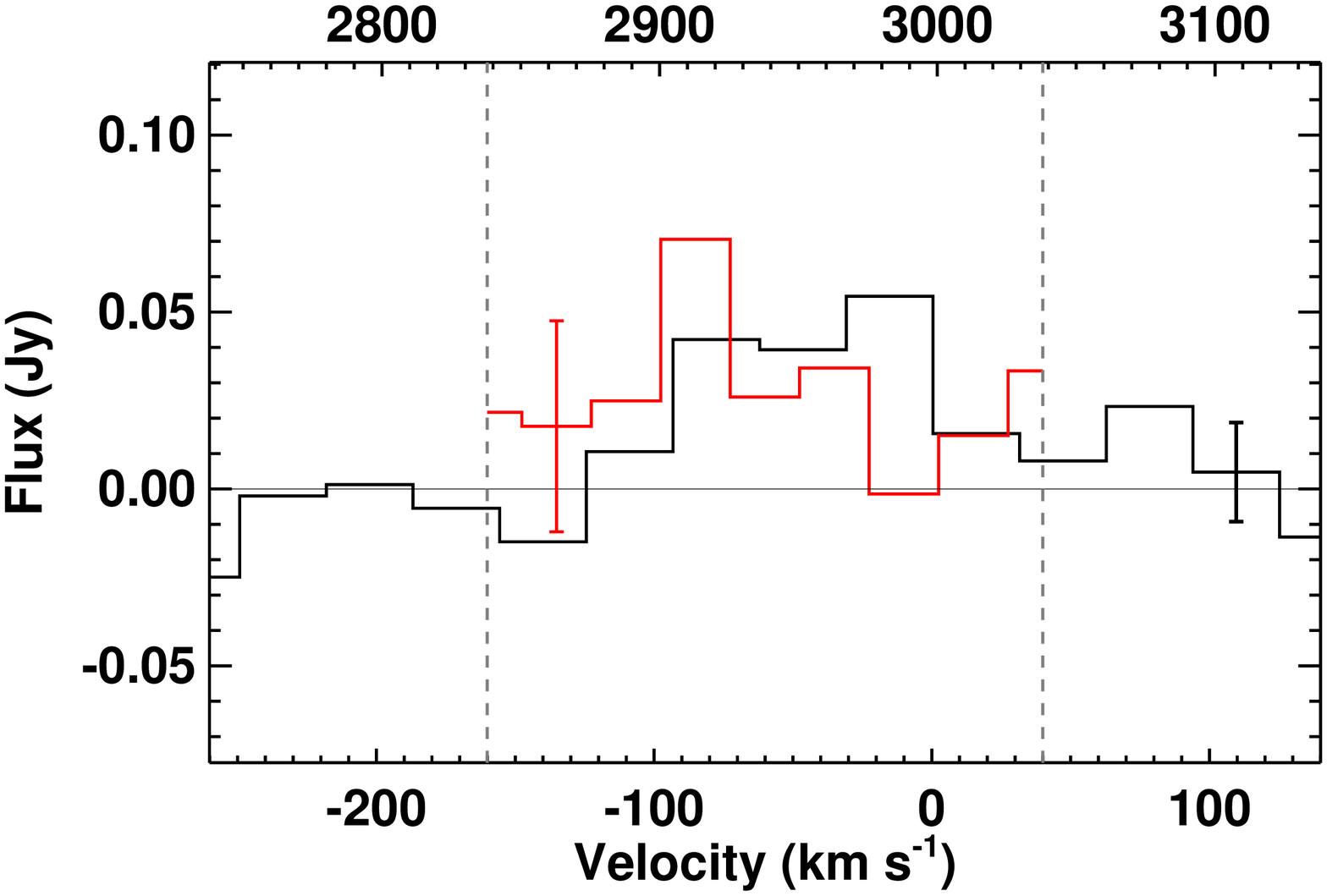}}
\end{figure*}
\begin{figure*}
\subfloat{\includegraphics[height=1.6in,clip,trim=0.1cm 1.4cm 0.4cm 2.4cm]{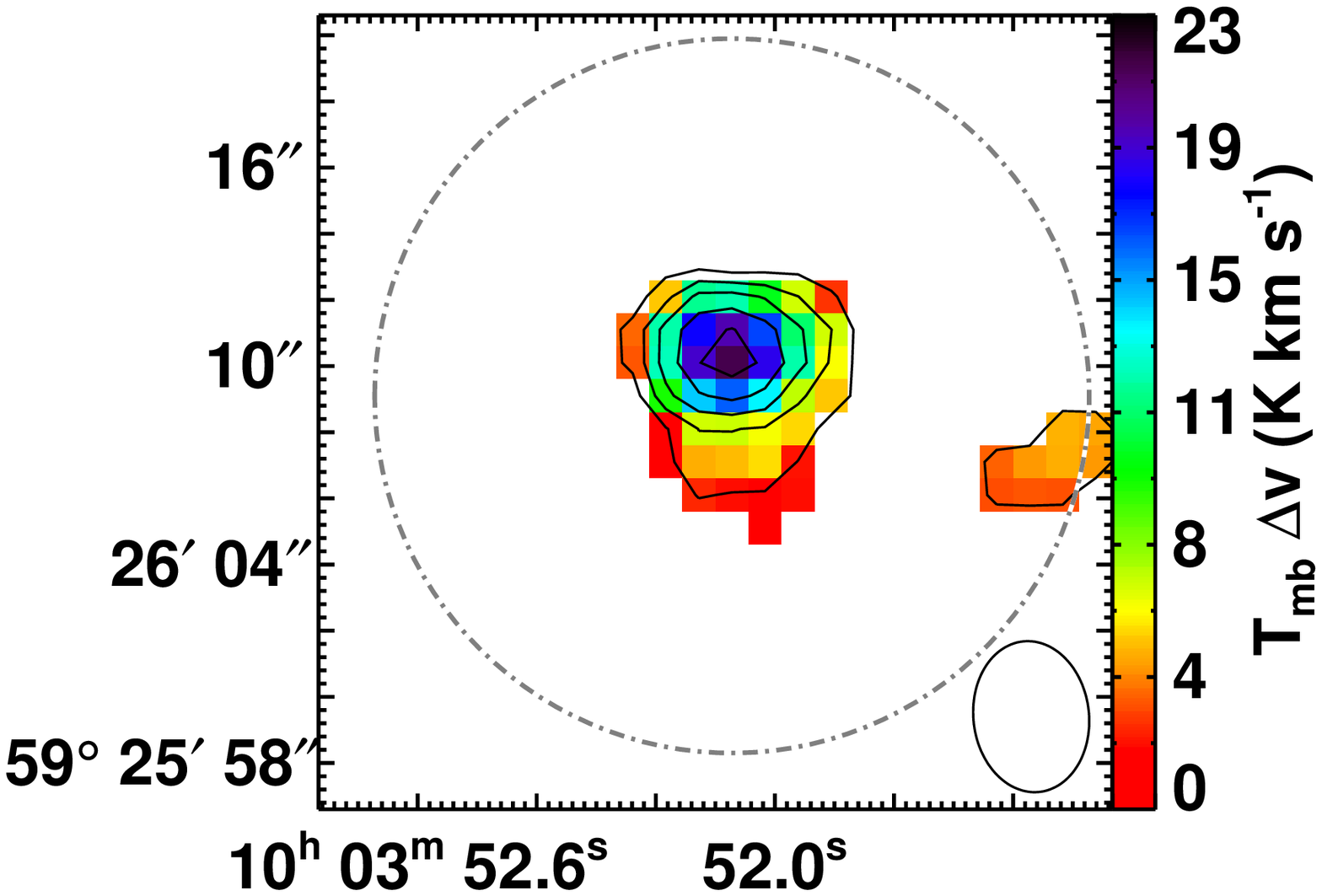}}
\subfloat{\includegraphics[height=1.6in,clip,trim=0.1cm 1.4cm 0.4cm 2.4cm]{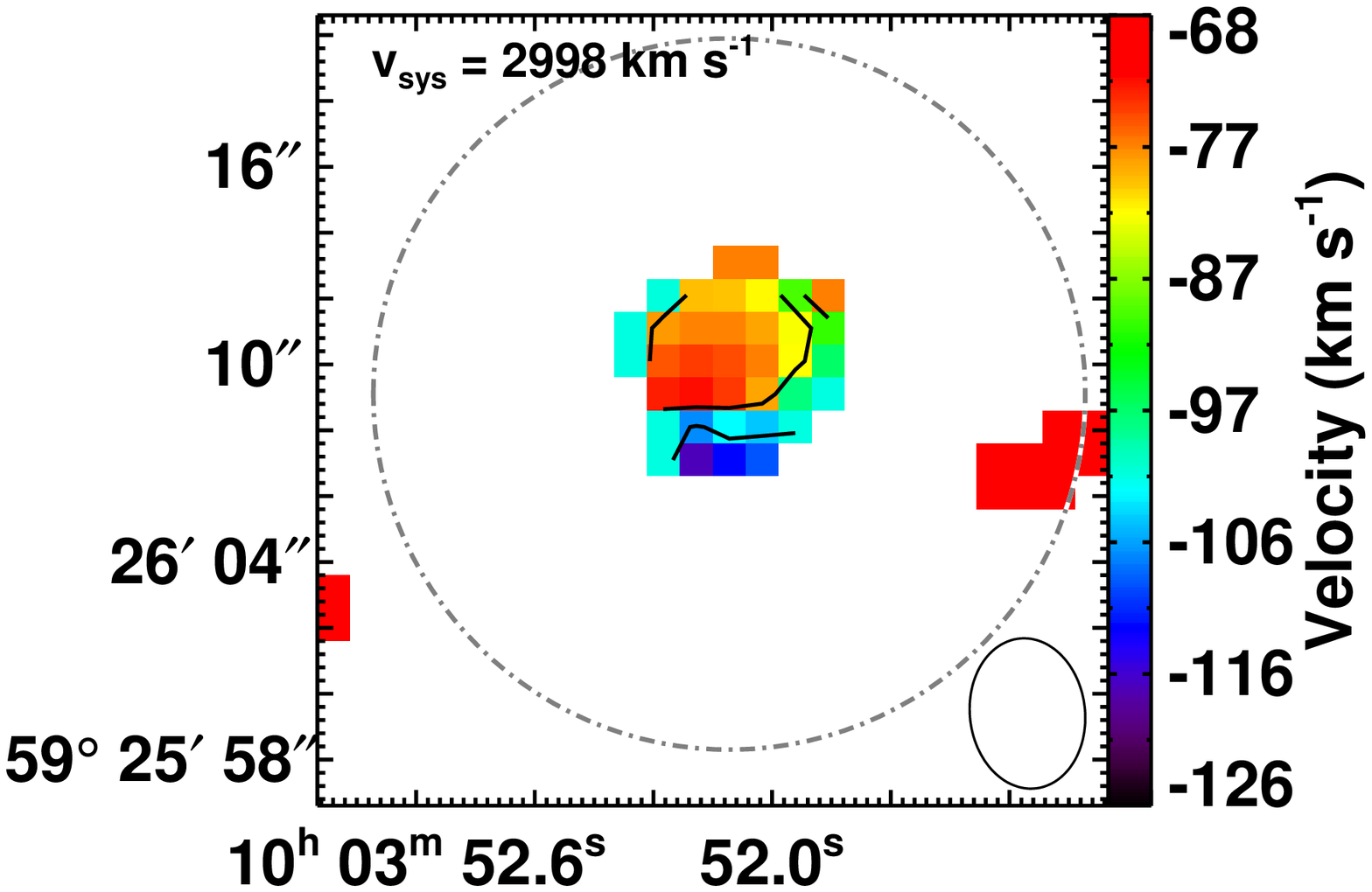}}
\subfloat{\includegraphics[height=1.6in,clip,trim=0cm 1.4cm 0cm 0.9cm]{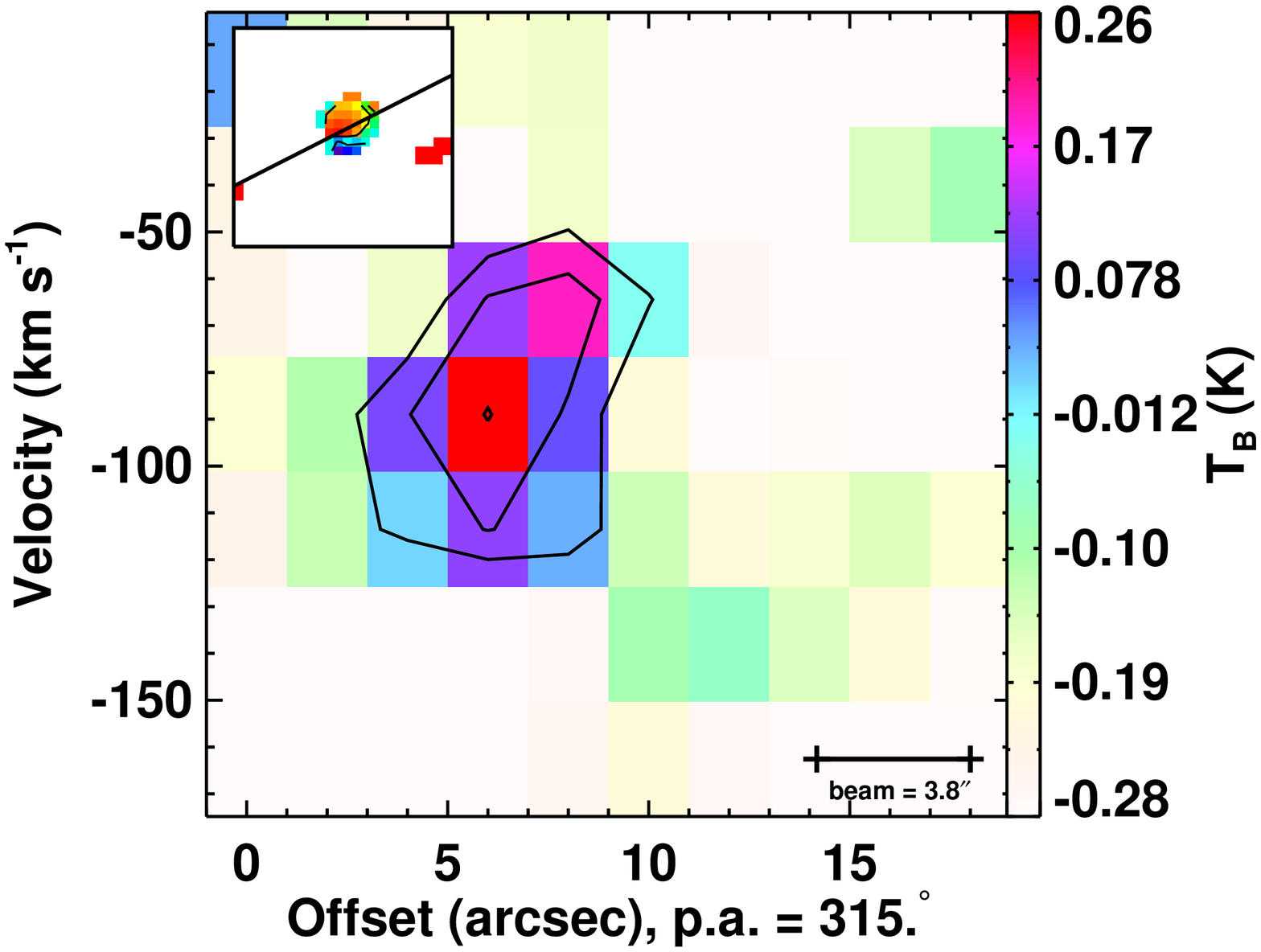}}
\end{figure*}
\begin{figure*}
\subfloat{\includegraphics[width=7in,clip,trim=0.2cm 0.8cm 0.1cm 1.3cm]{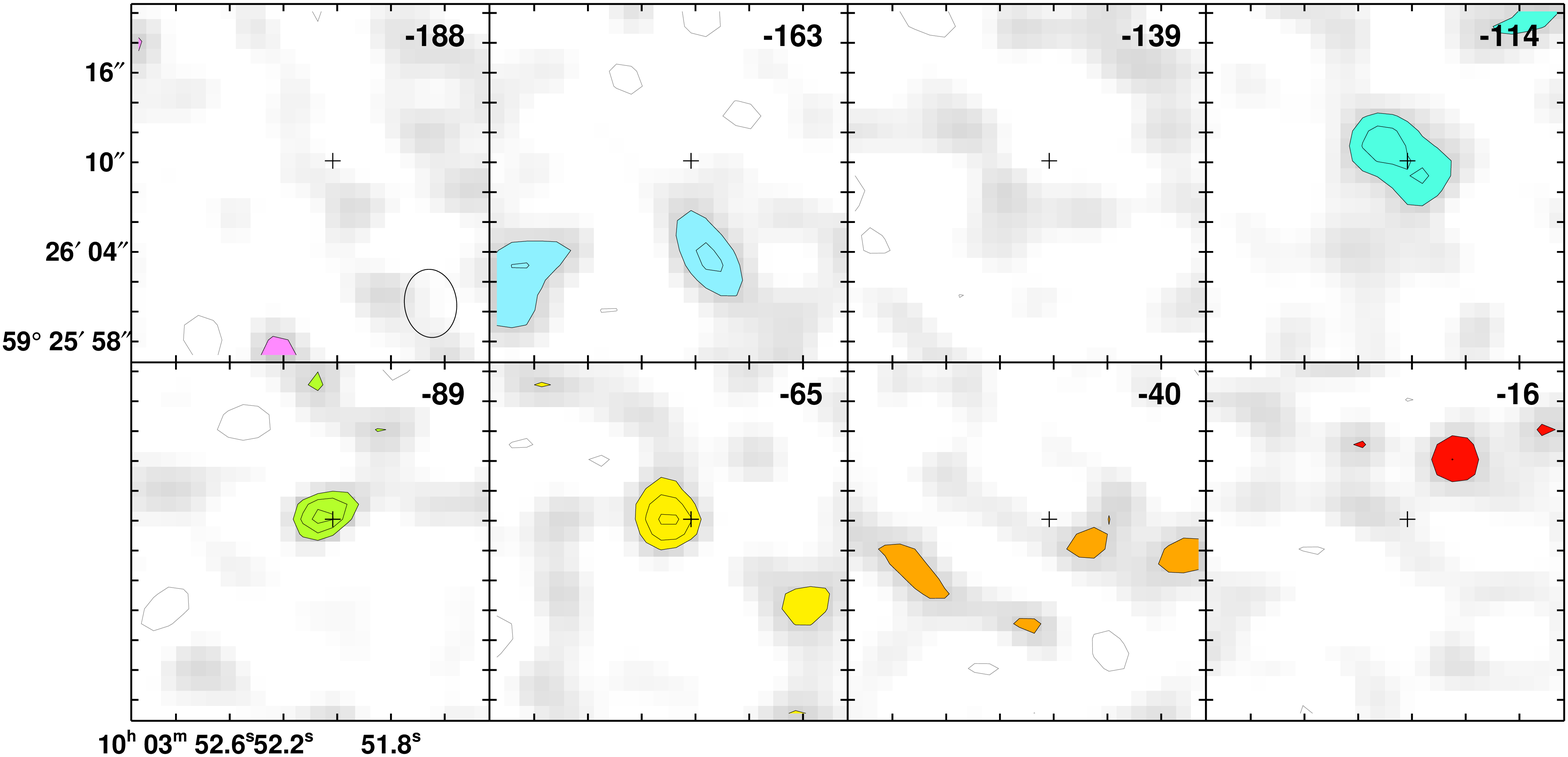}}
\caption{{\bf UGC~05408} is a field regular rotator ($M_K$ = -22.03) with a bar stellar morphology.  It contains a dust bar and filaments.  At a distance 45.8 Mpc, it is the most distant detection in the CARMA \atlas\ survey.  The moment0 peak is 3.5 Jy beam$^{-1}$ \kms.}
\end{figure*}

\clearpage
\begin{figure*}
\centering
\subfloat{\includegraphics[height=2.2in,clip,trim=2.2cm 3.2cm 0cm 2.7cm]{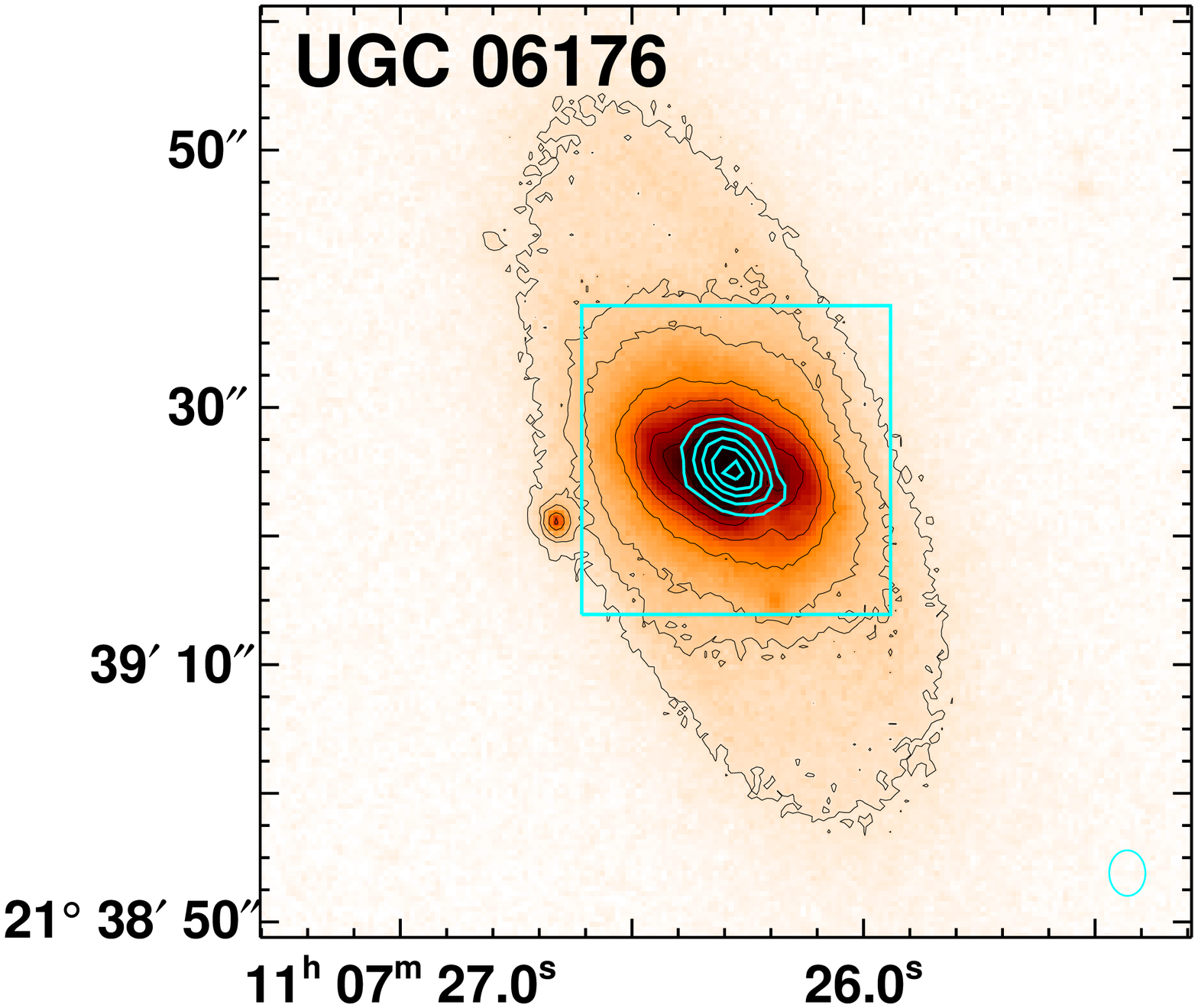}}
\subfloat{\includegraphics[height=2.2in,clip,trim=0cm 0.6cm 0.4cm 0.4cm]{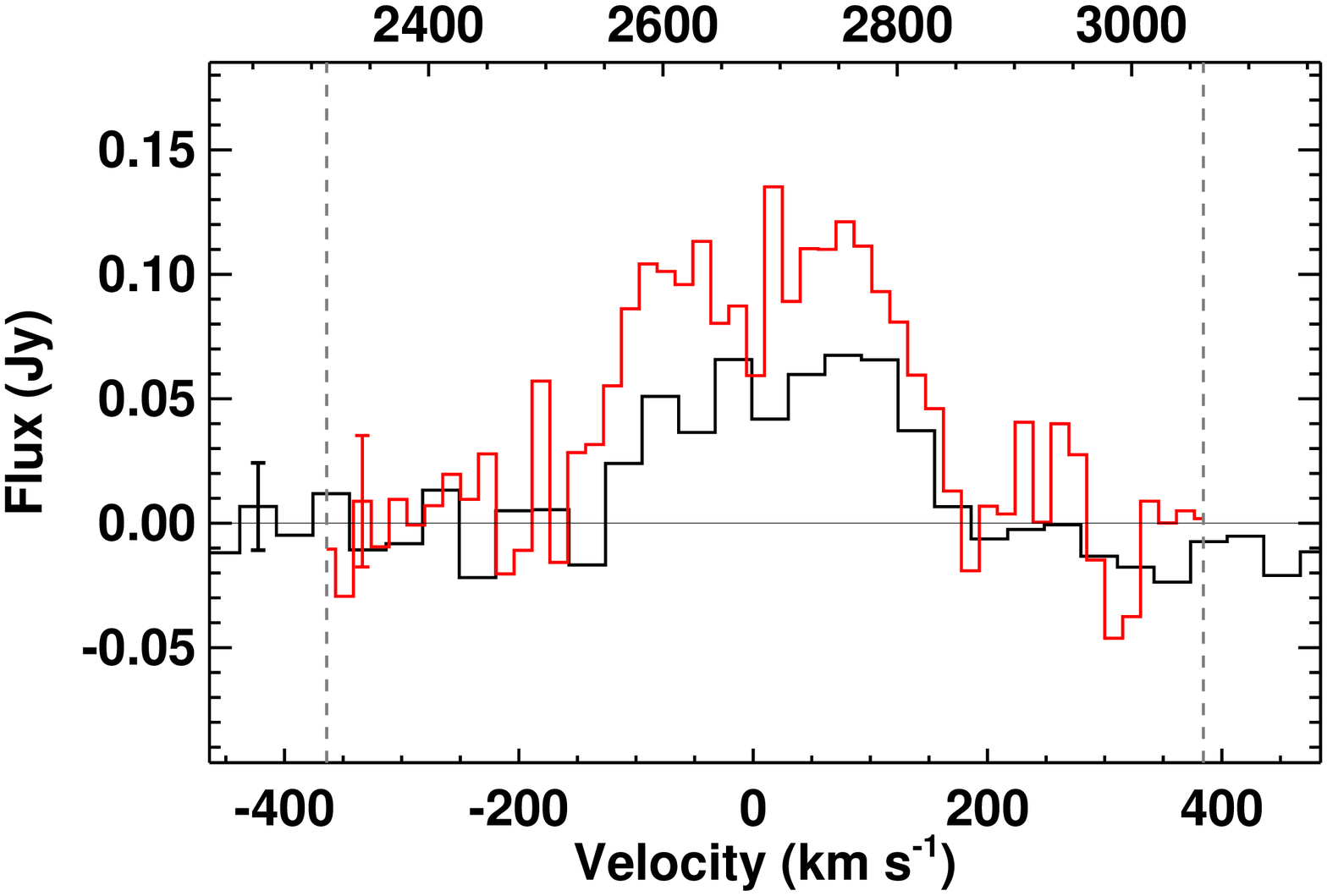}}
\end{figure*}
\begin{figure*}
\subfloat{\includegraphics[height=1.6in,clip,trim=0.1cm 1.4cm 0.6cm 2.4cm]{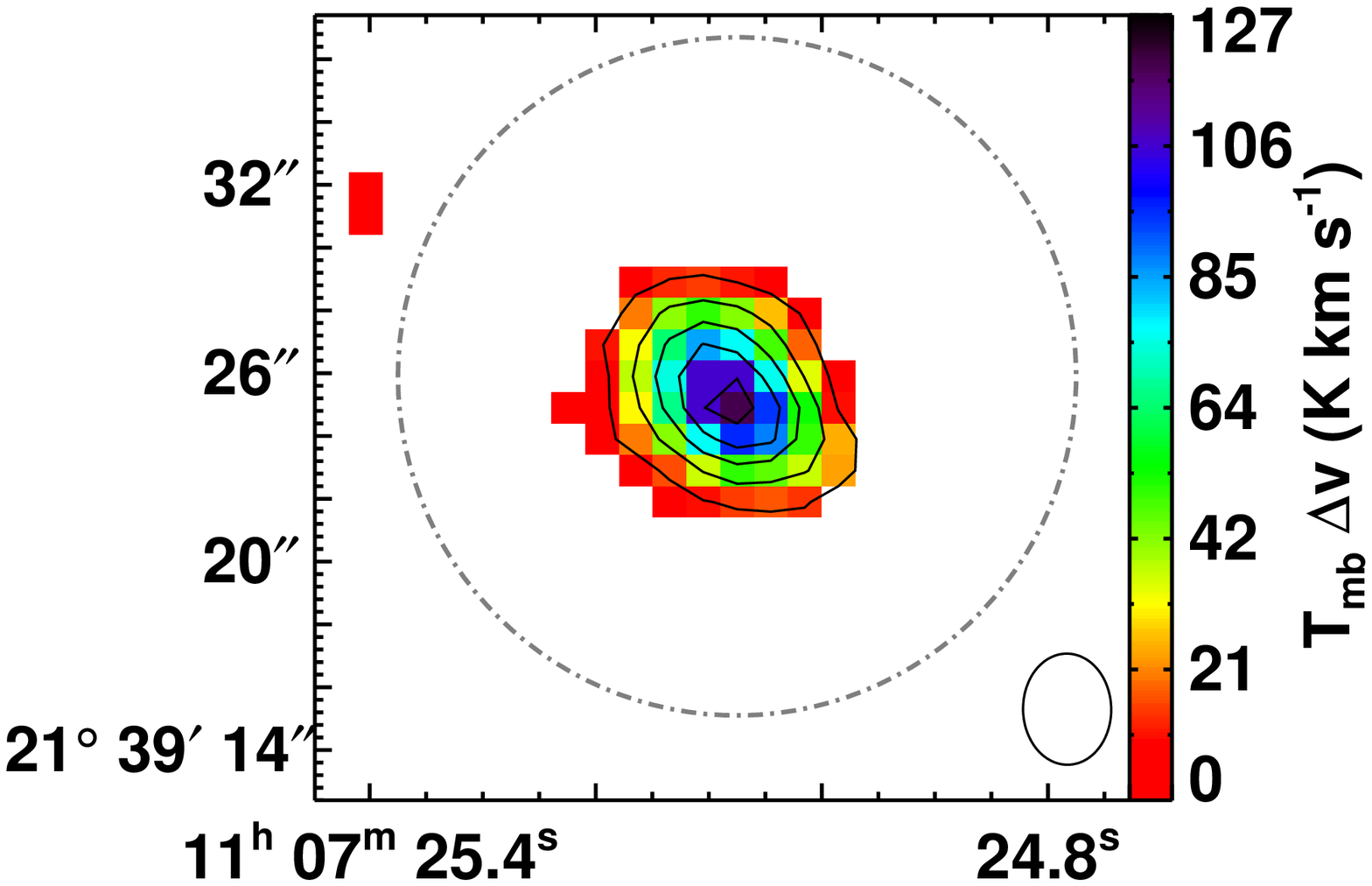}}
\subfloat{\includegraphics[height=1.6in,clip,trim=0.1cm 1.4cm 0cm 2.4cm]{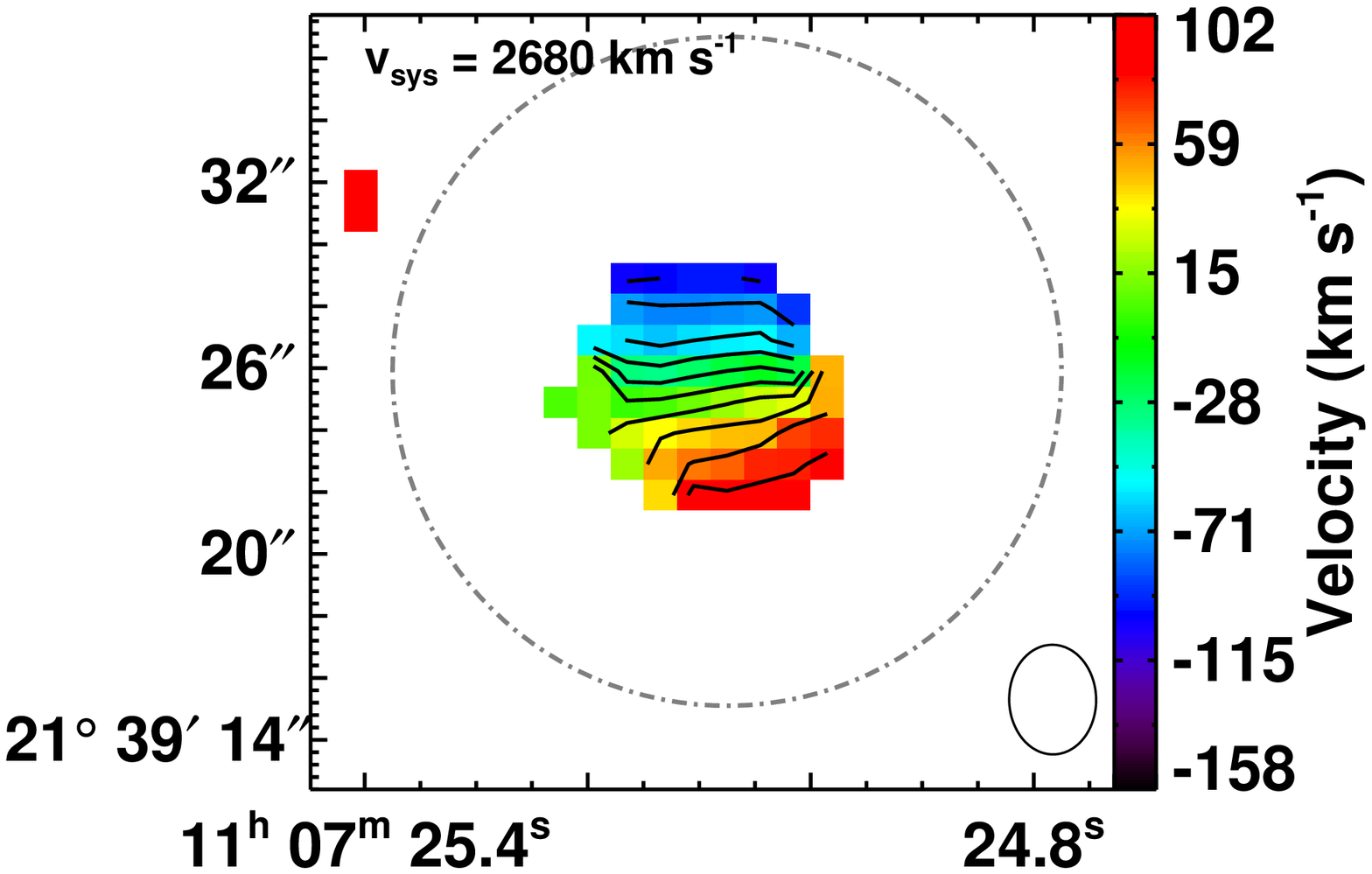}}
\subfloat{\includegraphics[height=1.6in,clip,trim=0cm 1.4cm 0cm 0.9cm]{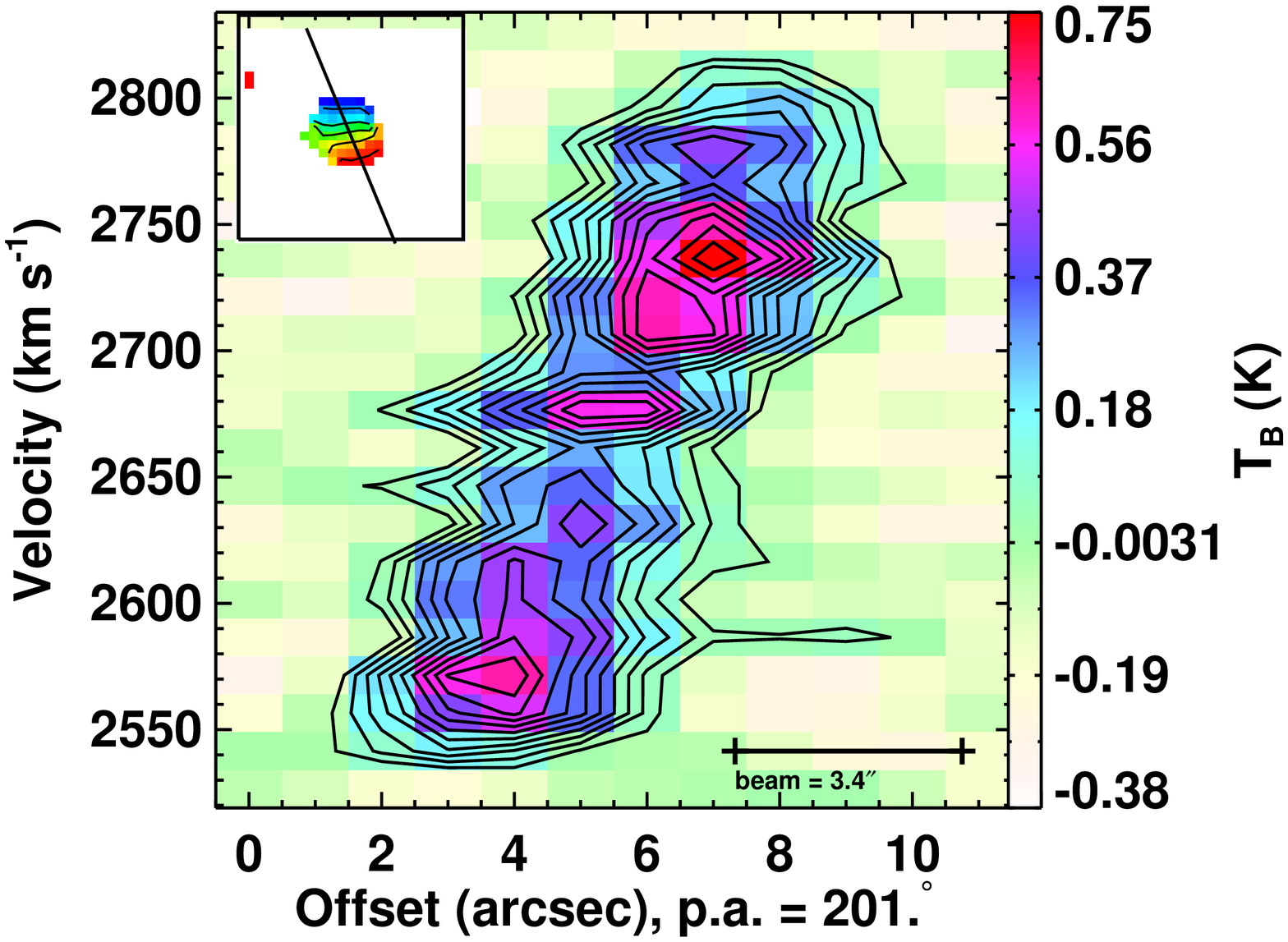}}
\end{figure*}
\begin{figure*}
\subfloat{\includegraphics[width=7in,clip,trim=0.3cm 0.8cm 0.1cm 1.3cm]{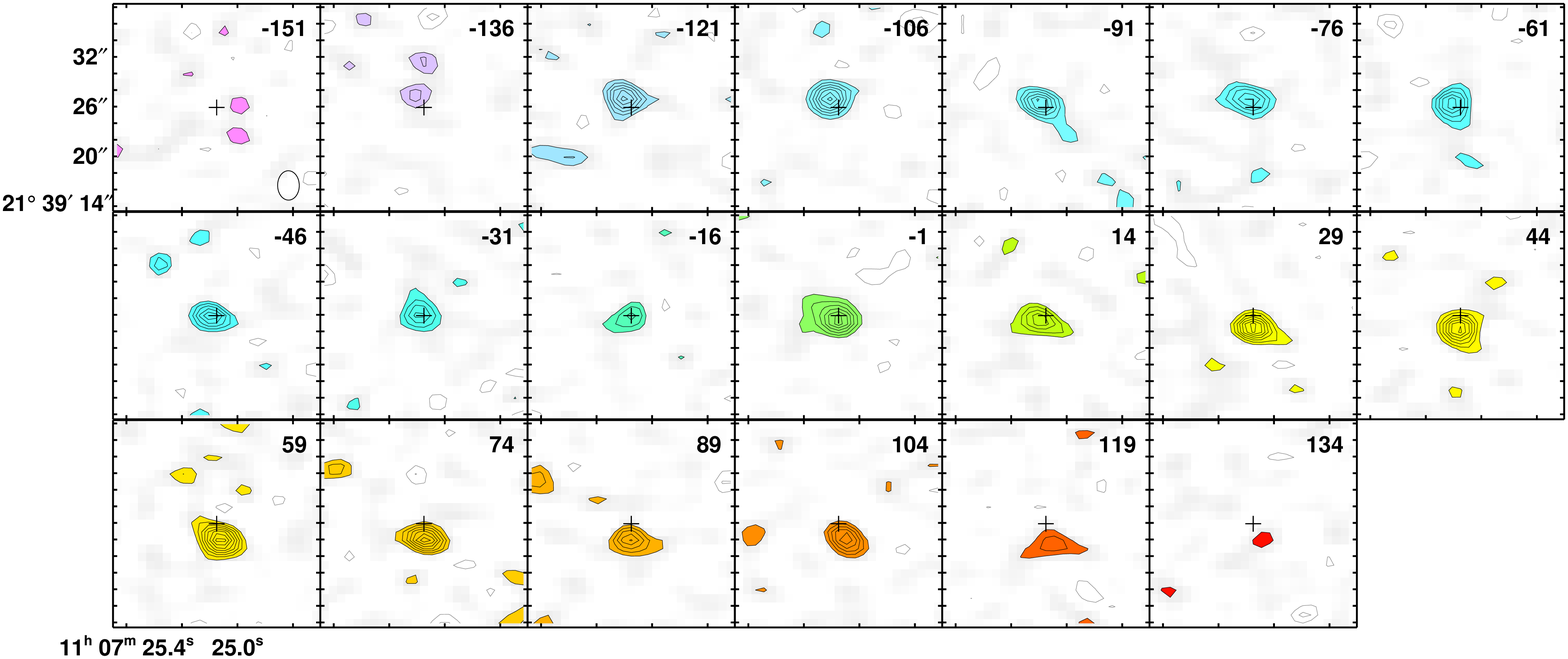}}
\caption{{\bf UGC~06176} is a field regular rotator ($M_K$ = -22.66) with bar and ring stellar morphology.  It contains a dust bar, ring, and filaments.  The moment0 peak is 14 Jy beam$^{-1}$ \kms.}
\end{figure*}

\clearpage
\begin{figure*}
\centering
\subfloat{\includegraphics[height=2.2in,clip,trim=2.2cm 3.2cm 0cm 2.7cm]{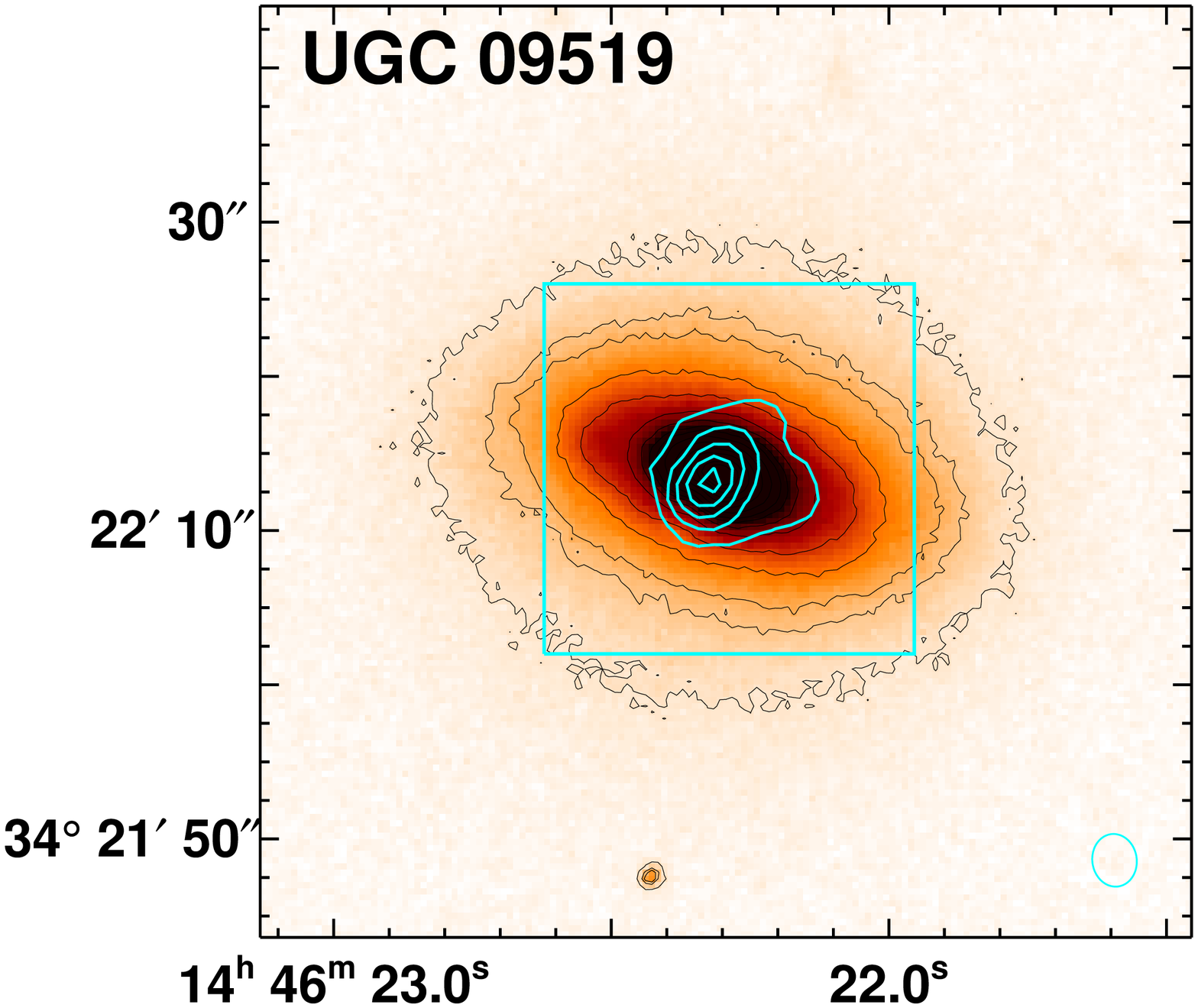}}
\subfloat{\includegraphics[height=2.2in,clip,trim=0cm 0.6cm 0.4cm 0.4cm]{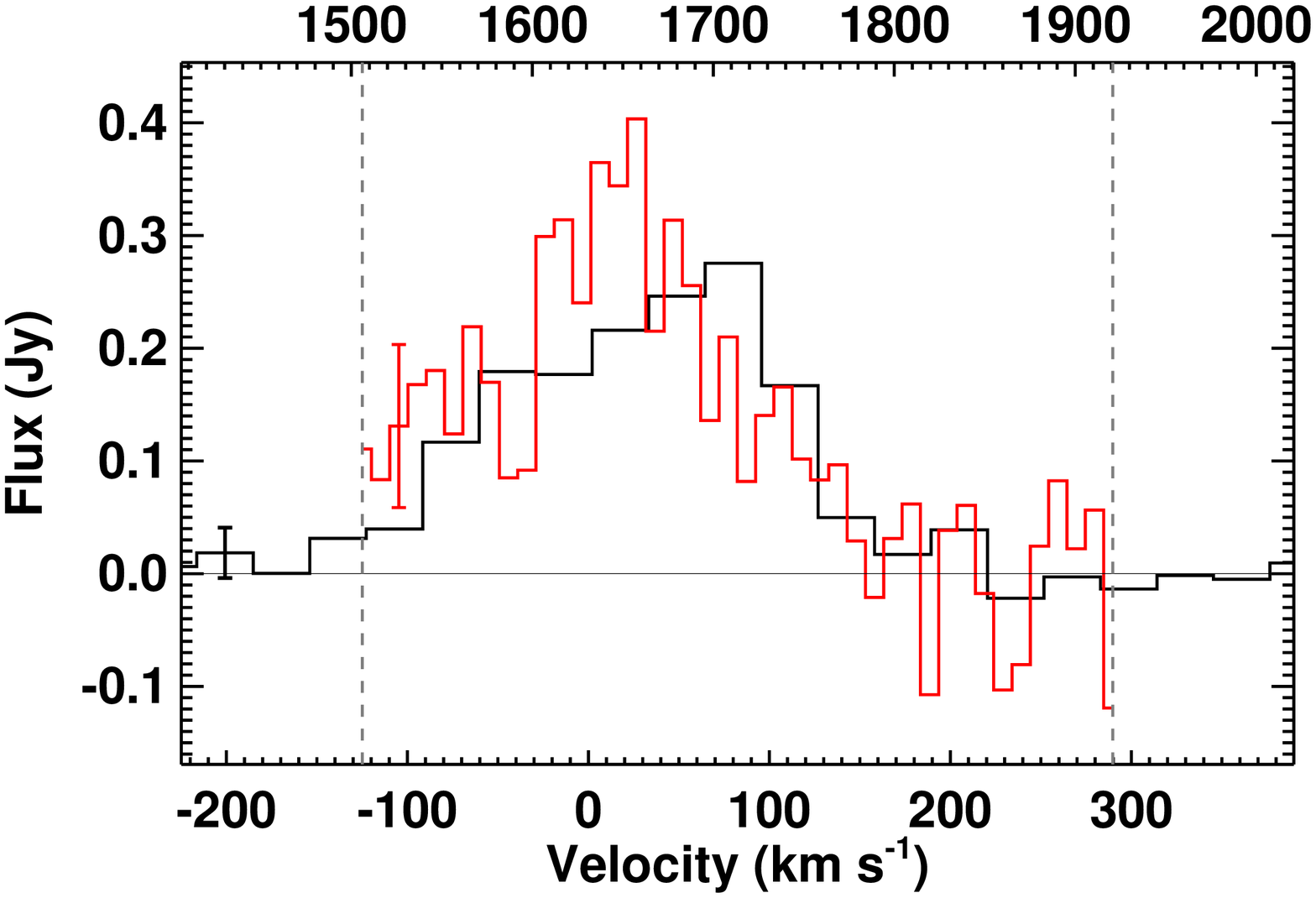}}
\end{figure*}
\begin{figure*}
\subfloat{\includegraphics[height=1.6in,clip,trim=0.1cm 1.4cm 0.6cm 2.4cm]{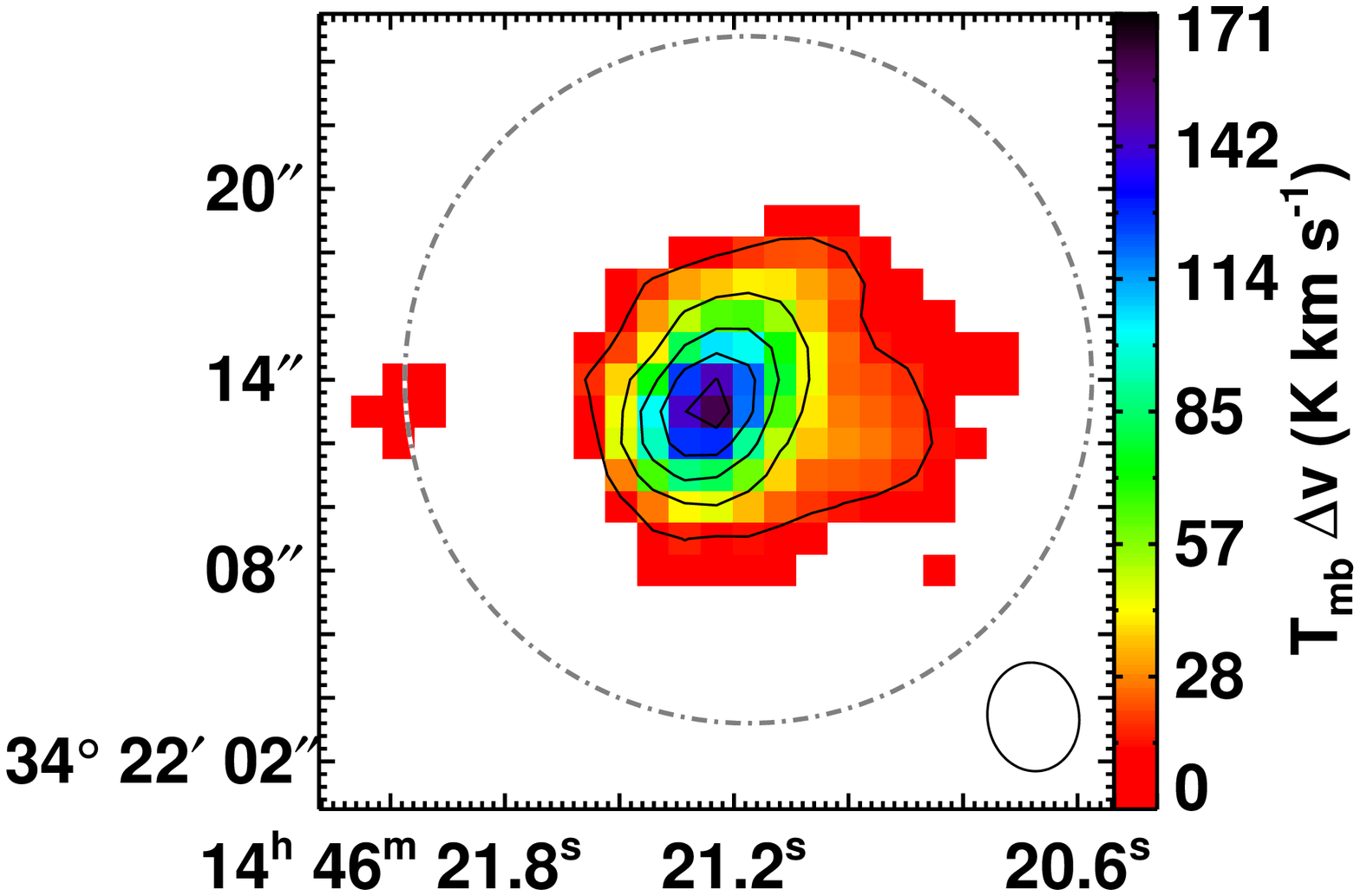}}
\subfloat{\includegraphics[height=1.6in,clip,trim=0.1cm 1.4cm 0cm 2.4cm]{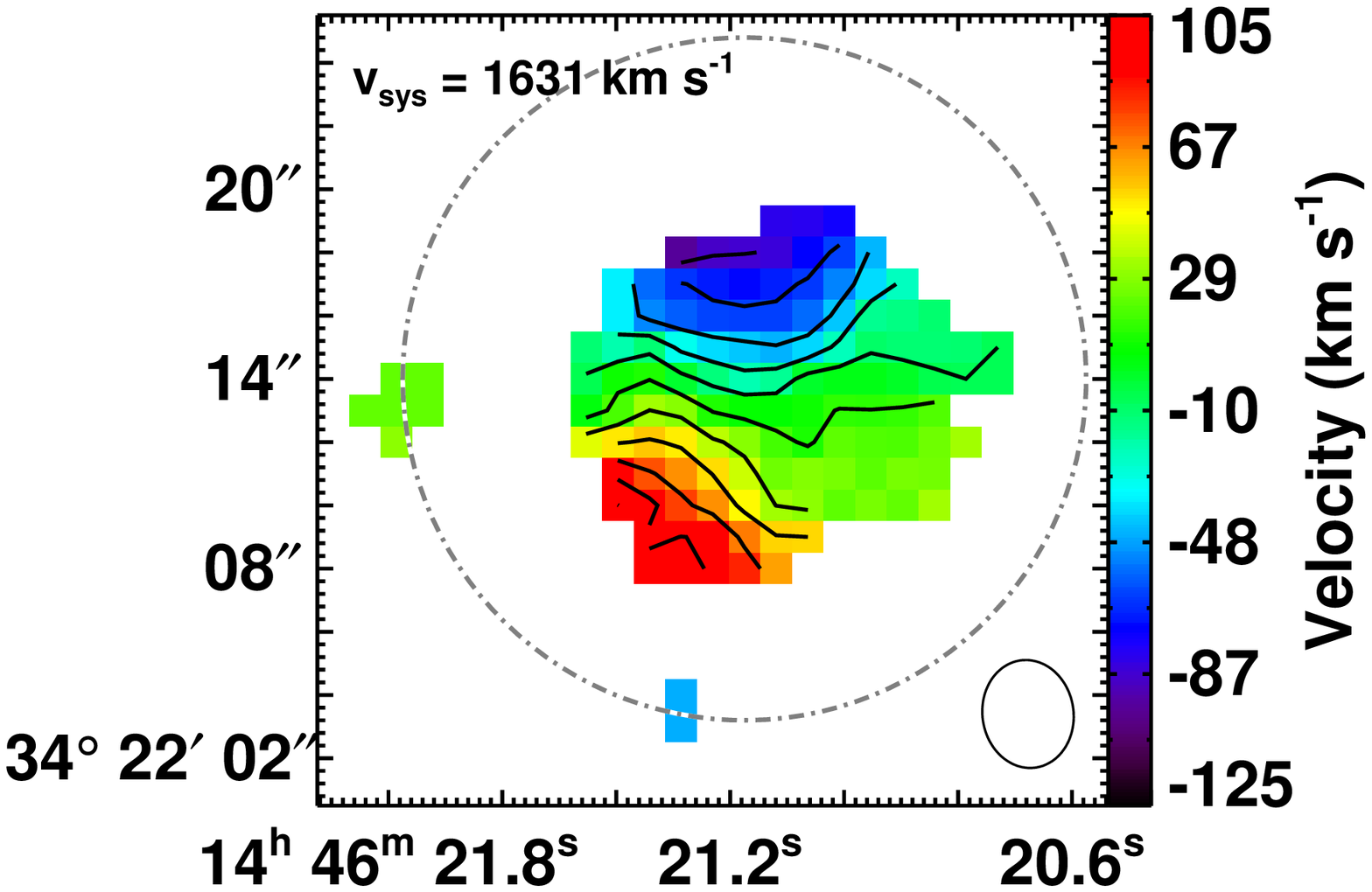}}
\subfloat{\includegraphics[height=1.6in,clip,trim=0cm 1.4cm 0cm 0.9cm]{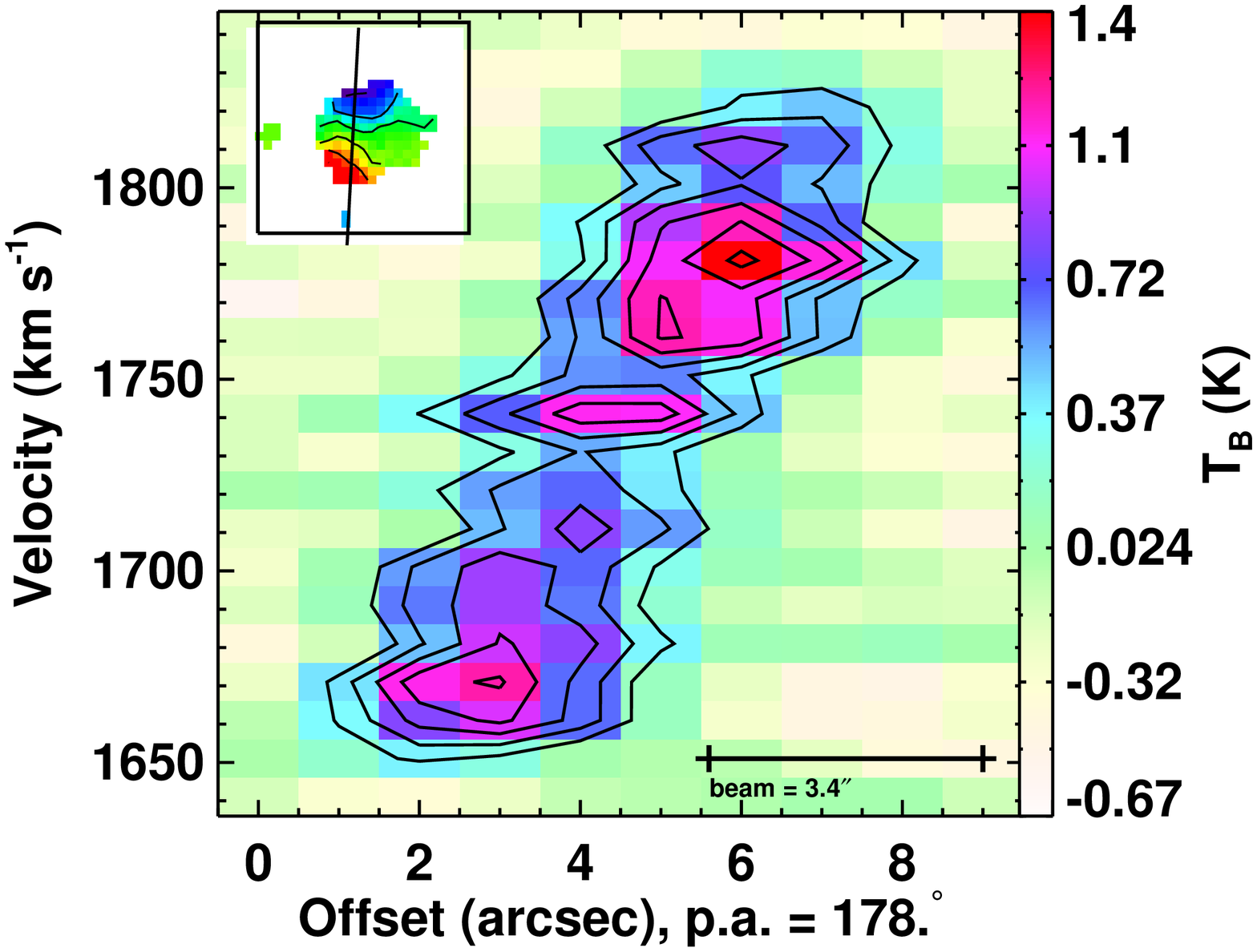}}
\end{figure*}
\begin{figure*}
\subfloat{\includegraphics[width=7in,clip,trim=0.6cm 0.8cm 0.1cm 1.3cm]{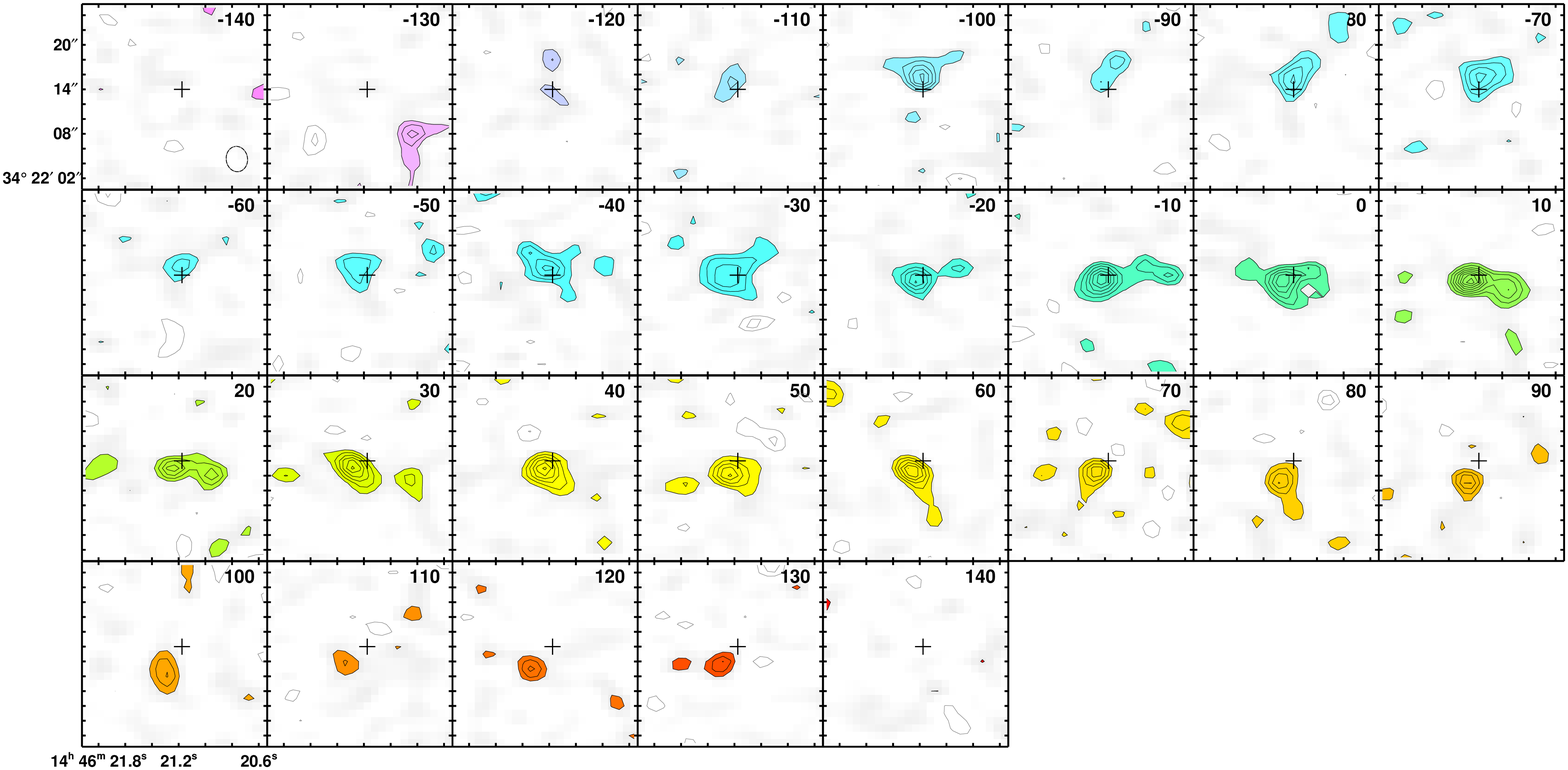}}
\caption{{\bf UGC~09519} is a field regular rotator ($M_K$ = -21.98) with normal stellar morphology.  It contains dust filaments.  The moment0 peak is 18 Jy beam$^{-1}$ \kms.}
\label{fig:lastgal}
\end{figure*}

\clearpage
\section{Interferometric CO data from the literature}
\label{app:litgals}

\begin{figure*}
\centering
\subfloat{\includegraphics[width=7in,clip,trim=0cm 0.2cm 0.5cm 0cm]{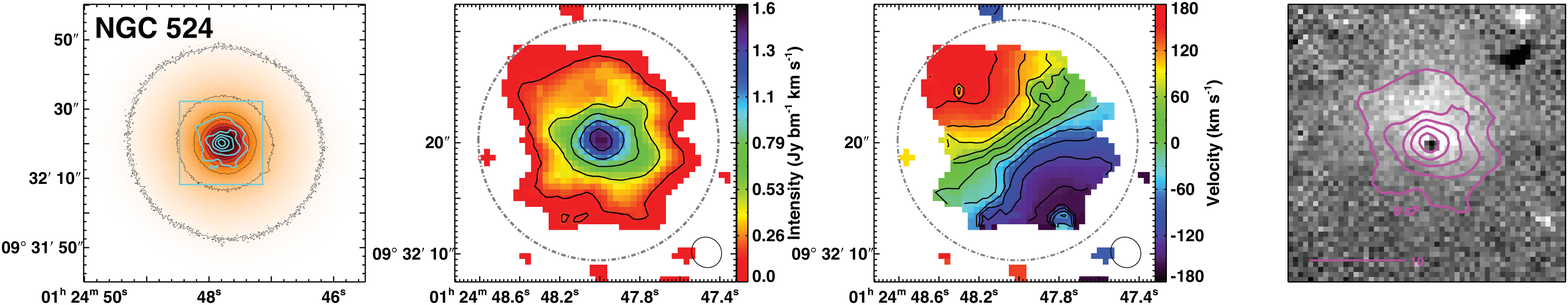}}\\
\subfloat{\includegraphics[width=7in,clip,trim=0cm 0.2cm 0.5cm 0.6cm]{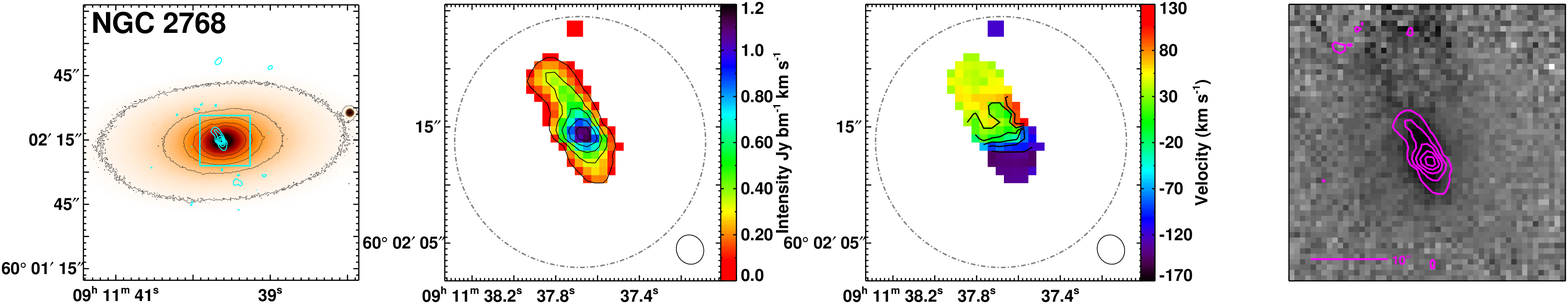}}\\
\subfloat{\includegraphics[width=7in,clip,trim=0cm 0.2cm 0.5cm 0.6cm]{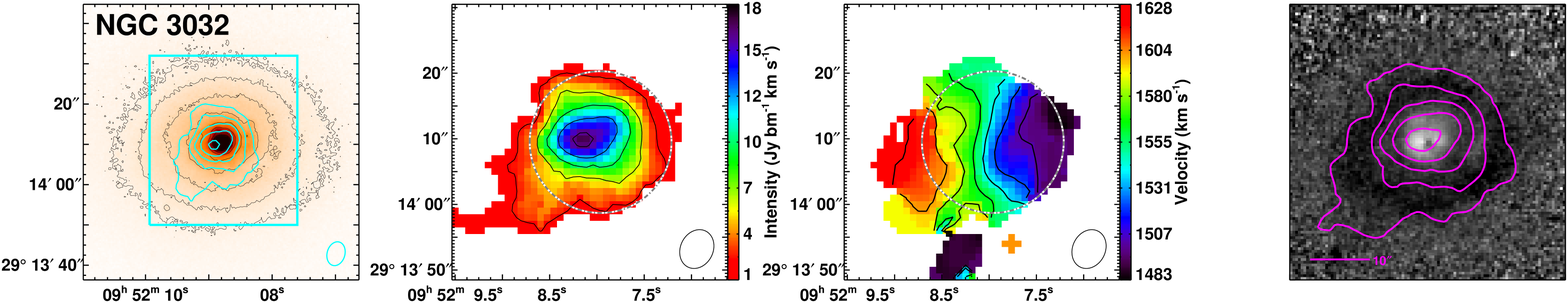}}\\
\subfloat{\includegraphics[width=7in,clip,trim=0cm 0.2cm 0.5cm 0.6cm]{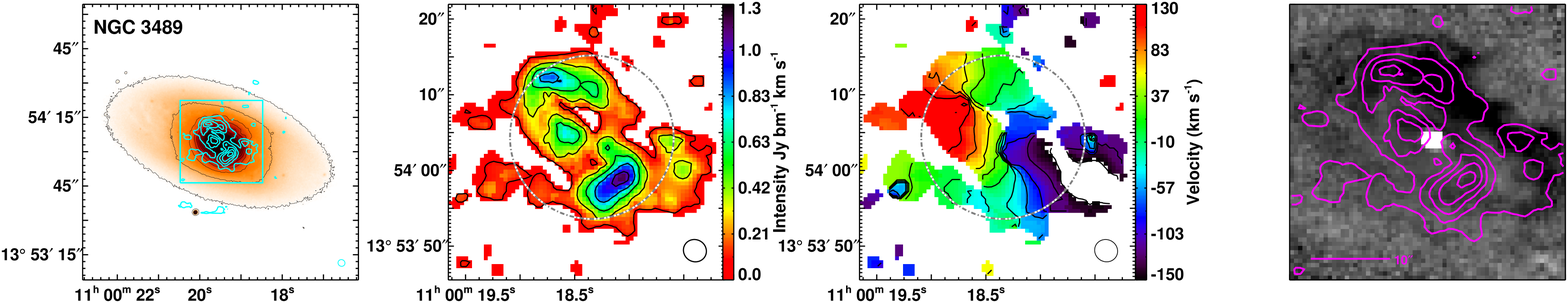}}\\
\caption{Interferometric CO(1--0) data from \atlas\ in the literature.  From top-to-bottom, NGC~524 \citep{crocker+11}, NGC~2768 \citep{crocker+08}, NGC~3032 \citep{ybc08}, NGC~3489 \citep{crocker+11}, NGC~4459 \citep{ybc08}, NGC~4476 \citep{young02}, NGC~4477 \citep{crocker+11}, NGC~4526 \citep{ybc08} and NGC~4550 \citep{crocker+09}.  For each galaxy, from left-to-right, the panels show the $r$-band image overlaid with the CO(1--0) integrated intensity (moment0) contours (cyan), the color-scale and contours of the CO integrated intensity (moment0) map overlaid with the IRAM 30m telescope beam (gray), the color-scale of the CO mean velocity (moment1) map overlaid with isovelocity contours and the IRAM 30m telescope beam (gray), and the $g-r$ colour image overlaid with the CO(1--0) moment0 contours (magenta).  For comparisons to unsharp-masked HST data of these galaxies, see the papers cited above.}
\label{fig:lit_gals}
\end{figure*}
\addtocounter{figure}{-1}
\begin{figure*}
\subfloat{\includegraphics[width=7in,clip,trim=0cm 0.2cm 0.5cm 0.6cm]{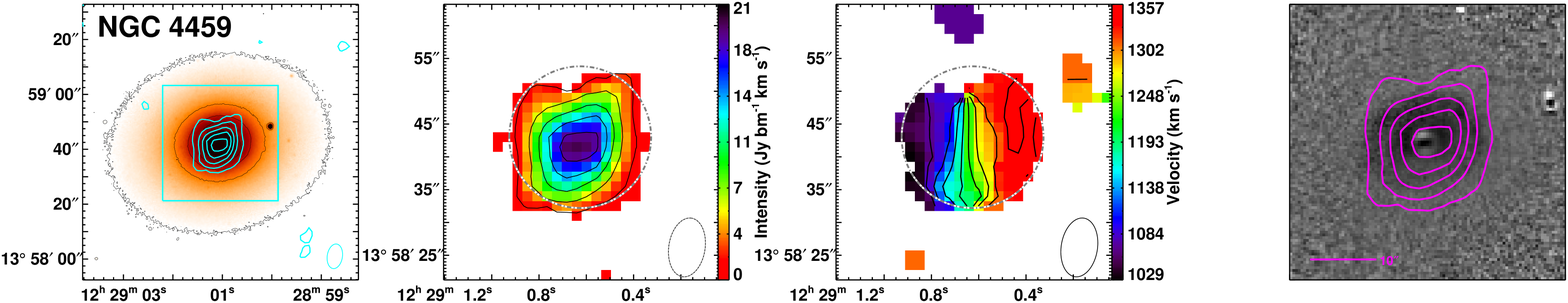}}\\
\subfloat{\includegraphics[width=7in,clip,trim=0cm 0.2cm 0.5cm 0.6cm]{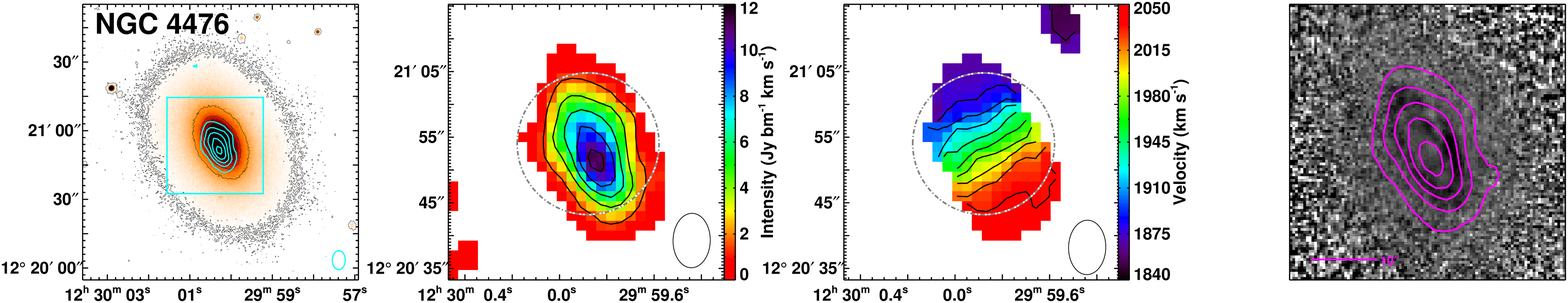}}\\
\subfloat{\includegraphics[width=7in,clip,trim=0cm 0.2cm 0.5cm 0.6cm]{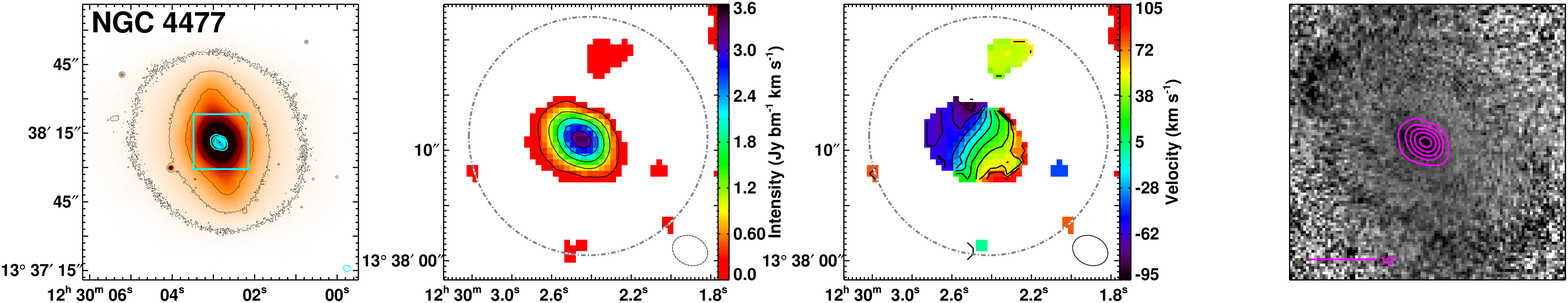}}\\
\subfloat{\includegraphics[width=7in,clip,trim=0cm 0.2cm 0.5cm 0.6cm]{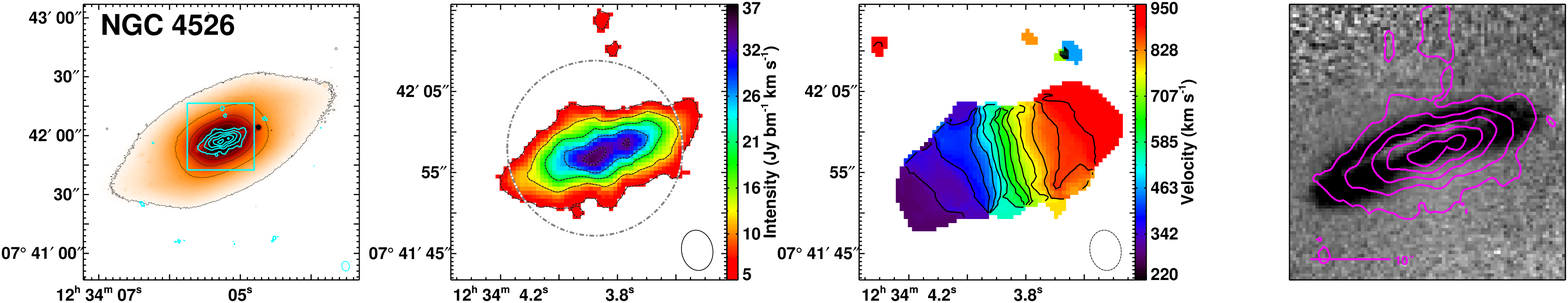}}\\
\subfloat{\includegraphics[width=7in,clip,trim=0cm 0.2cm 0.5cm 0.6cm]{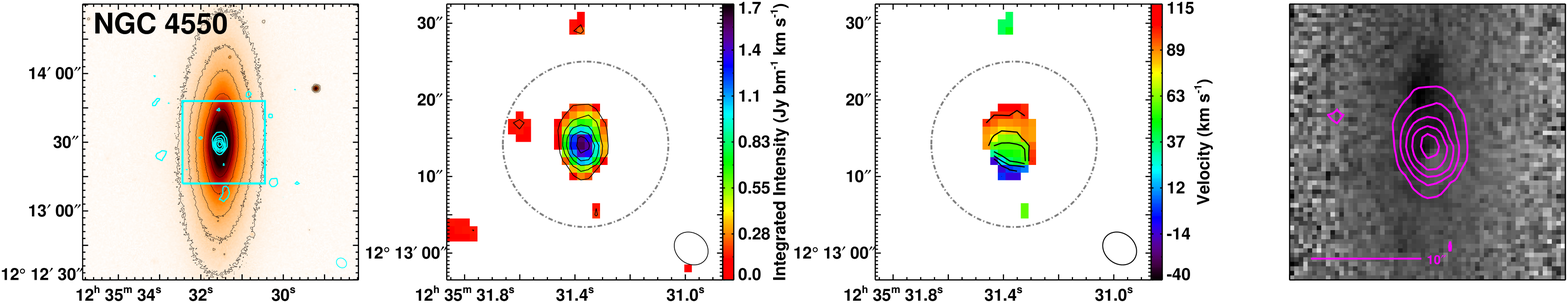}}
\caption{Continued}
\end{figure*}

\section{Galaxies observed with CARMA not in the \atlas\ survey}
\label{app:remgals}
\clearpage
\begin{figure*}
\centering
\subfloat{\includegraphics[height=2.2in,clip,trim=2.2cm 3.2cm 0cm 2.7cm]{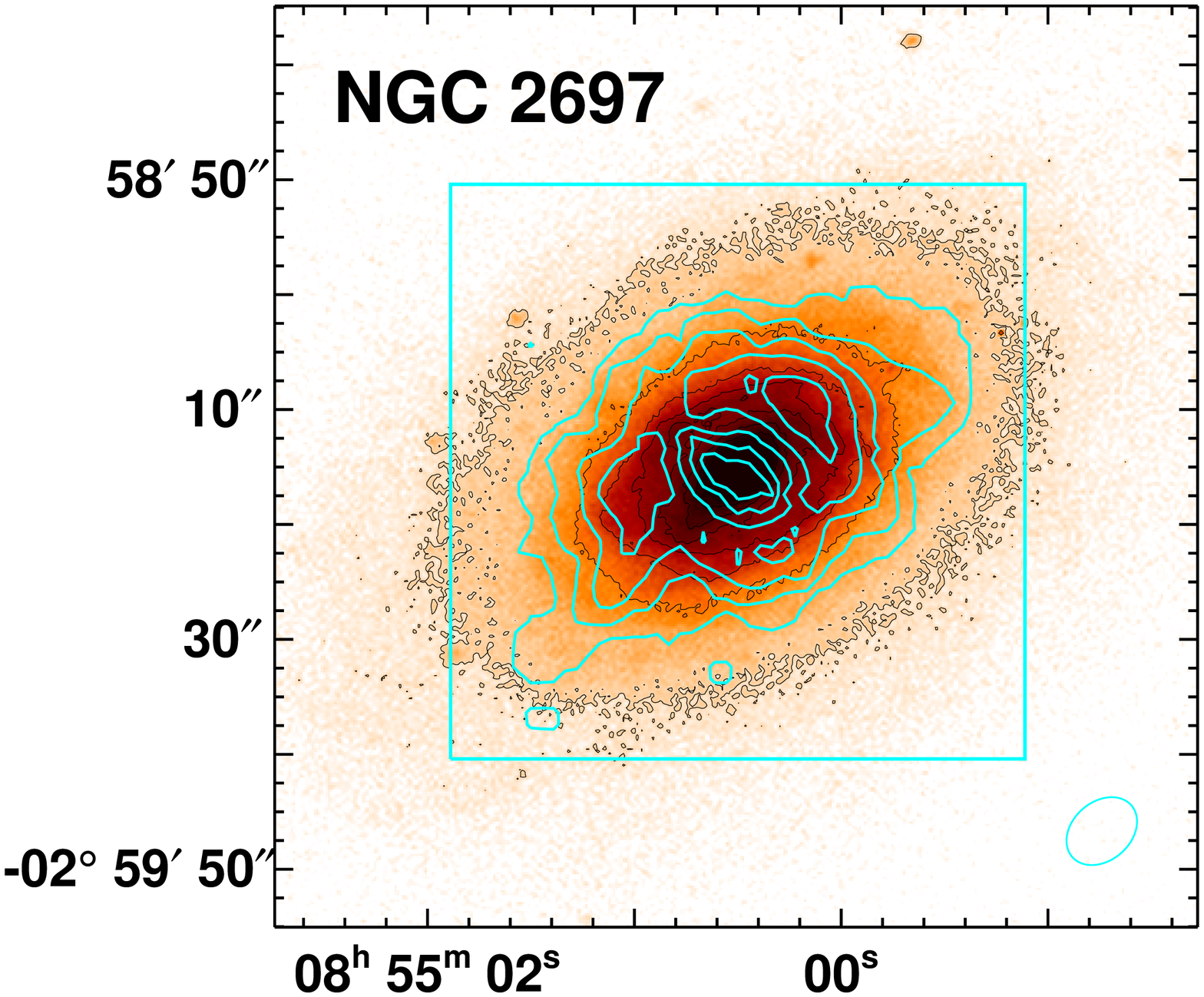}}
\subfloat{\includegraphics[height=2.2in,clip,trim=0cm 0.6cm 0.4cm 0.4cm]{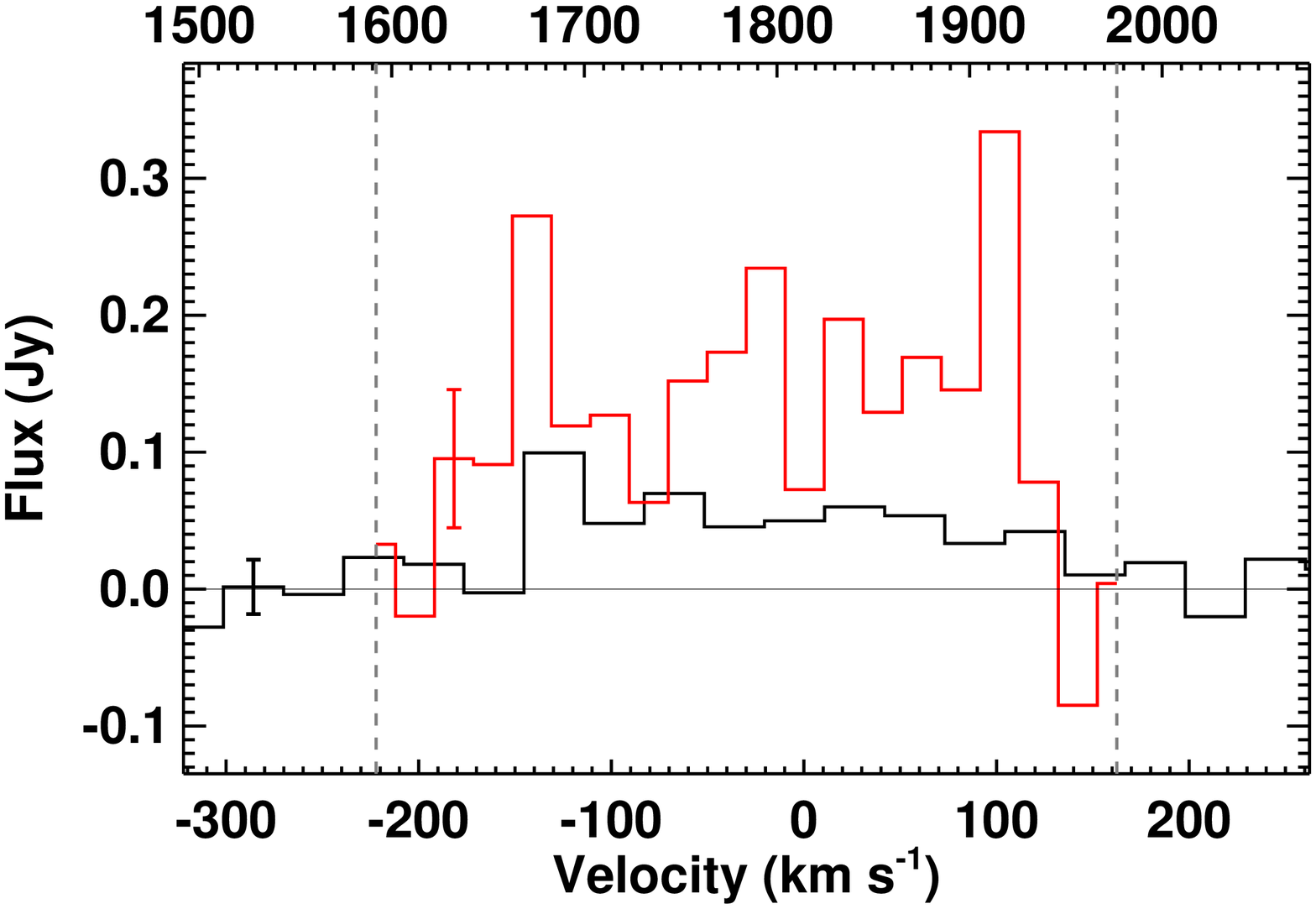}}
\end{figure*}
\begin{figure*}
\subfloat{\includegraphics[height=1.6in,clip,trim=0.1cm 1.4cm 0.6cm 2.5cm]{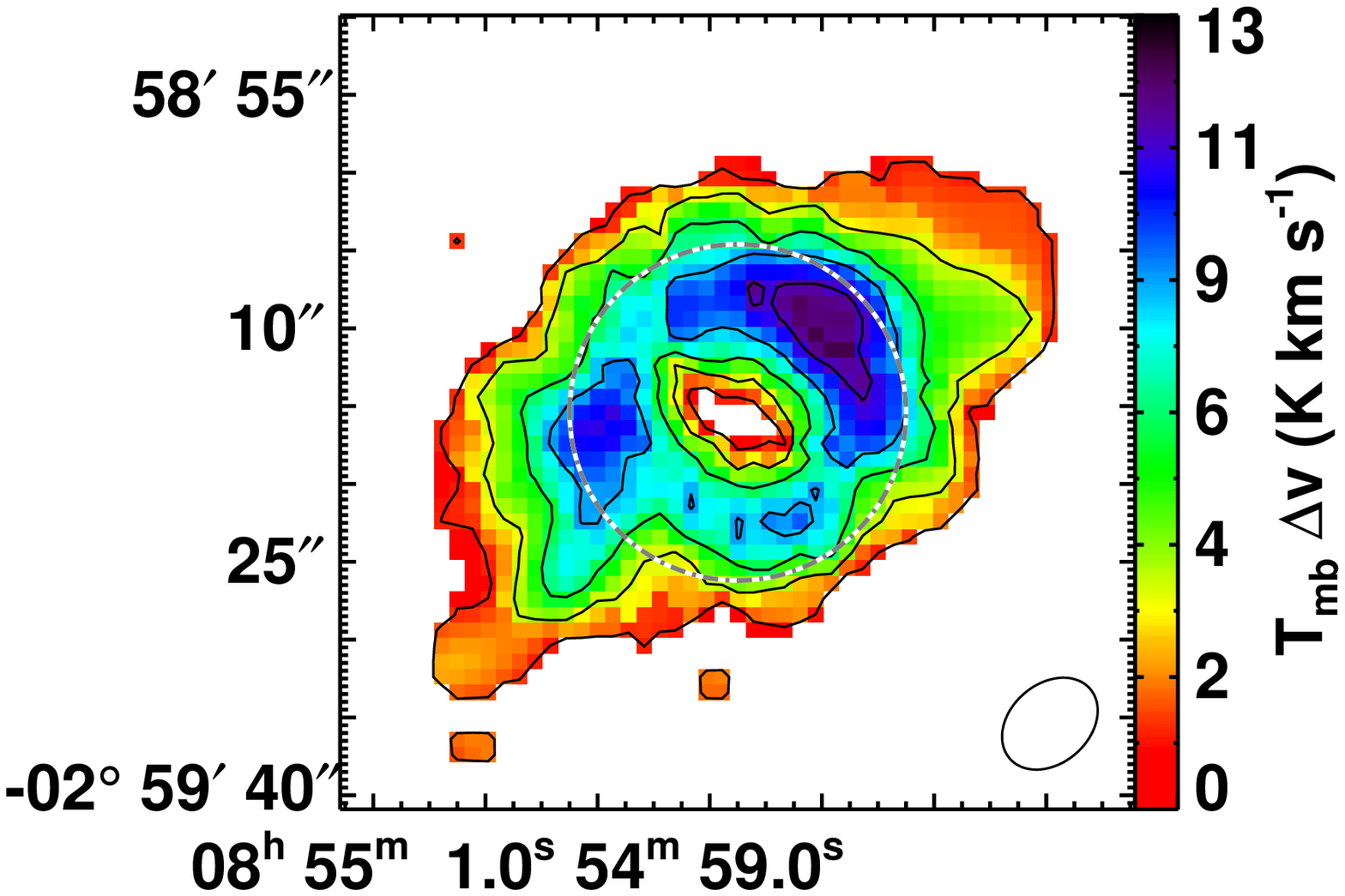}}
\subfloat{\includegraphics[height=1.6in,clip,trim=0.1cm 1.4cm 0cm 2.5cm]{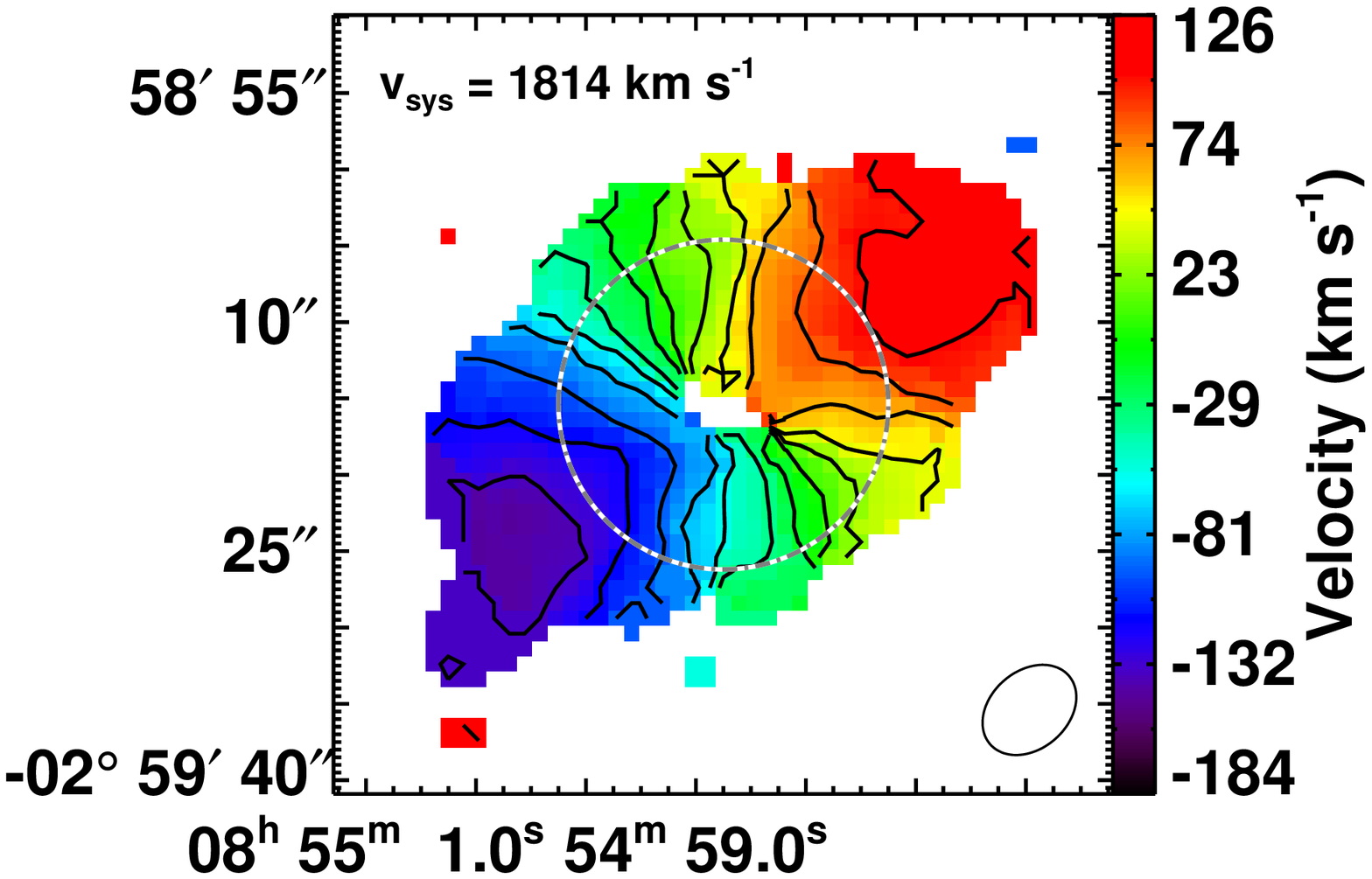}}
\subfloat{\includegraphics[height=1.6in,clip,trim=0cm 1.4cm 0cm 0.9cm]{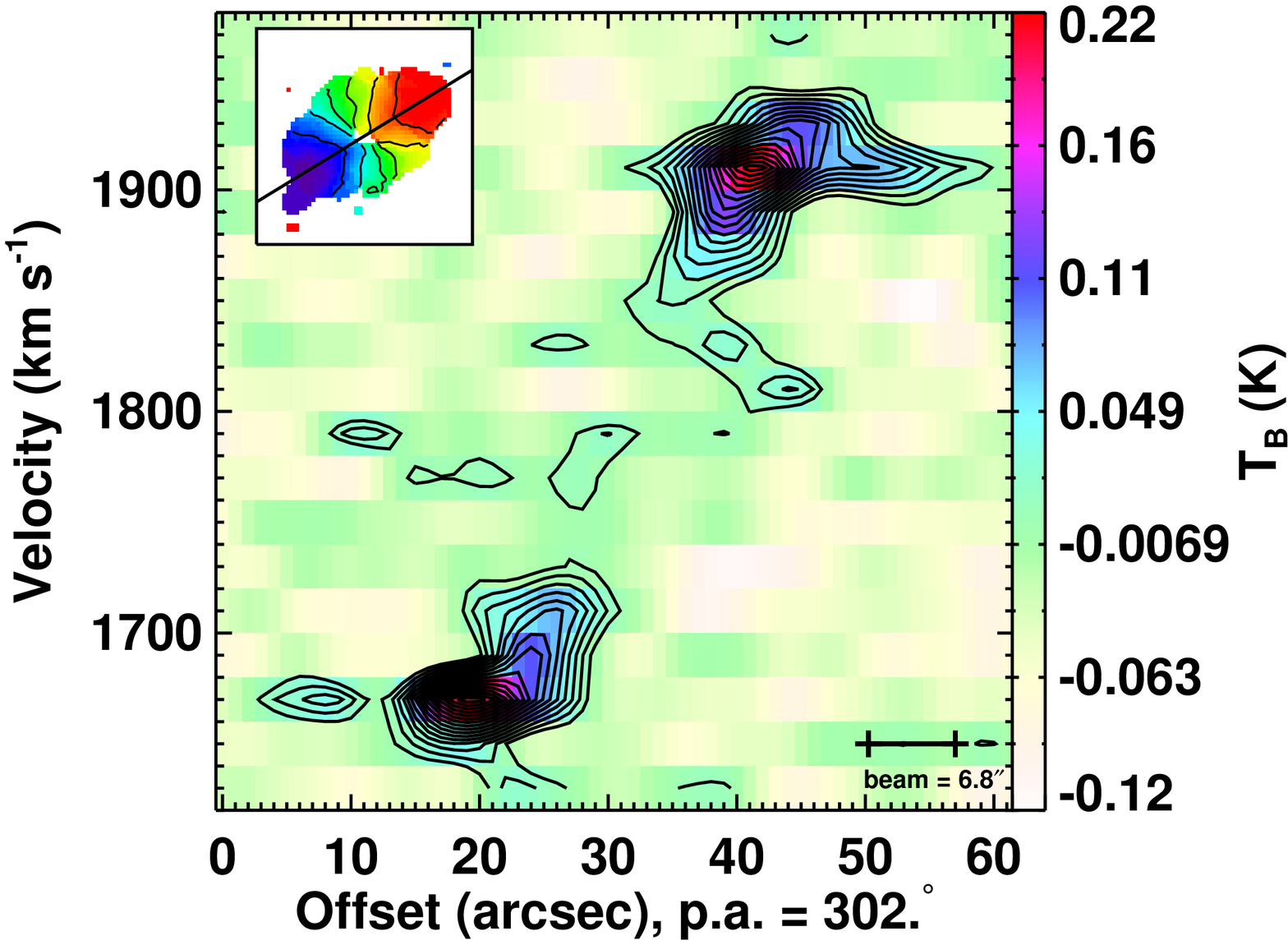}}
\end{figure*}
\begin{figure*}
\subfloat{\includegraphics[width=7in,clip,trim=1.2cm 1.5cm 1.2cm 4.6cm]{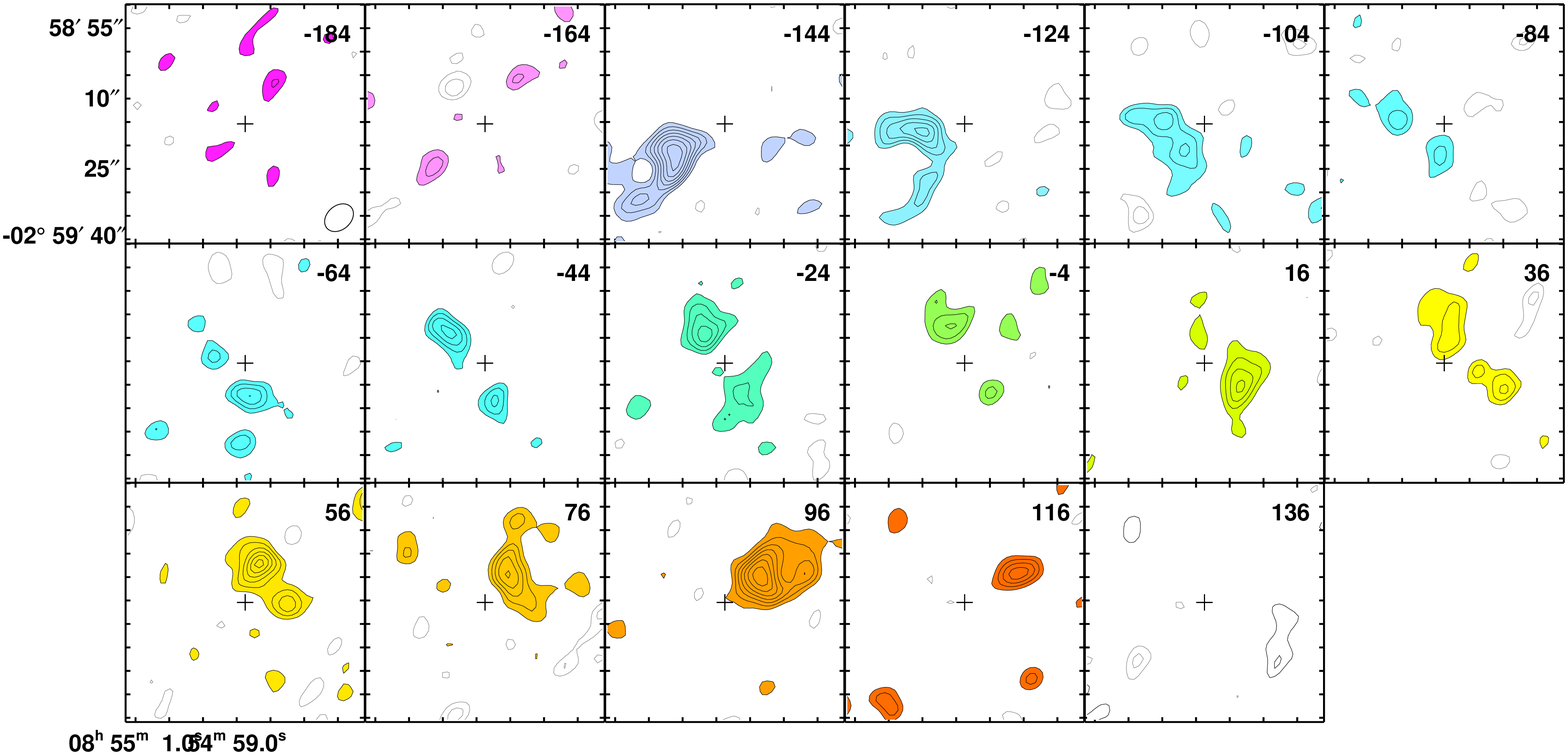}}
\caption{{\bf NGC~2697} ($M_K$ = -22.19) was removed from the sample as being mis-classified and containing spiral structure in the stellar isophotes.  
The moment0 peak is 4.1 Jy beam$^{-1}$ \kms.}
\end{figure*}

\clearpage
\begin{figure*}
\centering
\subfloat{\includegraphics[height=2.2in,clip,trim=2.2cm 3.2cm 0cm 2.7cm]{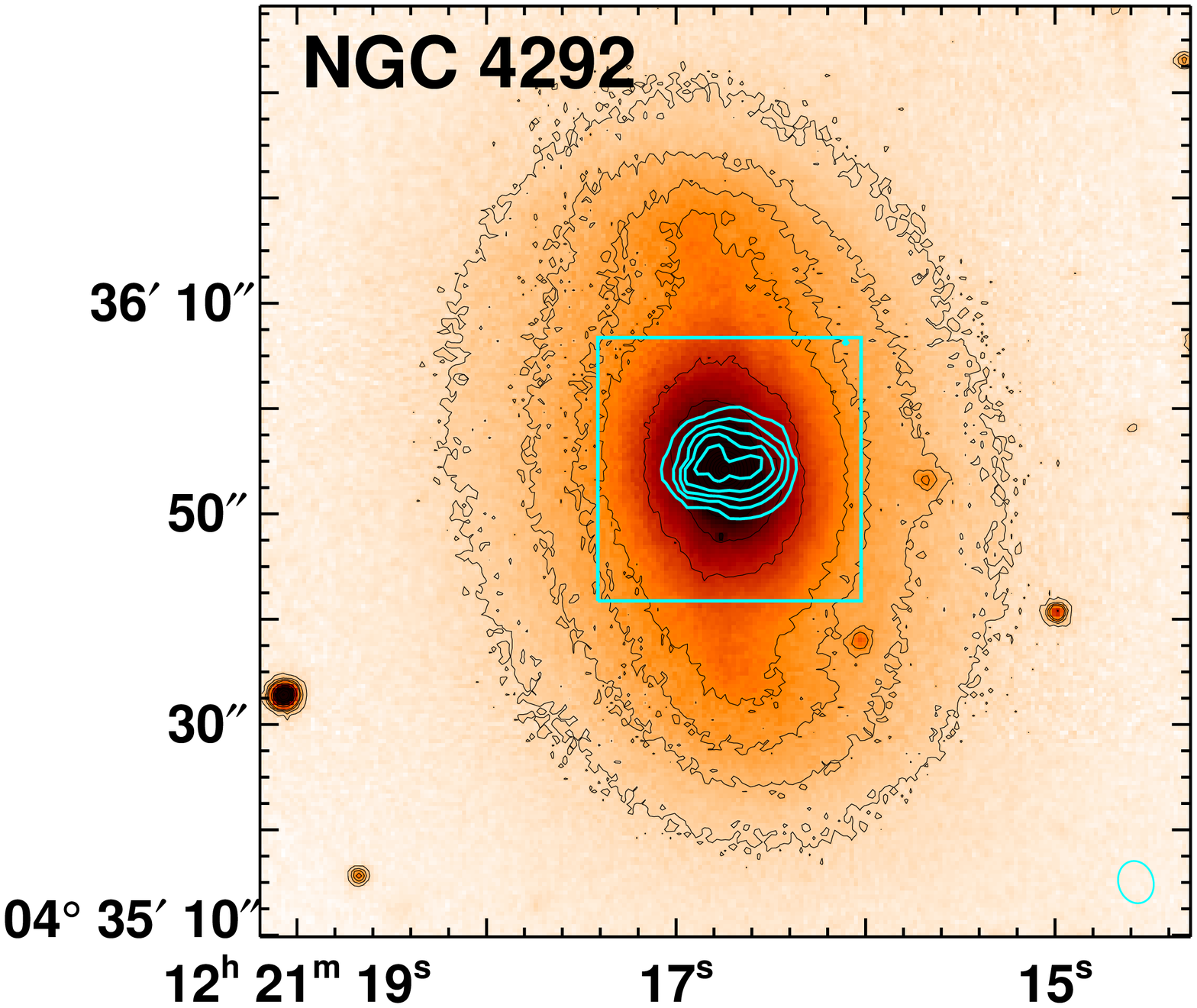}}
\subfloat{\includegraphics[height=2.2in,clip,trim=0cm 0.6cm 0.4cm 0.4cm]{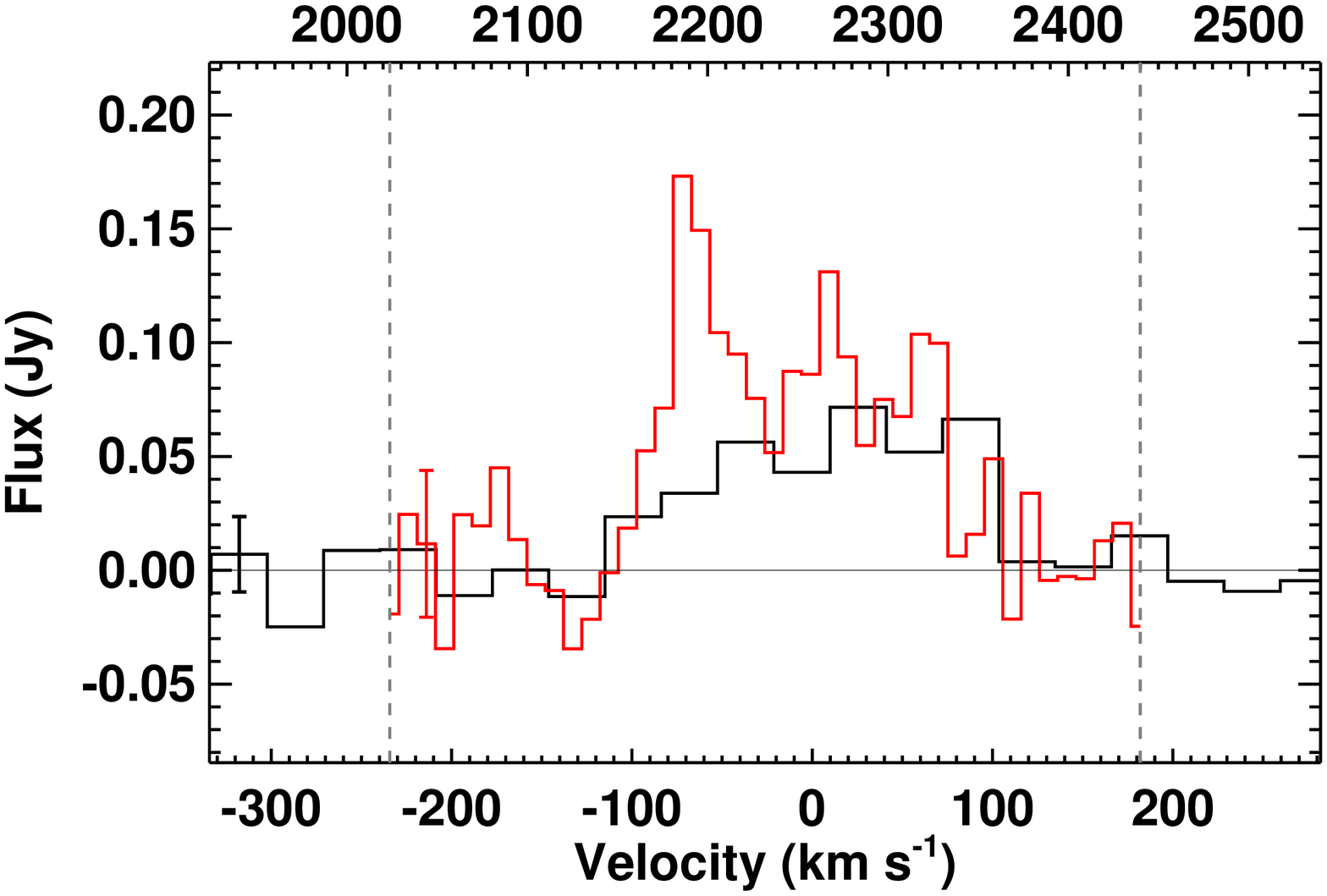}}
\end{figure*}
\begin{figure*}
\subfloat{\includegraphics[height=1.6in,clip,trim=0.1cm 1.4cm 0.6cm 2.4cm]{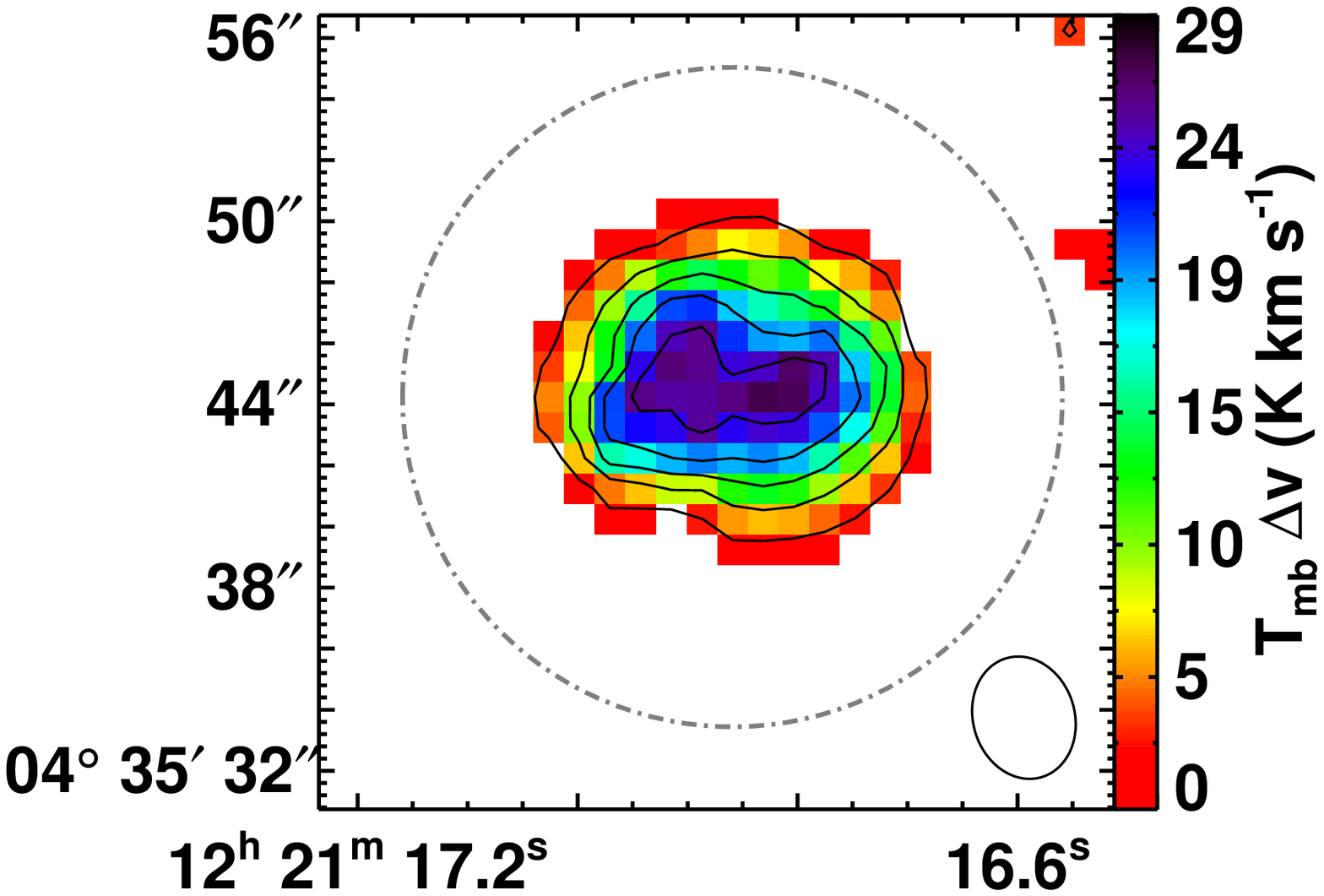}}
\subfloat{\includegraphics[height=1.6in,clip,trim=0.1cm 1.4cm 0cm 2.4cm]{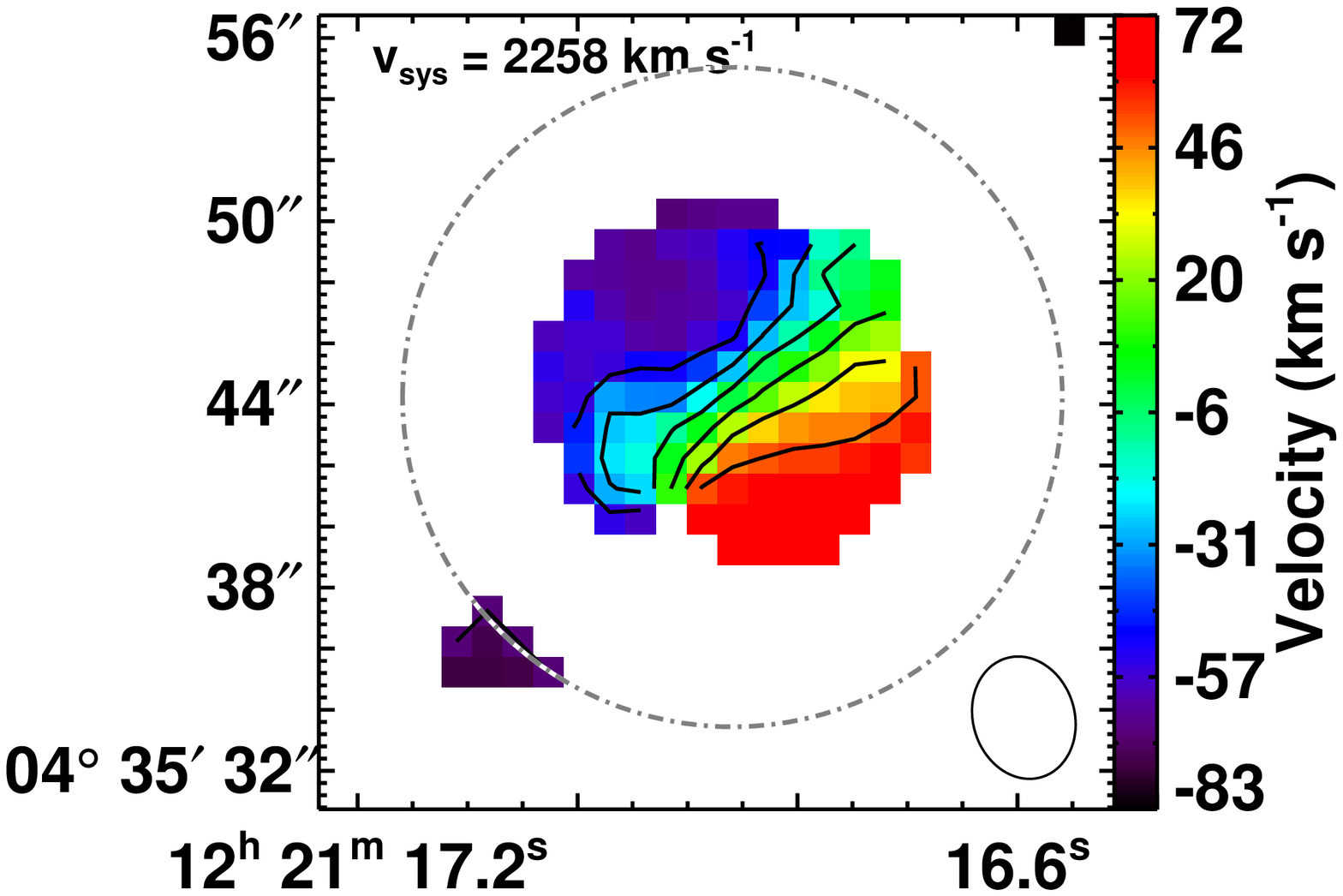}}
\subfloat{\includegraphics[height=1.6in,clip,trim=0cm 1.4cm 0cm 0.9cm]{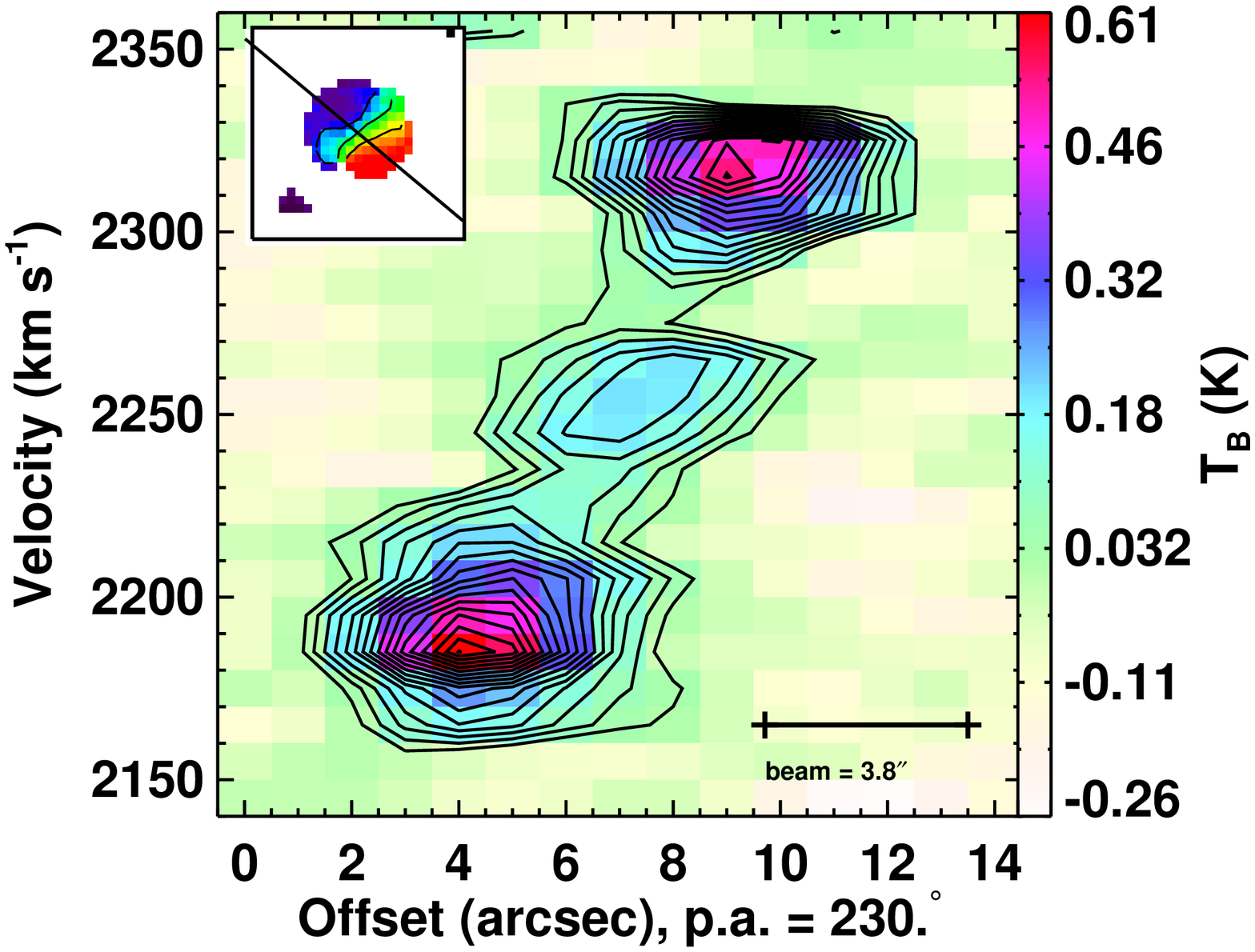}}
\end{figure*}
\begin{figure*}
\subfloat{\includegraphics[width=7in,clip,trim=1.6cm 6.2cm 6.8cm 0cm]{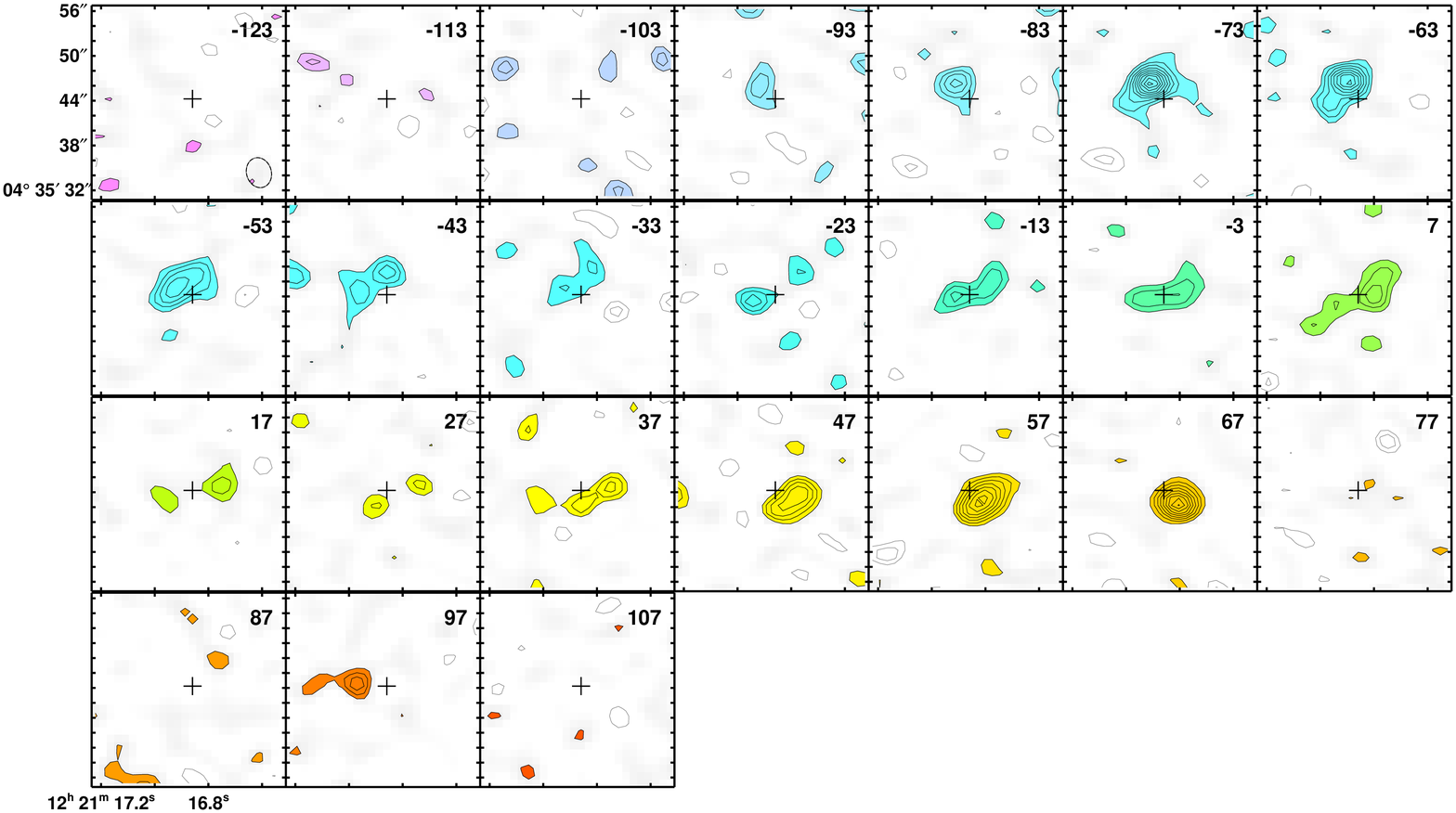}}
\caption{{\bf NGC~4292} is a Virgo cluster galaxy ($M_K$ = -23.2).  It was removed from the \atlas\ sample based on its not being observed with {\tt SAURON}, and thus lacks stellar kinematic data.  The moment0 peak is 4.3 Jy beam$^{-1}$ \kms.}
\end{figure*}

\end{document}